\documentclass{article}

\usepackage[numbers,sort&compress]{natbib}

\usepackage[utf8]{inputenc}
\usepackage{hyperref}
\hypersetup{
    colorlinks=true,
    linkcolor=blue,
    filecolor=magenta,      
    urlcolor=blue,
    citecolor=red,
}

\usepackage{amsmath,amsfonts,amssymb}
\usepackage{graphicx}
\usepackage{xfrac}   
\usepackage{comment}
\usepackage{authblk}

\usepackage{url}
\newcommand\be{\begin{equation}}
\newcommand\ee{\end{equation}}
\newcommand\bees{\begin{eqnarray}}
\newcommand\ees{\end{eqnarray}}
\newcommand{\dgw}{D_L^{\,\rm GW}}
\newcommand{\dem}{D_L^{\,\rm em}}

\newcommand{\ra}{\rightarrow}
\newcommand\eq[1]{Eq.~(\ref{#1})}

\newcommand\Eq[1]{Eq.~(\ref{#1})}

\newcommand{\lp}{\left (}
\newcommand{\rp}{\right )}
\renewcommand{\d}{{\rm d}}
\newcommand{\PBH}{{\rm PBH}}

\newcommand{\msun}{{\, \rm M}_\odot}
\newcommand{\mtot}{ {\, \rm M_{\rm tot}}  }
\newcommand{\degsq}{\, \rm deg^2}

\usepackage[margin=1in,footskip=0.25in]{geometry}

\usepackage{soul}

\date{}
\title{\hspace{12cm}\vspace{-2cm}\href{https://lisa.pages.in2p3.fr/consortium-userguide/wg_cosmo.html}{\includegraphics[height=2cm]{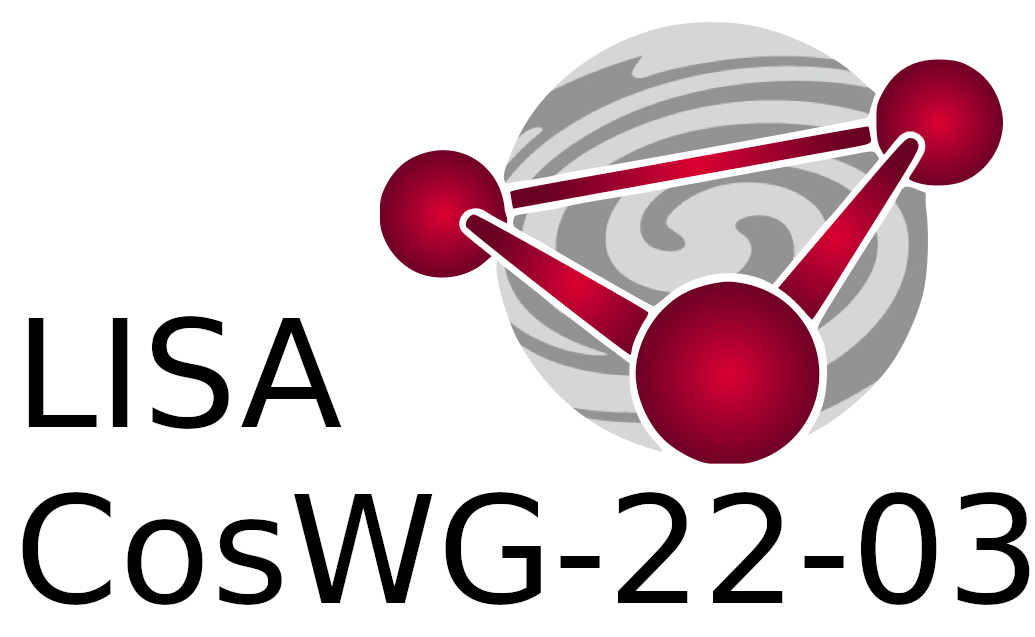}} \\[2cm] Cosmology with the Laser Interferometer Space Antenna
}

\author[1]{Pierre Auclair}
\affil[1]{Laboratoire Astroparticule et Cosmologie, Universit\'e de Paris, CNRS, 75013 Paris, France} 

\author[2]{David Bacon$^*$}
\affil[2]{Institute of Cosmology \& Gravitation, University of Portsmouth, PO1 3FX, UK}

\author[3]{Tessa Baker} 
\affil[3]{\footnotesize Department of Physics and Astronomy, Queen Mary University of London, Mile End Road, London E1 4NS, UK}

\author[4]{Tiago Barreiro} 
\affil[4]{Universidade Lus\'ofona de Humanidades e Tecnologias
Campo Grande, 376, PT1749-024, Lisboa, Portugal}

\author[5,6,7]{Nicola Bartolo}
\affil[5]{\footnotesize  Dipartimento di Fisica e Astronomia ``G. Galilei", Universit\`a degli Studi di Padova, via Marzolo 8, I-35131 Padova, Italy}
\affil[6]{\footnotesize INFN, Sezione di Padova, via Marzolo 8, I-35131 Padova, Italy}
\affil[7]{\footnotesize INAF - Osservatorio Astronomico di Padova, Vicolo dell'Osservatorio 5, I-35122 Padova, Italy}

\author[8]{Enis Belgacem}
\affil[8]{\footnotesize Institute for Theoretical Physics, Faculty of Science, Utrecht University, Princetonplein 5, 3584 CC Utrecht, The Netherlands}

\author[9]{Nicola Bellomo}
\affil[9]{\footnotesize Theory Group, Department of Physics, The University of Texas at Austin, Austin, TX 78712, USA}

\author[10,11,12]{Ido Ben-Dayan}
\affil[10]{\footnotesize Department of Physics, Ariel University, Ariel, POB 3, 4070000, Israel}
\affil[11]{\footnotesize Berkeley Center for Cosmological Physics, University of California, Berkeley, CA 94720, USA}
\affil[12]{\footnotesize Department of Physics, University of California, Berkeley, CA 94720, USA}

\author[5,6,7]{Daniele Bertacca}

\author[13]{Marc Besancon}
\affil[13]{\footnotesize CEA Paris-Saclay, Irfu/DPhP, Bat.141, 91191 Gif sur Yvette, France}

% ...

\begin{document}

\maketitle

%%%%%%%%%%%%%%%%%%%%%%%%%%%%%%%%%%%%%%%%%%%%%%%%%%%%%%%%%%%%%%%%%%%%
\begin{abstract}
\noindent The Laser Interferometer Space Antenna (LISA) has two scientific objectives of cosmological focus: to probe the expansion rate of the universe, and to understand stochastic gravitational-wave backgrounds and their implications for early universe and particle physics, from the MeV to the Planck scale. However, the range of potential cosmological applications of gravitational wave observations extends well beyond these two objectives. This publication presents a summary of the state of the art in LISA cosmology, theory and methods, and identifies new opportunities to use gravitational wave observations by LISA to probe the universe.

\end{abstract}
%%%%%%%%%%%%%%%%%%%%%%%%%%%%%%%%%%%%%%%%%%%%%%%%%%%%%%%%%%%%%%%%%%%%

\tableofcontents

\newpage
%%%%%%%%%%%%%%%%%%%%%%%%%%%%%%
%\include{TEXTinstructions}
%
%\newpage
%%%%%%%%%%%%%%%%%%%%%%%%%%%%%%%
%\include{TEXTintroduction}
% Here Sec. 1 starts
\section{Introduction}

\small \emph{Contributors: R.~Caldwell, G.~Nardini.}\\ \normalsize

The Laser Interferometer Space Antenna (LISA)~\cite{LISA:2017pwj} is a planned space-borne gravitational wave (GW) detector that will open a new frontier on astrophysics and cosmology in the mHz frequency band. This European Space Agency-led mission includes participation by ESA member countries and significant contributions from NASA and the US, as well as from several other Countries. Phase A work is on track for mission adoption in mid 2020s, and is compatible with a launch in the mid 2030s.

LISA will consist of a trio of satellites, located at the vertices of an equilateral triangle, in an Earth-trailing heliocentric orbit. The $2.5$-million km distances between the satellites will be monitored using precision laser interferometry to detect passing GWs. Here we consider a nominal mission of six years with a duty cycle of around 75\%, although we understand that the instruments will be engineered to a specification that will enable a possible extension.

LISA will be sensitive to GWs from a wide array of sources~\cite{LISA:2017pwj}. A primary target will be the inspiral and merger of massive binary black holes (MBBHs), ranging in masses $10^4 - 10^7~M_\odot$, at redshifts out to $z\sim 10$. A significant foreground signal will be the many galactic white dwarf binaries, each effectively a monotone source. 

By the time LISA launches, the state of GW observation will have evolved.  The extended Advanced  Laser Interferometer Gravitational Wave Observatory (LIGO), Advanced Virgo, and Kamioka Gravitational Wave Detector (KAGRA) family of ground-based GW detectors will have begun implementing third-generation technology demonstration upgrades. The network of pulsar-timing radio telescopes will have grown to include the Square Kilometer Array (SKA). Yet, LISA will be different from its predecessors. The size of the detector will enable access to a completely fresh part of the GW spectrum, leading to observations of new astrophysical sources as well as a new window on primordial stochastic gravitational wave backgrounds (SGWB). Many sources will produce overlapping signals, owing to the improved sensitivity. Extracting individual sources and events, and discriminating from an unresolved hum, will be part of the challenge.
 
According to the mission proposal~\cite{LISA:2017pwj}, LISA has two main scientific objectives of purely cosmological bearing. The first is to probe the expansion rate of the universe, with specific requirements to measure the dimensionless Hubble parameter by means of GW observations alone, and further to constrain cosmological parameters through joint GW and electromagnetic (EM) observations. The second such objective is to understand SGWBs and their implications for early universe and particle physics.  
This will entail the characterisation of the astrophysical SGWB, and subsequently a measurement or bound on the amplitude and spectral shape of a cosmological SGWB. There are further scientific imperatives to use LISA to explore the fundamental nature of gravity and to search for unforeseen sources with relevance for cosmology. There is a wealth of cosmological information that may be extracted from LISA observations.

We start with Secs.~\ref{sec:standard_sirens} and \ref{sec:gw_lensing} on standard sirens and weak gravitational lensing; these are ``sure bets" for LISA, based on our current understanding of source populations. These sections are directly related to  LISA science objective SO6 ``probe the rate of the expansion universe"~\cite{SciRD}. They also identify new opportunities to derive cosmological information from GW astrophysical sources, in connection with LISA science objectives SO1, SO2, SO3 and SO4, which are devoted to understanding the galactic and extragalactic astrophysical source populations~\cite{SciRD}. We follow with sections on more speculative topics, which are potentially profound and revolutionary. Sec.~\ref{sec:modified_gravity} discusses the constraints on modified gravity theories that may be achieved through measurement of GW sources at cosmological distances. Results on this research subject are aligned with LISA science objectives SO5 ``explore the fundamental nature of gravity and black holes" as well as the aforementioned SO6. Sec.~\ref{sec:SGWB} introduces the theoretical foundations, observables, and conventions relevant for subsequent sections. Sec.~\ref{sec:PTs}, Sec.~\ref{sec:CosmicStrings} and Sec.~\ref{sec:Inflation} describe predictions of SGWBs sourced by first-order phase transitions, cosmic strings, and inflationary processes.  Sec.~\ref{sec:nonStandard} explores how these diverse SGWB signals convey  unique information on the expansion rate of the universe at redshift $\sim\!1000$ or higher.  These latter four sections touch on topics that are crucial for LISA science objective SO7 ``understand SGWB and their implications". Inflation not only leads to a SGWB but also to density  perturbations which may give rise to the formation of primordial black holes (PBHs), which is the subject of Sec.~\ref{sec:PBH}. Finally, Sec.~\ref{sec:toolstrans} and Sec.~\ref{sec:pipesgwb} present existing or planned tools and methods to analyse the GW signals discussed in the previous sections. Such tools and methods potentially constitute key deliverables for SO1-SO7 as well as LISA science objective SO8 ``search for GW bursts and unforeseen sources".

Certain topics of cosmological interest are intentionally omitted from this document: dark matter particles, some tests of general relativity (GR), waveform uncertainties,  astrophysical backgrounds, and the astrophysics of discrete sources such as MBBHs. These and related topics are covered by the living reviews maintained by the LISA Astrophysics~\cite{AstroWP}, Data Challenge~\cite{LDCWP}, Fundamental Physics~\cite{FundWP} and Waveform Working Groups~\cite{WaveWP}. Such reviews complement the picture presented herein, by the Cosmology Working Group. The goal of all these documents is to both identify LISA science objectives and corresponding work packages, and to alert the scientific community about novel research opportunities, or potential gaps. 
The tests of general relativity of Sec.~\ref{sec:modified_gravity} and PBH science of Sec.~\ref{sec:PBH} are exemplary cases of why these living reviews are needed. The original LISA proposal~\cite{LISA:2017pwj} makes no mention of these science investigations. But in recent years, as the subjects have evolved, a set of new science objectives have been proposed to cover them. The possibility that similar situations arise again justifies the effort and interest for living reviews that report on the thrilling, blooming, and fast-evolving LISA science.

Hereafter, we include a table of the acronyms used in this document.

\begin{table}
\small
\begin{center}
$~$\\[-1cm]
\begin{tabular}{ |l|l| }
 \hline
Acronym & Definition \\
 \hline
 BAO & baryon acoustic oscillations \\
 BBH & binary black hole \\
 BBN & big bang nucleosynthesis \\
 BBO & Big Bang Observer \\
 BH & black hole \\
 BNS & binary neutron star \\
 BSM & beyond the standard model of particle physics \\
 CE & Cosmic Explorer \\
 CP & charge parity \\
 CMB & cosmic microwave background \\
 CPL & Chevalier-Polarski-Linder \\
 DECIGO & DECihertz Interferometer Gravitational wave Observatory \\
 DE & dark energy \\
 DES & Dark Energy Survey \\
 DESI & Dark Energy Spectroscopic Instrument \\
 EDM & electric dipole moment \\
 E-ELT & European-Extremely Large Telescope \\ 
 EFT & effective field theory \\
 EM & electromagnetic \\
 EMRI & extreme mass ratio inspiral \\
 EoS & equation of state \\
 ET & Einstein Telescope \\
 EWPT & electroweak phase transition \\
 FLRW & Friedmann–Lema\^itre–Robertson–Walker \\
 FOPT & first-order phase transition \\
 GB & galactic binary \\
 GW & gravitational wave \\
 GR & general relativity \\
 IMBBH & intermediate-mass binary black hole \\
 IMS & interferometry metrology system \\
 IR & infrared \\
 KAGRA & Kamioka Gravitational Wave Detector \\
 KiDS & Kilo-Degree Survey \\
 $\Lambda$CDM & cosmological constant plus cold dark matter \\
 LIGO & Laser Interferometer Gravitational Wave Observatory \\
 LISA & Laser Interferometer Space Antenna \\
 LSS & large scale structure \\
 % LSST & Legacy Survey of Space and Time \\
 MBBH & massive binary black hole \\
 MBH & massive black hole \\
 MCMC & Markov Chain Monte Carlo \\
 MHD & magnetohydrodynamic \\
 NG & Nambu Goto\\
 PBH & primordial black hole \\
 PISN & pair-instability supernova \\
 PLS & power law sensitivity \\
 ppE & parameterized post-Einsteinian \\
 PTA & pulsar timing array  \\
 RD & radiation domination \\
 QCD & quantum chromodynamics \\
 SGWB & stochastic gravitational wave background \\
 SKA & Square Kilometer Array \\
 SM & standard model of particle physics\\
SNR & signal-to-noise ratio \\
 SOBH & stellar-origin black hole \\
 SOBBH & stellar-origin binary black hole \\
 TDI & time domain interferometry \\
 UV & ultraviolet \\
 \hline
\end{tabular}
\end{center}
\caption{\small Commonly used acronyms.}
\end{table}
\normalsize

\newpage
%%%%%%%%%%%%%%%%%%%%%%%%%%%%%%%%%
%\include{TEXTsirens}
% Here Sec. 2 starts
\section{Tests of cosmic expansion and acceleration with standard sirens}
\label{sec:standard_sirens}

\small \emph{Section coordinators: J.M.~Ezquiaga, A.~Raccanelli,  N.~Tamanini. Contributors: D.~Bacon, T.~Baker, T.~Barreiro, E.~Belgacem, N.~Bellomo, D.~Bertacca, C.~Caprini, C.~Carbone, R.~Caldwell, H-Y.~Chen, G.~Congedo, M.~Crisostomi, G.~Cusin, C.~Dalang, W.~Del Pozzo, J.M.~Ezquiaga, N.~Frusciante, J.~Garc\'ia-Bellido, D.~Holz, D.~Laghi, L.~Lombriser, M.~Maggiore, M.~Mancarella, A.~Mangiagli, S.~Mukherjee, A.~Raccanelli,  A.~Ricciardone, O.~Sergijenko, L.~Speri, N.~Tamanini, G.~Tasinato, M.~Volonteri, M.~Zumalacarregui.}\\ \normalsize

\subsection{Introduction}
Broadly speaking, to learn about the universe and its cosmic expansion we need to measure distances and times. 
GW astronomy offers a unique perspective in this matter, since the signal emitted by a compact binary coalescence is well predicted by GR. Namely, the amplitude of the GW is inversely proportional to its luminosity distance and it only depends on the masses and orbital inclination of the binary system source. 
Since cosmological propagation at the background level (namely, excluding the effect of perturbations over the Friedmann–Lema\^itre–Robertson–Walker (FLRW) geometry) only changes the overall strain amplitude, one can use the frequency evolution of the GW to unveil the masses of the compact binary and the relative amplitude of the two GW polarisations to estimate the orbital inclination, obtaining thus a direct and absolute measurement of the luminosity distance. 
However, GW signals alone do not provide a way to relate the time (of merger, for example) in the observer frame to the one in the source frame. 
To access this information, one needs an independent determination of the redshift of the source. In such a case the GW signal from compact binary coalescence can be considered a standard siren \cite{Schutz:1986gp}, namely a cosmological event for which a distance measurement and complementary redshift information are both available.
For example, the binary neutron star (BNS) merger GW170817, observed by the Advanced LIGO and Advanced Virgo detectors jointly with several EM facilities which spotted associated EM emissions, has already been used as a proof-of-principles, low-redshift measurement of $H_0$ \cite{Abbott:2017xzu}.
On the other hand, massive black holes (MBHs) seen by LISA with an EM counterpart could be used to map the cosmic expansion up to high redshift \cite{Holz:2005df,Tamanini:2016zlh,Tamanini:2016uin,Belgacem:2019pkk}. 

In this section we will present the different standard sirens that LISA will detect and the information about the cosmological model that they will provide. 
The section is organized as follows. We begin by describing the concept of standard siren in Sec.~\ref{subsec:standard_sirens}, detailing the expected LISA bright and dark sirens.
We then consider the constraints that could be placed in the standard cosmological constant plus cold dark matter ($\Lambda$CDM) cosmological model in Sec.~\ref{subsec:LCDM_constraints}. 
Subsequently, we explore LISA capabilities to probe different dark energy (DE) models in Sec.~\ref{subsec:DE}. 
Next, we show the synergies of LISA with other EM and GW observatories in Sec.~\ref{sec:synergies}. 
Finally, we describe the benefit of cross-correlating LISA data with large-scale structure surveys in Sec.~\ref{sec:crosscorr}.

\subsection{Standard sirens}
\label{subsec:standard_sirens}

GW signals from compact binary coalescences are natural cosmic rulers because of the inverse dependence of the strain with the GW luminosity distance, 
$h \propto 1/D_{L}^{\mathrm{GW}}$. 
In GR and over a Friedman-Lema\^itre-Robertson-Walker (FLRW) background, the GW luminosity distance is given by
\begin{equation}
\label{eq:luminositydistanceCosmo}
D_{L}^{\mathrm{GW}}=\frac{c}{H_0}\frac{(1+z)}{\sqrt{\vert\Omega_k\vert}}\mathrm{sinn}\left[ H_0 \int_0^z \frac{\sqrt{\vert\Omega_k\vert}}{H(z')}dz'\right]\,,
\end{equation}
where $\mathrm{sinn}(x)=\sin(x),\,x\,,\sinh(x)$ for a positive, zero and negative spatial curvature respectively. Assuming a $\Lambda$CDM cosmology, the Hubble parameter is a function of the matter content $\Omega_m$, the curvature $\Omega_k$ and the amount of DE $\Omega_\Lambda$ (radiation at present time is negligible)
\begin{equation}
H(z)=H_0\sqrt{\Omega_m(1+z)^3+\Omega_k(1+z)^2+\Omega_\Lambda}\,.
\end{equation}
LISA will attempt to measure $H_0$, $\Omega_m$, $\Omega_k$ and $\Omega_\Lambda$ using sirens.

GW observations themselves, however, do not provide direct information about the redshift. 
Therefore, in order to be able to probe the cosmological evolution we need additional input. 
In the case in which the redshift of the GW source is directly obtained from an EM counterpart, we will refer to the source as a \emph{bright siren}.
A beautiful example of this kind of multi-messenger event was the LIGO/Virgo event GW170817 \cite{LIGOScientific:2017vwq}, which provided the first standard siren measurement of $H_0$ \cite{LIGOScientific:2017adf} (see \cite{LIGOScientific:2021aug} for an update on the measurement of $H_0$ from standard sirens). As we present in Sec.~\ref{sec:Bright_Sirens}, LISA will be sensitive to very different bright sirens, but the concept remains the same. 
On the other hand, when the redshift information is obtained from an analysis that does not include EM counterparts, we will refer to the GW sources as \emph{dark sirens}.
In Sec.~\ref{sec:Dark_Sirens} we will present different dark siren classes that LISA will detect and which can be used to obtain cosmological information by cross-matching the sources with galaxy catalogues and looking for correlated features in the mass distribution.

\begin{figure}[t!]
\includegraphics[width = 0.85
\textwidth]{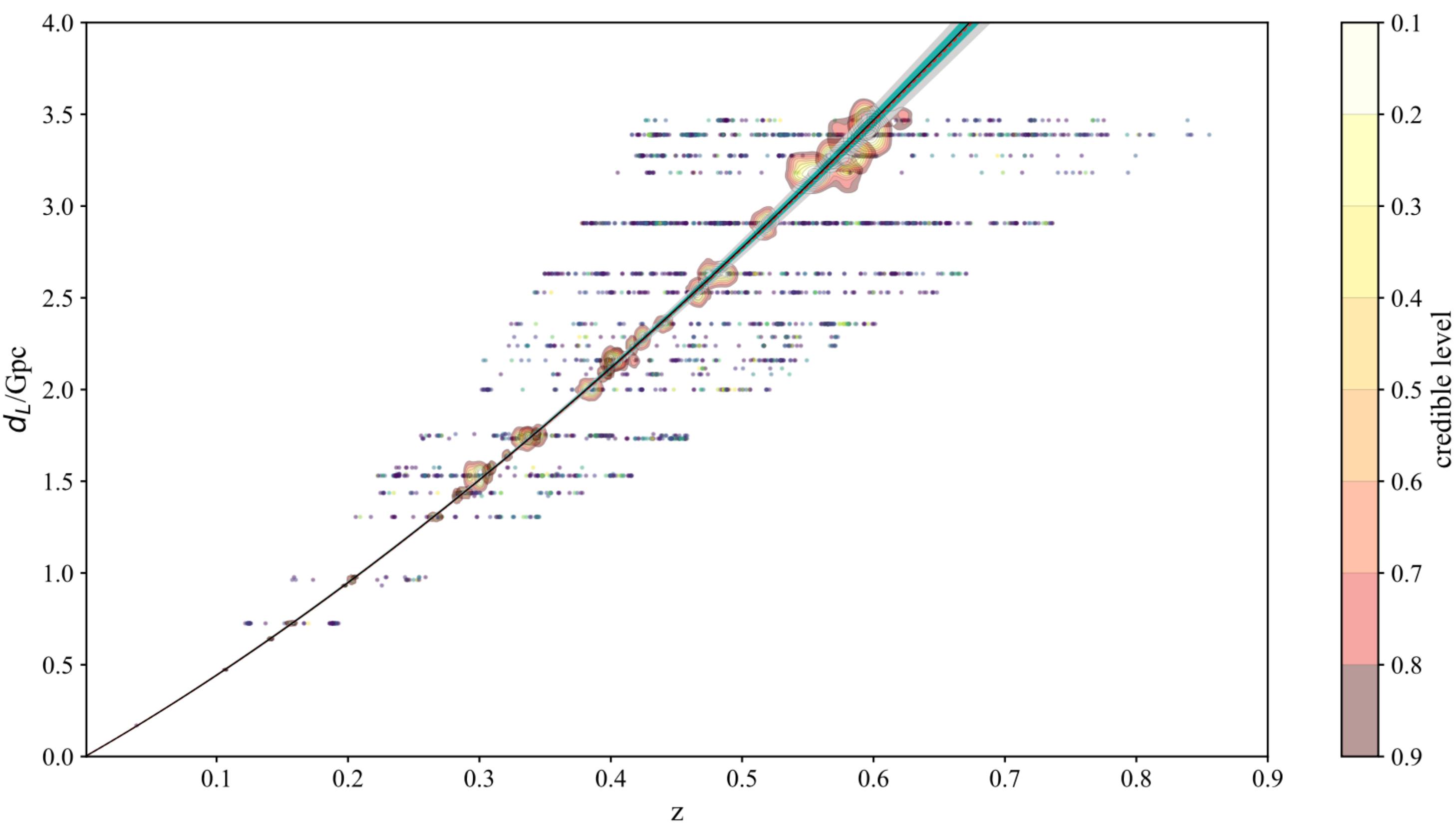}\\
\includegraphics[width = 0.49\textwidth]{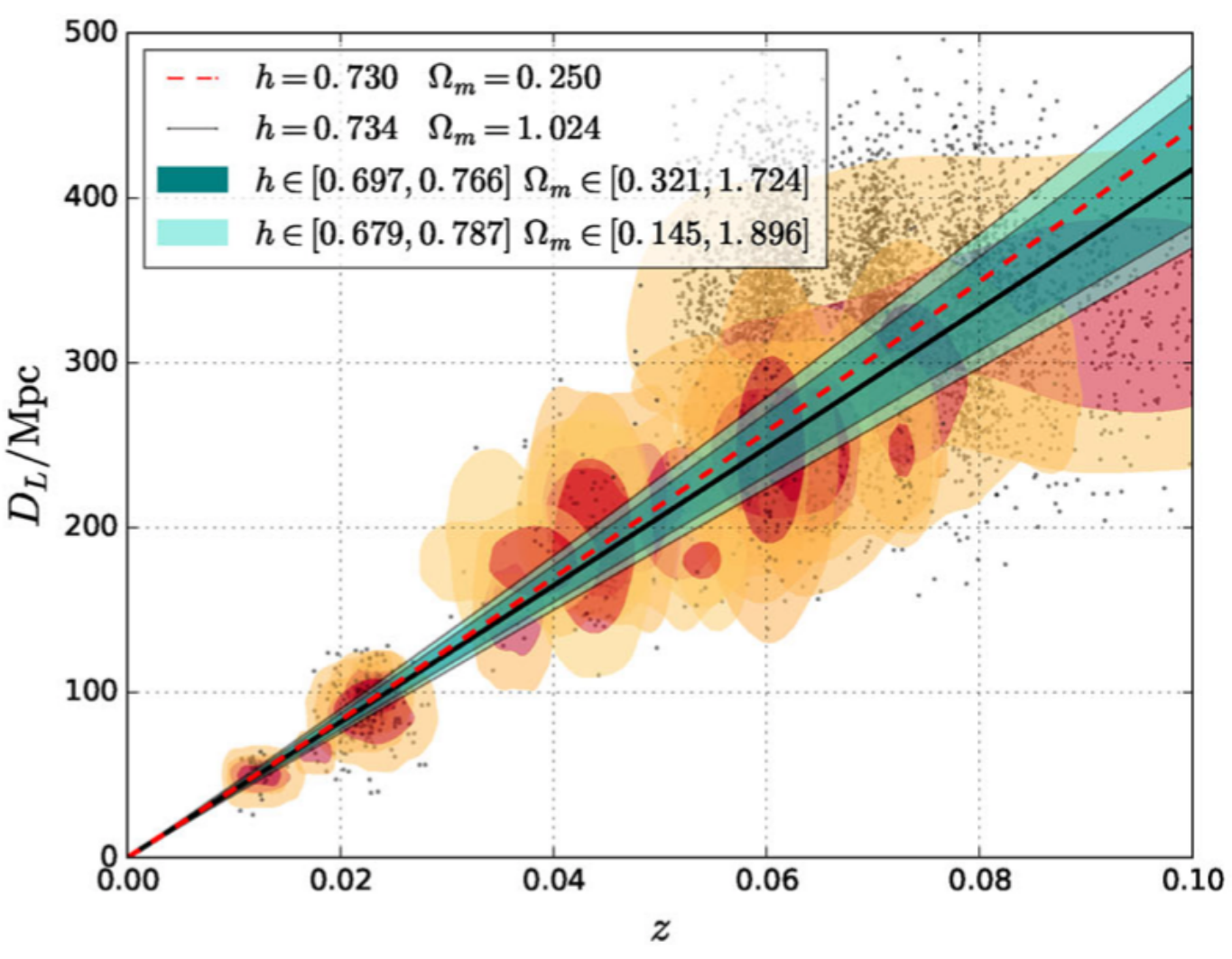}
\centering
\includegraphics[width = 0.49\textwidth]{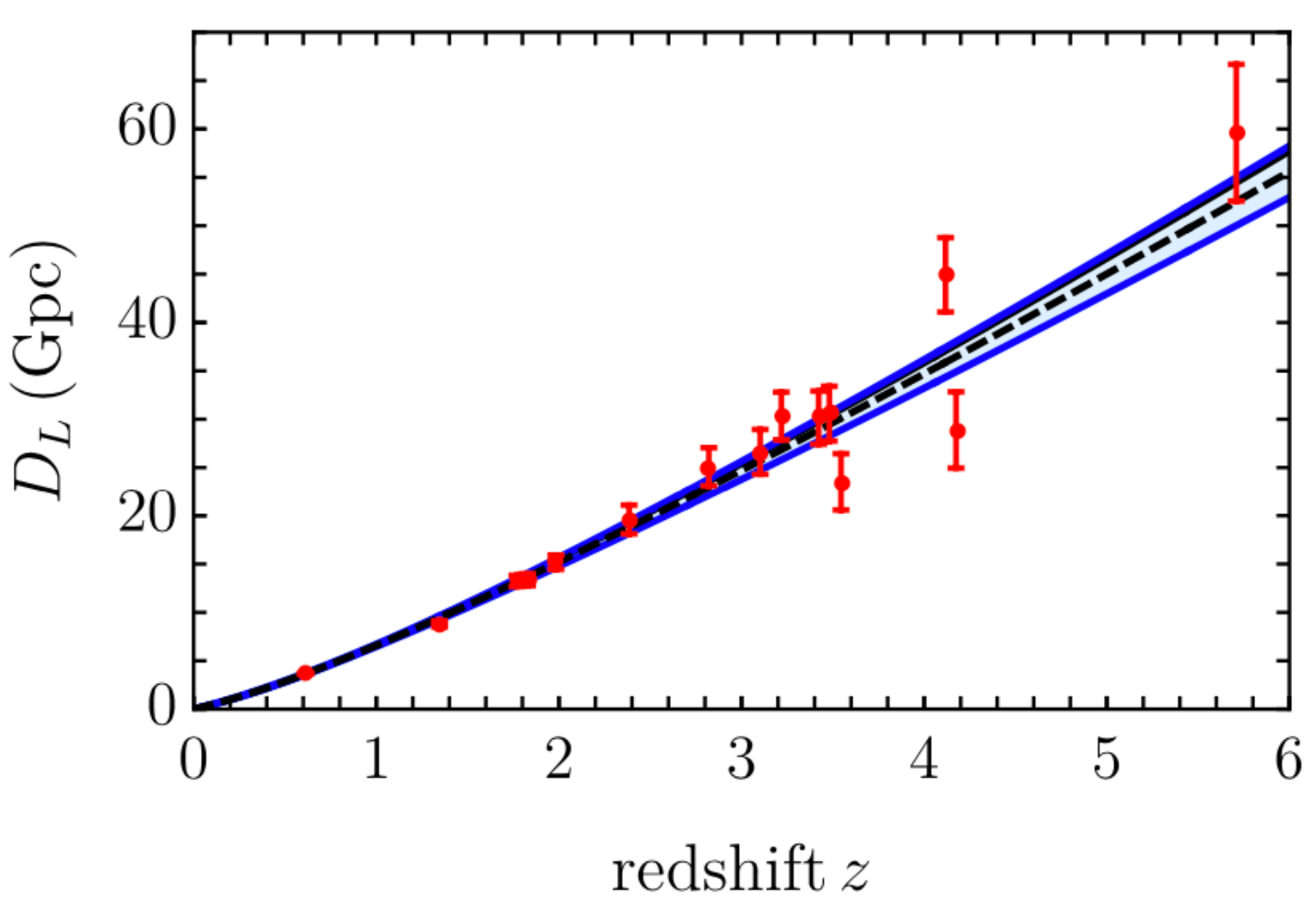}
\caption{\small 
Examples of LISA standard siren data sets at different redshift ranges.
Data from low-redshift dark sirens, defined by SOBBHs, are reported in the bottom-left plot (reproduced from Ref.~\cite{DelPozzo:2017kme}).
Data at intermediate redshifts correspond to EMRIs and are reported in the top plot (reproduced from Ref.~\cite{Laghi:2021pqk}).
High-redshift data are provided by standard sirens from MBBHs and are reported in the bottom-right plot (reproduced from Ref.~\cite{Speri:2020hwc}).
Note how SOBBHs and EMRIs, being dark sirens, can only provide broad likelihood regions in the Hubble diagram, while MBBHs, being bright sirens, provide precise data points thanks to the unique redshift association coming from the EM counterpart identification.
}
\label{fig:lisa_sirens}
\end{figure}

Modern analyses of standard sirens are based on Bayesian inference. We refer the reader to Sec.~\ref{sec:tools_standard_sirens} for a glimpse at the actual statistical tools LISA will use and a detailed discussion of their associated systematic uncertainties. In what follows we focus mostly on the different GW sources and their potential as standard sirens. 
LISA will detect three types of potential standard siren populations: MBBHs at $1 \lesssim z \lesssim 8$, extreme mass ratio inspirals (EMRIs) at $0.1 \lesssim z \lesssim 1$ and SOBBHs at $z \lesssim 0.1$. 
An example of the expected Hubble diagrams from these three different standard sirens populations can be found in Fig.~\ref{fig:lisa_sirens}.
The details regarding each of these populations will be presented in the following.

\subsubsection{Bright sirens: MBBHs with electromagnetic counterpart}
\label{sec:Bright_Sirens}

LISA will detect the coalescence of MBBHs up to redshift $z \simeq 15$--$20$. However, the mass and redshift distributions of the events are still uncertain. Currently, our knowledge of MBHs is limited to cases where either an active galactic nucleus is present or to quiescent MBHs in nearby galaxies \cite{Kormendy:2013dxa,Graham:2015kba,Greene:2019vlv}. The population of MBBHs accessible by LISA might be considerably different from our current expectations. Several groups have attempted to address this question with hydrodynamics simulations \cite{Salcido:2016oor,Katz:2019qlu,Volonteri:2020wkx} or semianalytic formation models \cite{Volonteri:2002vz,Barausse:2020mdt,Ricarte:2018mzn,2022MNRAS.511..616T,2018MNRAS.474.3825V}. While the former are able to handle more naturally hydrodynamical, thermodynamical, and dynamical processes, the latter are computationally efficient and can be used to explore a larger parameter space. 

Two main sources of uncertainties affect the expected redshift and mass distribution of merging MBBHs: black hole seeding and delay time prescription \cite{Klein:2015hvg}. If MBHs grow from the remnants of metal-poor population-III stars, the population of MBBHs accessible by LISA is expected to peak at the total mass  $\mtot \simeq 10^3 \msun$. However if MBHs arise from the monolithic collapse of gas in protogalaxies, the mass distribution is expected to range from $10^4 \msun$ up to few $10^7 \msun$. We note that additional formation mechanisms have been proposed and that they would further modulate the distribution of merging MBHs. 
Moreover, delay times, between the merger of two galaxies and the merger of their central MBHs, shape the redshift distribution, with short delays leading to more mergers at higher redshift. 
Further uncertainties arise from the gas inflow to the halo centre, its efficiency and the geometry of accretion.
Even if LISA will be able to distinguish different formation scenarios \cite{Sesana:2010wy}, the aforementioned uncertainties reflect in a broad interval for the number of events detected and their distributions (e.g.~mass, mass ratio, spins, redshift).

Combining these uncertainties, LISA should be able to detect between a few and several tens of MBBH events per year \cite{Klein:2015hvg}. 
Multiple-body interactions among a MBH binary and one or more intruder MBHs, arising naturally from the hierarchical galaxy formation process, might still produce $\simeq 10-20$ events per year \cite{Bonetti:2018tpf}.

LISA will also provide exquisite accuracy on MBBH parameters. For the search of a possible EM counterpart, the sky position accuracy is of paramount importance. Taking into account the full inspiral-merger-ringdown GW signal, MBBHs from few $10^5 \msun$ to few $10^6 \msun$ can be localised within $0.4 \degsq$ up to $ z \simeq 3$ \cite{Mangiagli:2020rwz}, but with high accuracy obtained only at merger. For these sources the posterior on the sky position is expected to be Gaussian; however, for more massive and distant sources, the recovered sky position is expected to present multimodalities \cite{Marsat:2020rtl}.
For heavy systems with total mass $\mtot > 10^7 \msun$ the ringdown portion of the signal might carry most of the information for the source localisation \cite{Baibhav:2020tma}. 
For cosmology applications, also the estimate on the luminosity distance plays a fundamental role: due to the typical large signal-to-noise ratio (SNR) value for these sources, LISA should be able to constrain the luminosity distance to better than $10\%$ for most of the events at $ z < 3$ \cite{Tamanini:2016zlh}.

If MBBHs evolve in gas-rich environment, EM radiation might be produced by the accretion of gas onto the MBHs close to or after merger. The orbital motion of the binary is expected to open a cavity in the circumbinary disk. Hydrodynamical simulations show that minidisks generally form around each BH from the stream of gas from the circumbinary disk \cite{Farris:2014zjo,Tang:2018rfm}. 

This leads to pre-merger EM emission across all wavelengths \cite{dAscoli:2018fjw}, which can be identified if the pre-merger sky localisation is good \cite{DalCanton:2019wsr}.
In the optical band, the Vera C. Rubin Observatory (formerly known as Legacy Survey of Space and Time) \cite{Abell:2009aa} will reach a magnitude limit of 24.5 in $30 \, \rm s$ of pointing over a field of view of $\simeq 10 \degsq$.
This survey speed enables the Vera C. Rubin Observatory to cover a sky area of $\simeq 100 \degsq$ allowing for possible pre-merger EM detection, though the number of detected counterparts is expected to be low \cite{Tamanini:2016zlh}.
At or after merger, several transients have been proposed, from spectral changes and brightening \cite{Schnittman:2008ez,Rossi:2009nk} to jets \cite{Palenzuela:2010nf,Khan:2018ejm,Yuan:2021jjt}.
In X-ray, Athena \cite{Nandra:2013jka}, with a field of view of $0.4 \degsq$  and a limiting flux of $\approx 3\times 10^{-16}$erg cm$^{-2}$ s$^{-1}$ in 100 ks, will be optimal to search for possible post-merger signatures. Similarly in radio, SKA \cite{2009IEEEP..97.1482D} 
will observe the launch of putative post-merger radio jets with an initial field of view of $ 1 \degsq$. 

A counterpart detection strategy has been proposed \cite{Tamanini:2016zlh} consisting in first localising the source in the radio with the SKA, and subsequently to proceed to a redshift determination of the host galaxy in the optical with the Extremely Large Telescope \cite{ELTcite}. This strategy is more promising than direct optical identification with the Vera C. Rubin Observatory. Depending on the seed and dynamical evolution models, LISA will detect between $\simeq 10$ and $\simeq 25$ MBBH events with EM counterpart during a mission assumed to be of five year \cite{Tamanini:2016zlh}.
These estimates are however affected by several astrophysical uncertainties that have as yet not been properly characterised.
It is, however, robust to expect that MBBH bright sirens will all be detected at relatively high redshift: one study finds standard sirens distributions peaking at around  $z\sim 2-3$ \cite{Tamanini:2016zlh,Tamanini:2016uin}. 
Therefore, MBBH bright sirens will be of great relevance to test the $\Lambda$CDM model, and possible deviations from it, in a redshift range so far scarcely probed by EM observations.
In Secs.~\ref{subsec:LCDM_constraints} and \ref{subsec:DE} we review the cosmological constraints that LISA will be able to impose both at low and high redshift.

%-FIGURE LISA STANDARD SIRENS-

\subsubsection{Dark sirens: SOBBH, EMRIs, IMBBHs}
\label{sec:Dark_Sirens}

In addition to observing the EM counterpart of individual events, there are other methods that one can employ to obtain information about the redshift of the GW source.
The most common and widely-used among these methods relies on statistical matching the inferred position of the GW source with a catalogue of galaxies with known redshift (see \cite{LIGOScientific:2021aug} for the application of this method to the Advanced LIGO-Advanced Virgo-KAGRA data). GW events for which an EM counterpart cannot be identified, but which can still be used to extract cosmological information statistically, are usually referred to as dark standard sirens.
In this section we outline how dark sirens are treated by correlating galaxy catalogues with the localisation of the source and using properties of their mass distribution. 
Generally speaking, dark standard sirens have the advantage to be applicable to all kind of GW sources for which a distance measurement can be retrieved (not exclusively those emitting EM counterpart signals), although a large number of them are necessary to achieve precise measurements through solid statistics.

In the absence of an EM counterpart, redshift information can be extracted by putting a prior on potential hosts from a galaxy catalogue, assuming that galaxies are good tracers of binary black hole (BBH) mergers.
In order to do that (see more details on statistical methods in Sec.~\ref{sec:tools_dark_sirens}), one associates the GW event with every galaxy within the 3D localisation error of the event, assigning to each of them a certain probability of being the true host galaxy of the GW event.
In this way, by stacking together the information gathered from several dark sirens, one can statistically infer the true values of the cosmological parameters.
For this method to be effective, one requires a large number of events to combine statistics or a very small localisation volume.
Errors in the luminosity distance scale with the SNR, $\Delta D_L/D_L\sim 1/\rho$, while the localisation depends also on the duration of the source, since LISA orbital modulation can be used to help disentangle the location in the sky. 
One complication is given by the fact that galaxy catalogues are in general not complete, especially at high redshift. 
This requires one to include information on the missing galaxies within the catalogue, and the catalogue prior must be supplemented by a suitable ``completion" term \cite{Chen:2017rfc,Fishbach:2018gjp,Gray:2019ksv,Finke:2021aom} (again see Sec.~\ref{sec:tools_dark_sirens} for more details). 
This aspect 
will remain a limiting factor until catalogues with very large completeness are available. 
The availability of complete galaxy catalogues, small GW localisation regions and accurate redshift determination (including characterisation of the uncertainty due to peculiar velocities \cite{Howlett:2019mdh,Mukherjee:2019qmm,Nicolaou:2019cip}), will be crucial in order for the statistical method to give competitive constraints on cosmological parameters.
At the time when LISA will be taking data, galaxy catalogues will be available from a plethora of current and future experiments, providing observations of different types of galaxies, with varying number density and sky coverage, over different redshift ranges. In particular, there should be available at that time the completed observations from Euclid and Dark Energy Spectroscopic Instrument (DESI), with the addition of redshift-deep catalogues from the Subaru Prime Focus Spectrograph project and the Roman Space Telescope. Additionally, the Vera C. Rubin Observatory should have observed hundreds of millions of galaxies, and deep and wide catalogues should be available from SPHEREx. Finally, on time-scales comparable with LISA, the full SKA2 and the ATLAS satellite should provide extremely deep and full-sky catalogues of the sky. Moreover, there are plans to build a next-generation billion-galaxies survey as a successor of DESI.

Stellar-origin black holes (SOBHs) are guaranteed dark sirens for LISA.
From the observations with current LIGO/Virgo detectors \cite{Abbott:2020gyp}, we know that there is a population of BBHs with masses between $\sim5M_\odot$ and $\sim50M_\odot$ that LISA will see in their early inspiral. 
Some of those events will be subsequently detected by ground-based detectors becoming in this way ``multi-band" events (see Sec.~\ref{sec:multiband}). 
SOBBHs with good enough localisation can be used as dark sirens \cite{Kyutoku:2016zxn,DelPozzo:2017kme}. 
Because of their low masses, cosmologically useful SOBHs could only be seen by LISA up to $z\sim0.1$, probing essentially the local expansion rate $H_0$. 
In practice, only those events with better than $20\%$ accuracy in the luminosity distance and with a typical sky-localisation error better than $\sim1\mathrm{deg}^2$  will be useful as dark sirens.
According to current forecasts, LISA could detect $\sim 10 - 100$ dark sirens \cite{DelPozzo:2017kme}, though uncertainties on the LISA noise level at the high-frequency end of its band \cite{Audley:2017drz}, on the detection threshold \cite{Moore:2019pke} and merger rate of these systems \cite{Abbott:2020gyp}, might well invalidate the most optimistic expectations. In Secs.~\ref{subsec:LCDM_constraints} and \ref{subsec:DE} we will review the cosmological constraints that LISA can impose with SOBBHs as dark sirens.

LISA will also be able to use EMRIs as dark sirens.
They will in fact be detected at cosmological distances, possibly in high numbers, though the rates are so far extremely uncertain.
EMRIs detected by LISA are expected to be broadly peaked around $0.5 < z < 2$, potentially reaching $z \sim 4$ \cite{Babak:2017tow}.
A recent investigation showed that events up to $z \sim 0.7$ can safely be used to estimate $H_0$ \cite{Laghi:2021pqk}, provided they can be detected with high SNR.
These events can reach relative uncertainties on $\Delta D_L/ D_L$ below $0.05$ and typical sky-localisation errors less than 1 deg$^2$, representing the best well-localised events suited for a statistical approach on the inference of the GW redshift.
As we will see in Secs.~\ref{subsec:LCDM_constraints} and \ref{subsec:DE}, according to the analysis of Ref.~\cite{Laghi:2021pqk}, LISA will be able to use from a few to several tens of EMRIs to extract useful cosmological information.

Redshift information on a GW source can also be obtained in a statistical way performing a population analysis when there are distinctive features in the source mass distribution. 
This is simply because GW observatories are only sensitive to the redshifted or detector frame masses, which directly relate to the source masses via
\begin{equation}
    m_z=(1+z)m\,.
\end{equation}
Therefore, if the mass distribution presents a known feature, e.g.~a peak or a drop, at a reference scale which is invariant under cosmic evolution, by observing this feature in the GW events at different luminosity distance bins one can infer their redshift. 
If such a feature exists, this would be a very convenient probe of the cosmic expansion because it only requires GW data.

A good example of such features occurs in the mass spectrum of SOBBHs. 
This is because as stars become more massive, a runaway process induced by electron-positron pair production known as pair-instability supernova (PISN)  is triggered \cite{Barkat:1967zz,Fowler:1964zz,Heger:2001cd,Fryer:2000my,Heger:2002by,Belczynski:2016jno}. 
These PISN result in complete disruption of the stars, preventing the formation of remnant BHs and thus inducing a gap in the mass spectrum starting at around $50M_\odot$. Nonetheless, for sufficiently massive stars the PISN process is insufficient to prevent direct collapse, and a population of BBHs with masses larger than $\sim120M_\odot$ could arise. 
Therefore, the theory of PISN predicts a gap in the BBH mass spectrum with two edges that act as reference scales.\footnote{Recent analyses have shown however that  both the lower end of the gap \cite{Farmer:2019jed} and its width \cite{Farmer:2020xne} are robust against ambient factors and nuclear reaction rates.}
While the lower edge of the gap lies within the main sensitivity of present LIGO/Virgo detectors and could lead to precise measurements of $H(z)$ \cite{Farr:2019twy, Ezquiaga:2022zkx}, 
LISA will be more sensitive to the upper edge if a population of ``far side" binaries in fact exists \cite{Ezquiaga:2020tns}.
These alternative methodologies are currently under development and will need to be further investigated in the future, especially in the framework of GW cosmology with LISA.

Finally, we mention another effect that in principle allows one to access the redshift information directly from the GW signal alone. 
The variation of the background expansion of the universe during the time of observation of the binary induces an effectively -4 post-Newtonian term in the waveform phase, whose amplitude directly depends on the redshift and on the value of the Hubble parameter both at the source and at the observer \cite{Seto:2001qf,Nishizawa:2011eq}. 
Unfortunately, redshift perturbations due to the inhomogeneous distribution of matter between the source and the observer also depend on time, and therefore also contribute to the extra terms in the phase \cite{Bonvin:2016qxr}. 
Among these, the time-varying peculiar velocity of the GW source centre of mass may dominate the signal, effectively preventing the extraction of the redshift information from the amplitude of the dephasing (but possibly allowing a measurement of the binary’s peculiar acceleration  \cite{Tamanini:2019usx,Inayoshi:2017hgw}).

\subsubsection{Systematic uncertainties on standard sirens} \label{sec:sirens_systematics}

Bright and dark sirens will suffer from some common systematic uncertainties. 
The measurements of the binary luminosity distances are affected by the detector calibration uncertainty \cite{2016RScI...87k4503K,Chen:2020zoq} and the accuracy of the waveform models (see Ref.~\cite{Abbott:2018wiz} for discussions in the context of LIGO/Virgo observations). 
Accurate waveforms will be particularly needed for the high SNR sources that LISA will detect. 
Moreover, high redshift sources will be affected by weak lensing uncertainties \cite{Holz:2004xx,Hirata:2010ba,Cusin:2020ezb} (see Sec.~\ref{sec:gw_lensing} for more details). 
In addition, the parameter estimation of the luminosity distance will be subject to degeneracies with the orbital plane inclination and other parameters, though for long duration signals or when higher harmonics are measured \cite{Baibhav:2020tma}, this degeneracy can be broken.
Finally, our understanding of the possible observational selection effect \cite{Chen:2020dyt} as well as the astrophysical rate evolution (see e.g.~Fig.~12 of Ref.~\cite{Finke:2021aom} for an application to LIGO-Virgo data) are critical to the accuracy of standard siren analysis as well.
Not many investigations have so far assessed the systematic uncertainties affecting standard siren measurements with LISA.
A thorough exploration of all these effects, needed to consolidate our confidence on LISA cosmological observations, will be necessary in the future.

\subsection{Constraints on $\Lambda$CDM }
\label{subsec:LCDM_constraints}

In this subsection we present how well LISA will be able to constrain the cosmological parameters of the standard $\Lambda$CDM model, by using different classes of standard sirens as presented in Sec.~\ref{subsec:standard_sirens}.
We first focus on the Hubble constant $H_0$ and consider constraints on additional parameters afterwards.
We  conclude the subsection with a discussion on consistency tests of $\Lambda$CDM at high-redshift with LISA MBBH standard sirens.

\subsubsection{$H_0$ tension and standard sirens}
\label{sec-H0tension}

The standard model of cosmology is extremely successful and allows one to describe the universe from the time of BBN to the present time of cosmic acceleration. Remarkably, it contains only six parameters, one of which, the present-day \textit{Hubble constant} $H_0$ describes the expansion rate of the universe. At small redshifts, it relates the luminosity distance and the redshift of a source such that $D_L(z\ll 1)= cz/H_0$. Consistency of the model requires the inferred value of $H_0$ to be independent of the probe and any deviations should be seen as a sign of unaccounted for systematics or more excitingly, new physics. 

In recent years, two sets of values for $H_0$ have emerged from so-called  \textit{early} or \textit{late} measurements of $H_0$ depending on the origin of the calibration. As the error bars shrink, it becomes increasingly clear that the two values are in tension, reaching $4-5 \sigma$ disagreements \cite{Verde:2019ivm}. The most precise measurement of $H_0$ from the early universe comes from the cosmic microwave background (CMB) (at recombination redshift $z \sim 1100$) with an inferred value of $H_0 = 67.4 \pm 0.5$ km s$^{-1}$ Mpc$^{-1}$ at 68\% C.L. assuming a flat $\Lambda$CDM cosmology \cite{Aghanim:2018eyx}.
In this case the $H_0$ value is inferred from the angle upon which the scale associated to the horizon at the last scattering surface is projected, which is obtained from the measurement of the density fluctuations.  Compatible values of $H_0$ are obtained also from 
the
Atacama Cosmology Telescope and WMAP5 for which $H_0 = 67.6 \pm 1.1$ km s$^{-1}$ Mpc$^{-1}$ at 68\% C.L. \cite{Aiola:2020azj} and from  the joint analysis of Dark Energy Survey
(DES) clustering and weak lensing data with baryon acoustic oscillations (BAO) and big bang nucleosynthesis (BBN), $H_0 = 67.2 ^{+ 1.2}_{-1.0}$ km s$^{-1}$ Mpc$^{-1}$ at 60\% confidence \cite{Abbott:2017smn}.

In contrast, several teams have measured a significantly higher  value of $H_0$ in the local universe with redshifts $z\leq 1$ in a model independent fashion. 
For example, the SH0ES team used Cepheid calibrated supernovae type Ia to measure $H_0 = 74.03 \pm 1.42$ km s$^{-1}$ Mpc$^{-1}$ \cite{Riess:2019cxk} (see also recent updates \cite{Riess:2021jrx}). The H0LiCOW collaboration used strong lensing time delays of background quasars to infer $H_0=73.3^{+1.7}_{-1.8}$ km s$^{-1}$ Mpc$^{-1}$ \cite{Wong:2019kwg}. The Megamaser Cosmology project used very long baseline interferometric observations of water masers in Keplerian orbits around MBHs to measure $H_0=73.9\pm 3.0$ km s$^{-1}$ Mpc$^{-1}$ \cite{Pesce:2020xfe}. The Carnegie-Chicago Hubble Program collaboration used tip of the red giant branch measurements in the large Magellanic cloud to calibrate 18 supernovae type Ia, instead of Cepheids, and found a slightly lower $H_0=69.8\pm 2.6$ km s$^{-1}$ Mpc$^{-1}$ \cite{Freedman:2019jwv}.
 
GWs offer an independent test of the tension using bright or dark sirens as described in Sec.~\ref{sec:Bright_Sirens} and Sec.~\ref{sec:Dark_Sirens}. The LIGO-Virgo collaboration used the first bright siren, namely GW170817, to infer $H_0= 70^{+12}_{-8}$km s$^{-1}$ Mpc$^{-1}$ at 1$\sigma$ \cite{Abbott:2017xzu}, which lies somewhat in between the early and late universe values but with a worse precision if compared to current EM results.
Furthermore, current dark siren measurements reached an inferred value of $H_0 = 75^{+25}_{-22}$ km s$^{-1}$ Mpc$^{-1}$ at 1$\sigma$ \cite{Fishbach:2018gjp,Finke:2021aom, Abbott:2019yzh}. The rather large error bars are expected to shrink as $1/\sqrt{N}$, where $N$ indicates the number of events. Percent-level precision on $H_0$ is expected in the 2020s with BNS mergers detected by Advanced LIGO-Virgo and their EM counterpart observations \cite{Nissanke:2013fka,Chen:2017rfc}. Bright sirens in the era of the third generation of ground-based GW detectors could further constrain other cosmological parameters, such as $\Omega_m$ and $w_0$ \cite{2010CQGra..27u5006S,Zhao:2010sz,Cai:2016sby,Jin:2020hmc,Chen:2020zoq} and usher us into the era of precise GW cosmology.

LISA will offer an alternative and complementary probe of $\Lambda$CDM which might as well provide useful information on the Hubble tension. In the following, we explore the potential of LISA to probe $H_0$ and beyond.

\subsubsection{LISA forecast for $H_0$}
\label{subsec:forecastH0}

LISA will be able to contribute measurements of $H_0$ coming from different classes of standard siren sources (see Sec.~\ref{subsec:standard_sirens}). For the time being, the literature has described only measurements coming from individual classes of sources. 
A complete analysis combining the constraining power of different LISA GW sources is still missing.

\begin{figure}
    \centering
\includegraphics[width=0.45\textwidth]{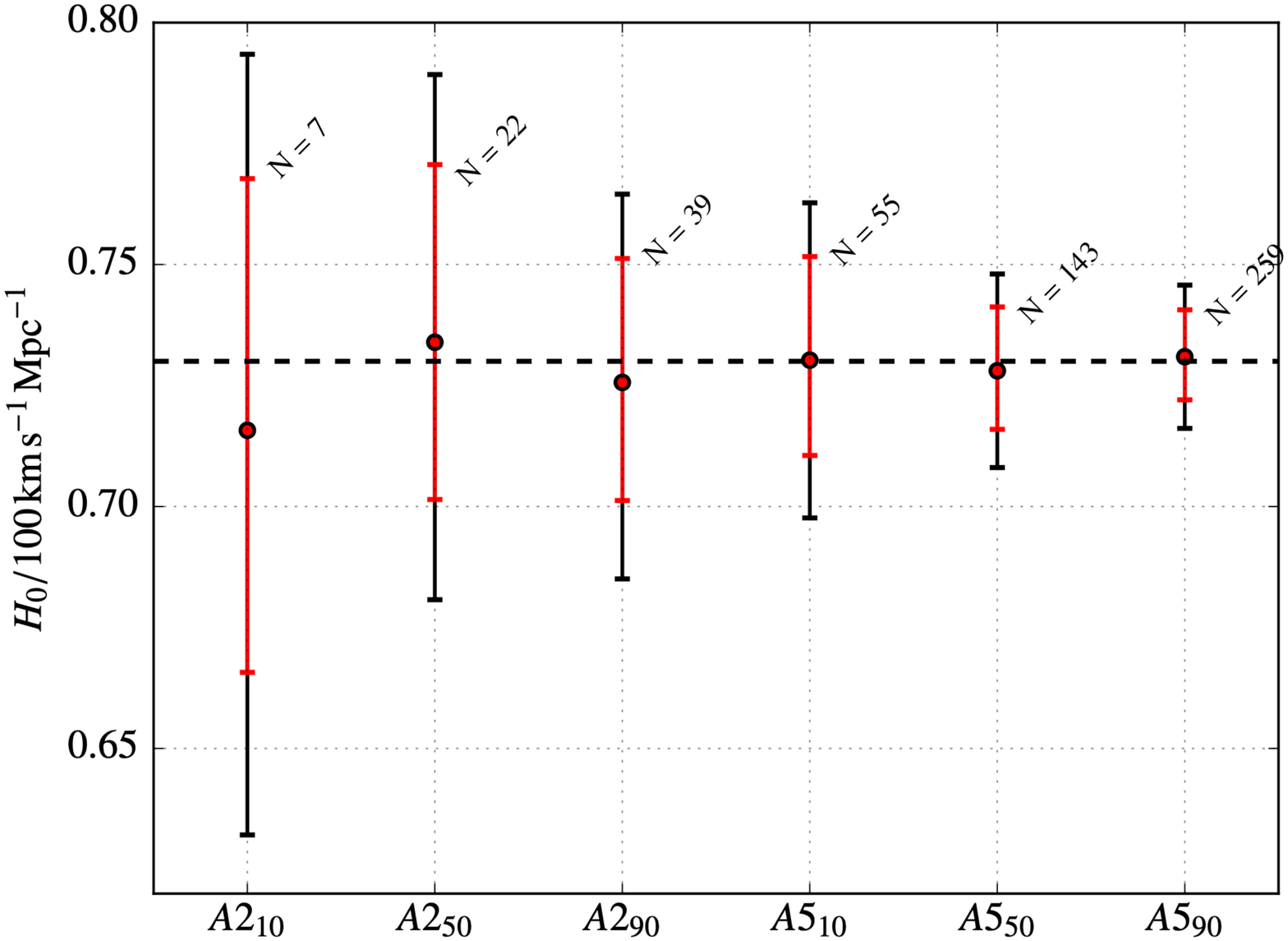}
    \includegraphics[width=0.45\textwidth]{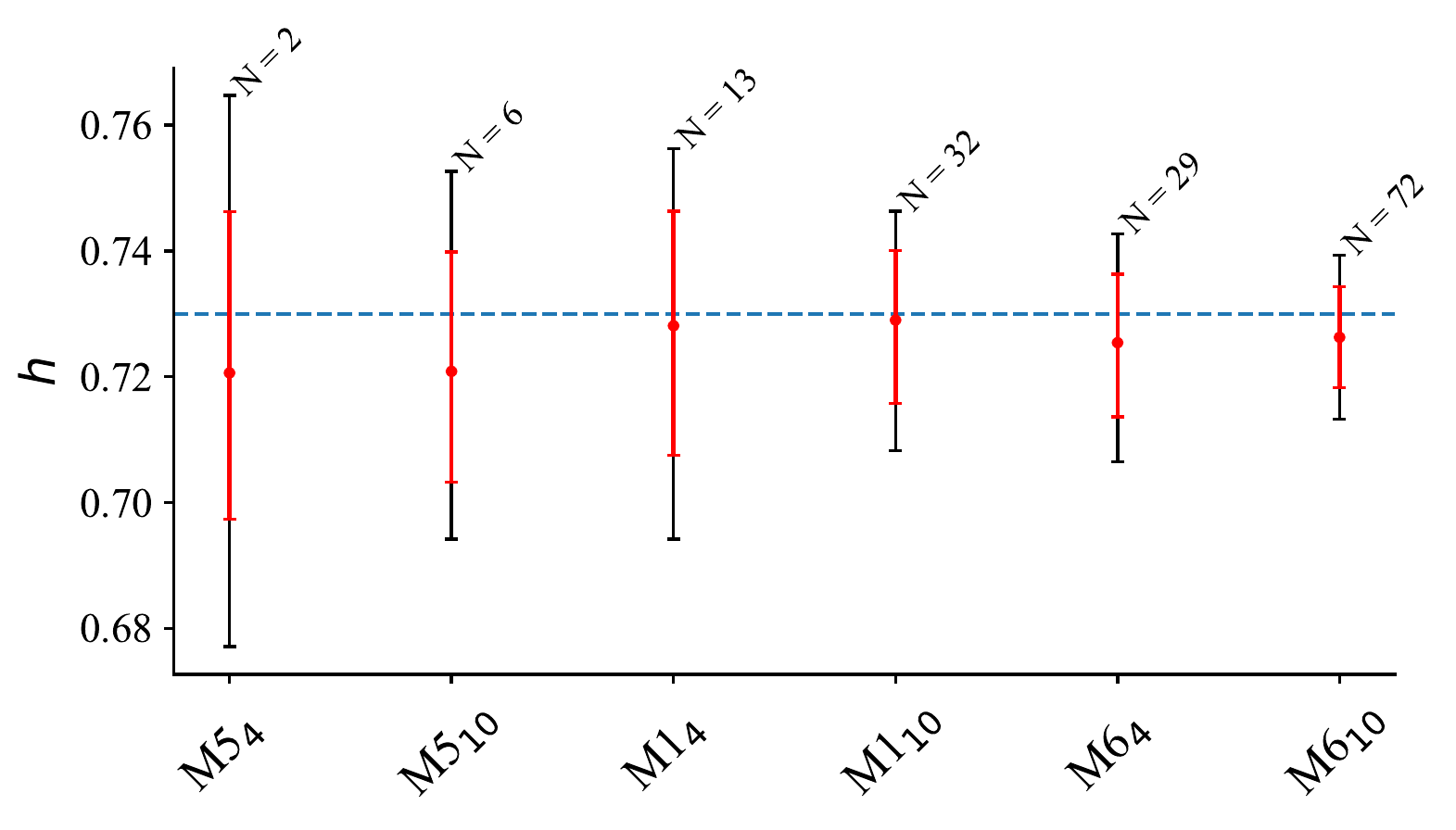}
    \caption{\small Left panel: $90\%$ (black) and $68\%$ (red) credible intervals for $H_0$/km s$^{-1}$ Mpc$^{-1}$ for each of the LISA configurations considered in Ref.~\cite{DelPozzo:2017kme}. The credible regions are averaged over the galaxy hosts realisations. Right panel: 90\% (black) and 68\% (red) percentiles, together with the median (red dot) of $h$ for the pessimistic, fiducial, and optimistic EMRI models (M5, M1, M6, respectively) and for two different LISA observational scenario (4 and 10 years). Here $h = H_0 / (100\, {\rm km\, s}^{-1} {\rm Mpc}^{-1})$ and the blue dashed horizontal line denotes the true cosmology (set at $h=0.73$ in Refs.~\cite{DelPozzo:2017kme,Laghi:2021pqk}). For each data point, we also report the average number $N$ of EMRIs considered in the analysis. Plots taken from  Refs.~\cite{DelPozzo:2017kme, Laghi:2021pqk}.}
    \label{fig:h0-summary}
\end{figure}

By considering SOBBHs, and by cross-matching with simulated galaxy catalogues, Refs.~\cite{Kyutoku:2016ppx,DelPozzo:2017kme} found that constraints on $H_0$ can reach the few \% level.
In particular, in the study presented in Ref.~\cite{DelPozzo:2017kme}, several different instrumental configurations for LISA were investigated, as well as several coalescence rate models within the range allowed by the LIGO-Virgo observations from O1. The SOBBHs entering the analysis were selected to have SNR $>8$, a cosmological redshift $< 0.1$ and an uncertainty on the luminosity distance smaller than $20\%$. No other selection criteria were applied. With the aforementioned selections, 
the number of SOBBHs considered ranged from a pessimistic case of 7, yielding an accuracy on $H_0$ of $7\%$ to a most optimistic case of 259, yielding an accuracy of $1\%$, see left panel in Fig.~\ref{fig:h0-summary}. 
 
A similar analysis can be done also with EMRIs.
A first investigation \cite{MacLeod:2007jd} pointed out that $\sim\!20$ EMRIs detected at $z\sim0.5$ could be enough to constrain $H_0$ at the 1\% level.
The analysis provided in Ref.~\cite{MacLeod:2007jd} employed however a simplified approach to estimate cosmological forecasts with LISA, and moreover assumed the more optimistic mission design considered at the time.
A recent, more detailed analysis has been performed with LISA EMRIs \cite{Laghi:2021pqk}. Using only the most informative, high-SNR ($> 100$) EMRIs up to redshift $z \leq 0.7$, cross-matching with the galaxy catalogue obtained from the simulated sky of Ref.~\cite{Henriques:2012ku}, it is shown that constraints at the few \% can be forecast for $H_0$. An analysis of three different EMRI population models taken from Ref.~\cite{Babak:2017tow}, representing a pessimistic, a fiducial, and an optimistic scenario, points out that in a 4-year LISA mission lifetime constraints are expected to be at $3.6\%$, $2.5\%$, and $1.6\%$ (68\% CL), respectively, while in case of a 10-year mission $H_0$ can be constrained at the $2.6\%$, $1.5\%$, and $1.1\%$ accuracy (68\% CL). 
The different accuracy in the various scenarios reflects the different number of useful EMRIs available in each model, which in case of 10 years of observation and after the SNR selection, ranges from $\sim\!5$ (in the worst scenario), passing to $\sim\!30$ (in the fiducial scenario), up to $\sim\!70$ (optimistic scenario), see right panel in Fig.~\ref{fig:h0-summary}.
 
Finally the last standard siren class that LISA can employ to constrain $H_0$ are MBBHs.
Although these events are expected to be detected at high-redshift ($z>1$), by assuming $\Lambda$CDM one can set bounds on $H_0$, i.e.~on the cosmic evolution at low-redshift.
By simulating different populations of MBBH mergers, performing (simple) parameter estimations over the expected GW signal, and by simulating the emission and detection of possible EM counterparts, recent works showed that constraints at the few \% level can be imposed on $H_0$ \cite{Tamanini:2016zlh,Tamanini:2016uin,Belgacem:2019pkk}.
These studies predict around four useful standard sirens with observed EM counterpart, per year.
Their redshift distribution peaks between redshift 2 and 4, with tails up to $z\simeq 8$.
The dominant contribution on the distance uncertainty of these events is not the LISA measurement error, but rather the systematic effect due to week lensing which dominates at high-$z$ providing an estimated average uncertainty of up to 5-10\% \cite{Hirata:2010ba,Cusin:2020ezb}.
Nevertheless even if LISA will detect only a few MBBH standard sirens, the fact that these will be at high redshift, with relatively precise distance determination, and of course with the single redshift value identified with the EM counterpart, will allow for interesting constraints on $H_0$ at the few percent level \cite{Tamanini:2016zlh,Tamanini:2016uin,Belgacem:2019pkk}.
This is comparable with what is expected for low-redshift, more numerous LISA dark sirens such as SOBBHs and EMRIs.
As we will show below, MBBHs will however be more interesting for cosmological analyses beyond the simple measurement of $H_0$.
 
The joint-inference on $H_0$ resulting from the combination of the analyses described above for SOBBHs, EMRIs and MBBHs, is expected to provide interesting constraints, possibly reaching the 1\% level, or better.
Such a combined investigation however has not yet been performed and will be the focus of future studies.
Note also that a further class of standard sirens for LISA could be provided by intermediate-mass binary black holes (IMBBHs), ranging from $\mathcal{O}(100)$ to $\mathcal{O}(1000)$ solar masses.
Although recent LIGO/Virgo observations may point towards the existence of this class of BBHs \cite{Abbott:2020tfl}, their merger rate and population properties are completely unknown at the moment \cite{Toubiana:2020drf}, making current LISA forecasts too uncertain to be seriously considered.
Given their high-mass and redshift range, IMBBHs could nevertheless well represent the best class of LISA dark sirens if they can be detected in high numbers.
Moreover the association with possible EM counterparts, which for these systems are being realistically considered \cite{Graham:2020gwr}, could well turn the LISA IMBBH detections that merge within the LIGO/Virgo band in a relatively short time, into useful multi-band bright sirens with a great potential to yield precise cosmological measurements \cite{Muttoni:2021veo}.

By surveying the considerations made above, we can expect LISA to deliver constraints on $H_0$ at the few percent level or better.
The robustness of these constraints against uncertainties in the overall analysis, including for example calibration and waveform modelling issues, will be something to carefully assess in the future.
In any case, a precise and accurate measurement of $H_0$ with LISA will provide useful insights on the Hubble tension, should the tension still persist.
On the other hand, a further independent and complementary measurement of $H_0$ will help strengthen our confidence in the value of the Hubble constant, especially if hints of physics beyond $\Lambda$CDM appear.

\subsubsection{$\Lambda$CDM beyond $H_0$}

\label{subsec:beyond_H0}

The relatively high-redshift reach for some of the standard sirens sources detected by LISA, will allow the inference on cosmological parameters beyond $H_0$.
In particular MBBHs will be extremely useful to test the cosmic expansion at high-redshift ($z\lesssim 8$) while EMRIs could be useful for cosmological applications up to $z\sim 1$, or more generally up to the redshift at which we will be able to employ fairly complete galaxy catalogues.

As shown in Refs.~\cite{Tamanini:2016zlh,Tamanini:2016uin}, MBBH mergers with an identified EM counterpart could be used by LISA to infer the values of $\Omega_m$ and $\Omega_k$, albeit with large uncertainties.
Assuming $\Lambda$CDM, constraints on $\Omega_m$ are forecast to reach the $\sim 10\%$ level (68\% C.L.) only in the most optimistic scenarios \cite{Tamanini:2016zlh,Belgacem:2019pkk,Spergel:2015sza}, while allowing for spatial curvature degrades these estimates for $\Omega_m$ to $\sim 25\%$ and yields a measurement of $\Omega_k$ at similar precision \cite{Tamanini:2016zlh}. Needless to say these results will certainly not be competitive with EM observations, but will at least provide an independent and complementary measurement.

Similarly, EMRIs will be able to provide information on $\Omega_m$ only in the most optimistic scenarios. In fact recent estimates \cite{Laghi:2021pqk} indicate that using the loudest $\mathcal{O}(70)$ events, as predicted in an optimistic population scenario, it is possible to get constraints with $\sim20\%$ accuracy (68\%CL) on $\Omega_m$ when jointly inferred with $H_0$. Even though this result is not expected to compete with current and future EM observations \cite{Amendola:2016saw,Scolnic:2017caz}, it is somehow not surprising since the EMRIs used in Ref.~\cite{Laghi:2021pqk} are distributed up to $z \lesssim 0.7$. Future analysis, possibly including a larger number of events, thus including low-SNR events which are typically coming from high redshifts, are expected to give more informative results on cosmological parameters beyond $H_0$. 
 
Preliminary investigations of the full $\Lambda$CDM cosmological model where the curvature term $\Omega_k$ is not fixed to 0, seem to indicate that, even for moderate redshift sources such as the loudest EMRIs (SNR $> 100$) at $z\lesssim 0.7$, LISA will provide some simultaneous constraints on all cosmological parameters.
In the fiducial scenario of Ref.~\cite{Laghi:2021pqk}, $\Omega_k$ can be constrained with an accuracy of $\sim 30\%$, while $\Omega_m$ can be constrained with an accuracy of $\sim 55\%$, while retaining an accuracy on $H_0$ of $\sim 2\%$ (all 68\% CL) \cite{Laghi:2021icm}. 
These results will need to be further investigated in the future, however they are suggestive that a possible simultaneous inference of different cosmological parameters with LISA standard sirens will indeed be possible.
In particular the full LISA cosmological analysis with results obtained from the combination of all classes of LISA standard sirens, namely SOBBHs, EMRIs and MBBHs, should substantially increase the accuracy of the estimates on $\Omega_m$ and $\Omega_k$ above, thanks especially to the combination of cosmological datasets from different redshift ranges which might break some of the degeneracy between the cosmological parameters.
A future combined investigation will thus be needed to thoroughly assess the ability of LISA to constrain parameters beyond $H_0$.

\subsubsection{Tests of $\Lambda$CDM at high-redshift}
\label{subsec:LCDM_highz_tests}

The most interesting class of LISA standard sirens are MBBHs, not only because they are expected to produce detectable EM counterparts (bright sirens) but also because they will be detected at high redshift and thus can be employed to test the cosmic expansion at early epochs largely unexplored by current EM cosmological surveys.
As we will show in Sec.~\ref{subsec:DE} and in Sec.~\ref{sec:modified_gravity}, MBBHs will have the potential to test different cosmological models alternative to $\Lambda$CDM.
Here we briefly mention how well LISA can test $\Lambda$CDM itself at high $z$, taking into account possible general deviations (to be discussed shortly) and by comparing with EM probes at similar redshifts.
Note also that in analogy to EM distance observations, LISA MBBHs can as well be employed to probe the fundamental assumptions of $\Lambda$CDM, e.g.~the cosmological principle \cite{Cai:2017aea}.

Let us first of all recall how well MBBHs can test $\Lambda$CDM.
As shown in the previous section, in the most optimistic cases $\Omega_m$ can be tested at the $\sim 10\%$ level, which reflects the fact that at $z\gtrsim 1$ the universe is expected to be matter dominated and thus to provide information mainly on $\Omega_m$.
Any deviation from the standard matter-dominated evolution at redshift $1\lesssim z \lesssim 4$ however will be constrained by LISA, at a level not attained by current EM observations.

To put in context the potential of LISA we can compare its constraining power with other EM measurements of the cosmic expansion at high $z$.
LISA will in fact provide independent and complementary data which will not only deliver useful and accurate information on deviations from $\Lambda$CDM but will also be used to check and cross-validate EM measurements, expected to be sparse and inaccurate at such high-redshift.
As a clear example, LISA can successfully compete with quasar cosmological observations at $z\sim 2$ and above \cite{Speri:2020hwc}.
Current quasar distance measurements indicate possible issues in the Hubble diagram at $z\gtrsim 2$, where low-significance deviations from $\Lambda$CDM have already been claimed \cite{Risaliti:2018reu} (see also Refs.~\cite{Velten:2019vwo,Yang:2019vgk,Banerjee:2021xln}).
To understand if such apparent deviations are due to systematic effects or are indeed due to new physics beyond $\Lambda$CDM, complementary measurements at the same redshift range will be needed.
As shown in Ref.~\cite{Speri:2020hwc}, four MBBH standard sirens detected by LISA are on average enough to unequivocally confirm or rule out the apparent deviation claimed in Ref.~\cite{Risaliti:2018reu}, and thus to reveal
if indeed this is due to systematics in the quasar Hubble diagram or to new physics.
Standard siren observations will play a crucial role to test deviations from the $\Lambda$CDM.

GW observations provide a more reliable distance luminosity measurement than other EM observations, such as quasars. Standard sirens rely in fact on fundamental predictions of GR and not on phenomenological relations between observed quantities.
Although the quasar Hubble diagram will become more precise and accurate in the time between now and when LISA will fly, for example with observations by the eROSITA \cite{Merloni:2012uf} and Euclid \cite{Amendola:2016saw,Barnett:2019rtg} missions, intrinsic systematics on the quasar cosmological measurement might still be unresolved.
LISA will thus provide a unique complementary test of the cosmic expansion at $z\gtrsim 2$, which not only will yield accurate measurement of deviations from $\Lambda$CDM but it will also be used to cross-validate any results obtained by other EM observations at the same redshift range, notably with quasars.

\subsection{Probing dark energy}
\label{subsec:DE}

This subsection discusses how LISA can probe the fundamental nature of DE through GW standard sirens.
Here only simple alternative DE models are considered, and in particular we  assume that the underlying gravitational theory remains GR.
Models of DE based on modification of GR are discussed in details in Sec.~\ref{sec:modified_gravity}.

\subsubsection{Equation of state of dark energy: $w_0$ and $w_a$}

Deviations from the standard cosmological model include the presence of a DE fluid with effective equation of state (EoS) given by: $p_{\rm DE}=w_{\rm DE}(a)\rho_{\rm DE}$, where $p_{\rm DE}$ and $\rho_{\rm DE}$ are the effective pressure and density of the DE fluid respectively. 
Although the functional dependence of $w_{\rm DE}(a)$ may be very non-trivial (see e.g.~self-accelerating cosmologies), for practical purposes
we consider here only a simple phenomenological parametrization introduced by Chevalier-Polarski-Linder (CPL) \cite{Chevallier:2000qy,Linder:2002et}: 
\be
 w_{\rm DE}(a)=w_0 + w_a(1-a) \,, \label{CPLeq}
\ee
where $w_0$ and $w_a$ are constants and indicate, respectively, the value and the time derivative of $w_{\rm DE}$ today. We refer to Sec.~\ref{sec:modified_gravity} about modified gravity for more complicated, model dependent forms.
    
In all the scenarios studied by Ref.~\cite{Belgacem:2019pkk} (which depend on the BH seeds and on the assumptions about the error on redshift), the GW sources considered are expected not to contribute relevantly to improve the knowledge on $w_0$, with respect to what is already known from current cosmological observations. More precisely, the 1$\sigma$ error $\Delta w_0$ only goes from $\Delta w_0=0.045$ (using CMB, type Ia supernovae and BAO data) to $\Delta w_0=0.044$ adding MBBH standard sirens to the datasets, even in the best scenario where LISA alone can only reach a 20\% relative 1$\sigma$ uncertainty on $w_0$ \cite{Tamanini:2016uin}.
It is important to remark, however, that these outcomes are based on a mission duration of 4 years and sources are limited to MBBHs with EM counterparts. Significant improvements are expected by extending the data taking time or by combining with information from other sources, notably EMRIs.

Indeed, recent investigations \cite{Laghi:2021pqk} suggest that EMRIs will deliver constraints on $w_0$ of the order of $\sim 10\%$, when inferred simultaneously with $w_a$. When assuming prior knowledge of $H_0$ and $\Omega_m$, constraints on $w_0$ are estimated at the $\sim7\%$ level in a realistic EMRI scenario, reaching $\sim 5\%$ in the best case (all 90\% C.L.). While relevant information on $w_0$ can be obtained with moderately low-redshift events, $w_a$ is expected to be measurable only with higher-redshift events, which in the joint cosmological inference of $w_0$ and $w_a$ of Ref.~\cite{Laghi:2021pqk} are not considered, and thus no relevant measurement of $w_a$ is obtained.

Although from these estimates it seems that LISA standard sirens will not be competitive with future EM observations, we stress that they will anyway provide independent and complementary measurements which will increase our confidence on any insight on the nature of DE.
This is strikingly important for modified gravity models of DE, where GWs can indeed provide orthogonal information with respect to EM observations; see Sec.~\ref{sec:modified_gravity}.

\subsubsection{Alternative dark energy models}

The  CPL phenomenological parametrization discussed in the previous subsection is appropriate for detecting deviations from the $\Lambda$CDM paradigm occurring at small redshifts. However, the theoretical description of DE may require more sophisticated models. Here, we briefly discuss such a possibility focusing on scenarios that change the background cosmological expansion with respect to $\Lambda$CDM. We assume the standard evolution equation for GWs. Possible modifications to GW propagation  through  cosmological spacetimes -- motivated by modified graviton dispersion relations, or  non-minimal couplings of the dark-energy sector with curvature -- are described in Sec.~\ref{sec:modified_gravity}. 
 
Models for DE can  include quite   a large  number of parameters, leading to rich dynamics for the DE sector as a function of redshift. Some investigations aim   to dynamically explain the puzzling small value for the present-day acceleration rate, leading to a time-dependent evolution of the DE density. These  include  quintessence scenarios \cite{Wetterich:1987fm,Zlatev:1998tr},  which can be generalised to  kinetically-driven \cite{ArmendarizPicon:2000dh} and kinetic-braiding models \cite{Deffayet:2010qz} without modifying the propagation properties of GWs. Other scenarios aim  to alleviate the coincidence problem relating DE with DM at intermediate redshifts, for example in  the  DE-DM interacting models \cite{Wetterich:1994bg,Amendola:1999er}, or in Chaplygin gas cosmology \cite{Kamenshchik:2001cp}. More exotic possibilities include  holographic  DE, associating  the present-day acceleration of the universe with the size of the particle horizon \cite{Li:2004rb}. See e.g.~Refs.~\cite{Copeland:2006wr,Li:2011sd,Ishak:2018his,Huterer:2017buf} for  reviews, including  analysis of  cosmological implications and observational prospects of DE scenarios. A recent resurgence of interest on  DE model building, based on the previous approaches,  has been motivated by the $H_0$ tension discussed in Sec.~\ref{sec-H0tension}; see \cite{DiValentino:2021izs} for a review.  Among many examples, such tension can be alleviated in scenarios with DM-DE interactions or with features at small or intermediate redshifts \cite{DiValentino:2017iww,DiValentino:2016hlg,Keeley:2019esp,Raveri:2019mxg}, in ranges that might be probed with GW sirens. See e.g.~Ref.~\cite{Knox:2019rjx} for a comprehensive discussion on this topic, including comparison between theoretical ideas and existing cosmological constraints. 
 
These theoretical models suggest that  distinctive DE  effects  can occur at different redshifts, from very small to relatively large values of $z$.
The capability of LISA to probe cosmological expansion in a large range of redshifts, as discussed in the previous sections, provides invaluable opportunities for building independent cosmological tests (see Sec.~\ref{subsec:LCDM_highz_tests}) and thus to probe different DE scenarios, in a complementary way with respect to EM probes.
A clear example is given by the investigations of Refs.~\cite{Cai:2017yww,Caprini:2016qxs} where LISA forecasts for testing cosmological models allowing for DE-DM interactions or for early DE have been produced using MBBH as bright sirens.
Similar analyses using LISA dark sirens are still missing in the literature and will constitute material for future explorations.
By considering the results obtained with both SOBBHs and EMRIs for standard cosmological models (see Sec.~\ref{subsec:LCDM_constraints}),
it would be very interesting to further develop these studies by designing an efficient, unified method to reconstruct the redshift dependence of DE with GWs, similar for example to what already done with EM observations; see e.g.~Ref.~\cite{Sahni:2006pa}.
Assumptions going beyond the linear CPL parameterisation discussed in the previous section might be better suited to test specific alternative scenarios, or for improving the DE reconstruction in a wider redshift interval, e.g.~through polynomial fitting \cite{Alam:2003sc}.
Alternatively, methods based on a principal component analysis of a  binned parametrization of the signal as a function of redshift, as proposed in Ref.~\cite{Huterer:2002hy}, might be adapted and applied to GW observations with LISA. 

Besides the alternative DE models already considered in the literature, there are plenty of others that can still be tested by LISA and for which detailed analyses will be needed in the forthcoming years in order to understand how well LISA will constrain them and thus to better assess and expand the science case of the mission.
We conclude by mentioning again that the nature of DE can be further investigated if this is connected to an underlying gravitational theory beyond GR.
In this case new observational signatures might appear, giving rise to a richer phenomenology and to more promising LISA results.
Such models and analyses will be discussed in Sec.~\ref{sec:modified_gravity} in the context of modified theories of gravity~\cite{Baker:2019ync}.

\subsection{Synergy with other cosmological measurements}
\label{sec:synergies}

GW observations by LISA will provide a wealth of cosmological information as we have seen in the previous sections. 
But LISA will not be alone in the quest of understanding the cosmic history. It is thus of great importance to assess how LISA could complement with other facilities. 
In what comes next, we discuss LISA synergies with EM observatories and other GW detectors.

\subsubsection{Integration with standard electromagnetic observations}
\label{subsec:synergy_EM}

To date, most studies constraining cosmological parameters with GWs have focused on the Hubble parameter, see Sec.~\ref{subsec:forecastH0}. This is because the cosmological information in GW amplitudes is primarily held in the luminosity distance of the source, $D_L$, which for low redshifts ($z\ll1$) can be approximated to $D_L\simeq c z /H_0$. 
The majority of LISA sources will exist at higher redshifts where this approximation breaks down. Instead the full expression for the luminosity distance must be used (see Eq.~(\ref{eq:luminositydistanceCosmo})) and so GW amplitudes in principle have sensitivity to further cosmological parameters traditionally measured electromagnetically, such as the fractional densities $\Omega_m$ and $\Omega_\Lambda$, and the DE EoS parameters $w_0$ and $w_a$, see Secs.~\ref{subsec:beyond_H0} and \ref{subsec:DE}. 

A key outstanding question is whether specific combinations of EM and GW probes have the ability to improve constraints on these parameters and break degeneracies between them. Details about the galaxy surveys co-temporal with LISA are not available at present, but as a conservative approach we can consider `Stage IV' experiments planned over the next decade, such as DESI, Euclid, the Vera C. Rubin Observatory, and the Roman Space Telescope \cite{Aghamousa:2016zmz,Laureijs:2011gra,rubinobservatory, Spergel:2015sza}. The corresponding CMB data will come from the LiteBIRD mission \cite{2012SPIE.8442E..19H}. Direct cross-correlation of GW sources with galaxy catalogues will be considered in Sec.~\ref{sec:crosscorr}. Here we focus instead on probe combination, though we note that there is a lack of comprehensive studies  on this topic in the current literature.

One advantage leveraged by Stage IV galaxy surveys is the ability to combine multiple probes from the same instrument. Most commonly the main probes are shear power spectra from galaxy weak lensing, BAO measurements from galaxy clustering, supernovae, and, additionally, strong gravitational lenses in some analyses. As examples of the expected constraints,  Ref.~\cite{Zhan:2017uwu} provides forecasts for standard cosmological parameters using weak lensing, BAO and supernovae data from the Vera C. Rubin Observatory, combined with current Planck CMB data. The resulting 68\% confidence intervals on $\left\{w_0, w_a\right\}$ and $\omega_m = \Omega_m h^2$ are $\left\{7.05\times 10^{-2},1.86\times 10^{-1}\right\}$ and $7.73\times 10^{-4}$ respectively, with $h$ being the normalised Hubble parameter.
Similar results are expected from the Euclid space mission, which should yield 68\% confidence intervals around $\left\{3\times 10^{-2}, 1.3\times 10^{-1}\right\}$ for $\left\{w_0, w_a\right\}$.
Comparing these values to the forecasts using LISA EMRI detections presented in Ref.~\cite{Laghi:2021pqk}, it seems unlikely that LISA will be able to offer competitive constraints on $\Omega_m$. However, in mildly optimistic scenarios they may offer moderate constraints on the DE parameters, with Ref.~\cite{Laghi:2021pqk} forecasting $95\%$ confidence intervals on $\left\{w_0, w_a\right\}$ of approximately $ \left\{ 6.5\times 10^{-2},6.5\times 10^{-1}\right\}$ for a four-year LISA mission and optimistic MBH population models. 

An alternative strategy is to find probe combinations where GW data can break degeneracies existing between EM probes. One such example is put forwards in Ref.~\cite{Qi:2021iic}, which determines forecasts for the combination BNS data from the DECihertz Interferometer Gravitational wave Observatory (DECIGO) with measurements of redshift drift (the Sandage-Loeb effect \cite{Loeb:1998bu}) from the SKA and the European-Extremely Large Telescope (E-ELT). Redshift drift measurements use high-resolution spectroscopy of the HI emission line (SKA) or Lyman-$\alpha$ absorption lines in quasar spectra (E-ELT) over long time frames (10 years+) to directly measure tiny shifts the line frequencies. Although experimentally challenging, this constitutes a \textit{direct} measurement of $H(z)$, as opposed to the integrated effect of $H(z)$ probed by luminosity distances; see Eq.~\eqref{eq:luminositydistanceCosmo}.

GWs can likewise offer a direct measurement of $H(z)$ through the dipole of the luminosity distance. Eq.~\eqref{eq:luminositydistanceCosmo} gives the luminosity distance-redshift relation in a perfectly homogeneous and isotropic universe; in fact small corrections to this are induced by gravitational lensing and peculiar velocities. As shown in Refs.~\cite{Bonvin:2005ps, Bonvin:2006en} for the EM case (see e.g.~Ref.~\cite{Nishizawa:2010xx} for the GW case), the peculiar motion of our Galaxy with respect to the CMB frame induces a dipole mode in the luminosity distances. This dipole moment is given by
\begin{eqnarray}
D_L^{(1)}(z)=\frac{|{\bf v_0}|\,(1+z)^2}{H(z)} \,,
\end{eqnarray}
where ${\bf v_0}$ is the dipole anisotropy in the CMB, estimated to be approximately $369.1\pm 0.9$~km/s. Combining measurements of this dipole from DECIGO BNS sources with the redshift drift measurements above, Ref.~\cite{Qi:2021iic} finds substantial breaking of degeneracies in the $H_0-\Omega_m$ plane, leading to 1-$\sigma$ bounds on $\{H_0, \Omega_m\}$ of $\{0.78\, {\rm km\, s}^{-1}\mathrm{Mpc}^{-1}, 0.006\}$, competitive with current bounds. Mild improvements on the constant DE EoS were also obtained ($\sigma_{w_0}\sim 0.03$), though no meaningful constraint on $w_a$ was possible. 
Further analyses are needed to properly understand if such method can equally be applied to LISA.

On cosmological scales, a bias prescription is used to model how GW events trace the large-scale DM distribution. Analogously to galaxy bias, the GW bias is modelled as scale-independent on large scales. Using the parameterisation $b_{\textrm{GW}}(z)=b^0_{\textrm{GW}}\,(1+z)^\alpha$, Ref.~\cite{Mukherjee:2020hyn} finds the parameters $b^0_{\textrm{GW}}$ and $\alpha$ to be uncorrelated with $H_0$ and $\Omega_m$ for a $\Lambda$CDM model ($w_0=-1$ fixed). This suggests cautious optimism that  combined GW and EM constraints on cosmological parameters should not be strongly sensitive to the GW bias prescription.

\subsubsection{Complementarity with other gravitational wave observatories}
\label{sec:multiband}

\begin{figure}[t!]
\centering
\includegraphics[width = 0.48\textwidth]{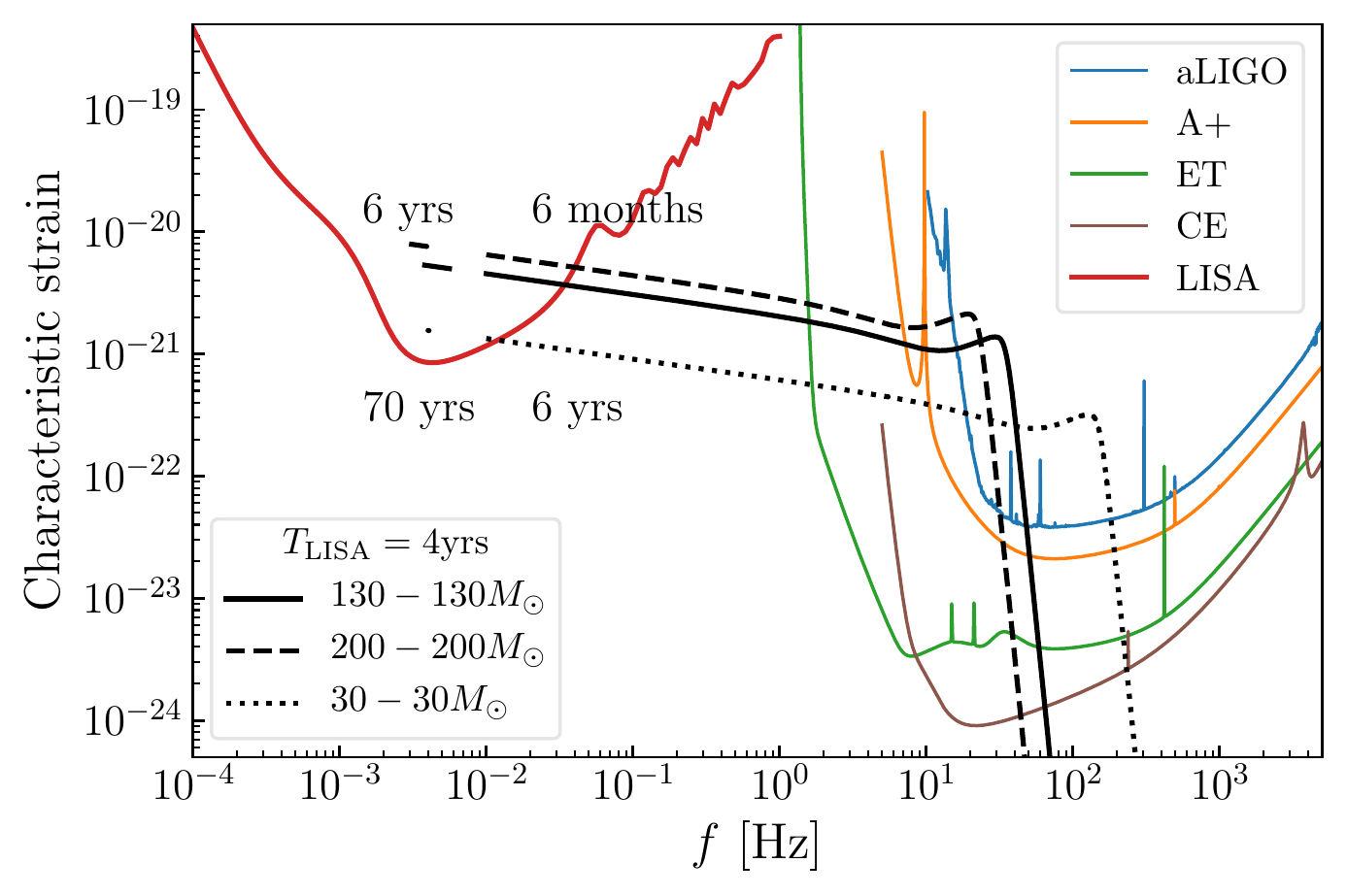}
\includegraphics[width = 0.48\textwidth]{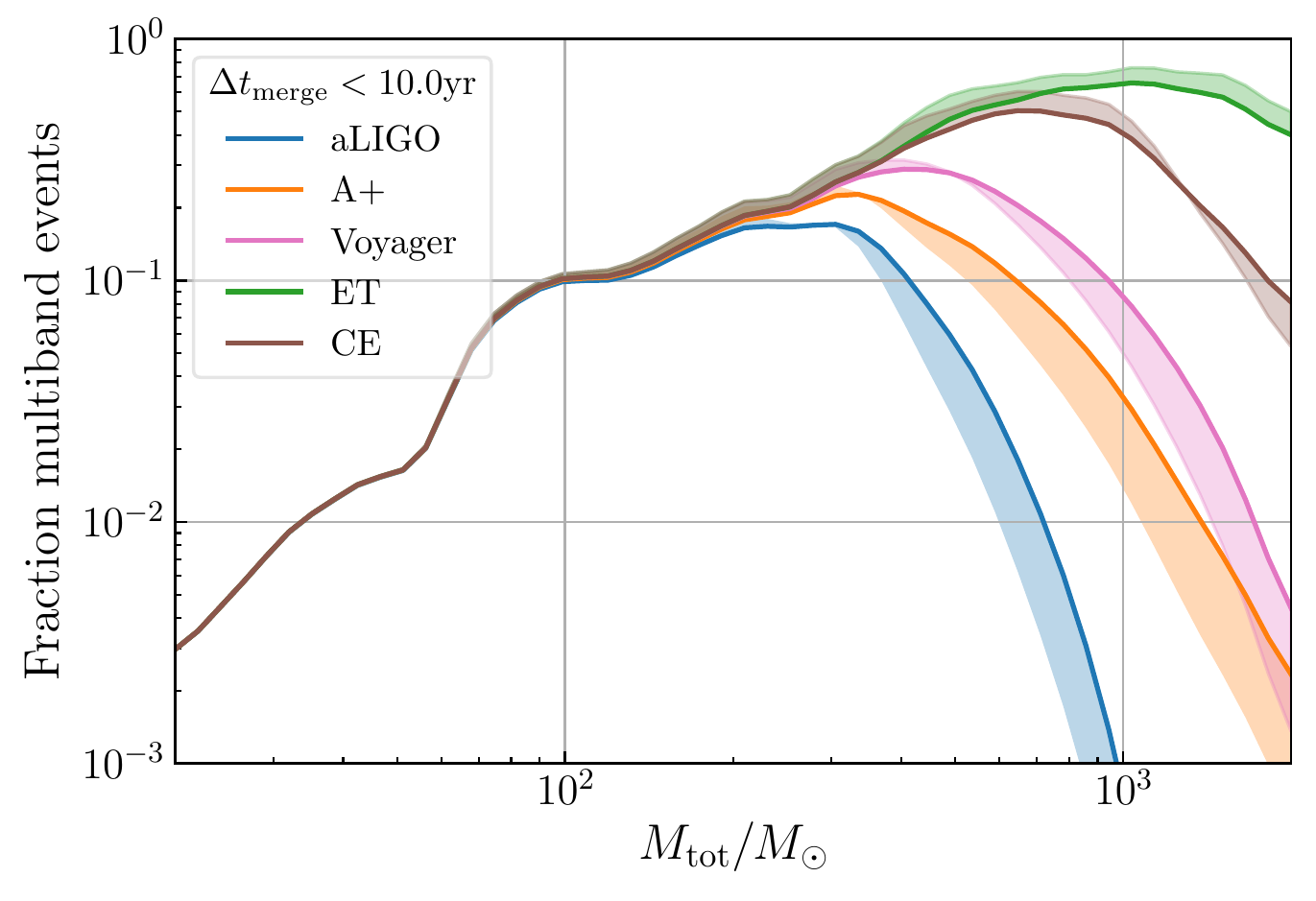}
\caption{\small On the left, GW signals as seen by LISA and different ground-based detectors. The times indicate the time to merge which depends on the initial frequency. On the right, fraction of events, observed by LISA and that will merge within 10 years, being detected by a ground-based detector. Right panel reproduced from Ref.~\cite{Ezquiaga:2020tns}.}
 \label{fig:multi_band}
\end{figure}

In addition to the synergies with other cosmological surveys, LISA will also build new synergies with other GW detectors. 
In particular, LISA will be able to detect the early inspiral phase of compact binary coalescences that later merge within the frequency bands of Earth-based interferometers. 
Some example signals are displayed in the left panel of Fig.~\ref{fig:multi_band}. 
Because the same signal is detected across different frequencies, these events are known as multi-band. 
Different populations of BBHs could become multi-band sources. 
Most notably, the SOBBHs that current LIGO/Virgo detectors are observing could have been seen if LISA was online a few years before these detections \cite{Sesana:2016ljz}. 
Nonetheless, LISA high frequency sensitivity limits their number \cite{Moore:2019pke}.
If present in nature, IMBBHs would be more promising candidates \cite{AmaroSeoane:2009ui,Jani:2019ffg,Sedda:2019uro}, and possibly yield interesting cosmological results \cite{Muttoni:2021veo}.
There is however a limit to their masses, because if they are too heavy they will merge before reaching the frequencies of ground-based detectors. (This limitation could be removed with a deci-Hertz observatory \cite{Sedda:2019uro}.)  
The right panel of Fig.~\ref{fig:multi_band} displays the \emph{fraction of multi-band events}, defined as the subset of LISA detections merging within 10 years and being detected by a ground-based instrument \cite{Ezquiaga:2020tns}. For concreteness we consider Advanced LIGO (aLIGO), its possible upgrade (A+), Voyager and the third-generation detectors Cosmic Explorer (CE) and Einstein Telescope (ET). 
Interestingly, the multi-band fraction peaks where the upper end of the PISN mass gap is expected to be found, implying that far-side binaries could be promising multi-band sources \cite{Ezquiaga:2020tns}. As noted in Ref.~\cite{Gerosa:2019dbe}, for $M_\mathrm{tot}\lesssim100M_\odot$, there is no difference for the multi-band ratio between 2G and 3G detectors because the fraction is limited by LISA high-frequency range.
On the contrary, for $M_\mathrm{tot}>200M_\odot$ the difference among ground-based detectors are sizeable and depends mostly on their low frequency sensitivity.

Besides individual sources, LISA and ground-based detectors could share SGWBs.
For example the background of unresolved SOBBHs could be within the reach of LISA detectability \cite{Sesana:2016ljz}. 
Similarly, binaries above the PISN gap could leave an additional background \cite{Mangiagli:2019sxg}. 

Finally, LISA could also complement other space-based detectors. In particular, there are several proposals such as Taiji \cite{Guo:2018npi} and TianQin \cite{Wang:2019tto} that could potentially fly at the same time as LISA. 
These additional detectors would help improve the cosmological inference~\cite{Wang:2021srv,Wang:2020dkc,Baral:2020mzs,Zhao:2019gyk}. 
Moreover,  any further detector in the deci-Hertz frequency band would be an excellent addition to LISA science~\cite{Sedda:2019uro}.

\subsection{Cross-correlation and interaction with large scale structure}
\label{sec:crosscorr}

GW maps of resolved events or SGWB can be cross-correlated with other large-scale-structure (LSS) tracers to perform a variety of astrophysical and cosmological measurements. Here we focus on the use of the SGWB measurements while technicalities about this signal are postponed to Sec.\ref{sec:SGWB}.

\subsubsection{Cross-correlations with resolved events}
The analysis of the cross-correlations between galaxy surveys and resolved GW events from compact object mergers has a rich scientific potential~\cite{Laguna:2009re}. From binary mergers with an EM counterpart, such an analysis can constrain DE and modified gravity models~\cite{Camera:2013xfa}. A similar analysis for sources as the first ones detected by LIGO/Virgo, yields constraints on the distance-redshift relation, the Hubble constant and other cosmological parameters~\cite{Oguri:2016dgk}. PBH scenarios as well as different astrophysical models can also be tested~\cite{Raccanelli:2016cud, Raccanelli:2016fmc, Scelfo:2018sny, Scelfo:2020jyw}.

The cross-correlation of the resolved GW sources with galaxies is a further  promising avenue. It enables the measurement of the redshift of the GW sources that do not have EM counterpart. 
This technique makes it possible to measure the value of Hubble constant, the matter density, the DE EoS and its redshift evolution from the GW sources. Along with probing the expansion history of the universe, it also provides evidence on the relation between the GW sources and DM distribution through the redshift-dependent galaxy bias parameter $b(z)$ \cite{Oguri:2016dgk, Raccanelli:2016cud, Scelfo:2018sny, Mukherjee:2018ebj, Mukherjee:2020hyn}.

The observable considered for such studies is the (3D) angular power spectrum:
\begin{equation}
    C_\ell^{\rm GW,\,LSS} (z_1,z_2) = \frac{2}{\pi} \int dk k^2 P(k) \, \Psi_\ell^{\textrm{GW}}(k) \Psi_\ell^{\textrm{LSS}}(k)\,.
\end{equation}
The two $\Psi_\ell^X(z)$ kernels encapsulate the physical processes in place:
\begin{align}
  \Psi_\ell^{X}(k) &= \int N^X(z) b^X(z) D(z) W(z) \Delta_\ell^X(k,z) \, dz \, ,
\end{align}
where $X$ stands for either GW or LSS when considering one or the other observable, and $W(z)$ are the observational window functions related to the experiment specifications.
The $\Delta_\ell (k,z)$ terms are the observed overdensities including effects from the intrinsic clustering, velocity effects (redshift-space distortions and Doppler), lensing and gravitational potentials, respectively, and they include information on galaxy clustering, gravity, and of course cosmological and astrophysical parameters:
\begin{equation}
\Delta_\ell(k,z) = \Delta^\mathrm{den}_\ell(k,z) + \Delta^{\mathrm{vel}}_\ell(k,z) +  \Delta^{\mathrm{len}}_\ell(k,z) +  \Delta^{\mathrm{gr}}_\ell(k,z) \, ;
\label{eq:relative contr}
\end{equation}
for the complete expression of those terms, see e.g.~Refs.~\cite{Bonvin:2011bg, Scelfo:2018sny}. In particular, this correlation can constrain DE and modified gravity models~\cite{Camera:2013xfa, Raccanelli:2016fmc}. In this case, the modifications due to different models of gravity and DE enter in the $\Psi_\ell$ kernels, and the advantage of using the GW-LSS cross-correlation comes from the complementarity with other measurements as well as the potentially higher redshift range for the GW bin.

The fact that GW events trace the LSS allows us to also test and constrain astrophysical models~\cite{Scelfo:2020jyw, Libanore:2020fim}. In this case, the change in the model $C_\ell$ will be in the merger rate, the redshift distribution and the bias of the compact objects' hosts.

Merging of BHs that are the endpoint of stellar evolution happens almost exclusively in galaxies that had sufficient star formation, and therefore in halos with a relatively large galaxy bias ($b_g>1$).
Conversely,  PBHs preferentially merge in lower biased objects, and thus have a lower cross-correlation with luminous galaxies~\cite{Bird:2016dcv}. 
Therefore, the cross-correlation of GW maps with galaxies, which can measure the bias of the BBH hosts, provides information on the abundance of PBHs \cite{Raccanelli:2016cud, Scelfo:2018sny} (for more details on PBHs, see Sec.~\ref{sec:PBH}).

Further correlations contain other useful information. The tomographic shear maps and the number density distribution of GW sources, combined with shear-shear and GW-GW auto-correlations, also constrain the cosmological parameters~\cite{Osato:2018mtm}. Thanks to GW-LSS cross-correlations with LISA and future galaxy survey data,   the detection of GW weak lensing seems viable~\cite{Mukherjee:2019wcg}. The cross-correlation between GW weak-lensing and CMB-lensing might also allow to test fundamental predictions of GR~\cite{Mukherjee:2019wfw}.

\subsubsection{Cross-correlations with the stochastic gravitational wave background}

A complementary study to the cross-correlation between GW resolved sources and LSS is the cross-correlation with the astrophysical SGWB. There are at least two astrophysical backgrounds that LISA will detect: one contribution generated by the galactic binary (GB) mergers, which is expected to dominate at low frequencies (up to $\sim 10^{-3} $Hz), and one coming from extragalactic BBH inspirals, expected to be relevant at larger frequencies ($\sim 10^{-3} - 10^{-2}$Hz). Several phenomena in the early universe also source a stochastic signal whose strength  is poorly predicted (see Sec.~\ref{sec:SGWB} and subsequent sections). The SGWB is then expected to be a combination of an astrophysical and a cosmological component, and a priori any of the two can dominate the SGWB signal in the LISA band. In order to be sensitive to both of them, it is of fundamental importance to find ways to disentangle the two signals. The spectral shape of each contribution is a standard tool to try to disentangle the components, however, due to the richness of sources expected in the LISA band, it is worth finding other ways to characterise the SGWB contributions. A promising approach consists in cross-correlating  the SGWB with matter distribution at late times. Since, as we will see, the  GW energy density depends on cosmological perturbations (besides astrophysical dependencies), it correlates with other cosmological probes. Some analyses and forecasts of the cross-correlation signals between GW observatories and future galaxy surveys, as e.g.~Euclid and SKA, are presented in Refs.~\cite{Contaldi:2016koz, Cusin:2017fwz, Cusin:2017mjm, Cusin:2018rsq,Jenkins:2018kxc,Jenkins:2018lvb, Jenkins:2018uac, Cusin:2019jpv,  Cusin:2019jhg, Jenkins:2019uzp, Jenkins:2019nks, Bertacca:2019fnt, Pitrou:2019rjz, Mukherjee:2019oma, Alonso:2020mva}). These cross-correlations not only can be useful to disentangle the origin of the SGWB, but represent completely new observables to infer cosmological information. For instance, along with the spatial fluctuation of the astrophysical SGWB, its temporal fluctuation provides a measurement of the high redshift merger rate of the astrophysical sources contributing to the SGWB~\cite{Mukherjee:2019oma}.

As for the case of resolved events, the observable quantity is the angular cross power spectrum, between the the galaxies overdensity and the energy density of the astrophysical SGWB, hereafter labelled AGWB:
\begin{equation}
C^{\rm AGWB\,\times\,LSS}_\ell = \frac{2}{\pi} \int k^2 dk P_\zeta(k) \Delta^{\rm AGWB}_\ell(k) \Delta^{\rm LSS}_\ell(k),
\end{equation}
where~$P_{\zeta}(k) = A_s \left(k/k_\mathrm{pivot}\right)^{n_s-1}$ is the scale-invariant curvature power spectrum, with $A_s$ and~$n_s$ the amplitude and tilt respectively, while~$k_\mathrm{pivot}$ is the pivot scale. The two transfer functions, $\Delta^{\rm AGWB}_\ell(k)$ and $\Delta^{\rm LSS}_\ell(k)$, contain astrophysical and cosmological information (see the previous section).
The astrophysical information can be included focusing on the anisotropies of the astrophysical SGWB energy density. The total GW energy density per logarithmic frequency  $f_{\rm o}$ and solid angle~$\Omega_o$ along the line-of-sight~${\bf n}$ of a SGWB is~\cite{Cusin:2017fwz, Bertacca:2019fnt}
\begin{equation}
\Omega^{\rm TOT}_{\rm GW} \left(f_{\rm o}, {\bf n}\right)= \frac{f_{\rm o}}{\rho_{{\rm c}}} \frac{ {\rm d} \rho^{\rm tot}_{\rm GW}}{ {\rm d} f_{\rm o} {\rm d} \Omega_{\rm o} } \, .
\end{equation}
It contains both a background (monopole) contribution in the observed frame ${\bar{\Omega}}_{\rm GW}(f_o)$, which is homogeneous and isotropic, and a direction-dependent contribution $\Omega_{\rm GW}(f_{\rm o}, {\bf n}  )$. Starting from these, we can define the total relative fluctuation as~$\Delta^\mathrm{TOT}_\mathrm{AGWB}(f_o, \hat{\mathbf{n}})= \left(\Omega^\mathrm{TOT}_\mathrm{AGWB}-\bar{\Omega}^\mathrm{TOT}_\mathrm{AGWB}\right)/\bar{\Omega}^\mathrm{TOT}_\mathrm{AGWB}$. 
The contributions to the astrophysical SGWB energy density fluctuation are computed in Refs.~\cite{Contaldi:2016koz, Cusin:2017fwz, Cusin:2017mjm, Cusin:2018rsq, Cusin:2018avf, Jenkins:2018kxc, Jenkins:2018lvb, Cusin:2019jpv,Cusin:2019jhg,Jenkins:2019uzp, Jenkins:2019nks, Bertacca:2019fnt, Pitrou:2019rjz, Alonso:2020mva}. Here, following Ref.~\cite{Bertacca:2019fnt}, we report its expression in the Poisson gauge
\begin{equation}
\begin{aligned}
&\Delta\Omega_{\rm AGWB}(f_o, \hat{\mathbf{n}}, \mathbf{\theta}) =   \sum_{[i]}  \int  dz \, W(z) \, \mathcal{F}^{[i]}(f_o, z, \mathbf{\theta})
\Bigg\{ b^{[i]} D  \\& + \left(b^{[i]}_\mathrm{evo} - 2 - \frac{\mathcal{H}'}{\mathcal{H}^2}\right) \hat{\mathbf{n}}\cdot\mathbf{V} - \frac{1}{\mathcal{H}}\partial_\parallel(\hat{\mathbf{n}}\cdot\mathbf{V}) - (b^{[i]}_\mathrm{evo}-3) \mathcal{H} V  \\
&\quad + \left(3 - b^{[i]}_\mathrm{evo} + \frac{\mathcal{H}'}{\mathcal{H}^2}\right)\Psi + \frac{1}{\mathcal{H}}\Phi' + \left(2 - b^{[i]}_\mathrm{evo} + \frac{\mathcal{H}'}{\mathcal{H}^2}\right) \int_0^{\chi(z)} d\tilde{\chi} \left(\Phi'+\Psi'\right)  \\
&\quad + \left(b^{[i]}_\mathrm{evo} - 2 - \frac{\mathcal{H}'}{\mathcal{H}^2}\right) \left( \Psi_{o} - \mathcal{H}_0 \int_{0}^{\tau_0} d\tau \left.\frac{\Psi(\tau)}{1+z(\tau)}\right|_{o} -  \left(\hat{\mathbf{n}}\cdot\mathbf{V} \right)_{o} \right) \Bigg\}\,,
\end{aligned}
\label{eq:fluctuation_poissongauge}
\end{equation}
where the \textit{density}, \textit{velocity}, \textit{gravity} and \textit{observer} terms, in the first, second, third and fourth line, respectively contain all the cosmological information. On the other hand, the function $\mathcal{F}^{[i]}(f_o, z, \mathbf{\theta})$ contains all the astrophysical dependencies: e.g.~the mass and spin distribution of the binary, the emitted GW energy spectrum, the clustering properties of GW events and the details of the GW detectors. The functions $b^{[i]}(z,\mathbf{\theta})$ and~$b^{[i]}_\mathrm{evo}(z,\mathbf{\theta})$
are the bias and the evolution bias of the $i$-th type of GW source, which specify the clustering properties of GW sources and characterise the formation of sources.

The cross-correlation analysis of the astrophysical SGWB (from sources at all redshifts along the line of sight) with galaxy number counts at a given redshift leads to a tomographic reconstruction of the redshift distribution of the sources ~\cite{Cusin:2018rsq, Mukherjee:2019oma, Canas-Herrera:2019npr, Alonso:2020mva, Cusin:2019jpv,Yang:2020usq}. Subtleties about the noise and other characteristics of the detector play an important role, so that cross-checks on the detector performances are possible by means of this analysis \cite{Bertacca:2019fnt, Alonso:2020mva}.
Overall, the cross-correlation analysis shows that the combination of galaxy surveys with the astrophysical SGWB can be a powerful probe for GW physics and  multi-messenger cosmology.

\subsubsection{Large-scale structure effects on gravitational-wave luminosity distance estimates}

Here we discuss the effect of cosmological perturbations and inhomogeneities on estimates of the luminosity distance of compact object binary mergers through GWs.
It is important to account for such effects on GW propagation to obtain robust measurements for precision cosmography.

The main attempts to investigate perturbation effects on GWs involve  the integrated Sachs-Wolfe effect~\cite{Laguna:2009re}, peculiar velocities or accelerations~\cite{Kocsis:2005vv, Bonvin:2016qxr,Tamanini:2019usx, Mukherjee:2019qmm, Nicolaou:2019cip}, lensing~\cite{Pyne:1995iy, Wang:1996as, Takahashi:2003ix, Sathyaprakash:2009xt, Dai:2016igl, Takahashi:2016jom, Baker:2016reh, Dai:2017huk, Haris:2018vmn, Dai:2018enj, Oguri:2018muv, Contigiani:2020yyc, Ezquiaga:2020dao, Ezquiaga:2020spg} and 
environmental effects~\cite{Barausse:2014tra}.
Coherent peculiar velocity of the binaries at low redshift and weak gravitational lensing by intervening inhomogeneities can affect the identification of the hosts' redshift. This introduces changes of typically a few percent (but occasionally much larger) in the flux, while not significantly affecting the redshift, and thus provides a source of noise in the $D_L(z)-z$ relation~\cite{Sathyaprakash:2009xt, Hirata:2010ba}.
Using the local wave zone approximation to define the tetrads at source position~\cite{Maggiore:1900zz},  the corrections to the luminosity distance read~\cite{Bonvin:2016qxr, Bertacca:2017vod}\footnote{For simplicity,  we have dropped all  contributions evaluated at the observer, assuming concordance background model, and work in Poisson Gauge.}
 \begin{eqnarray}
\label{DL_2}
\frac{ \Delta D_L}{\bar D_L} &=& \left(1-\frac{1}{{\cal H} \bar \chi} \right)  v_\| -  \int_0^{\bar \chi} d \tilde \chi \, \frac{ \left(\bar \chi-\tilde \chi\right)}{\tilde \chi \bar \chi} \,  \triangle_\Omega  \Phi  + \nonumber\\
&+&\frac{1}{{\cal H} \bar \chi} \Phi - 2\left(1-\frac{1}{{\cal H} \bar \chi} \right) \int_0^{\bar \chi} d\tilde \chi\, \Phi' - 2\Phi + \frac{2}{\bar \chi} \int_0^{\bar \chi} d\tilde \chi \, \Phi \; ,
\end{eqnarray} 
where prime denotes the derivative with respect to $\eta$ and ${\cal H} = a'/a$, 
$ \triangle_\Omega\equiv\bar \chi^2 \bar \nabla^2_\perp = \bar \chi^2 (\bar \nabla^2
- \bar \partial_\parallel^2 - 2 {\bar \chi}^{-1} \bar \partial_\parallel)= (\cot \partial_\theta + \partial_\theta^2 +\partial_\varphi/\sin^2\theta)$.
We can recognise in~\eq{DL_2} the presence of a velocity term (the first term), followed by a lensing contribution, and the final four terms account for the 
Sachs-Wolfe, Integrated
Sachs-Wolfe, volume and Shapiro time-delay effects.

The additional $D_L$ uncertainty due to the inclusion of perturbations is expected to  peak at low-$z$ due to velocity contributions; however, velocity effects rapidly decrease and lensing takes over~\cite{Bertacca:2017vod}. Those results indicate that the amplitude of the corrections could be important for future interferometers such as LISA.

In presence of DE or modifications of gravity, the GW luminosity distance might differ from the EM signals also for large-scale fluctuations~\cite{Garoffolo:2019mna, Dalang:2019rke, Jana:2020vov, Dalang:2020eaj}. In particular, linearised fluctuations of the GW luminosity distance contain contributions directly proportional to the clustering of the DE field~\cite{Garoffolo:2019mna, Garoffolo:2020vtd} and by combining luminosity distance measurements from GW and supernovae sources, it is possible to uncover field inhomogeneities detecting them directly. See e.g.~Ref.~\cite{VanDenBroeck:2010fp} for an analysis of weak lensing effects in the measurement of the DE EoS with LISA.

\newpage
%%%%%%%%%%%%%%%%%%%%%%%%%%%%%%%%%
% Here Sec. 3 starts

\section{Gravitational lensing of gravitational wave signals}
\label{sec:gw_lensing}

\small \emph{Section coordinators: D.~Bacon, M.~Zumalacarregui. 
Contributors: D.~Bacon, G.~Congedo, G.~Cusin, J.M.~Ezquiaga, S.~Mukherjee, M.~Zumalacarregui. } \\

\subsection{Introduction}
Similarly to EM radiation, GWs feel the gravitational potential of both massive objects and also the LSS while traveling across the universe. This opens the possibility of both probing the distribution of structure in the cosmos and the fundamentals of the underlying gravitational interactions. LISA will offer a unique perspective since it will detect high-redshift GWs in a lower frequency band compared to ground-based detectors, increasing the lensing probabilities and the detectability of diffraction effects.
 
GW lensing phenomena are characterized by two properties: the convergence $\kappa$\footnote{Because GW sources are effectively point-like, no image distortions are observable and shear influences GW observations only through its effect on the magnification. This is very different from lensing of galaxies, where shear distortions are directly observable.}
(governing whether there is weak or strong gravitational lensing) and a dimensionless frequency $w$ (governing whether wave optics or geometric optics is relevant). The convergence, $\kappa$, is defined as the integral along the line of sight of the redshift-weighted second derivative of the Newtonian potential.
Depending on the strength of the gravitational potential, two gravitational lensing regimes exist: weak ($\kappa \ll 1$) or strong ($\kappa\simeq 1$). In the first case, the main observable effect is magnification of the observed flux (or equivalently a change in the inferred luminosity distance). In the second case the effect can be both a magnification/demagnification and also production of multiple signals with a time delay that is a function of the lens properties. We will examine both regimes in Secs.~\ref{sec:lensing_weak} and \ref{sec:lensing_strong} below.

The low frequency and phase coherence of GWs allows the observation of wave effects. It is convenient to define the dimensionless frequency
\begin{equation} \label{eq:lensing_freq_dimensionless}
 w = 8\pi G M_L f = 4\pi \left(\frac{M_L}{10^8 M_\odot}\right)\left(\frac{f}{\rm{mHz}}\right) \;, 
\end{equation}
as well as a magnification factor $F(w) = h(w)_\text{lensed}/h(w)_\text{unlensed}$, where $M_L$ is the redshifted lens mass. GW lensing is accurately described by geometric optics only if $w\gg 1$ (more precisely, $w\Delta \tilde t_j\gg 1$ for all images). Then, the amplification factor (the ratio between the lensed and unlensed waveform) is a sum over multiple images $j$:
\begin{equation}\label{eq:lensing_geometric_optics}
F(w) = \sum_j |\mu_j|^{1/2}\exp(i w \Delta \tilde t_j - i\pi n_j)\,,
\end{equation}
where $\mu_j$ is the magnification, $\Delta \tilde t_j$ is the time delay (in units of $4GM_L$) evaluated on the $j$-th image position. Here $n_j$ is the Morse phase of the image, respectively $0$, \sfrac{1}{2}, 1  for minima, saddle point and maxima of the time delay function \cite{Schneider:1992,Takahashi:2003ix}. These image types are also known as type I, II and III respectively.

For low ($w$ considerably less than 1) or intermediate ($w\sim1$) frequencies it is necessary to consider the wave optics regime, where the amplification factor is 
\begin{equation} \label{eq:lensing_wave optics}
    F(w) = \frac{w}{2\pi i}\int dx^2 \exp(i w \Delta \tilde t(\vec x, \vec b))\,,
\end{equation}
where the time delay $\Delta \tilde t(\vec x, \vec b)$ is now a function of the (normalized) lens plane coordinate $\vec x$ and source location $\vec b$. (See Ref.~\cite{Takahashi:2003ix} for details). The low frequency limit $w\ll 1$ corresponds to $F\sim 1$, or no magnification. This can be understood as a wave not being sensitive to an object whose effective size is smaller than its wavelength. The high frequency limit corresponds to the geometric optics result Eq.~(\ref{eq:lensing_geometric_optics}), as can be obtained from a Gaussian expansion around the images, which correspond to the extrema of the time delay $\delta\tilde t$ (sub-dominant corrections to geometric optics can be computed \cite{Takahashi:2004mc}). We will discuss wave effects in gravitational lensing and LISA opportunities in Sec.~\ref{sec:lensing_wave}.

\subsection{Weak lensing }\label{sec:lensing_weak}

The first regime that GWs will undergo is weak lensing. Because of the relatively deep observations LISA will be able to achieve (median $z\sim2$, reaching out much deeper depending on source types) most of the observed events (if not all) will be subjected to lensing by the large scale structure.
In recent years weak lensing has become one of the primary tools in cosmology to study the distribution of DM and the nature of DE (through the evolution with redshift).
Galaxy lensing surveys such as KiDS \cite{Heymans:2020gsg} and DES \cite{Troxel:2017xyo} are now providing tighter constraints on cosmological parameters, such as geometry parameters ($H_0$, $\Omega_m$, and $\Omega_\Lambda$) and matter clustering parameters ($\sigma_8$ and $n_S$) to a few percent level.
Soon Euclid \cite{Blanchard:2019oqi} and the Vera C. Rubin Observatory \cite{Abell:2009aa} will further constrain those parameters, extending the analysis to DE ($w_0$ and $w_a$) and various models beyond $\Lambda$CDM, down to the percent level.
Most likely, the study of systematic errors affecting those measurements will be the topic of the next decade or so.
GWs have the potential to revolutionize the field with virtually bias-free measurements of the luminosity distance.
In fact, LISA will give us the first deep luminosity distance measurements that will be both accurate (as this relies on assuming GR is the correct theory) and also precise with typical errors up to a few percent depending on source type and position in the sky.
At the same time, the typical root-mean-squared  error due to lensing is 0.02 for $z<1$ and ramping up quite rapidly with redshift~\cite{Cusin:2020ezb}.
This will be a source of unwanted bias and extra noise in the Hubble diagram inference, which must be accounted for in the analysis. However, if modelling of the lensing is introduced, then this systematic error can become a new piece of useful information.
For these reasons LISA will also establish the groundwork for planned second generation space-based detectors.

\subsubsection{As a source of noise and bias for standard sirens}

Magnified sources are easier to detect than de-magnified sources. This simple fact biases the distribution of lensing magnifications of an observed source sample. Lensing selection effects are often neglected when estimating the lensing-induced uncertainty on the cosmological distance measurement from high-redshift GW sources. However, when selection effects are included, flux conservation is no longer enforced, and the  mean  of  the magnification  distribution is shifted  from 1 for sufficiently high-redshift sources. This introduces an irreducible (multiplicative) bias on the distance reconstruction, independent of the sample size. In Ref.~\cite{Cusin:2020ezb} the effect of selection bias on a population of MBBHs is examined, for different scenarios of MBH formation. It is shown that, while the effect of the bias on the distance estimator for sources with an EM counterpart is typically below the variance threshold, it becomes relevant for high redshift sources when the statistics of detectable events is large (e.g.~when MBHs form at high redshift from remnants of Population III stars), and it should be taken into account in population studies. 

\subsubsection{As a probe complementary to standard sirens}

Thanks to the variety and large number of sources, LISA will be able to probe a wide range of redshifts with sufficient statistical significance to constrain cosmological parameters.
Given that LISA will be able to measure luminosity distances to within a few percent (with optimistic estimates pushing this limit down to one percent) and the vast majority of its sources (e.g.~MBBHs, IMBBHs and EMRIs) at a median redshift of about 2, weak lensing by the LSS will be expected to dominate the error budget of any luminosity distance measure. This has to be taken into account in geometric analyses, such as the luminosity distance-redshift relationship as illustrated in the previous sections.
Here we  explore an avenue to extract additional cosmological information: it is well known that the weak lensing signal for background objects is correlated with foreground large scale structure. Foreground over- (under-) densities will magnify (demagnify) the observed luminosity distance to background sources by a small factor. This can be expressed in terms of the convergence via
\begin{equation}
D_L'=(1-\kappa) D_L~,
\end{equation}
where $D_L$ is the (unknown) unlensed distance, $\kappa$ is the projected 2D convergence, and $D_L'$ is the (observed) lensed distance.
The observed luminosity distance now incorporates information about the convergence field, which in turn is cosmology dependent.
In fact its power spectrum has cosmological information imprinted into it that can be extracted from data. It was long ago speculated that this signal could be exploited if large enough statistics (very high number densities and very precise distance measurements) were available in the future with second generation space detectors such as BBO \cite{Cutler:2009qv}.
This approach was then applied to third-generation ground based detectors and deci-Hz space detectors to prove that the method would in principle be applicable to DE and modified gravity in a tomographic analysis \cite{Camera:2013xfa}.
More recently, it was shown that a joint analysis of lensing and distance-redshift on a combined dataset of LISA and third generation ground detectors would allow us to jointly constrain geometry parameters (such as $H_0$ and density) and clustering parameters ($\sigma_8$ and $n_s$) at the same time with percent precision ($\Lambda$CDM with curvature, nominal distance errors, and nominal number density)~\cite{Congedo:2018wfn}. This is because a geometry probe such as distance-redshift can act as a strong prior to break degeneracies that are usually present in lensing. 
LISA will be the first detector to have good enough statistics to prove that such analysis is indeed possible, bringing in independent constraints on additional cosmological parameters such as $\sigma_8$. In doing this, LISA will pave the way for a combined analysis with other third-generation ground-based detectors such as ET, and also in the future for new purposely-built detectors operating in other bands.
(One such proposal is a deci-Hz detector currently under evaluation by ESA \cite{Baker:2019ync}).
Not only will LISA provide the first independent measurement of weak lensing, but it will open the field of statistical cosmology to GWs which could dominate cosmology in the decades to come.

\subsection{Strong lensing}\label{sec:lensing_strong}

The probability that a GW is lensed depends on the redshift of the source and the distribution of lenses. The merger rate at high redshift depends on the astrophysical formation channels which can affect the event rate of well detected events and the sub-threshold lensed events observed by LIGO/Virgo \cite{Mukherjee:2021qam} or, in future, by CE and ET \cite{
Piorkowska:2013eww, Biesiada:2014kwa, Ding:2015uha}. One of the robust ways to predict the lensing event rate is by using the detection or an upper bound on the amplitude of the SGWB \cite{Mukherjee:2020tvr, Buscicchio:2020cij}. For high-redshift GWs, the intervening matter between the source and the observer magnifies the GW signal, thereby introducing a systematic error in the luminosity distance determination.   Correcting for this weak lensing effect is relevant when inferring cosmological parameters from GW standard sirens~\cite{Holz:2002cn,Hirata:2010ba}; see Sec.~\ref{sec:lensing_weak}. 
But a  fraction of these GW events pass close enough to the lens so that lensing  produces multiple images of the same event. 
LISA could detect a few strongly-lensed massive BBH during its mission \cite{Sereno:2010dr}. 
The detection of multiple lensing events could provide new means for cosmography with LISA \cite{Sereno:2011ty}. 

In the strong lensing regime the time delay between the images increases linearly with the mass of the lens, reaching delays of months for a $10^{12}M_\odot$ galaxy lens. Each of the lensed images has a different magnification, but they also acquire a fixed phase shift depending on how many times the image has crossed a lens caustic \cite{Schneider:1992}; recall the $i\pi n_j$ term in Eq.~(\ref{eq:lensing_geometric_optics}). 
The origin of this phase shift can be also understood from the folding of the wavefront, which in strong-lensing configurations produces a phase of the waveform associated with the properties of the lensing potential \cite{Schneider:1992,Dai:2017huk}. 

In the case of minima of the lensing potential (type I images), there is no phase shift and the lensed waveform reproduces exactly (modulo magnification) the shape of the emitted signal. For maxima (type III), there is an overall change of the sign, $e^{-i\pi}=-1$, but the morphology of the signal remains intact. 
However, for saddle points (type II), the associated phase shift can distort the waveform, since the frequency-independent phase shift will introduce a frequency-dependent time delay that will differentially transform the GW signal \cite{Ezquiaga:2020gdt}. 
For non-precessing, quasi-circular orbits of equal mass binaries the emitted GWs are dominated by a single quadrupole radiation mode (22). In that case, the lensing phase shift accounts for a $\pi/4$ shift of the coalescence phase \cite{Dai:2017huk,Ezquiaga:2020gdt}. 
Nonetheless, whenever higher modes, precession or eccentricity are relevant, the lensed waveform is modified so that it does not conform with expectation from (unlensed) templates in GR \cite{Ezquiaga:2020gdt}. 
This sets up the possibility that type II images near the detection threshold might be missed with standard template bank searches, but also offers the opportunity of identifying strongly lensed events with a single type II image if the SNR is high enough. 
Explorations of the capabilities of third generation ground-based detectors to identify type II images are performed in Ref.~\cite{Wang:2021kzt}.

Precise measurement of the phase of long duration signals by LISA could be key in distinguishing different types of lensed images \cite{Ezquiaga:2020gdt}.
The observation of very high SNR MBBHs and sources with very asymmetric masses, such as IMRIs and EMRIs, could give LISA a unique opportunity to identify strongly lensed events, even when detecting a single image, i.e.~one term in Eq.~(\ref{eq:lensing_geometric_optics}), if the other images are received outside of the observing window or are too faint to be recovered. 

Another very strong lensing effect can exist for LISA MBBH sources. If we observe these in EM during the inspiral, we expect
self-lensing flares to occur when the two BHs are aligned with the
line-of-sight (to within about an Einstein radius). In a recent pair
of papers, \cite{Davelaar:2021eoi, Davelaar:2021gxx}, a binary emission model was ray-traced, and a
distinct feature was found in the light curve imprinted by the BH shadow from
the lensed BH.  A dip occurs in the lightcurve when the
foreground BH lenses the shadowed part of the background
BH. This could make it possible to extract BH shadows that are
spatially unresolvable by high-resolution very-long baseline interferometry.  These shadows are
another probe of the metric around the MBH.

\subsection{Wave effects} \label{sec:lensing_wave}

\begin{figure}
    \centering
    \includegraphics[width=0.5\textwidth]{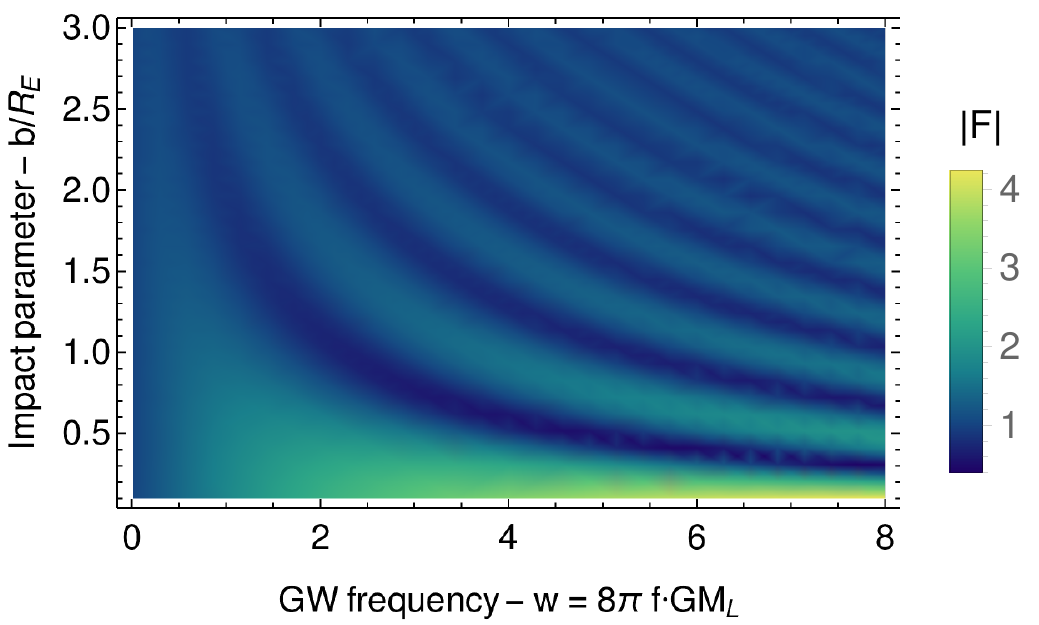} \hfill
    \includegraphics[width=0.45\textwidth]{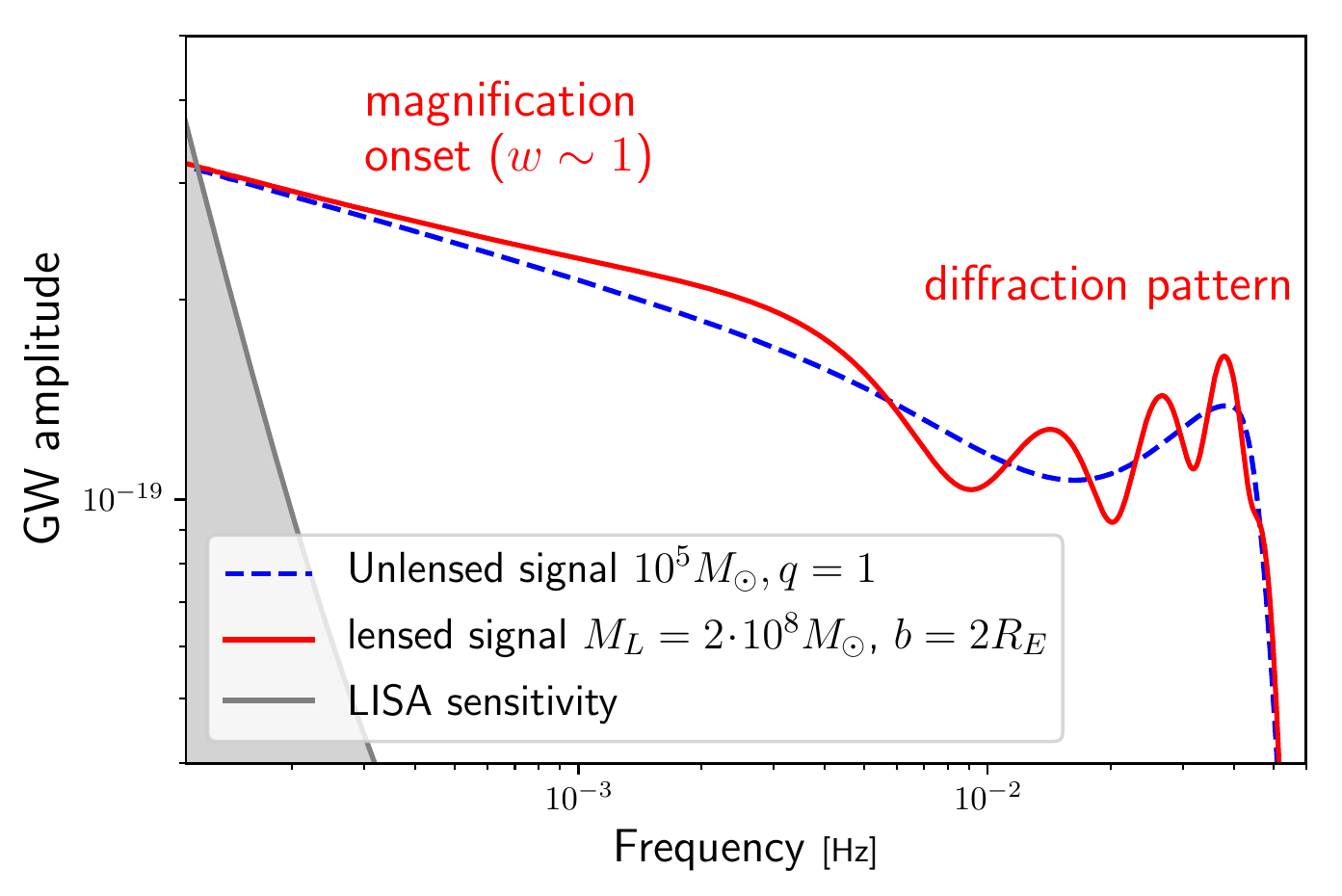}
    \caption{\small GW lensing and wave effects. 
    Left panel: Amplification $|F(w)|$ caused by a point lens in the wave optics regime. 
    Low frequency signals $w\ll 1$ undergo no magnification. An oscillatory pattern emerges at higher frequencies.
    Right panel: Imprint of strong lensing on a MBBH source, as observed by LISA. The frequency dependence (onset of magnification, diffraction pattern) carries information about the gravitational lens.
    \label{fig:lensing_wave_effects}}
\end{figure}

GWs can be emitted at low frequencies ($w\lesssim1$), allowing the observation of wave diffractive phenomena.
For typical LISA sources, wave optics as in Eq.~(\ref{eq:lensing_wave optics}) needs to be considered 
for lenses with masses $M_L\sim 10^6-10^9 M_\odot$, cf.~Eq.~(\ref{eq:lensing_freq_dimensionless}). In contrast, wave effects in gravitational lensing are not observed for EM sources, due to the finite size of these sources and their high EM frequencies ($w\gg1$), cf.~Figs.~12, 13 of Ref.~\cite{Oguri:2019fix}.
 
Wave effects produce a frequency-dependent amplification factor $F$. At very low frequencies $w\ll 1$ no magnification occurs, $F\sim 1$, as a wave is not affected by an object smaller than its wavelength. As the frequency increases, an oscillatory pattern emerges, caused by the superposition of different images. This pattern persists for GWs even in the geometric optics limit  $w\gg 1$ through the exponents in Eq.~(\ref{eq:lensing_geometric_optics}). We do not see this pattern for optical EM sources due to incoherent emission and lack of time resolution.
Note that diffraction prevents the magnification for GWs from diverging, even for perfectly aligned, symmetric lenses with normalized impact parameter $|\vec b|/R_E\to 0$. 
An inspiralling binary sweeps through a range of frequencies before coalescence; a gravitational lens imprints the frequency-dependent amplification factor into the waveform. These effects are shown in Fig.~\ref{fig:lensing_wave_effects}.

Several studies address the detectability of wave effects by LISA. 
Ref.~\cite{Sereno:2010dr} estimates that a handful of strongly lensed events ($\lesssim 5$) will be observed in a 5 year mission. 
Because the lensing probabilities are dominated by galactic halos with $M_L \sim 10^{12}M_\odot$, wave effects are likely to be unimportant for most strongly lensed LISA sources.
However, frequency-dependent diffraction effects in the weak lensing regime may be detectable for a small fraction of sources: Ref.~\cite{Gao:2021sxw} find that $0.1 - 1.6\%$ of LISA's MBBH with total mass $10^5 - 10^{6.5}M_\odot$ and redshift $4 - 10$ may present detectable wave-optics.
LISA provides new opportunities to observe these frequency-dependent effects and use them to probe the large-scale structure of the universe at the scale of sub-halos. 

Wave effects can help identify GW signals (or groups of them) as lensed events and characterize gravitational lenses. We  discuss both possibilities below. 

\subsubsection{Identification of lensed signals}

Diffraction-induced features in the gravitational waveform could be used to identify the event as lensed.  The main target for ground detectors is the first diffraction peak, appearing at low frequencies \cite{Dai:2018enj}. These diffraction effects can be detected even for impact parameters comparable to the Einstein radius $b\gtrsim R_E$, for which there is only one image in the geometric optics limit.
Orbital parameters are expected to have some degeneracies with diffraction, but it is in principle possible to distinguish them. Spin precession creates a modulation that affects the amplitude more than the phase (while diffraction affects both equally). Orbital eccentricity can be distinguished by higher harmonics in GWs, which are not induced by gravitational lensing.

Wave effects may produce systematic errors in different analyses. Diffraction-induced distortions make the GW phase appear to arrive earlier than an unlensed EM signal \cite{Takahashi:2016jom, Morita:2019sau}, although this is only an apparent superluminality \cite{Suyama:2020lbf,Ezquiaga:2020spg}. This effect has to be taken into account \cite{Ezquiaga:2020spg} when, for example, inferring constraints on the speed of gravity from the possible pre-merger modulated EM brightness of a MBB~\cite{Haiman:2017szj}.

The interplay between wave optics and gravitational polarizations may provide alternative means to characterize lensed GWs. This has been investigated using low-frequency corrections to geometric optics. These corrections cause the polarization plane defined in geometric optics to become blurred due to diffraction effects, which leads to the rise of apparent vector and scalar polarization modes \cite{Cusin:2019rmt}. 
Moreover, GWs in curved space propagate on different geodesics depending on their circular polarization (an effect known as the gravitational spin hall effect) \cite{Yamamoto:2017gla,Andersson:2020gsj}. 
These effects might provide means to characterize lensed GWs. As they are suppressed by factors of the GW frequency, these phenomena are more important for LISA than for ground-based detectors.

\subsubsection{Reconstruction of large scale structure and lens properties}

GW diffraction is sensitive to the power spectrum of matter density fluctuations on very small physical scales \cite{Oguri:2020ldf}. 
Ground detectors are sensitive to this power spectrum at wavenumbers where baryonic effects are smallest,  $k\sim 10^6 h\text{Mpc}^{-1}$ for $f\sim 0.1-1$Hz.
Individual high SNR sources at high redshift, such as LISA observed MBBHs, may be sensitive to frequency-dependent modulations of the GW phase and amplitude (SNR $\gtrsim 500$ at mHz frequency, $z_S\sim 3$, cf.~Fig.~12 of \cite{Oguri:2020ldf}). 
Lensing induced diffraction can be enhanced substantially in non-standard scenarios that affect the small-scale power spectrum (e.g.~PBHs).

Wave effects can then improve the chances of detection and provide additional information about the lens.
Inspiralling binaries detectable by LISA sweep through a range of frequencies that might include the onset of diffraction for $w\sim 1$ and/or the characteristic diffraction pattern for $w>1$; see Eq.~(\ref{eq:lensing_freq_dimensionless}). 
For the case of LISA MBBHs lensed by point masses and singular isothermal spheres, diffraction allows for a determination of lens mass and source position at the level of $10\%$ to $\sim 0.1\%$ for $M_{L}\lesssim 10^7 M_\odot$ and $M_{L}\gtrsim 10^8 M_\odot$ for SNR $\sim 1000$ \cite{Takahashi:2003ix}. Point lenses can be identified for large impact parameters $b/R_E\sim \mathcal{O}(10)$ (rather than $\sim 1$), increasing significantly the likelihood of detection $\propto b^2$.
Diffraction is also sensitive to cuspy (or singular) matter distributions, producing an image with magnification $\mu\sim (GMf)^{\alpha-3}$ for a profile with central density $\rho\sim r^{-\alpha}$ ($0<\alpha\leq 2$) \cite{Takahashi:2004mc}.
Frequency-dependent wave effects can be used to individually detect DM subhalos with $M\lesssim 10^7 M_\odot$ and even measure their mass profiles, although LISA might not provide enough events with sufficiently high SNR \cite{Choi:2021jqn}.

Investigations of microlensing (the effect of small-scale structure, such as stars, on the images produced by a large lens) have focused on LIGO-Virgo sources. Microlenses have masses such that wave diffraction effects are important at LISA frequencies, and would normally not leave an observable imprint (magnification $\sim$1). However, they can be observed if they are present in regions of high magnification of the macrolens, such as caustics \cite{Diego:2019lcd}. Nevertheless, in most situations the macrolens model will be a sufficient description \cite{Cheung:2020okf}. 
GW microlensing is also sensitive to the stellar mass function, particularly in regions of high magnification \cite{Mishra:2021xzz}.

\newpage
%%%%%%%%%%%%%%%%%%%%%%%%%%%%%%%%%
% Here Sec. 4 starts

\section{Constraints on modified gravity theories}
\label{sec:modified_gravity}

 \small \emph{Section coordinator: D.~Vernieri. Contributors: T.~Baker, E.~Belgacem, G.~Calcagni, M.~Crisostomi, N.~Frusciante, K.~ Koyama, L.~Lombriser, M.~Maggiore, S.~Mukherjee,  M.~Sakellariadou, I.~D.~Saltas, D.~Vernieri, F.~Vernizzi, M.~Zumalacarregui.} \normalsize

\subsection{Models and theories}\label{sec:models}

To test GR  using the propagation of GWs, it is useful to  compare it to other models that predict observational deviations. The simplest case that one can consider 
is the  presence of an additional scalar interaction, e.g.~a so-called scalar-tensor theory of gravity. 
Over the years, increasingly more general theories devoid of Ostrogradsky instabilities~\cite{Ostrogradksi} have been introduced with the aim to account for the most general modification in this setup.
The advent of the Galileon field theory~\cite{Nicolis:2008in} quickly led to its covariant generalization~\cite{Deffayet:2011gz} and so to the reappraisal of the Horndeski theory~\cite{Horndeski:1974wa}, i.e.~the most general scalar-tensor theory with second order field equations. Relaxing this last assumption in a suitable way, has resulted in the
beyond-Horndeski~\cite{Zumalacarregui:2013pma} or Gleyzes--Langlois--Piazza--Vernizzi model~\cite{Gleyzes:2014dya,Gleyzes:2014qga}, and subsequently
degenerate higher-order scalar-tensor theories~\cite{Langlois:2015cwa,Crisostomi:2016czh,BenAchour:2016fzp}.   

At low energy, those relevant for cosmology, the Lagrangian of these theories schematically reads~\cite{Pirtskhalava:2015nla}
\begin{equation}
\label{lagrWBG}
{\cal L} = \sum_{n=0}^{\infty}  \sum_{m=0}^{3} c_{n,m} (\phi) \Lambda_2^4 \left( \frac{( \partial \phi)^2}{\Lambda_2^4} \right)^n \left(  \frac{\square \phi}{\Lambda_3^3} \right)^m + \ldots\;,
\end{equation} 
where $c_{n,m} (\phi)$ are dimensionless coefficients that depend mildly on $\phi$, $\Lambda_2$ corresponds to the spontaneous  Lorentz breaking scale, $\Lambda_3$  is related to the ultraviolet (UV) cut-off of these scalar-field theories while the ellipses refer to suitable gravitational terms necessary to keep the equations of motion of the propagating degrees of freedom of second order. 
Higher derivative terms with $m>0$ have been introduced in cosmology to  explain the accelerated expansion without a cosmological constant and simultaneously pass Solar System tests~\cite{Luty:2003vm,Nicolis:2008in} using the so-called Vainshtein screening mechanism~\cite{Vainshtein:1972sx,Babichev:2013usa}. 
In this case, the two scales must be chosen as $\Lambda_2 \sim (M_{\rm Pl} H_0)^{1/2}$ and $\Lambda_3 \sim (M_{\rm Pl} H_0^2)^{1/3}$, where $M_{\rm Pl}$ is the Planck mass and $H_0$ the value of the current Hubble constant.

The above models represent low energy effective field theory  (EFT) whose cut-off scale can be chosen according to the certain type of phenomenology they aim to describe, e.g.~DE.
An alternative path to follow is to work out cosmological phenomena in a top-down way~\cite{Calcagni:2020edt} starting from a fundamental theory of gravitational and matter interactions, such as quantum gravity.

Quantum gravity is a generic name applied to any theory where the gravitational interaction is consistent with the laws of quantum mechanics. Exploring the cosmological implications of these theories is important to assess how LISA can contribute to our knowledge of fundamental physics.

The corrections to cosmological dynamics and evolution of the universe arising in quantum-gravity scenarios are expected to be very small since they dominate at Planck scales $\ell_{\rm Pl}$ or high curvature, compared to horizon scales $H^{-1}_0\sim 10^{60}\ell_{\rm Pl}$ or low curvature. However, this may hold true for perturbative curvature corrections to GR, while other mechanisms may enhance deviations from classical gravity either by inflation of metric fluctuations to cosmic scales at early times or by non-perturbative cumulative effects on the propagation of GWs. The former mechanism will be explored in Sec.~\ref{sec:sgwbqg}, while the latter will be developed in this section in parallel with the other models. In particular, the dimension of spacetime in quantum gravity changes with the scale probed and this can affect both the dispersion relation of GWs and the luminosity distance of GW astrophysical sources~\cite{Calcagni:2019kzo,Calcagni:2019ngc}.

 Quantum effects can also manifest themselves as non-local terms in the quantum effective action, and these can affect the long-distance dynamics of the theory. In particular, assuming  a non-local term that corresponds to a dynamical generation of a mass for the conformal mode leads to a predictive model that passes all current cosmological constraints~\cite{Maggiore:2013mea,Belgacem:2017cqo,Belgacem:2020pdz} and predicts modified GW propagation. In this model the effect can be quite large, leading, at the redshifts accessible to LISA, to deviations from GR that can be as large as $80\%$~\cite{Belgacem:2019lwx,Belgacem:2020pdz}.

The gravitational interaction can be modified by considering a Lagrangian constructed with a general function of the scalar associated with non-metricity $Q$, i.e.~$f(Q)$~\cite{Nester:1998mp,Jimenez:2019ovq,Bajardi:2020fxh}. In this theory  at least two additional scalar modes are introduced with respect to GR. The equivalent of GR is obtained when the Lagrangian coincides with the non-metricity scalar~\cite{BeltranJimenez:2019tjy}.  The $f(Q)$ theory modifies the propagation equation of GWs due to an additional scale-independent  friction term which is related to a time-dependent effective Planck mass~\cite{BeltranJimenez:2017tkd}. As a consequence,  the  GW luminosity distance, $D_L^{GW}$, differs from the standard EM  luminosity distance, $D_L^{em}$~\cite{Frusciante:2021sio}.  For a specific model it has been shown~\cite{Frusciante:2021sio} that at the redshifts relevant for LISA, the deviations with respect to the standard cosmological model can be relevant depending on the value of the free parameter and that GW detectors such as LISA show a strong power in constraining it.  

\subsection{General expression for modified cosmological gravitational wave propagation } \label{sec:twelve}

%%%%%%%%%%%%%%%
\subsubsection{Modified cosmological gravitational wave propagation}
%%%%%%%%%%%%%%%

The GW propagation over cosmological distances for generalised theories of gravity and other extensions of $\Lambda$CDM can be parametrised by the wave equation~\cite{Saltas:2014dha,Gleyzes:2014rba,Lombriser:2015sxa,Nishizawa:2017nef,Belgacem:2017ihm,Ezquiaga:2018btd,Lombriser:2018guo,Belgacem:2019pkk}
\begin{equation} \label{gwprop_eq:gw}
 h''_A+ 2\left[1-\delta(\eta,k) \right]{\cal H}\, h'_A+\left[ c_T^2(\eta,k)\,{k^2}+m_T^2(\eta,k) \right] h_A\,=\,\Pi_A
\,,
\end{equation}
where the primes indicate derivatives with respect to conformal time $\eta$. Here, we assume that the GW propagates far away from the source in the form of a plane wave. 
The modified friction ($\delta$), speed ($c_T^2$), mass ($m_T^2$) and the source ($\Pi_A$) are in general assumed to be functions of both time and wave-number $k$.\footnote{The function  $\delta$ is also indicated with other names in the literature, such as $\nu$ and $\alpha_M$, related to $\delta$  by  $\nu\,=\,\alpha_M\,=\,-2 \delta$. Here we follow the notation of Ref.~\cite{Belgacem:2019pkk}.} Furthermore, parity-violating theories could also introduce a dependence of these quantities on the polarization index $A$.

As regards the friction term, within theories beyond GR next to the standard Hubble friction, one may encounter an extra contribution described through $\delta$. The latter can for instance parametrise an effective Planck mass evolution rate or the impact of extra dimensions, or the effect of other modifications of GR at the cosmological scale.
The wave propagation may further be modified by a deviation in its speed $c_{\rm T}$ and the mass of the graviton $m_T$. 
Finally, in  standard GR the source term is due to the anisotropic stress tensor (a typical mechanism that can generate it is neutrino free streaming, see e.g.~Sec.~19.5.3 of Ref.~\cite{Maggiore:2018sht}), and in modified gravity can in general get further contributions
(as is the case for instance in bimetric theories~\cite{Hassan:2011zd}).
In several modified gravity theories 
the only modification is in the friction term and is wavenumber-independent,  so that Eq.~(\ref{gwprop_eq:gw}) reads
\begin{equation}
h''_A  +2 {\cal H}[1-\delta(\eta)] h'_A+k^2h_A=0\, .
\label{prophmodgrav}
\end{equation}
In that case, the effect of the modification is such that, from the measurement of the waveform of the inspiral  of a coalescing binary, we do not get the standard luminosity distance obtained from EM signals (that, in this context, we will denote by $\dem$), but rather a GW luminosity distance, $\dgw:=1/h_A$ defined as the inverse of the strain amplitude, up to a constant coefficient. In the case of \Eq{prophmodgrav}, this quantity is related to $\dem$ by~\cite{Belgacem:2017ihm}
\begin{equation}
D_L^{\,\rm GW}(z)=D_L^{\,\rm em}(z)\exp\left[-\int_0^z \,\frac{dz'}{1+z'}\,\delta(z')\right]\, .
\label{dLgwdLem}
\end{equation}
For the purpose of comparison with the data, it is convenient to have a simple parametrization of this effect in terms of a few parameters, rather than a full function of redshift (similarly to  the $(w_0,w_a)$ parametrization of the DE EoS). In practice, just as for the DE EoS, a parametrization is useful only if it is quite economical, with at most two parameters, otherwise it will be difficult to extract them from the data. For modified GW propagation, a convenient choice is~\cite{Belgacem:2018lbp}   
\begin{equation}
\label{eq:fit}
\frac{D_L^{\,\rm GW}(z)}{D_L^{\,\rm em}(z)}=\Xi_0 +\frac{1-\Xi_0}{(1+z)^n}\, ,
\end{equation}
which  depends on the two parameters
$(\Xi_0,n)$.
This parametrization reproduces the fact that, as $z\ra 0$, the ratio $D_L^{\,\rm GW}/D_L^{\,\rm em}$ must go to one  since as the redshift of the source goes to zero, there can be no effect from modified propagation. In the opposite limit of large redshifts, in \eq{eq:fit} $D_L^{\,\rm GW}/D_L^{\,\rm em}$ goes to a constant value $\Xi_0$. This reproduces what happens in  typical DE models, where deviations from GR only appear in the recent cosmological epoch, so that $\delta(z)$ goes to zero at large redshift and $D_L^{\,\rm GW}(z)/D_L^{\,\rm em}(z)$  saturates to a constant. \Eq{eq:fit} indeed fits the explicit results from a large class of modified gravity models~\cite{Belgacem:2019pkk}. The modification in the luminosity distance shown in \eq{eq:fit} can be tested from observations by combining with EM observational probes such as the BAO and CMB. By combining the BAO angular scale $\theta_{BAO}(z)$ with the measured luminosity distance from the GW sources and the sound horizon scale from CMB $r_s$, one can write a unique relation to measure the parameters $\Xi_0$ and $n$~\cite{Mukherjee:2020mha}:
\begin{equation}\label{modfandbao}
  D^{GW}_L(z)\theta_{BAO}(z)= \bigg[\Xi_0 +\frac{1-\Xi_0}{(1+z)^n} \bigg](1+z) r_s.
\end{equation}
If GR is the correct theory of gravity, the above equation indicates that the product of $\theta_{BAO}(z)$ and $D^{GW}_L(z)$ should scale as $(1+z)r_s$ as a function of redshift. Otherwise, it will exhibit a departure from $(1+z)r_s$ which can be explored   to reconstruct the parameters $\Xi_0$ and $n$ for both bright and dark standard sirens as a function of redshift~\cite{Mukherjee:2020mha}.   

The $(\Xi_0,n)$ parametrization is the one in terms of which have been written explicit codes and Markov Chain Monte Carlo (MCMC), in order to assess the sensitivity of LISA to modified GW propagation, including the degeneracies of $\Xi_0$ with other parameters such as $w_0$, see below. Other two-parameter parametrizations can be found in Refs.~\cite{Belgacem:2019pkk,Calcagni:2019ngc}.

%%%%%%%%%%%%%%%
\subsubsection{The damping term (standard sirens)}
%%%%%%%%%%%%%%%

Late-time constraints on the modified damping term have been forecasted in Refs.~\cite{Lombriser:2015sxa,Belgacem:2017ihm,Amendola:2017ovw, Belgacem:2019pkk, Baker:2020apq, Mukherjee:2020mha, Frusciante:2021sio} based on standard siren tests~\cite{Schutz:1986gp,Holz:2005df}, and early-time modifications can be constrained by CMB B-modes~\cite{Amendola:2014wma}. However, the dark sirens detected by LIGO and Virgo already impose some actual limits on $\Xi_0$~\cite{Arai:2017hxj,Belgacem:2018lbp,Lagos:2019kds, Finke:2021aom}. 
The modified friction term can also be constrained with GW data only analyzing the mass distribution of compact binaries~\cite{Ezquiaga:2021ayr}, similarly to the methods developed to bound $H_0$ and other cosmological parameters~\cite{Farr:2019twy,Ezquiaga:2020tns,Mastrogiovanni:2021wsd}.
So far there has not been much exploration of a possible frequency dependence of $\delta$. (See, however, Ref.~\cite{Belgacem:2019pkk} for forecasts on oscillations in the GW amplitude.)

In specific models of modified gravity, the parameters effecting the GW luminosity distance also impact EM observables. For the Horndeski family of theories, constraints of this type can be found in Refs.~\cite{Alonso:2016suf, SpurioMancini:2019rxy,Bonilla:2019mbm,Allahyari:2021enz}, and the combination of EM+GW constraints in Refs.~\cite{Noller:2018wyv,Baker:2020apq}.

In certain theories GWs can decay via interactions. This effect is induced by 
non-linear couplings from $m=2$ in Eq.~\eqref{lagrWBG}, 
as the spontaneous violation of Lorentz symmetry allows GWs to decay into two scalar excitations perturbatively~\cite{Creminelli:2018xsv} or with resonant enhancement~\cite{Creminelli:2019nok} 
for which, importantly, LISA will probe a different and complementary region of the parameter space already constrained by LIGO/Virgo. 
 
%%%%%%%%%%%%%%%
\subsubsection{The gravitational wave speed} \label{sec:mg_gw_speed}
%%%%%%%%%%%%%%%

The GW speed $c_T$ is constrained by a variety of measurements.
The detection of ultra high energy cosmic rays implies a strong constraint on gravitational Cherenkov radiation from subluminal propagation of the waves, as otherwise the radiation would decay away at a rate proportional to the square of their energy $\mathcal{O}(10^{11}~\textrm{GeV})$ before reaching us~\cite{Moore:2001bv,Caves:1980jn,Kimura:2011qn}.
For galactic $\mathcal{O}(10~\textrm{kpc})$ or cosmological $\mathcal{O}(1~\textrm{Gpc})$ origin, the relative deviation in $c_{\rm T}$ is constrained to be smaller than $\mathcal{O}(10^{-15})$ or $\mathcal{O}(10^{-19})$, respectively.
This bound, however, only applies for subluminal propagation, redshifts of $z\lesssim0.1$, and modifications in the high-energy regime.

Another constraint on $c_T$ at the subpercent level can be inferred from the energy loss in binary pulsar systems~\cite{Jimenez:2015bwa,Brax:2015dma}.
Constraints on $c_{\rm T}$ have also been discussed as forecasts for potential arrival time comparisons with nearby supernovae emissions~\cite{Nishizawa:2014zna,Saltas:2014dha,Lombriser:2015sxa,Brax:2015dma}, which are however very rare, for LISA eclipsing binary systems~\cite{Bettoni:2016mij}, or the weak bounds that can be inferred without counterpart emissions from BBH mergers~\cite{Cornish:2017jml}, from the comparison of the GW arrival times between the terrestrial detectors~\cite{Blas:2016qmn}, or the CMB B-mode power spectrum~\cite{Raveri:2014eea} for early-time modifications.
A stringent and prominent direct constraint on deviations of $c_T/c=1$ of $\lesssim\mathcal{O}(10^{-15})$ was obtained from the arrival times of the GWs from the LIGO/Virgo event GW170817~\cite{TheLIGOScientific:2017qsa,Monitor:2017mdv} and its EM counterparts.
As anticipated, the measurement left a strong impact across a wide range of cosmic acceleration models~\cite{Lombriser:2015sxa,McManus:2016kxu,Lombriser:2016yzn,Creminelli:2017sry,Sakstein:2017xjx,Ezquiaga:2017ekz,Baker:2017hug}.
This can be observed from Eq.~\eqref{lagrWBG}: for $\dot \phi^2 \sim \Lambda_2^4$ and $H \dot \phi \sim \Lambda_3^3$, a $c_{ T}$ very close to $c$ implies a fine tuning of the coefficients $c_{n,2}$ and $c_{n,3}$ for all $n$. 

Importantly, however, the constraint only applies to low redshifts of $z\lesssim0.01$ and the LIGO/Virgo frequency range~\cite{Battye:2018ssx,deRham:2018red}.
In particular, it was argued that UV completion terms for modified gravity theories naturally recover a luminal speed of gravity in the high-energy limit tested by GW170817 while allowing deviations at lower energies relevant to modifications that could drive cosmic acceleration~\cite{deRham:2018red}.
LISA will provide a threefold improvement over a GW170817-like bound.
With the detection of GWs from massive BBHs up to $z\sim10$ and their EM counterparts $c_T$ will be tested across much larger distances, tightening the current constraint.
Additionally, a more robust measurement of arrival-time delays can be achieved~\cite{Mangiagli:2020rwz, Haiman:2017szj, Tang:2018rfm}.
Furthermore, LISA tests the frequency range below the expected UV transition of modified gravity models~\cite{deRham:2018red}.
Besides providing a new test of GR at different energy scales, a measurement of the speed of GWs with LISA is thus of particular relevance to cosmic acceleration models.
%FV%
Another way to relax the constraints on $c_T$ is by considering larger values of $\Lambda_3$  or smaller values of $\dot \phi^2$. For instance, it was noted that by raising  $\Lambda_3$ in Eq.~\eqref{lagrWBG} by three orders of magnitudes is enough to agree with most of the  constraints discussed in this section~\cite{Noller:2019chl}. Although this means that cosmological effects of these theories become irrelevant,  their consequences  could become manifest in compact astrophysical objects, potentially probed by LISA.
%

%%%%%%%%%%%%%%%
\subsubsection{Dispersion}
%%%%%%%%%%%%%%%

An energy $E$ dependent GW velocity is also introduced with non-vanishing graviton mass $m_g=\mu h/c^2\neq0$, where the group velocity becomes $(v_g/c) \simeq 1 - (m_gc^2/E)^2$.
The combination of current LIGO/Virgo sources yields a constraint of $m_g\leq1.76\times10^{-32}~{\rm GeV}/c^2$~\cite{Abbott:2020jks}.
More distant and more massive sources are generally more effective in constraining the dispersion relation and the lower energy range of LISA will be favourable for tightening the bounds on $m_g$.
Currently, the strongest bound on the mass of the graviton with $m_g\leq7\times10^{-41}~{\rm GeV}/c^2$ is inferred from weak gravitational lensing~\cite{Choudhury:2002pu}.

Frequency dependent modifications in the velocity term of Eq.~(\ref{gwprop_eq:gw}) can also more generally be parametrised with the group velocity $(v_g/c) \simeq 1 + (\alpha - 1) A_{\alpha} E^{\alpha-2} + \mathcal{O}(A_{\alpha}^2)$, where $\alpha$ and $A_{\alpha}$ may parametrise quantum-gravity effects, extra dimensions, or Lorentz invariance violations~\cite{Mirshekari:2011yq}.
The parametrisation recovers for instance the effect of a massive graviton for $\alpha=0$ with $A_0 = m_g^2 c^4$. In quantum gravity, the parameter $\alpha$ depends on the dimension of spacetime and of momentum space~\cite{Calcagni:2019ngc}.

%%%%%%%%%%%%%%%
\subsubsection{Interaction terms and gravitational wave  oscillations} \label{sec:mg_gw_interactions}
%%%%%%%%%%%%%%%

Finally, the presence of a source term $\Pi\neq0$ further modifies the GW amplitude with an oscillatory correction of order $\Pi\gamma_{ij}/(H k)$~\cite{Nishizawa:2017nef}. 
This effective source term appears naturally in theories with multiple tensor fields. These tensor fields could have a fundamental origin like in bigravity~\cite{deRham:2010kj,Hassan:2011zd} or they could be an effective field arising from a combination of multiple vector fields as in Yang-Mills~\cite{Cervero:1978db,Galtsov:1991un,Darian:1996mb}, Abelian multi gauge fields in a gaugid configuration~\cite{Piazza:2017bsd} and multi Proca fields~\cite{ArmendarizPicon:2004pm,Hull:2014bga,Allys:2016kbq,Jimenez:2016upj}. (See Ref.~\cite{Jimenez:2019lrk} for a survey of the theory landscape.) 
The induced GW oscillations have been studied in concrete examples such as Refs.~\cite{Max:2017flc,Max:2017kdc} and gauge field DE~\cite{Caldwell:2016sut,Caldwell:2018feo}. 
The presence of additional tensor modes can leave an imprint in the luminosity distance, introduce waveform modulations and chiral effects~\cite{Jimenez:2019lrk}. 

The effect of GW oscillations on the luminosity distance can be constrained with LISA standard sirens~\cite{Belgacem:2019pkk} (see Sec.~3.3). In the coherent regime, GW oscillations appear as frequency dependent modulations of the waveform and can thus be tested without the need of redshift information, cf.~Fig.~11 in Ref.~\cite{Belgacem:2019pkk}. GW oscillations in bigravity depend on a mixing angle (defining the interaction between the two spin-2 fields) and are suppressed by the graviton mass as $\tilde m_g^2/f^2$. The lower frequency range and high SNR will allow LISA to improve constraints on $m_g$ to $m_g\lesssim 2\cdot10^{-25}$eV,  an improvement by 3 orders of magnitude  with respect to current LIGO bounds ($m_g\lesssim 10^{-22})$~eV~\cite{Max:2017flc}.

%%%%%%%%%%%%%%%
\subsubsection{Polarisation effects, gravitational wave  lensing and triple systems}
%%%%%%%%%%%%%%%

GW propagation beyond GR can lead to effects on standard ($+,\times$) or novel GW polarizations (e.g.~scalar and vector waves). Due to Lorentz symmetry and parity, on the homogeneous FLRW background $h_\times,h_+$ interacts only with additional tensor polarizations, either fundamental (bigravity, multigravity) or composite (multiple vector fields), leading to the effects discussed in Sec.~\ref{sec:mg_gw_interactions}. Theories including parity-violating interactions can lead to propagation effects on left/right polarized GWs~\cite{Alexander:2007kv,Okounkova:2021xjv}.

Interactions between standard and novel polarizations in parity-preserving theories occur beyond the homogeneous background, leading to new phenomena in GW lensing beyond GR~\cite{Ezquiaga:2020dao}. Locally, the propagating field is a superposition of the $+,\times$ and novel polarizations, known as propagation eigenstates. Propagation eigenstates evolve independently with a well defined speed that can depend on the position and direction. In general, each propagation eigenstate has a different speed, leading to birefringence (polarization dependent arrival time and deflection angles). Birefringence causes very distinct effects when the (total) time delay between different polarizations is shorter than the duration of the signal in the observable band. Long signals at high redshift (e.g.~MBBHs) will allow LISA to probe birefringence in ways complementary to ground detectors.

GW lensing beyond GR can be particularly powerful in configurations in which the source and the lens are in close proximity. Interesting targets for LISA are stellar mass binaries in the vicinity of a MBH (a scenario suggested by the possible EM counterpart to GW190521~\cite{Graham:2020gwr}). GW birefringence effects can be strongly enhanced by the strong gravity, potentially allowing tests of Horndeski theories at even higher precision than GW170817 (cf.~Sec.~\ref{sec:mg_gw_speed} and Fig.~16 in Ref.~\cite{Ezquiaga:2020dao}). 
GW emission by a binary near a MBH can be affected by the scalar hair of the central body~\cite{Brax:2020ujo}.
A stellar-mass binary orbiting a MBH undergoes a characteristic Doppler shift and (possibly) a strong lensing pattern that can be used to identify the triple system~\cite{Toubiana:2020drf}. Those systems can excite the quasi-normal modes of the central BH during the inspiral, with stellar-mass binaries detected by LISA sensitive to central masses $M\sim 10^6M_\odot$~\cite{Cardoso:2021vjq}. 
These configurations will allow LISA to perform novel tests of GR, either through the confirmation of the triple-nature of the system or by direct detection of quasi-normal modes.
 
%%%%%%%%%%%%%%%
\subsubsection{Relation to ppE framework}
%%%%%%%%%%%%%%%

The modifications in Eq.~(\ref{gwprop_eq:gw}) can also be cast into the parameterized post-Einsteinian (ppE) framework~\cite{Yunes:2009ke}, introduced to parametrise the effects of GR departures in the dynamical strong-field regime on the gravitational waveforms from the binary coalescence of compact objects,
\begin{equation}
 h(f) = \left( 1 + \sum_j \alpha_j u^j \right) e^{i\sum_k \beta_k u^k} h_{\rm GR}(f) \,,
\end{equation}
where $u\equiv(\pi \mathcal{M} f)^{1/3}$ with chirp mass $\mathcal{M}$ and frequency $f$. 
The GR waveform $h_{\rm GR}(f)$ is reproduced for vanishing $\alpha_j$ and $\beta_k$.
The series in $\alpha_j$ can be expressed as an integral over $\delta$ whereas the series in $\beta_k$ may be expressed as an integral involving $c_T$ and $m_T$~\cite{Nishizawa:2017nef}.
In the ppE subclass of the generalised inspiral-merger-ringdown waveform model $h(f) = e^{i\delta\Phi_{\rm gIMR}} h_{\rm GR}(f)$~\cite{Li:2011cg} (see e.g. Ref.~\cite{Nishizawa:2017nef} for the deatils of this parameterization) the modification is hence independent of $\delta$.
The above parametrization has been further extended to include parity violating effects and then used to put constraints on parity violation in gravity by using the LIGO/Virgo O1/O2 catalog~\cite{Yamada:2020zvt}.

%%%%%%%%%%%%%%%
\subsection{Open problems}
%%%%%%%%%%%%%%%

An interesting open problem is the mixture of propagation and source effects for extended theories of gravity and the role of screening mechanisms.
A post-Newtonian expansion for the source emission in screened regimes can be performed using a scaling relation~\cite{McManus:2017itv,Renevey:2020tvr}.
For some GR extensions the relevant modification in $\nu$ are in fact determined by the screened environments of emitter and observer rather than the cosmological background whereas for $c_T$ screening effects may safely be neglected.

In addition to the effects on the propagation of GWs, modified gravity theories can also affect the generation of GWs.
Shift-symmetric scalar-tensor theories are particularly suited for describing DE since they mediate a long range (massless) interaction. However, 
in these theories, there is a no-hair theorem that proves that the only BH solution is locally isometric to Schwarzschild~\cite{Hui:2012qt, Creminelli:2020lxn}. One way to avoid the no-hair theorem is to consider a time dependent scalar field solution and a number of hairy BH solutions have been found in Horndeski, beyond-Horndeski as well as degenerate higher-order scalar-tensor theories~\cite{Babichev:2016rlq}. Another possibility for hairy BH solutions is to involve the coupling with the Gauss-Bonnet invariant~\cite{Sotiriou:2013qea}. A no-hair theorem also exists for neutron stars~\cite{Barausse:2015wia, Lehebel:2017fag}. Again the time dependence of the scalar field can evade this no-hair theorem~\cite{Sakstein:2016oel, Ogawa:2019gjc}. It is still an open question whether these hairy solutions can be formed dynamically and how they modify the generation of GWs such as BH ringdown~\cite{Noller:2019chl}. We note that even if modified gravity theories predict the same BH solutions in static environments, this is not necessarily the case in dynamical situations, and also their perturbations are different in different theories~\cite{Barausse:2008xv, Tattersall:2017erk}.  

An interesting possibility to evade such no-hair theorems, at least for neutron stars, is to allow for a conformal coupling with matter. The conformal coupling breaks the shift symmetry, but so long as this is Planck suppressed, this breaking will be soft. In this case non-trivial solutions featuring screening exist~\cite{Babichev:2009ee} and, most importantly, pass the test of non-linear numerical evolution~\cite{terHaar:2020xxb}. 

These modified gravity theories admit the existence of scalar GWs. A detection of scalar GWs is a smoking gun for the deviation from general relativity~\cite{Scharre:2001hn}. Screening mechanisms are expected to suppress the generation of scalar GWs~\cite{Dar:2018dra}, however, very recently, numerical relativity simulations have been performed, which point at a partial breakdown of the screening in BH collapse~\cite{terHaar:2020xxb,Bezares:2021yek} and in the late inspiral and merger of BNSs~\cite{Bezares:2021dma}. In more detail, stellar collapse seems (quite surprisingly) to produce a very low frequency signal potentially detectable by LISA, while waveforms from BNSs seem to deviate from their GR counterparts at the quadrupole (but not dipole) multipole order.  

Finally, there are various theoretical constraints on modified gravity models considered in this section. A notable one is the decay of GWs into scalar field perturbations~\cite{Creminelli:2018xsv, Creminelli:2019nok}. This is due to the low strong-coupling scale of the theory, which is close to the energy scale probed by LIGO/Virgo.
Other constraints can be placed in terms with $m=1$ and $m=2$ in Eq.~\eqref{lagrWBG},
and come from avoiding instabilities in the scalar field sector that can be induced by passing GWs~\cite{Creminelli:2019kjy}.
If new states are present at a scale parametrically below the cut-off, the theory is of no use for the GWs predictions at LIGO/Virgo~\cite{deRham:2018red}. LISA will detect GWs at much lower frequencies and will play an essential role in constraining these theories.

\newpage
%%%%%%%%%%%%%%%%%%%%%%%%%%%%%%%%%
% Here Sec. 5 starts

\section{Stochastic gravitational wave background  as a probe of the early universe}
\label{sec:SGWB}
\small \emph{Section coordinator: S.~Kuroyanagi. Contributors: 
N.~Bartolo, C.~Caprini, G.~Cusin, D.G.~Figueroa, S.~Kuroyanagi, M.~Peloso, S.~Renaux-Petel, A.~Ricciardone.}

\subsection{Introduction}

The SGWB contains information on the GW events that occurred across the whole history of the universe. Its measurement has an immense value for cosmology. 
%On one hand, the SGWB encloses  knowledge of all the binary events that are to faint to be individually resolved.  On the other, it  
It gives access to stages of the early universe that cannot be directly probed by either EM or neutrino observations. The universe before the CMB epoch was indeed not transparent to photons, and neutrinos produced in the very early universe are too weak to be detected at current or forthcoming experiments. We now introduce some key concepts about the SGWB and its observables, while we focus on the physics of its cosmological sources in the subsequent sections.

\label{subsec:GWprobeIntro}

\subsubsection{What is a stochastic gravitational wave background?}

Early universe phenomena that emit GWs typically lead to the production of a background of GWs of stochastic nature. This means that the tensor perturbation $h_{ij}({\mathbf x},t)$ that defines the background is a random variable with different realisations everywhere in space. SGWBs can therefore be characterised only at a statistical level, by means of ensemble averages. As there is only one observable universe, what is customary in cosmology is to invoke the {\it ergodic} hypothesis, stating that either spatial or temporal averages are equivalent to an ensemble average of the underlying statistical distribution. In other words, we interpret that, by observing today a large region of the universe (or for this matter a given region for long enough time), we have access to many realisations of the system. This holds under two conditions: 1) if the universe is almost homogeneous and isotropic at the time of the GW production, so that the initial condition for the generation of GWs is the same (in a statistical sense) everywhere; 2) if the GW source respects causality. Under these circumstances,  the properties of SGWBs from the early universe can be studied by means of the ergodic hypothesis. 
 
Due to causality, a cosmological GW source acting at a given time in the early universe cannot produce a signal correlated at length and time scales larger than the cosmological horizon at the moment of GW generation. If we denote with a subscript $p$ the time of production, the (physical) correlation scale of the emitted GWs must satisfy $\ell_p\leq H_p^{-1}$.\footnote{Let us note that we have used the inverse Hubble rate $H^{-1}_p$ as the cosmological horizon, as that is a good approximation for most of the cosmological evolution of the universe, except during inflation.} Equivalently, the GW signal cannot be correlated at time scales larger than $\Delta t_p\leq H_p^{-1}$. At the present time we have access to much larger length and time scales than today's redshifted scale associated with $H^{-1}_p$, so a SGWB signal in our detectors is perceived as the superposition of many signals uncorrelated in time and space. The number of independent signals in a given region today can be actually counted, knowing the evolution of the universe and the details of the GW generation mechanism. 

We note that the above arguments remain valid also for causal sources continuously emitting GWs over a long period, say during several Hubble times. The paradigmatic example of this is the case of a cosmic string network, which emits GWs continuously, all the way from the moment of the phase transition that created it, till today. The SGWB signal in this case is mostly dominated (at least in the Nambu Goto (NG) approach to strings, see Sec.~\ref{sec:CosmicStrings}) by the superposition of the GWs emitted by sub-horizon string loops. This GW signal is perceived today as a background formed by the superposition of many GW emissions at different times of cosmic history, and from many different regions. Therefore the observation of this signal today cannot be resolved beyond its stochastic nature, exactly for the same reasons discussed above. The main difference in fact between a continuously sourced background and one arising from a source localised in time, is rather that the former extends over a long frequency range, precisely because its source has been emitting during many Hubble times.
 
In the case of inflation, the above arguments do not apply, as the causal horizon grows exponentially during the inflationary phase. Yet, it is well known that inflation produces a SGWB, as we discuss in Secs.~\ref{sec:Inflation} and \ref{sec:nonStandard}. The inflationary background of GWs is actually considered to be stochastic because of the intrinsic quantum nature of the generating process. In particular, this background originates during inflation due to the quantum vacuum fluctuations of tensor metric perturbations, which become a random variable. The tensors become effectively classical  during the inflationary accelerated expansion as their wavelengths are exponentially stretched to super-Hubble scales, where they acquire very large occupation numbers. (See the discussion e.g.~in Ref.~\cite{Caprini:2018mtu}.) This transition after Hubble-crossing renders the tensor perturbation a stochastic variable. When the tensor perturbations re-enter the Hubble radius during cosmic evolution after inflation, they form a GW signal that is intrinsically stochastic.

Finally, let us notice that the superposition of astrophysical GWs also forms a SGWB, which will act as a foreground for cosmological SGWBs. We will briefly review possible astrophysical background contributions expected in the mHz band in Sec.~\ref{Sec:AstroSGWB}.\\

\subsubsection{How do we characterize a stochastic gravitational wave background?} 

Since we speak about random tensor variables, the Fourier mode $h_r(\mathbf{k}, t)$ are also considered to be random variables. For statistically homogeneous and isotropic, unpolarised and Gaussian GW backgrounds, we define the tensor amplitude power spectrum as
\be
\label{powerspec}
\langle h_r(\mathbf{k}, t) \, h^*_{p}(\mathbf{q}, t) \rangle = \frac{8\pi^5}{k^3} \,
\delta^{(3)}(\mathbf{k} - \mathbf{q}) \, \delta_{r p} \, h_c^2(k, t)\,,
\ee
with $h_c$ a dimensionless real function, depending only on the time $t$ and the comoving wave-number $k = |\mathbf{k}|$. The previous factor $8\pi^5$ has been chosen so that we can write
\be
\label{hcketa}
\langle h_{ij}(\mathbf{x}, t) \, h_{ij}(\mathbf{x}, t) \rangle = 2 \, \int_0^{+\infty} \frac{dk}{k} \, h_c^2(k, t)\,,
\ee
with the factor $2$ in the RHS as a convention motivated by the fact that the LHS involves the contribution from two independent polarisations. It is then clear that $h_c(k, t)$ represents the characteristic tensor amplitude per logarithmic wave-number interval and per polarisation state, at a given time $t$. 

Another relevant quantity to characterise a SGWB is the spectrum of GW energy density per logarithmic wave-number interval, $d \rho_{\rm GW} / d \mathrm{log} k$, defined as
\be
\label{rhogw}
\rho_{\rm GW} \, = \, \frac{\langle \dot{h}_{ij}(\mathbf{x}, t) \, \dot{h}_{ij}(\mathbf{x}, t) \rangle}{32 \pi G} \, = \, 
\frac{\langle h'_{ij}(\mathbf{x}, \eta) \, h'_{ij}(\mathbf{x}, \eta) \rangle}{32 \pi G \, a^2(\eta)} \, = \, 
\int_0^{+\infty} \frac{dk}{k}\,\frac{d \rho_{\rm GW}}{d \mathrm{log} k}\,,
\ee
where in the second equality we have converted the derivatives with respect to the physical time $t$ into derivatives with respect to the conformal time $\eta$.
An expression for the GW energy density power spectrum $d \rho_{\rm GW} / d \mathrm{log} k$ valid for free waves inside the Hubble radius, can be found using the simple relation ${h'_c}^{2}(k, \eta) \simeq k^2 \,h_c^{2}(k, \eta)$, where the prime symbol ($'$) represents the derivative with respect to the conformal time, and ${h'_c}^{2}(k, \eta)$ characterises the expectation value $\langle h'_{ij}(\mathbf{x}, \eta) \, h'_{ij}(\mathbf{x}, \eta) \rangle = 2 \, \int_0^{+\infty} \frac{dk}{k} \, {h'_c}^{2}(k, \eta)$. Using this, we arrive at~\cite{Caprini:2018mtu}
\be
\label{drhodlogk}
\frac{d \rho_{\rm GW}}{d \mathrm{log} k} = \frac{k^2\,h_c^2(k, \eta)}{16 \pi G \, a^2(\eta)} \, .
\ee
As $h_c(k, \eta) \propto 1/a(\eta)$ for sub-Hubble modes, the GW energy density is diluted as radiation with the expansion of the universe, $\rho_{\rm GW} \propto a^{-4}$, as expected for massless degrees of freedom.

Finally, in order to connect the above expressions with experimental observables, we need to re-express the GW spectra today in terms of the present-day physical frequency, $f = k / (2 \pi \, a_0)$, associated with the comoving wave-number $k$. In order to do this, we first write the characteristic tensor amplitude per logarithmic frequency interval today as $h_c(f) = h_c(k, t_0)$. We then define the one-sided spectral density of a SGWB as
\be
\label{Shf}
S_h(f) = \frac{h_c^2(f)}{2 f} \, ,
\ee
which has dimensions $\mathrm{Hz}^{-1}$. The reason to define this quantity is that it is directly comparable to the noise in a detector, parametrised by $S_n(f)$. On the other hand, it is convenient to normalise the spectrum of the GW energy density per logarithmic frequency interval like
\be
\label{Omegagw}
\Omega_{\rm GW}(f) = \frac{1}{\rho_c}\,\frac{d \rho_{\rm GW}}{d \mathrm{log} f}\,,
\ee
where $\rho_c =3 H^2/(8 \pi G)$ is the critical energy density at time $t$. The quantity traditionally considered by cosmologists is then
\be
\Omega_{\rm GW}^{(0)}(f) = \frac{4 \pi^2}{3 H_0^2} \, f^3 \, S_h(f) \, ,
\label{OmandSh}
\ee
which corresponds to the normalised energy density spectrum today. In terms of the dimensionless amplitude $h_c = \sqrt{2 f\,S_h}$, we can also write the following relations
\begin{eqnarray}
\label{Omandhc}
S_h(f) = 7.98 \times 10^{-37} \, \left(\frac{\mathrm{Hz}}{f}\right)^3 \, h^2\,\Omega_{\rm GW}(f)\,\frac{1}{\rm Hz}\,, \\
h_c(f) = 1.26 \times 10^{-18} \, \left(\frac{\mathrm{Hz}}{f}\right) \, \sqrt{h^2\,\Omega_{\rm GW}(f)} \, .
\end{eqnarray}\\

\subsubsection{Current constraints by other observations.}

There are several ways to place constraints on the SGWB energy density. We briefly review different types of observations which provide constraints on the SGWB at different frequencies. 

BBN places a bound on a primordial (i.e.~prior to BBN) SGWB as the latter contributes to the total energy density of extra relativistic species and affects the expansion rate of the universe during BBN. In order not to spoil BBN by changing the resulting light-element abundances, the energy density of the SGWB should satisfy \cite{Caprini:2018mtu, Allen:1996vm}
\begin{equation}
\int d (\ln f) \Omega_{\rm GW}(f) h^2 \le 5.6 \times 10^{-6} \left( N_{\rm eff}^{\rm (upper)} - 3.046 \right) \,,
\end{equation}
where $N_{\rm eff}^{\rm (upper)}$ is the upper bound on the effective number of relativistic degrees of freedom $N_{\rm eff}$. Applying the 2$\sigma$ upper limit for $N_{\rm eff}$ from BBN (obtained by observations of $^4$He  and D) as $N_{\rm eff}^{\rm (upper)} = 3.41$ \cite{Cyburt:2015mya}, we obtain $\Omega_{\rm GW} h^2 < 2.3 \times 10^{-6}$ for a logarithmic frequency bin. This constraint is valid for GWs generated before the BBN epoch $T\sim 1$ MeV.

The CMB is also a powerful tool to constrain the primordial SGWB. CMB anisotropies on large scales, both in temperature and in polarisation (E and B modes), can be induced by the tensor metric perturbations. Non-detection of such anisotropies provides a bound on the tensor-to-scalar ratio $r$ and the current best bound is given by the combined Planck and BICEP2/Keck data up to 2018: $r<0.036$ at $k=0.05 {\rm Mpc}^{-1}$ \cite{Akrami:2018odb,Ade:2018gkx,Tristram:2020wbi,BICEP:2021xfz}, which corresponds to $\Omega_{\rm GW} h^2 <  1.5 \times 10^{-16}$ at $f=7.7 \times 10^{-17}$~Hz.
On small scales $f>10^{-15}$~Hz, GW modes are well inside the cosmological horizon and behave like massless neutrinos, contributing to $N_{\rm eff}$, thus affecting the growth of density perturbations as well as the expansion rate at recombination \cite{Smith:2006nka,Sendra:2012wh,Clarke:2020bil}. The updated constraint from the temperature anisotropy is $\Omega_{\rm GW} h^2 < 1.7 \times 10^{-6}$ for adiabatic initial conditions and $\Omega_{\rm GW} h^2 < 2.9 \times 10^{-7}$ for homogeneous (non-adiabatic) initial conditions at the $95$\% confidence level \cite{Clarke:2020bil}. In the case of adiabatic initial conditions, GW perturbations evolve in the same way as neutrino perturbations. Most known sources of a SGWB produce an unperturbed background and we should impose homogeneous initial conditions, which assume no initial density perturbation.

Ground-based interferometer experiments, which have sensitivity to the frequency range of $10-100$Hz, are rapidly improving the upper limit on a SGWB. The latest constraint by the O3 run of LIGO/Virgo gives $\Omega_{\rm GW} < 5.8 \times 10^{-9}$ (with $h=0.679$) at the $95$\% confidence level for a flat (frequency-independent) SGWB \cite{Abbott:2021xxi}. The limit is obtained by combining data from the earlier O1 and O2 runs and 99\% of the sensitivity comes from the frequency band $20-76.6$~Hz.

Pulsar timing arrays probe low-frequency GWs at a frequency range of $10^{-9}$ and $10^{-6}$~Hz. Millisecond pulsars are known to have an extremely stable pulse frequency. GWs affect the pulse propagation and change the pulse times of arrival, thus their presence can be tested by regularly monitoring pulse frequencies. Radio telescope projects, such as NANOGrav \cite{Alam:2020laa}, EPTA \cite{Desvignes:2016yex}, and PPTA \cite{Kerr:2020qdo}, have been updating the upper limit on the GW amplitude, and recently 
%. Recently, the NANOGrav and EPTA 
% collaborations of these projects have 
reported the possible detection of a SGWB~\cite{Arzoumanian:2020vkk, Chen:2021rqp, Goncharov:2021oub}. The strain amplitude for a $f^{-2/3}$ power-law spectrum inferred by the two experiments has central value at
$h_c=1.92\times 10^{-15}$ and $h_c=2.95\times 10^{-15}$ at $f=1\,{\rm yr}^{-1}$, respectively, which correspond to $\Omega_{\rm GW} h^2 \simeq  2$ -- $5 \times 10^{-9}$ at $f= 3.2 \times 10^{-8}$~Hz. A confirmation of quadrupolar spatial correlations in the signal is needed to establish the detection.

A SGWB at $\mu$Hz frequencies can in principle be observed in binary pulsars as well as in Earth-Moon-like or Earth-Satellites-like systems. The binary system's orbits indeed exhibit resonant interactions with the SGWB in some regimes. Current estimates show that such binaries permit to rule out the existence of a SGWB with $\Omega_{\rm GW}\gtrsim 6\times 10^{-6}$ at  frequencies  $f\sim 1\ \mu$Hz,  with improvement to $\Omega_{\rm GW}\gtrsim 5\times 10^{-9}$ by late 2030's~\cite{Blas:2021mqw,Blas:2021mpc}. 

\subsection{Generation mechanisms}
\subsubsection{Cosmological}
Among the targets of the LISA mission there is the cosmological SGWB, which includes different sources active in the early universe.
Such a background is characterised by the spectral energy density $\Omega_{\rm GW} (f)$ defined in Eq.~\eqref{Omegagw} and by some peculiar features, which will be described in next sections, that can be useful in the process of characterisation and disentanglement from the astrophysical background, like chirality, non-Gaussianity, anisotropies and so on. A very well known example of a cosmological SGWB is the irreducible GW background due to quantum vacuum tensor fluctuations produced during inflation, which spans a large range of frequencies with an almost scale-invariant spectrum. For the simplest realisation of inflation, it has an  amplitude that is too small to be detected by LISA. Besides this, there are several primordial mechanisms which can lead to a non-flat cosmological-SGWB frequency profile at the scales probed by GW interferometers (from $10^{-5}$ to $10^{2}$ \rm{Hz}): from models where the inflaton is coupled with extra (gauge) fields (Sec.~\ref{sec-GWaddf}) \cite{ Barnaby:2010vf,Cook:2011hg,Sorbo:2011rz, Barnaby:2011qe,Dimastrogiovanni:2016fuu,Peloso:2016gqs, Domcke:2016bkh} to models with features in the scalar power spectrum (Sec.~\ref{sec:features}) \cite{Flauger:2009ab,Braglia:2020eai,Fumagalli:2020nvq}, models where spacetime symmetries are broken during inflation (Sec.~\ref{sec:supersolid}) \cite{Endlich:2013jia,Koh:2013msa,Cannone:2014uqa,Cannone:2015rra,Bartolo:2015qvr,Ricciardone:2016lym,Bartolo:2015qvr,Cannone:2014uqa,Akhshik:2014gja,Akhshik:2014bla}, or scenarios where non-attractor phases characterise the universe evolution \cite{Leach:2001zf, Namjoo:2012aa, Mylova:2018yap}. GWs sourced by second-order scalar fluctuations (Sec.~\ref{subsec:second-order-SGWB}) can further be associated with PBH formation (Sec.~\ref{sec:PBH}). These models are characterised by an amplitude which, still respecting the CMB bounds, have a large amplitude and a peculiar frequency shape which may enable detection by LISA. A dedicated analysis for the potential of the LISA space-based interferometer to detect the SGWB produced from different inflationary models has been performed in Ref.~\cite{Bartolo:2016ami}. This analysis has shown how LISA will be able to probe inflationary scenarios, in a complementary way to CMB experiments. Besides these, there are some post-inflationary mechanisms which can also generate GWs with a large amplitude at LISA scales: for instance several setups beyond the standard model of particle physics (BSM) exhibit a first-order phase transition (FOPT) around the TeV energy scale that  peaks in the LISA frequency window (see Sec.~\ref{sec:PTs}). A dedicated analysis for the detection of a cosmological SGWB from FOPTs have been done in Refs.~\cite{Caprini:2015zlo,Caprini:2019egz}.
Also cosmic defects can generate a cosmological SGWB which crosses the frequency window of the LISA detector. More precisely the GW signal from cosmic defects can be detected if the energy of the phase transition that created the defects is at the right scale (see Sec.~\ref{sec:CosmicStrings}). A recent analysis to probe the ability of LISA to measure this background, considering leading models of the string networks has been done in Ref.~\cite{Auclair:2019wcv}. In the most optimistic case, LISA might be able to probe cosmic strings with tensions $G\mu \gtrsim \mathcal O (10^{-17})$. It has been recently pointed out \cite{Boileau:2021gbr} that, depending on different assumptions on the astrophysical background and the galactic foreground, LISA will be able to probe cosmic strings with tensions  $G\mu \gtrsim \mathcal O(10^{-16} - 10^{-15})$.

The detection of any of these SGWBs from the early universe, would allow us to test high energy scales beyond the reach of particle colliders, like the Large Hadron Collider (LHC).

\begin{figure}
    \centering
	\includegraphics[width = 1.0\textwidth]{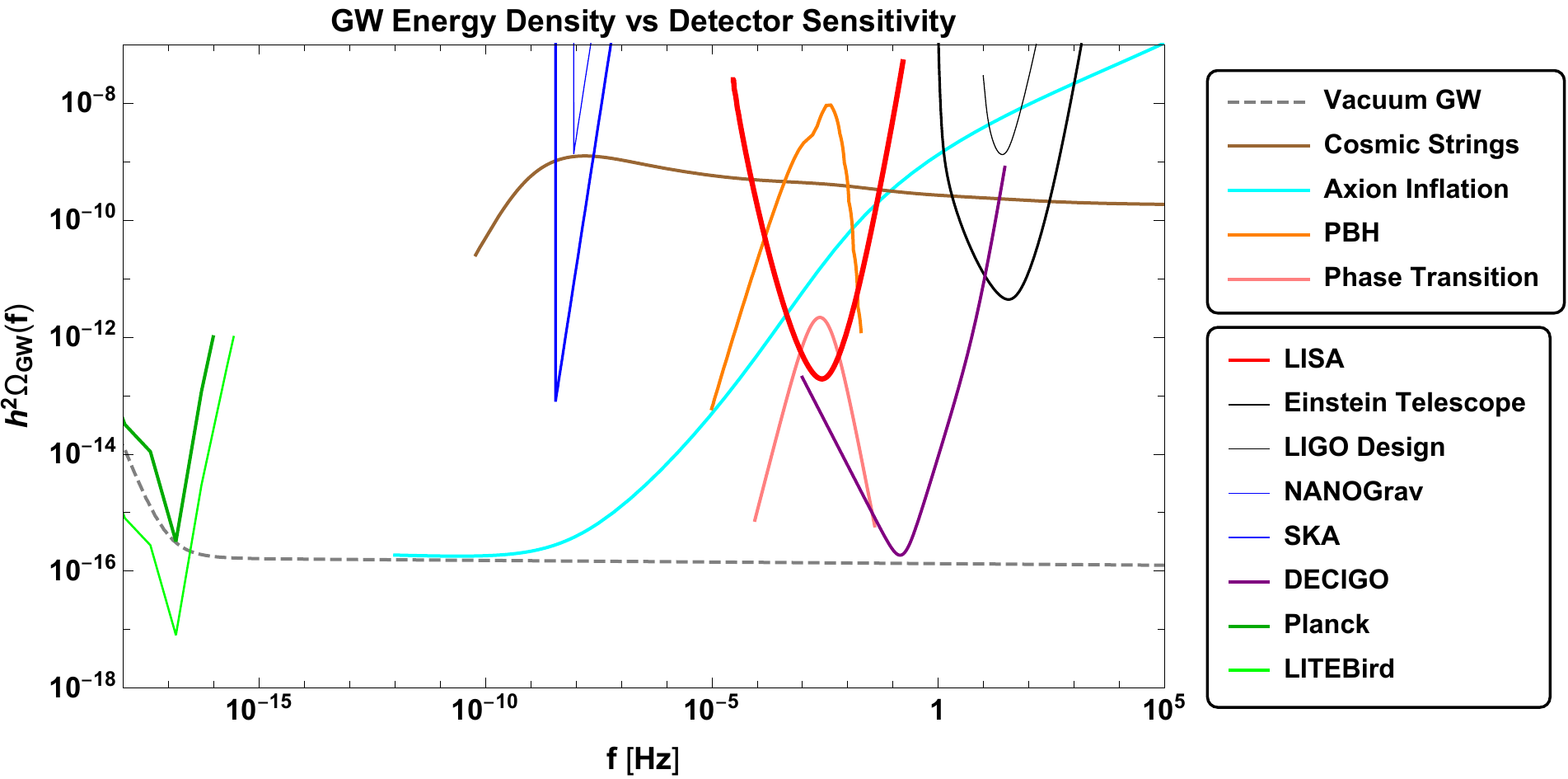}    \caption{\small SGWB  energy density $h^2\Omega_{\rm GW}$ for different cosmological sources compared to the sensitivity of different GW detectors. As cosmological signals we have the vacuum GW contribution coming from inflation (grey dashed line) with $ r=0.044$ and $n_T=-r/8$, the signal expected in axion inflation models (cyan), the signal generated by cosmic string networks with $G\mu=10^{-10}$ (brown), the signal generated by a FOPT with $v_w=0.9$, $\alpha=0.1$, $\beta/H_*=50$, $g_*=100$, $T_*=200 \rm {GeV}$ (pink) and the signal generated at second-order
    by the formation mechanism of PBHs with $f_{PBH}=1$, $\sigma = 0.5$, $k_* = k_{\rm LISA}$ (orange). For GW detectors we report the sensitivity of Planck (darker green), LITEBird (green), EPTA (blue), SKA (darker blue), LISA (red), DECIGO (purple), LIGO Design (black) and ET (darker black).}
 \label{fig:SGWBsources}
\end{figure}

In Fig.~\ref{fig:SGWBsources} we collect GW cosmological signals expected to peak in the LISA frequency band and we compare them with the sensitivity of present and future GW detectors.

\subsubsection{Astrophysical}
\label{Sec:AstroSGWB}
The astrophysical  SGWB results from the incoherent superposition of signals emitted by numerous 
unresolved astrophysical sources from the onset of stellar activity until today. As any other background of radiation, 
the astrophysical  SGWB is quantifiable through its isotropic energy density level and through the spatial angular power spectrum encoding its anisotropy. Many different astrophysical sources contribute to the astrophysical  SGWB, including SOBBHs and BNSs \cite{TheLIGOScientific:2016wyq, Regimbau:2016ike, Mandic:2016lcn, Dvorkin:2016okx, Nakazato:2016nkj, Dvorkin:2016wac, Evangelista:2014oba}, merging MBBHs~\cite{Kelley:2017lek}, rotating neutron stars \cite{Surace:2015ppq, Talukder:2014eba, Lasky:2013jfa}, stellar core collapse \cite{Crocker:2017agi, Crocker:2015taa} and population III binaries \cite{Kowalska:2012ba}. The astrophysical information that can be extracted from the intensity and polarisation maps of the astrophysical  SGWB, are the collective properties of a given population of astrophysical sources (redshift and mass distribution, local properties of galactic environment, ...).

The astrophysical  SGWB from BBHs is expected to be dominant in the LISA band \cite{Dvorkin:2016okx} and below,  and may become a source of confusion noise for other sources and cosmological background emissions. Observations with LISA will allow for the study of some aspects of BBH populations that are difficult to observe with ground-based interferometers. For example, at the mHz frequencies accessible to LISA, some of the binaries may not be fully circularised, and their residual eccentricities may provide an indication to their formation channel. In particular, binaries formed through dynamical processes in dense stellar clusters can have measurable eccentricities. These can be constrained for the subset of resolved merger, and in addition the distribution of eccentricities of the entire population may also affect the resulting astrophysical  SGWB.

The detection of the BNS merger by the LIGO/Virgo network \cite{2017PhRvL.119p1101A,2017ApJ...848L..12A,LIGOScientific:2020aai} and the estimated rate of mergers in the local universe of $R=13 - 1900$ Gpc$^{-3}$ yr$^{-1}$ \cite{LIGOScientific:2021psn} led to the conclusion that in the Hz band these sources may have a comparable contribution to the astrophysical SGWB relative to BBHs \cite{Abbott:2017xzg,Abbott:2021xxi}. We may therefore expect that their contribution to the anisotropies of the astrophysical  SGWB will also be important also for LISA. While it will be difficult to disentangle the relative contributions of BBHs and BNSs to the overall astrophysical  SGWB, especially in view of the large modelling uncertainty in the BNS merger rates \cite{2018MNRAS.474.2937C,2019MNRAS.482.2234G}, it is interesting to note that their host galaxies are expected to have different properties. In the isolated BH formation scenario discussed e.g.~in Ref.~\cite{Cusin:2019jpv}, BH masses are heavily influenced by the metallicity of their progenitor stars. Specifically, metal-poor stars retain most of their mass throughout their evolution and collapse to form heavier BHs. As a consequence, these BHs form preferentially in high-redshift and/or low-mass galaxies \cite{2016MNRAS.463L..31L,2018MNRAS.474.4997C,2018MNRAS.481.5324M,Artale:2019doq}.  In contrast, NSs can also form in metal-rich environments. In view of the different clustering properties of the host galaxy populations, BBHs and BNSs can in principle give rise to very different anisotropic components of the astrophysical  SGWB \cite{Cusin:2019jpv}. 

Finally, LISA will also allow one to study the astrophysical  SGWB from other types of sources such as close white dwarf binaries (see e.g.~Ref.~\cite{Vecchio:2002ca}), which may also produce anisotropies in the galactic plane  \cite{2001PhRvD..64l1501U,2005PhRvD..71b4025K}. We refer the reader to Ref.~\cite{AstroWP} for an in-depth discussion on the astrophysical populations leading to this and the aforementioned astrophysical SGWBs.

\subsection{Characteristics of the stochastic gravitational wave background}
\subsubsection{Frequency profile of the stochastic gravitational wave background}
As described above, there are many different mechanisms to generate a SGWB both of cosmological and astrophysical origin. The question that arises is, how to determine the origin of the SGWB, once it is detected. Identifying the SGWB source is a challenging but crucial task, essential to extract physical information from the SGWB detection. 
One of the most useful properties at this aim is the spectral shape of the SGWB. More specifically, one aims at reconstructing $\Omega_{\rm GW}$ as a function of frequency, since the frequency profile of the SGWB depends on the generation mechanisms, thus providing a way to disentangle different SGWB sources (see for example  Ref.~\cite{Kuroyanagi:2018csn} for a list of sources, and examples on their frequency profiles). Reconstructing the frequency profile in detail also allows to explore the possibility that several sourcing components contribute to the SGWB within the same frequency band, leading to a complicated spectral shape arising from their superposition.

While the spectral shape of the primordial signal depends very much on the detail of the source properties, some fairly general considerations on the expected frequency profile of the cosmological SGWB are still possible, based essentially on causality \cite{Caprini:2009fx}.
Sources active on a (conformal) time-scale $\Delta t\lesssim \mathcal{H}_p^{-1}$, where $\mathcal{H}_p^{-1}$ denotes the Hubble time when the source starts operating (we assume here that this occurs during the radiation dominated era), typically lead to peaked SGWB signals in terms of the variable $\Omega_{\rm GW}(k)$. On (comoving) wavenumbers $k\lesssim 1/\Delta t$, in fact, the SGWB source is expected to be uncorrelated both in time and in space, since $k\lesssim 1/\Delta t \sim v / L < 1/L$ for $v$ and $L$ the typical speed and size of the anisotropic stresses. These latter are the part of the source energy momentum tensor leading to the GW generation.  
Since $\Omega_{\rm GW}(k)$ is the GW energy density per logarithmic wave-number, white-noise anisotropic stresses (flat in $k$) lead to $\Omega_{\rm GW}(k)$ increasing as $k^3$. This increase is expected to change at around $k_p\sim 1/\Delta t \sim v/L$, and the subsequent behaviour depends on the detail of the time and space structure of the source \cite{Caprini:2009fx}.
The SGWB spectrum  either peaks around $k_p$, and decreases for $k>k_p$ with a slope depending on the source characteristics; or it takes a shallower, but positive slope, in which case the peak occurs at a higher wave-number. This latter can correspond to an inverse length $L^{-1}$, though other time and length scales can also be relevant and show up in the $\Omega_{\rm GW}(k)$ spectrum: for example, the characteristic {\it time}-scale over which the source is coherent $\tau<\Delta t$ (as opposed to the source duration $\Delta t$), or different {\it length}-scales, relevant in the space distribution of the anisotropic stresses, other than $L$. 

Notable peaks in the spectrum are then expected for SGWB sources characterised by a finite, short duration $\Delta t \lesssim \mathcal{H}^{-1}_p$.
In particular, one example of sources giving rise to peaked spectra that are relevant for LISA (since they are typically peaking in the LISA band), are FOPTs related to the electroweak symmetry breaking, as described in Sec.~\ref{sec:PTs}. On the other hand, for generation mechanisms that turn on at a given time $\mathcal{H}^{-1}_p$ in the radiation era, but continue to source GWs throughout the universe evolution $\Delta t\gg \mathcal{H}^{-1}_p$, the region of $k^3$ increase is less and less relevant, being pushed towards the horizon today. 
One expects a wide frequency region over which the signal features a slower increase with $k$ or, in some cases, is constant. The most noteworthy example for LISA is the SGWB produced by topological defects in the scaling regime, such as a NG cosmic strings network, which is exactly scale-invariant in the LISA band for wide regions of the model parameter space (see e.g.~Ref.~\cite{Figueroa:2012kw} and the discussion in Sec.~\ref{sec:CosmicStrings}).  

Another almost scale-invariant SGWB is the one generated by slow roll inflation. This constitutes, however, an exception with respect to the cases described above. The SGWB is in fact generated as the tensor perturbations reenter the horizon during the radiation and matter eras, and therefore the spectral shape does not depend on the causal evolution properties of some source anisotropic stresses, but on the amplification of vacuum tensor perturbations during inflation. In some scenarios in which the inflaton is coupled to an external field, the SGWB is indeed produced by the field anisotropic stresses, and one can obtain blue-tilted spectra whose tilt depends on the model. GWs actively generated at second order in perturbation theory from large scalar fluctuations could also have peculiar features depending on the inflaton dynamics. The scenarios pertaining to these categories that are relevant for LISA are presented in detail elsewhere in this paper (c.f. Secs.~\ref{sec:Inflation}, \ref{sec:PBHsGWsources}, \ref{subsec:second-order-SGWB} and, for a review see Ref.~\cite{Caprini:2018mtu} and references therein).

\subsubsection{Anisotropies and propagation effects}
\label{Anisotropies and propagation effects}
Angular anisotropies in the energy density of the SGWB can be an efficient way to characterise its physical origin and properties. They provide a further tool to help in disentangling a SGWB of cosmological origin from an astrophysical one, besides the exploitation of their different frequency dependence. Angular anisotropies can be imprinted both at the epoch of the SGWB generation and at later times, during its propagation across cosmological perturbations. As such the anisotropies in the SGWB can provide a new way to characterise and distinguish various generation mechanisms of primordial SGWB and they allow one to probe the evolution of cosmological perturbations. Because the universe is transparent to GWs for sub-Planckian energies, the case of a cosmological SGWB represents a privileged observable to probe the physics of the early universe, and its anisotropies can preserve the memory of the initial conditions of the universe right after inflation.

We are interested in anisotropies and inhomogeneities in the energy density of the SGWB, therefore we allow the monopole Eq.~(\ref{OmandSh})  to be dependent on space and direction of observation 
\begin{equation}
 \Omega_{\textrm{GW}} = \frac{1}{4 \pi} \int d^2 {\hat n} \, \omega_{\textrm{GW}} ( \eta ,\, \vec{x} ,\, q ,\, {\hat n} )\, ,
\label{Omegaanisto}   
\end{equation}
thus defining the energy density contrast as 
\begin{equation}\label{dcontrast}
\delta_{\textrm{GW}} \equiv  \frac{\delta \omega_{\textrm{GW}} ( \eta ,\, \vec{x} ,\, q ,\, {\hat n} )}{\bar \omega_{\textrm{GW}} ( \eta ,\, q )}
=\frac{\omega_{\textrm{GW}} ( \eta ,\, \vec{x} ,\, q ,\, {\hat n} ) -\bar{\Omega}_{\textrm{GW}}(q,\eta)}{\bar{\Omega}_{\textrm{GW}}(q,\eta)}\, ,
\end{equation}
where $q=2\pi f$. Various approaches have been adopted to compute the angular anisotropies and their statistics (such as the angular power spectrum) both for the cosmological and astrophysical SGWB \cite{Alba:2015cms,Contaldi:2016koz,2018PhRvL.121t1303G,Bartolo:2019oiq,Bartolo:2019yeu,Cusin:2017fwz,Cusin:2017mjm, Cusin:2018avf, Pitrou:2019rjz, Bertacca:2019fnt,Cusin:2018rsq,  Jenkins:2018uac, Jenkins:2018kxc, Cusin:2019jpv, Cusin:2019jhg,Bartolo:2019zvb,DallArmi:2020dar}. A way to compute the SGWB anisotropies is to adopt a Boltzmann equation approach, similarly to CMB anisotropies \cite{Contaldi:2016koz,Bartolo:2019oiq,Bartolo:2019yeu,Cusin:2018avf, Pitrou:2019rjz}. In such an approach one considers the generation of high-frequency GW modes and their propagation across a background of lower frequency (large-scale) cosmological perturbations (which can be either scalar or tensor in nature). As for CMB photons, therefore, the propagating GWs become the cosmological carrier of the underlying cosmic inhomogeneities. Such an approach allows one to put in evidence at least two distinguishing features for a cosmological
SGWB \cite{Bartolo:2019oiq,Bartolo:2019yeu}: first the anisotropies imprinted at the production epoch can be characterised by a strong frequency dependent contribution; secondly, if primordial non-Gaussianity are present in the background large-scale cosmological perturbations, then they will be left imprinted into the SGWB anisotropies. The bispectrum of the angular anisotropies of $\delta_{\textrm{GW}}$ turns out therefore to be a new probe of primordial non-Gaussianity, potentially measurable at interferometers, beyond the CMB and large-scale structure measurements \cite{Bartolo:2019oiq,Bartolo:2019yeu}. 
For these reasons, besides the information they provide for a SGWB from inflation, anisotropies can be a new probe for a whole series of phenomena. 
They can be produced at the epoch of generation of GWs from a phase transition \cite{2018PhRvL.121t1303G,Kumar:2021ffi}, and they can characterise also the SGWB which is unavoidably produced by second-order curvature perturbations in PBH formation scenarios \cite{Bartolo:2019zvb}. Specific imprints in the SGWB anisotropies can be also generated by decoupled relativistic particles in the early universe, thus reinforcing the SGWB as a new window into the particle physics content of the universe \cite{DallArmi:2020dar}.

For a SGWB of astrophysical origin, the analytic derivation of energy density anisotropies can be found in Refs.~\cite{ Contaldi:2016koz,Cusin:2017fwz,Cusin:2017mjm, Cusin:2018avf, Pitrou:2019rjz, Bertacca:2019fnt}. When adopting a Bolzmann-like description, one needs to add an emissivity term to the Vlasov equation for the graviton distribution function, accounting for the generation process at galactic scales \cite{Cusin:2018avf, Pitrou:2019rjz}. For extragalactic background components, the primary contribution to the energy density anisotropy comes from clustering (sources are embedded in the cosmic web), while a secondary source of anisotropy is due to line of sight effects (e.g.~lensing, kinematic and volume distortion effects). Predictions for the energy density angular power spectrum have been presented in Refs.~\cite{Cusin:2018rsq,  Jenkins:2018uac, Jenkins:2018kxc, Cusin:2019jpv, Cusin:2019jhg, Bertacca:2019fnt} in the Hz band and in Ref.~\cite{Cusin:2019jhg} in the mHz band. Anisotropies are typically suppressed by a factor $10^{-1}-10^{-2}$ with respect to the monopole, the range of variability depending on the underlying astrophysical model for star formation and collapse, and the angular power spectrum scales as $\ell^{-1}$ on large scales. Different physical choices for the process of BH collapse and mass distribution lead to differences up to 50\% on the angular power spectrum in the mHz band, non degenerate with a global scaling~\cite{Cusin:2019jhg}. With LISA it may be possible to constrain the dipole and quadrupole components of the angular power spectrum, for sufficiently high SNR detection (i.e.~sufficiently high monopole)~\cite{Alonso:2020rar, Contaldi:2020rht}.

 The study of the cross-correlation of the SGWB energy density with the LSS (e.g. galaxy distribution) is an interesting subject to examine to distinguish the origin (cosmological versus astrophysical) of a given background component. Unlike a cosmological SGWB, the extragalactic astrophysical background is expected to be highly correlated with the large scale structure. Ways to exploit this feature are discussed in Sec.~\ref{sec:gw_lensing}.

\subsubsection{Polarisations}
As any background of radiation, a SGWB is fully characterised in terms of Stokes parameters, intensity (proportional to the background energy density), and $Q$, $U$, $V$ parameters describing polarisation. 
Classical diffusion of GW radiation from massive objects can generate a net polarisation out of an unpolarised flux, playing a role analogue to Thomson scattering for CMB photons~\cite{Cusin:2018avf}. The amount of polarisation that can be generated depends on the GW frequency, and it is more effective for large wavelength modes, for which wave effects are expected to be more important in an astrophysical context. An order of magnitude estimate of the effect gives that, in the mHz band, the net amount of polarisation generated by diffusion is suppressed by several orders of magnitude with respect to anisotropies in the intensity \cite{Cusin:2018avf}. As polarisation cannot be effectively generated during propagation and astrophysical background components are expected to be statistically unpolarised at emission, the detection of a highly polarised background component is a smoking gun of its cosmological origin.  

As we review in Sec.~\ref{inflation-chirality}, several inflationary mechanisms have been proposed that could produce a net circularly polarised SWGB, which is characterised by Stokes $V$ parameter.  A net chiral polarisation can be measured with a network of ground-based \cite{Seto:2007tn,Seto:2008sr,Crowder:2012ik,Smith:2016jqs} or space-based \cite{Orlando:2020oko} interferometers.~\footnote{The detection of a circularly polarised SWGB at CMB scales was studied in Refs.~\cite{Gluscevic:2010vv,Smith:2016jqs,Gerbino:2016mqb,Thorne:2017jft}.} 

The measurement is more problematic in the case of a single planar instrument such as LISA. In this case, a left-handed GW with wave-vector $\vec{k}$ produces the same effects as a left-handed GW with wave-vector $\vec{k}_p$, where $\vec{k}_p$ has been obtained from $\vec{k}$ with a reflection on the plane of the detector. Therefore a difference between the two polarisations cannot be detected in the case of an isotropic SGWB. 

A net polarisation can however be detected also by a planar instrument if the SGWB is not isotropic. As discussed in the previous subsection, It is natural to expect that a SGWB of cosmological origin has a dominant monopole component, with large-scale anisotropies of magnitude comparable to that of the CMB ones. This statement is, however, frame-dependent, and the most natural expectation is that the SGWB is isotropic in the CMB rest-frame. As seen in the CMB, the motion of the Solar System in this frame, with a velocity $v \simeq 10^{-3}$, produces a dipole anisotropy, with an amplitude suppressed by a factor $v$ with respect to that of the monopole. 
Refs.~\cite{Seto:2006hf,Domcke:2019zls} studied how the dipole signal might allow one to measure a net chirality with LISA. This can be done through the cross-correlation between the A and E channels, which vanishes both in the case of isotropic and of unpolarised SGWB. As estimated in Ref.~\cite{Domcke:2019zls}, the SNR associated with this measurement is 
\begin{equation}
{\rm SNR} \simeq \frac{v}{10^{-3}} \, \left\vert \frac{\sum_\lambda \, \lambda \, \Omega_{\rm GW}^\lambda \, h^2}{1.4 \cdot 10^{-11}} \right\vert \, \sqrt{\frac{T}{3 \, {\rm years}}} \;, 
\end{equation}
where $\lambda = \pm 1$ refers to the right and left chirality, respectively, and where $T$ is the observation time. 

\subsubsection{Non-Gaussianity}

There are two types of non-Gaussianity discussed in the context of GW observations. One is the non-Gaussianity of inhomogeneities, which is defined in position or momentum space (see e.g.~Sec.~\ref{Anisotropies-inflation}). GWs generated at sub-horizon scales cannot produce correlation across the horizon due to causality, thus the SGWB is Gaussian. Non-Gaussianity typically appears in GWs generated in the context of inflation, which could produce non-trivial spatial correlations stretched over the horizon (see Secs.~\ref{sec:Inflation} and \ref{sec:non-gaussianity}). See \cite{Bartolo:2019oiq, Bartolo:2019yeu} for a detailed derivation of the non-Gaussianity expected in the SGWB, which is simply generated by the evolution through the background large-scale underlying inhomogeneities, similarly to what happens for CMB photons. This is computed through the angular bispectra (i.e.~the three point function) of the graviton energy density.

The other type of non-Gaussianity is the one in the time signal (sometimes referred to as a SGWB in the ‘‘popcorn’’ or ‘‘shot noise’’ regime), which could be a useful statistical measure for a SGWB formed by overlapped short-duration events, such as the astrophysical background. If GW events are not frequent enough to overlap in time, the observed strain has a non-Gaussian distribution. Among the cosmological sources, the SGWB from cosmic strings could show this non-Gaussian feature \cite{Regimbau:2011bm} (see Method II in Sec.~\ref{ssec:SGWB-from-loops}). For non-Gaussianity of astrophysical sources, see Sec.~\ref{sec:sgwb:foregrounds}.

\newpage
%%%%%%%%%%%%%%%%%%%%%%%%%%%%%%%%%
% Here Sec. 6 starts

\section{First order phase transitions} 
\label{sec:PTs}

\small \emph{Section coordinators: J.~Kozaczuk, M.~Lewicki. Contributors: M.~Besancon, C.~Caprini, D.~Croon, D.~Cutting, G.~Dorsch, O.~Gould, R.~Jinno, T.~Konstandin, J.~Kozaczuk, M.~Lewicki, E.~Madge, G.~Nardini, J.M.~No, A.~Roper Pol, P.~Schwaller, G.~Servant, P.~Simakachorn.}

\subsection{Introduction}

Cosmological FOPTs are one of the most attractive sources of GWs in the early universe~\cite{Witten:1984rs,Hogan:1986qda}.
A FOPT can occur when the Higgs or any other scalar fields are trapped in a metastable vacuum in the early universe. 
As the universe cools down, thermal or quantum fluctuations drive the field over or through the potential barrier, resulting in bubbles of the stable phase nucleating in the sea of metastable phase.
These bubbles then expand and collide with each other to complete the transition.
The collision of the bubbles and the fluid motion around them produce a SGWB.
Since the generated GWs propagate to the present nearly without interaction, they maintain  information about the processes that produced them at the time of generation. Their detection may therefore reveal some of the properties of the high-energy universe.

In the standard model of particle physics (SM) there are in principle two phase transitions, at two energy scales: the scale of the electroweak gauge symmetry breaking, and that of the chiral symmetry breaking in quantum chromodynamics  (QCD). 
Given the coupling constants and the mass of the Higgs boson in the SM, it is known that the electroweak symmetry breaking is a crossover~\cite{Kajantie:1995kf,Csikor:1998eu}.
Moreover, at almost zero quark chemical potential, the QCD phase transition cannot be first-order~\cite{Stephanov:2007fk}.
However, FOPTs are predicted in a number of extensions of the SM aimed at addressing open questions such as the origin of the observed baryon asymmetry~\cite{Kuzmin:1985mm,Rubakov:1996vz}, the nature of DM, and the hierarchy problem.
 
The millhertz-band GWs that LISA will measure are in the best frequency band to explore  FOPTs occurring between the electroweak and the multi-TeV scales.
Therefore, it is important to understand how FOPTs source GWs, how such transitions arise from microphysics, and consequently to appreciate the implications that the detection of GWs from FOPTs has for particle physics.
We provide an up-to-date understanding on these topics in the following.

\subsection{Determination of the relevant parameters}

The main features of the SGWB spectrum can be determined from knowledge of four parameters related to the FOPT: the temperature $T_*$ of the plasma when the bubbles percolate, the transition strength $\alpha$, the inverse duration of the phase transition $\beta$, and the wall velocity $v_w$. These macroscopic parameters crucially depend on the underlying microscopic model, linking particle physics and cosmology. We dedicate this section to a discussion of how to compute the parameters entering the SGWB spectrum from the underlying microphysics.

The temperature at which the FOPT occurs is determined by the interplay between the nucleation of bubbles of the true-vacuum phase  and the expansion of the universe.
The nucleation rate per unit volume is given by $\Gamma = A(t) \exp[-S_c(t)]$, where $S_c$ is the critical bubble action. For thermally-induced nucleation, we have $A\sim T^4$ and $S_c = S_3/T$ with $S_3$ being the three-dimensional Euclidean action of the $O(3)$ symmetric bounce solution.\footnote{For quantum tunnelling, $S_c = S_4$ is the Euclidean action of the $O(4)$  symmetric bounce in four dimensions.}
The onset of the FOPT then occurs at the nucleation temperature $T_n$, at which on average one bubble is nucleated per horizon volume.
It is roughly given by the temperature at which $S_c \sim 140$. 
The characteristic temperature for GW production is however the temperature $T_*$ at which the bubbles percolate, when approximately one third of the comoving volume has transitioned to the true vacuum. 
For moderately strong transitions, these temperatures are typically sufficiently close to take $T_* \sim T_n$.
More precise formulas for the determination of the nucleation and percolation temperatures in from the critical action can be found in Ref.~\cite{Caprini:2019egz}.

The transition strength parameter $\alpha$ describes the amount of energy released as a fraction of the radiation energy $\rho_{\rm rad}$.
Different definitions are used throughout the literature, corresponding to different ways of mapping the particle physics model to the bag EoS commonly used for determining the efficiency of converting the released energy into bulk motion of the fluid~\cite{Espinosa:2010hh}. 
A simple but reasonably accurate definition of $\alpha$ in terms of a given particle particle physics model is obtained from the difference of the trace $\theta$ of the energy-momentum tensor between the phases, 
\begin{equation}
    \alpha = \frac{\Delta \theta}{\rho_{\rm rad}}\,,
    \qquad
    \Delta \theta = \Delta V(T) - \frac{T}{4} \frac{\partial \Delta V(T)}{\partial T} \,,
\end{equation}
where $V(T)$ is the thermal effective potential.
See Refs.~\cite{Giese:2020rtr,Giese:2020znk} for a comparison of other conventions regarding the transition strength as well as generalisations beyond the bag model.
The efficiency factors for the conversion of the energy released in the transition into bulk motion can then be calculated from the transitions strength $\alpha$ and the wall velocity~\cite{Espinosa:2010hh,Giese:2020znk}.

The calculation of the thermal effective potential, starting from an underlying particle physics model, is itself a nontrivial task.
Due to the infrared (IR) sensitivity of bosonic fields at high temperature, perturbation theory must be re-summed.
As a consequence, the effective expansion parameter increases $g^2 \to g$ (or even $g^2 \to \sqrt{g}$), and the expansion converges more slowly.
Typical one-loop calculations utilise daisy re-summation~\cite{Parwani:1991gq,Arnold:1992rz} to include the leading $\mathcal{O}(g^2)$ and $\mathcal{O}(g^3)$ contributions to the potential, while methods to include still higher order contributions have been developed~\cite{Parwani:1991gq,Arnold:1992rz,Farakos:1994kx,Braaten:1995cm,Kajantie:1995dw,Kajantie:2002wa,Ekstedt:2020abj,Gould:2021dzl}.
In addition to the slow convergence of the perturbative expansion, there are true IR divergences at $\mathcal{O}(g^6)$ (four-loop order) which re-summation does not resolve~\cite{Linde:1980ts}, underlying the importance of nonperturbative calculations~\cite{Gould:2019qek,Kainulainen:2019kyp,Niemi:2020hto,Halverson:2020xpg,Huang:2020mso}.

The bubble nucleation rate can be expanded around the percolation time $t_*$ as
\begin{equation}
    \Gamma \sim T^4 e^{-S_3/T} = \Gamma_* e^{-\beta(t-t_*)},
\end{equation}
and adiabaticity of the expansion of the universe yields $dT/dt = -TH$, so
\begin{equation}
    \dfrac{\beta}{H_*} = T_* \left.\dfrac{d(S_3/T)}{dT}\right|_{T_*}.
\end{equation}
This parameter determines whether the transition will complete mostly by the expansion of a few nucleated bubbles or by the nucleation of new bubbles everywhere in space.
Indeed, the larger $\beta/H_*$ is, the faster the nucleation rate increases in time, the more bubbles will nucleate inside a Hubble horizon before the completion of the phase transition, and the smaller will their radii be when they collide, since there had not been much time for them to expand before meeting a neighbour.
Thus one can expect an inverse relation between $\beta$ and the amplitude of the GW spectrum. Moreover, $\beta$ enters in the determination of the SGWB peak frequency.
Note that, for very strong transitions, the definition above may become inappropriate and should be reformulated~\cite{Huber:2007vva,Jinno:2017ixd}.

As with $\alpha$, the calculation of the bubble nucleation rate, $\Gamma$, is affected by the IR sensitivity of bosonic fields at high temperatures.
The spatial inhomogeneity of bubbles, as well as the time dependence of their creation, raises additional challenges.
While much effort has gone into overcoming these challenges~\cite{Gleiser:1993hf,Bodeker:1993kj,Berges:1996ib,Surig:1997ne,Strumia:1998nf,Moore:2000jw,Garbrecht:2015yza,Gould:2021ccf}, the theoretical uncertainties present in calculations of the bubble nucleation rate are less well understood than those for the effective potential.

The determination of the wall velocity from first principles is a much more daunting task. The passage of the wall perturbs the equilibrium of particles in the plasma, so that the temperature, velocity and chemical potential of each species are different from the background. The distribution function of each species can be written as $f_i = f_i^\text{eq}+\delta f_i$, and the perturbations can be calculated by solving the corresponding Boltzmann equation~\cite{Moore:1995si, Konstandin:2014zta},
\begin{equation}
    (p^\mu \partial_\mu + K_\mu \partial_{p_\mu})\, f + C[f] = 0,
    \label{eq:Boltzmann}
\end{equation}
where $K_\mu$ is (related to) the force term felt by the particles due to the passage of the bubble, and $C[f]$ are collision terms that must be computed from the microphysics of the model under consideration, taking into account the mutual interactions of particles present in the plasma at the electroweak phase transition (EWPT); see e.g.~Refs.~\cite{Moore:1995si, Kozaczuk:2015owa}. The Higgs equation reads~\cite{Konstandin:2014zta}
\begin{equation}
    \Box\phi + \dfrac{dV (\phi,T)}{d\phi} + \sum_i \dfrac{dm_i^2}{d\phi} \int \dfrac{d^3p}{(2\pi)^3 2E_p} \delta f_i = 0,
\end{equation}
and the term with fluid perturbations $\delta f_i$ acts as a friction countering the bubble expansion. From this equation one obtains the wall velocity and the wall width, typically by using an ansatz for the Higgs profile
\begin{equation}
    \phi(z) = \dfrac{\phi_0}{2}\left[1-\tanh\left(\dfrac{z}{L_w}\right)\right]
\end{equation}
and solving for the width $L_w$ and the velocity $v_w$. Although the wall velocity has been computed from first principles for a few models, such as in SUSY extensions~\cite{John:2000zq, Kozaczuk:2014kva}, Higgs plus singlet setups ~\cite{Kozaczuk:2015owa,Lewicki:2021pgr} and two Higgs doublet models~\cite{Dorsch:2016nrg}, the determination of $v_w$ as a function of the parameters of the underlying theory remains an open issue for most of the interesting models discussed in the literature. Crucially, the  collision terms appearing in the Boltzmann equation~(\ref{eq:Boltzmann}) have been calculated only for a few models, such as the SM~\cite{Moore:1995si} and singlet extensions~\cite{Kozaczuk:2015owa}, and even then are known only at a leading-log approximation. For most models, additional collision terms involving interactions of new particles among themselves and with the SM sector have not yet been computed. Altogether, it is fair to say that our current tools allow only for an estimate of the wall velocity, and there is still a large room for improvement on this front\footnote{In this respect, a promising alternative direction is the use of a string theory-inspired method~\cite{Maldacena:1997re} known as gauge/string duality, or holography. The power of this tool is that it allows for a first-principle determination of out-of-equilibrium observables in a four-dimensional quantum field theory. It was used in~\cite{Bea:2021zsu,Bigazzi:2021ucw,Bea:2022mfb} to compute the bubble wall velocity in a family of strongly-interacting, four-dimensional gauge theories.}.

\subsection{GW Sources}

A FOPT in the early universe occurs when at least one scalar field gets trapped at the symmetric metastable vacuum.
As the universe cools down, the scalar field tunnels through or thermally fluctuates over the barrier into the lower-energy vacuum, and at a certain temperature bubbles in the broken phase start to nucleate in the sea of the symmetric phase.
The latent heat released in the transition drives these bubbles to expand further, and the transition completes when these bubbles collide and merge with each other.
During bubble expansion, the latent heat is converted into the energy of the bubble wall and the thermal and kinetic motion of the surrounding plasma. However, due to spherical symmetry of the bubbles, no GWs are generated during this expansion period. It is only when the bubbles collide, at the end of the FOPT, that GWs are produced from the kinetic energy of the scalar field and the fluid. Since there are numerous independent collision regions today within the typical resolution of GW detectors, the resulting spectrum is stochastic. 

\begin{figure}
    \centering
    \includegraphics[scale=1.]{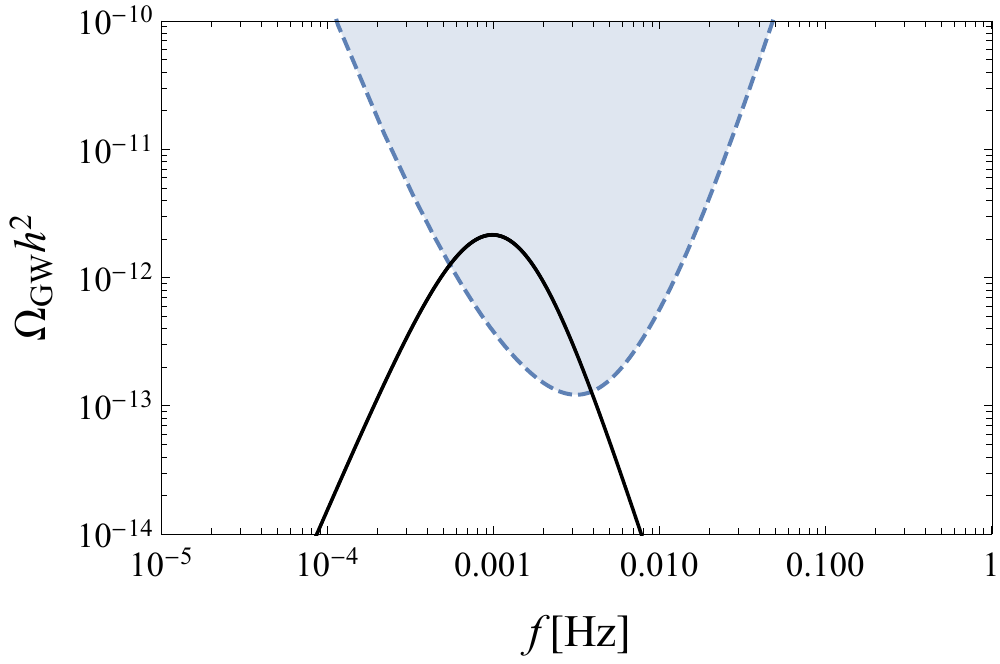}
    \caption{\small An example of the GW power spectrum from a FOPT, along with the LISA power-law-sensitivity (PLS)  curve of SNR\,$=10$. The model parameters used in this example are $v_w = 0.9$,
$\alpha = 0.1$, $\beta/H_* = 50$, $T_* = 200$ GeV, $g_* = 100$.
    %This plot was created using data from PTPlot version 1.0.1.
    }
    \label{fig:shape}
\end{figure}

The amount of GWs generated during a cosmological FOPT is quantified by the fractional energy density of these waves in the universe, $\rho_\text{GW}$, compared to the critical energy density, $\rho_c\equiv 3H^2/(8\pi G)$, where $H$ is the Hubble constant. One can rephrase the GW density parameter in Eq.~\eqref{Omegagw} as
\begin{equation}
    \Omega_\text{GW} \equiv \frac{1}{\rho_c}\frac{d \rho_\text{GW}}{d \ln k}\,.
\end{equation}
For a stochastic source, the spectrum is proportional to the two-point correlation of the metric perturbations $\langle \dot{h}(\mathbf{k},t) \dot{h}(\mathbf{k}^\prime, t)\rangle \equiv (2\pi)^3 \delta^{(3)}(\mathbf{k}-\mathbf{k}^\prime)P_{\dot{h}}(k)$, as
\begin{equation}
    \Omega_\text{GW} = 
    \frac{k^3}{24\pi^2 H^2} P_{\dot{h}}(k). 
\end{equation}
Since sufficiently distant regions could not have exchanged information with one another during the FOPT, the correlation function for low $|\mathbf{k}|$ corresponds to white noise, i.e.~$P_{\dot{h}}(\mathbf{k},t)\sim$ constant, and the spectrum grows as $k^3$. This growth cannot continue forever, so this pattern must eventually be broken and the spectrum must decrease at some point, for the total energy in GWs must be finite. Therefore, the SGWB spectrum must contain a peak. A typical shape is illustrated in Fig.~\ref{fig:shape} created using data from PTPlot version 1.0.1.~\footnote{
    Since the release of PTPlot version 1.0.0 there have been several modifications of PTPlot. This includes taking into account the recent erratum from Ref.~\cite{Hindmarsh:2017gnf} which led to the amplitude of the GW signal being overestimated by a factor of 10. For a more complete list of changes in version 1.0.1, see~\href{https://www.ptplot.org/ptplot/}{ptplot.org}.
    }

There are three main sources of GWs from bubble collisions: (i) the kinetic energy of the scalar field along the bubble walls, (ii) coherent motion of the plasma generated by the bubble expansion, also known as ``sound waves'', and (iii) turbulent motion of this fluid, 
typically expected to occur at late times after the fluid develops vortical motion. The total contribution to GW energy from FOPTs in the early universe is therefore the sum
\begin{align}
\Omega_{\rm GW}
&= 
\Omega_{\rm GW}^{\rm (coll)} + \Omega_{\rm GW}^{\rm (sw)} + \Omega_{\rm GW}^{\rm (turb)}.
\end{align}
How these three contributions arise in principle depends on the detail of the model.
However, as discussed in the previous section, it is known that a few parameters determined from microphysics are enough to discuss cosmological consequences: transition strength $\alpha$, nucleation speed $\beta$, wall velocity $v_w$, and the transition temperature $T_*$. 
These parameters determine the fluid profile of the expanding bubbles and the energy budget of the transition~\cite{Espinosa:2010hh,Giese:2020rtr}.

The transition dynamics can vary dramatically for different $\alpha$.
In typical transitions, in which $\alpha$ is moderate, the friction on the wall balances against the pressure from the released energy before bubbles collide, and as a result the walls approach a constant terminal velocity.
In this case most of the energy released in the transition is transferred to the surrounding plasma, and only a negligible fraction remains in the scalar field.
Macroscopically the energy transferred to the plasma can be classified into heat and bulk kinetic motion of the fluid.
The bulk motion takes the form of compression waves, or sound waves.
As long as the linearity of the fluid equation holds, i.e.~the kinetic energy in play is sufficiently low, sound waves from different bubbles continue to propagate even long after the transition completes, and they overlap to create random velocity field with typical length scale of the sound shell $l_{\rm shell}$.
This phenomenon is found in numerical simulations~\cite{Hindmarsh:2013xza,Hindmarsh:2015qta,Hindmarsh:2017gnf}, and  Refs.~\cite{Hindmarsh:2016lnk,Hindmarsh:2019phv} provide a corresponding analytic model  based on the picture of an ensemble of overlapping shells.
Interestingly, the velocity field sources GWs at the constant scale $l_{\rm shell}$ even well after the completion of the transition (provided that the transition is not too strong). Therefore, it works as an efficient source of GW production.
The resulting GW spectrum $\Omega_{\rm GW}^{\rm (sw)}$ contains two important characteristic scales of the bubble size and the fluid shell thickness.
A hybrid scheme to calculate the GW spectrum from this phenomena has been proposed recently~\cite{Jinno:2020eqg}.

The sound waves are expected to decay through the formation of non-linearities in the flow, after which the fluid enters a turbulent phase. The non-linearities develop on a time scale $\tau_{\rm sh}$, which is inversely proportional to the kinetic energy in the fluid. It has been shown that, even for weak transitions, $\tau_{\rm sh}$ is typically smaller than a Hubble time~\cite{Ellis:2020awk}, with $\tau_{\rm sh}$ becoming smaller as $\alpha$ increases. The decay of sound waves produced during a FOPT and the subsequent development of a turbulent flow has not been studied so far using direct numerical simulations and remains a major source of uncertainty when determining the GW spectrum. This holds in particular for strongly FOPTs, when the fluid kinetic energy is expected to be high, and the contribution of kinetic turbulence to the SGWB signal is expected to be relevant. Within the sound wave regime, typically the larger the transition strength the greater the production of GWs. FOPTs with smaller $\beta$ which produce fewer, larger bubbles are also preferred. For a detailed discussion on the dependence of the SGWB spectrum from sound waves on $\alpha$ and $\beta$, see Ref.~\cite{Caprini:2019egz}.

For deflagrations with $\alpha\sim \mathcal O(0.1)$ the sound shell picture breaks down~\cite{Cutting:2019zws}. In such transitions, substantial vorticity can be generated during the initial collision of sound shells and bubbles. Furthermore, hot high pressure regions of the metastable phase can form during the transition.
These hot regions suppress further bubble nucleation and decrease the wall velocity at the initial stage of the transition~\cite{Konstandin:2010dm,Megevand:2017vtb}, while they form droplets at the later stage, extending the duration of the transition and suppressing the production of kinetic energy. Both the generation of vorticity and the formation of droplets can modify the GW spectrum, either by speeding up the decay of sound waves or by reducing the fluid kinetic energy and thus GW spectrum amplitude. Further study is required in this regime to map out the production of vorticity and study the evolution of flows with mixed modes, as well as to find the suppression of kinetic energy from droplets in a wide range of parameter space. It is however worth noting that the wall velocity still needs to be computed as a function of parameter space for many BSM models. It is therefore unclear if there are many realistic models that would give rise to the strong transitions with subsonic wall velocities which would be affected by these results. 

Another interesting scenario is extremely strong transitions, in which $\alpha \gg 1$ is realized~\cite{Randall:2006py,Espinosa:2008kw,Konstandin:2011dr,Hambye:2013sna,Jaeckel:2016jlh,Jinno:2016knw,Marzola:2017jzl,Iso:2017uuu,Chiang:2017zbz,vonHarling:2017yew,Bruggisser:2018mus,Bruggisser:2018mrt,Hambye:2018qjv,Baldes:2018emh,Hashino:2018wee,Prokopec:2018tnq,Brdar:2018num,Marzo:2018nov,Breitbach:2018ddu,Baratella:2018pxi,Fairbairn:2019xog} but the bubble walls still reach a terminal velocity because of the higher-order friction terms on the wall~\cite{Bodeker:2017cim,Hoeche:2020rsg}.
In this case the relevant hydrodynamical solution is strong detonation.
While the dominant GW source comes from fluid kinetic motion, we must note several things:
first, the onset of turbulence is earlier in this type of model~\cite{Ellis:2020awk};
second, the overlap of sound shells, which is one of the requirements for the linear growth of the GW spectrum, may be somewhat delayed~\cite{Jinno:2019jhi}. Finally, in typical models there exists a maximum $\alpha$ for which transitions complete. Above this, vacuum energy dominates and the metastable state begins to inflate, prohibiting the percolation of the bubbles of the true vacuum~\cite{Ellis:2018mja}\footnote{In cases where the phase transition through thermal bubble nucleation fails other mechanisms such as vacuum fluctuations~\cite{Lewicki:2021xku} or spinodal instability~\cite{Bea:2021zol} can lead to completion of the transition which would have a very different phenomenology and clearly distinguishable GW spectra}.

When $\alpha$ exceeds a certain threshold, or simply in vacuum transitions, the bubble walls continue to accelerate until they collide (``run away").
While recent studies of the next-leading friction terms on bubble walls~\cite{Bodeker:2017cim,Hoeche:2020rsg} suggest that the parameter space for runaway is much smaller than previously thought~\cite{Bodeker:2009qy}, such behaviour can still occur in extremely strong transitions~\cite{Ellis:2019oqb,Ellis:2020nnr}.
In this case the only contribution is from the bubble wall $\Omega_{\rm GW}^{\rm (coll)}$.
The expanding and colliding bubble walls are highly relativistic and much thinner compared to the typical bubble size.
This observation motivates the so-called envelope approximation~\cite{Kosowsky:1992rz,Kosowsky:1992vn}, in which the shear stress is approximated to be localised in an infinitesimally thin shell at the bubble wall and disappears upon collision~\cite{Huber:2008hg,Jinno:2016vai}. However, recent studies revealed that for runaway and vacuum transitions the collided region of intersecting bubbles has a rich structure and cannot be neglected~\cite{Weir:2016tov,Cutting:2018tjt,Cutting:2020nla,Jinno:2019bxw}. The GW spectrum that takes collided regions into account has been actively studied numerically for two bubble ~\cite{Lewicki:2019gmv,Lewicki:2020jiv} and many bubble collisions~\cite{Cutting:2018tjt,Cutting:2020nla} as well as (semi)-analytically~\cite{Jinno:2017fby,Konstandin:2017sat}.  One of the implications is that, while the peak of the spectrum does not grow in amplitude after collisions as occurs for sound waves, the spectrum may have a growing structure towards the IR, which enhances the detection prospect by LISA.

In addition to the aforementioned sources, magnetic fields could be present prior
to or generated during a thermal FOPT.
In the presence of a magnetic field, the
turbulent motion of the primordial plasma would become dynamically coupled to
the magnetic field, leading to MHD turbulence
\cite{Brandenburg:2017neh, Brandenburg:1996fc, Christensson:2000sp}.
Moreover, the hydrodynamic turbulent motion induced by the expansion of the 
bubbles arises in an ionised plasma, and this fact in itself can lead to MHD turbulence.
The dynamical coupling between the velocity and the magnetic field
plays an important role in determining the shape of the GW signal.
The latter is affected by the presence of the
magnetic field both through the aforementioned dynamical coupling,
and because the magnetic anisotropic stresses produce GWs on their
own, as studied in early analytic
works \cite{Deryagin:1986qq, Kosowsky:2001xp, Caprini:2006jb, Gogoberidze:2007an,Caprini:2009yp}.
So far, the hydrodynamical simulations of thermal FOPTs have neglected magnetic fields.
The early analytic works on the subject require one to make
assumptions on the temporal correlation functions of the turbulence
velocity and the magnetic field, which strongly affect the SGWB spectral shape \cite{Caprini:2009fx}.
Previous analytical works were extended  with updated
modelling of the MHD turbulence in Ref.~\cite{Niksa:2018ofa}.
Numerical simulations  avoid resorting to analytical modelling of the MHD turbulence evolution, since they compute directly
the solution to MHD equations \cite{Pol:2019yex, Kahniashvili:2020jgm}.
Recent numerical simulations and previous analytic estimates agree on the
shape of the GW spectrum at high frequencies, scaling as $f^{-8/3}$.
However, the dynamical evolution of the magnetic field during the GW production
can affect this slope~\cite{Pol:2019yex}.
At intermediate frequencies, the numerical simulations of
Ref.~\cite{Pol:2019yex} predict a spectrum  $\Omega_{\rm GW}(f)\propto f$,
that eventually is expected to turn to $f^3$ at frequencies 
in the super-horizon range, due to causality \cite{Caprini:2009fx}.
$\Omega_{\rm GW}(f)$ evaluated from the numerical simulations grows as $f^3$ at early times, shifting to a linear increase $\propto f$  as the time progresses and GWs are building up.
The detail of the transition from $f$ to $f^3$ is an
active topic of research.
Combining numerical simulations with our theoretical understanding of
the MHD turbulence dynamics will help to obtain an accurate prediction of the spectral shape of the SGWB produced by MHD turbulence in the future.
As is the case  for other sources, the SGWB by MHD turbulence also depends on the parameters
of the FOPT.
The strength of the FOPT and the size of the bubbles can be related
to the kinetic energy density, and to the characteristic scale of turbulence,
respectively.

This picture can get more complicated in the case of magnetic field production
during the FOPT, and it would require an appropriate treatment
of the magnetic field in both the false and true vacuum regions, which
are dynamically converting through the bubble-driven FOPT.
During the FOPT, inhomogeneities of the Higgs field give
rise to the production of magnetic fields \cite{Vachaspati:1991nm, Ahonen:1997wh,Stevens:2007ep,Vachaspati:2020blt,Zhang:2019vsb,Yang:2021uid}.\footnote{Current studies consider vacuum FOPTs.
The extension to a thermal FOPT is not straightforward, and it requires further analysis.}
The previous mechanism does not require  CP violations,
however these are expected in relation to baryogenesis
through spontaneous lepton number symmetry 
breaking at a FOPT \cite{Cohen:1991iu}.
This parity violation
leads to the production of helical magnetic fields \cite{Vachaspati:1991nm, Cornwall:1997ms,Vachaspati:2020blt,GarciaBellido:2002aj}.
Such parity-violating turbulent sources lead to the 
production of circularly polarised GWs, studied analytically in Refs.~\cite{Kahniashvili:2005qi,Kahniashvili:2008pe,Kisslinger:2015hua,Ellis:2020uid}, and computed numerically in Ref.~\cite{Kahniashvili:2020jgm}.
Further studies are required in this direction for a clear understanding of
the GW and the polarisation signals produced from MHD turbulence
as a function of the FOPT parameters, and a computation of turbulence
in the plasma from first principles.

\subsection{Discovery prospects Beyond the Standard Model} 
\begin{figure}
    \centering
    \includegraphics[width=.95\textwidth]{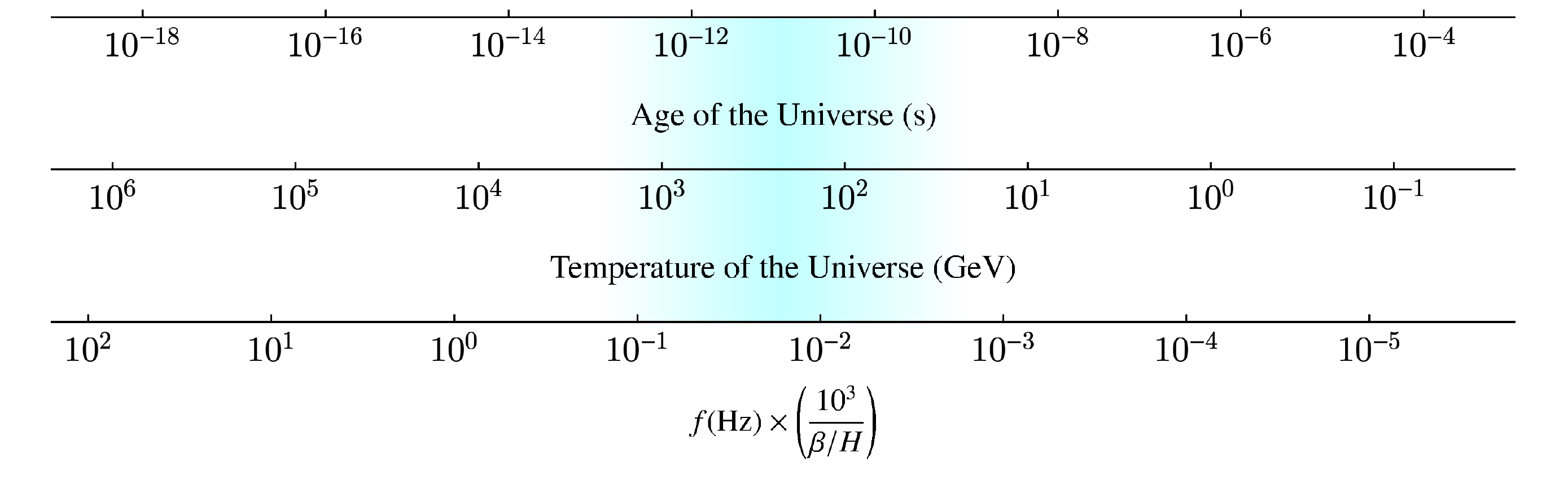}
    \caption{\small LISA sensitivity to the SGWB frequency peak of  FOPTs in the early universe. Here the fiducial values $v_w = 1$ and $g=106$ are used to relate the frequency peak of the acoustic spectrum \cite{Caprini:2019egz} to a time and temperature in the early universe, as a function of the transition rate parameter $\beta/H$.}
    \label{fig:cosmictimeline}
\end{figure}

GWs are a unique probe of the physics of the early universe. Unlike photons, gravitons were not in thermal equilibrium at early times, and may therefore directly encode information about events long before the time of last scattering. 
The characteristic broken power-law spectrum from a FOPT peaks at a frequency which can be related to a temperature and a time in the early universe via its redshift. Fig.~\ref{fig:cosmictimeline} shows this relation using the frequency peak of the acoustic spectrum, described in Ref.~\cite{Caprini:2019egz}. It is seen that for appropriate fiducial values, a FOPT at the electroweak scale temperature, $T\sim 10^2$ GeV, sources a SGWB spectrum peaking at frequencies to which LISA has its best sensitivity~\cite{Grojean:2006bp, Caprini:2015zlo, Caprini:2019pxz}.
The sensitivity to weak scale physics implies an opportunity to study FOPTs of several different kinds. In particular, several proposals for FOPTs that supply the out-of-equilibrium circumstance for baryogenesis are anchored to the weak scale. 
However, LISA is also sensitive to energy scales much beyond the electroweak scale: for exceptionally strong FOPTs the peak can fall outside the LISA sensitivity region but the tail of the  signal still has a high SNR. In this case, FOPTs even at the MeV or PeV scale can be detected.
This feature opens up potential complementary access in other experiments, as will be described in the next section. 

\subsubsection{First-order phase transitions in the LISA window}
FOPTs are studied in the context of a variety of models, which can be divided into two main categories: the spontaneous breaking of a gauge symmetry, and confinement due to the strong coupling of a gauge symmetry (and the related chiral symmetry breaking). The EWPT, which is particularly interesting in light of electroweak baryogenesis (see Sec.~\ref{sec:baryogenesis}), is an example of the former. The transition temperature associated with electroweak symmetry breaking implies that the peak frequency of the GW spectrum falls within the LISA sensitivity window, as seen in Fig.~\ref{fig:cosmictimeline}. As is well known, the EWPT is a crossover transition in the SM.  
Therefore, new physics is required to induce a FOPT that can produce a SGWB.
FOPTs can be studied in a model-independent effective-field-theory approach, adding non-renormalisable operators to the Higgs potential. These analyses show that for the EWPT to be sufficiently strongly first-order, the cut-off scale must be rather low, below the TeV scale~\cite{Zhang:1992fs,Grojean:2004xa,Delaunay:2007wb,Chala:2018ari}, therefore studies of the EWPT in SM extensions generically require the introduction of  at least one new degree of freedom at the $\mathcal{O}(100)$ GeV scale.

A weak-scale hidden sector featuring the spontaneous breaking of a gauge symmetry could also lead to a SGWB in the LISA window. An observable amplitude of the signal implies a significant fraction of the radiative degrees of freedom must have participated, hinting at a connection with DM \cite{Schwaller:2015tja,Jaeckel:2016jlh,Croon:2018erz,Bertone:2019irm}. The spectral form of such a signal is unlikely to give away much of its origin, though it can be related to particle spectra in specific models \cite{Croon:2018erz}. Complementary search strategies may be applied if such a spectrum is detected -- more about this in Sec.~\ref{sec:complementarity}.

Confining gauge theories are thought to feature a FOPT if the confined degrees of freedom either only include the gauge fields themselves, or if confinement implies the breaking of a chiral symmetry ${\rm SU}(N_f)\times {\rm SU}(N_f)$ and the number of chiral fermions (dynamical at the transition temperature) exceeds $N_f \geq 3$. The foundation for this insight dates back to an argument based on the linear sigma model \cite{Pisarski:1983ms} and has since been supported in part by lattice simulations \cite{Iwasaki:1995ij}. Studying the phenomenology of the transition is challenging, as effective theories either apply to the low-temperature (confined) phase and the high-temperature free phase. Moreover, an important role is played by the instantons of the strongly coupled theory. Initial explorations based on the linear sigma model indicate stronger SGWB are found for gapped spectra \cite{Croon:2019iuh}. However, the predictions depend on the chiral symmetry breaking model used \cite{Helmboldt:2019pan}. 

Another alternative is solitosynthesis, a FOPT resulting from the growth of non-topological solitons stabilised by a conserved global charge \cite{Griest:1989bq}. Unlike phase transitions resulting from the nucleation of critical bubbles, stable sub-critical populations of such solitons -- Q-balls -- exist, and accumulate charge until they reach a critical size and grow.
This slow process may imply supercooled transitions still complete \cite{Kusenko:1997hj,Croon:2019rqu}.
This scenario typically relies on the existence of a global symmetry under which the universe is asymmetrically charged. It has therefore been studied in the context of supersymmetry \cite{Kusenko:1997hj,Postma:2001ea,Pearce:2012jp} and asymmetric DM \cite{Croon:2019rqu}. 

As described in the previous section, the SGWB due to a cosmological FOPT can be described in terms of a small number of thermal parameters: characteristic temperature, bubble wall velocity, latent heat and a dynamical parameter such as a nucleation rate. These thermal parameters then predict a spectrum with a shape in most cases primarily defined by just two parameters: the peak frequency and peak amplitude. Therefore, the SGWB-inverse problem typically features degeneracies. However, resolution of the full spectrum and in particular the peak gives a unique picture of a crucial stage of the early universe, and may play an important role in answering some of the most fundamental questions.
Further resolution of the underlying BSM physics can also be accomplished through complementary experimental observation.

\subsubsection{Connection with baryogenesis}\label{sec:baryogenesis}

An  (electroweak) FOPT may also be responsible for the generation of the observed baryon asymmetry. This provides an appealing connection between baryogenesis and the direct probe of the electroweak epoch in the early universe through GWs especially since the corresponding signals would peak in the LISA band. At such a first order EWPT, satisfying simultaneously the three Sakharov conditions for baryogenesis (baryon number violation, CP violation and a departure from thermal equilibrium) becomes possible: baryon number is not an exact symmetry of the SM, violated via non-perturbative processes involving gauge and Higgs fields.\footnote{More precisely, the baryonic current $J^\mu_B$ is anomalous, and the anomaly $\partial_\mu J^\mu_B = (N_c g^2 ) / (32\pi^2) F^{\mu\nu}\widetilde{F}_{\mu\nu}$ allows for the creation of baryons through the dynamics of the non-abelian gauge fields.} In the presence of CP violation, these processes are biased towards producing more baryons than anti-baryons, and because the FOPT introduces a departure from thermal equilibrium, the reverse mechanism is suppressed, ensuring that the generated asymmetry is not washed-out and remains to the present day.

Among these \emph{electroweak baryogenesis} mechanisms, those most widely studied are \emph{non-local} ones, involving charge transport (see~\cite{Morrissey:2012db, Konstandin:2013caa} for reviews): CP violation in the scatterings of plasma particles with the phase transition boundary yields an excess of fermion handedness, which diffuses into the electroweak symmetric phase and is there converted to a baryonic excess via SM baryon number violating processes known as \emph{sphalerons}; the expansion of the FOPT bubbles then leads the net baryonic excess inside the bubble, where the asymmetry is frozen-out up to the present day. This last condition depends on the sphaleron rate inside the bubble, $\Gamma_\text{sph} \sim e^{-E_\text{sph}/T}$ (with $E_\text{sph}$ the sphaleron energy). Sufficient suppression of $\Gamma_\text{sph}$ for successful baryogenesis (i.e.~$\Gamma_\text{sph}$ much smaller than the Hubble expansion rate) yields the condition of a strong EWPT, $v/T \gtrsim 1$, where $v$ is the electroweak vacuum expectation value after the phase transition.

Up to very recently it has been assumed that these electroweak baryogenesis scenarios, as they rely on diffusion of particles in the plasma, were not effective for fast moving bubble walls (i.e.~faster than the speed of sound of the plasma, with the bubbles consequently expanding as detonations). 
On the other hand, GWs are more favourably produced by stronger transitions resulting in faster bubble walls.
This seemed to put a tension between the generation of a sizeable GW background from the FOPT (e.g.~large enough to be observable by LISA) and successful baryogenesis. However, recent works~\cite{Cline:2020jre,Laurent:2020gpg,Dorsch:2021ubz} indicate that while diffusion is not as efficient for supersonic bubble walls, it still allows for successful electroweak baryogenesis. It has also been shown that 
the production of a large GW signal can be made compatible with electroweak baryogenesis for bubbles expanding as subsonic deflagrations~\cite{Dorsch:2016nrg}, yet in this case the plasma friction against the moving wall must be large enough that even quite strong transitions remain subsonic, which requires the presence of sub-TeV particles with relatively large coupling to the Higgs --- a tightly constrained scenario by now.

The strongest bound on non-local electroweak baryogenesis comes from the non-observation of an electric dipole moment (EDM) of the electron by highly precise experiments~\cite{Andreev:2018ayy}. The additional sources of CP violation required by these transport mechanisms typically impact the electron EDM already at 2-loop order via Barr-Zee diagrams~\cite{Engel:2013lsa} --- two orders above the SM CP violation contribution, which manifests itself only when all three families of quarks are involved, thus affecting EDMs at 4-loop only, being therefore highly suppressed and unconstrained by EDMs. Whether electroweak baryogenesis is still viable under such tight bounds remains to be investigated. 
There are, however, models in which it might be possible to avoid those constraints. One possible example involves coupling the CP violation to a vacuum expectation value of an additional scalar field which vanishes after the transition rendering the model safe from EDM constraints~\cite{Espinosa:2011eu,Cline:2012hg,Vaskonen:2016yiu,Cline:2021iff}.
Alternatively, CP violation can be secluded in a dark sector, and  
communicated to the visible sector in a way that the corresponding EDM contribution is suppressed or absent~\cite{Cline:2017qpe,Carena:2018cjh,Carena:2019xrr}.
Yet another way is to increase the temperature of the EWPT  from symmetry non-restoration phenomena~\cite{Baldes:2018nel,Glioti:2018roy,Matsedonskyi:2020mlz,Biekotter:2021ysx,Carena:2021onl}. The observable effect is a GW  peak position shifted towards
higher frequencies.

Finally, there are also \emph{local baryogenesis} mechanisms where baryon production and CP violation occur at the same point in space, involving non-trivial gauge and Higgs field dynamics~\cite{Tranberg:2010af, Konstandin:2011ds}. 
However, these mechanisms are highly inefficient at electroweak scale temperatures~\cite{Lue:1996pr}, although they can still be relevant in more exotic scenarios of very supercooled FOPTs. The main advantage of such cold electroweak baryogenesis mechanisms is their ability to generate the observed baryon asymmetry with just the SM CP violation from the Cabibbo-Kobayashi-Maskawa matrix~\cite{Tranberg:2010af}, or from the strong CP phase~\cite{Servant:2014bla}, thus avoiding the tight bounds from EDMs~\cite{Andreev:2018ayy}. Since the thermal plasma is very diluted at the end of the transition due to the supercooling, the GWs are sourced mainly by the kinetic energy of the bubbles.

\subsection{LISA complementarity to other experimental tests}
\label{sec:complementarity}
The potential of LISA to explore FOPT is
complementary
to other experimental tests such as searches for beyond
SM physics at colliders, tests from flavour
physics including neutrinos physics
and B mesons physics or searches for signals from the 
dark sector.
LISA detection of SGWB from FOPTs can not only explore the electroweak energy
scale but could
explore higher energy scales which could be inaccessible to
present and future colliders. The other GW observatories available at the time that LISA flies, can also strengthen and complement the LISA findings.

\begin{figure}
    \centering
    \includegraphics[scale=.5]{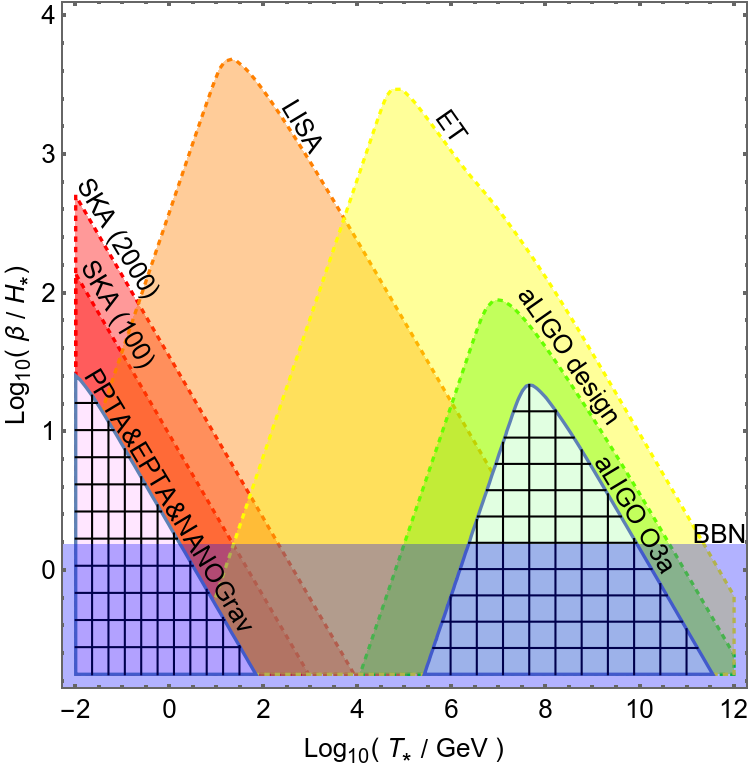}
    \caption{\small The parameter reach of present and future GW observational network for phase transitions in a runaway-like regime with $v_w \simeq 1$ and $\alpha \gg 1$.  In each shaded area the corresponding experiment (see labels) detects the FOPT signal with SNR\,$>\!10$.
The region labelled BBN, PTA and aLIGO O3 are ruled out. Figure adapted from Refs.~\cite{Figueroa:2018xtu, Megias:2020vek}.}
    \label{fig:FOPTforecast}
\end{figure}

Fig.~\ref{fig:FOPTforecast}  sketches the FOPT energy scales that LISA can probe with SNR\,$>\!10$ as well as its complementarity with other GW observatories expected in the early 2040's. The figure also displays the parameter region ruled out by current BBN, pulsar timing array (PTA) and LIGO/Virgo data.\footnote{The figure assumes that the SGWB hints arising in the present PTA analyses, are not of primordial origin. The energy scales detectable by LISA result broader than in Fig.~\ref{fig:cosmictimeline} since scenarios with the SGWB frequency outside the LISA band can still fulfil the SNR\,$>\!10$ requirement; see Refs.~\cite{Figueroa:2018xtu, Megias:2020vek} for details.}

The EWPT in the SM is a second order phase transition.
However, there is a wealth of BSM scenarios in
which the EWPT is of the first
order leading to sizeable SGWB signals. Without claiming to be exhaustive, one can for example mention the extension of the scalar sector of the SM with
extra singlet(s)~\cite{Craig:2013xia,Profumo:2014opa,Craig:2014lda,Hashino:2016xoj,Beniwal:2017eik,Kang:2017mkl,Matsui:2017ggm,Alves:2018oct,Beniwal:2018hyi,Ahriche:2018rao,Chala:2019rfk,Chiang:2019oms,Alves:2019igs,Alves:2020bpi,Shajiee:2018jdq,Vieu:2018zze,Morais:2019fnm} (although some could come without
collider traces~\cite{Ashoorioon:2009nf}), some having been
discussed in the context of LIGO~\cite{Balazs:2016tbi}, or
doublet(s)~\cite{Fromme:2006cm, Dorsch:2013wja,Dorsch:2014qja,Laine:2017hdk, Wang:2019pet,Zhou:2020xqi} (see also~\cite{Fujikura:2018duw} for an example from twin
Higgs models), 
composite Higgs models~\cite{Chala:2016ykx, Bruggisser:2018mrt,DeCurtis:2019rxl,Bian:2019kmg,Xie:2020bkl}, so-called
fermionic extensions of the SM involving for 
example $SU(2)_L$  doublet and/or two $SU(2)_L$ singlet
vector-like leptons strongly coupled
to the Higgs boson~\cite{Angelescu:2018dkk} or the various supersymmetric
extensions of the SM~\cite{Davies:1996qn,Apreda:2001us,Huber:2007vva,Huang:2014ifa,Kozaczuk:2014kva,Huber:2015znp,Demidov:2017lzf}.

There is also a wealth of theoretical BSM setups  in which
a FOPT can occur at a higher energy scale than
the EW energy
scale. These setups for example encompass extensions of the
scalar sector~\cite{Jinno:2015doa,Alanne:2019bsm,Baldes:2018nel},
classical conformal approaches~\cite{Brdar:2018num} or nearly conformally
invariant field theories
in which the generation of neutrino masses is linked to
spontaneous scale 
symmetry breaking~\cite{Agashe:2019lhy}, gauge extensions of the SM such as
models including 
$U(1)_{B-L}$~\cite{Madge:2018gfl,Ellis:2019oqb,Hasegawa:2019amx,Dev:2019njv,Zhou:2020idp,Ellis:2019tjf,Ellis:2020nnr} or left-right symmetries~\cite{Brdar:2019fur,Fornal:2020ngq},
grand unified theories, 
non-supersymmetric~\cite{Croon:2018kqn,Huang:2020bbe,Buchmuller:2019gfy,Coriano:2020kyb,Okada:2020vvb} or supersymmetric~\cite{Haba:2019qol},  
extra-dimension models and in particular the one including
warped 
spacetimes~\cite{Randall:1999ee,Goldberger:1999uk,Rattazzi:2000hs,Creminelli:2001th,Garriga:2002vf,Randall:2006py,Nardini:2007me,Konstandin:2010cd,Konstandin:2011dr,Bunk:2017fic,Megias:2018sxv,Megias:2018dki} (often coming together with a stabilisation mechanism such as the Goldberger-Wise mechanism~\cite{Goldberger:1999uk} and holographic duals of the Composite Higgs models mentioned above) or gauge-Higgs unification approaches~\cite{Adachi:2019apm}.

Many of these BSM models are being explored at current
particle colliders
such as the LHC. Searches performed at the LHC experiments
such as
ATLAS and CMS using the data from Run1 and Run2 did not
find any
evidence for BSM signals and have allowed one to put
constraints on
masses and couplings of their corresponding predicted
new particles.

\begin{figure}
    \centering
    \includegraphics[scale=.5]{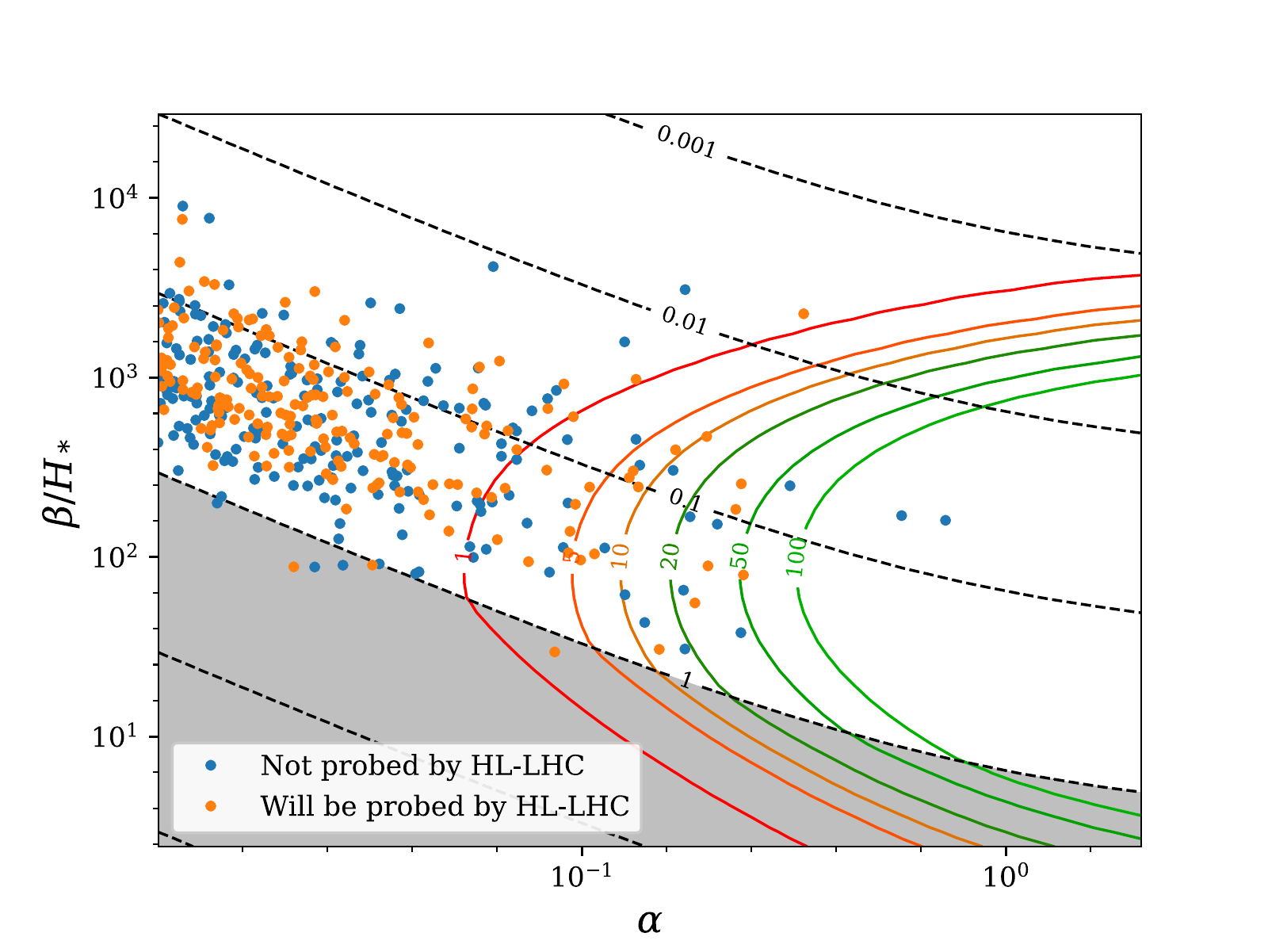}
    \caption{\small Typical values of $\alpha$ and $\beta/H_*$ for the general singlet model, superimposed with LISA SNR curves for $T_*=50$ GeV (solid lines). The models in blue (orange) are unlikely (likely) to be probed by the high-luminosity LHC. The dotted straight lines are the contours of the fluid turnover time quantifying the effect of turbulence. In the grey shaded region the decay of sound waves into turbulence is less important than the Hubble damping and the SNR curve reflects this effect. See Ref.~\cite{Caprini:2019egz} for details.}
    \label{fig:HL-LHC}
\end{figure}

These BSM models could possibly be further explored at
future colliders
such as the high-luminosity phase of the LHC or
the Future 
Circular Collider \footnote{Although many of them can
be elusive 
to collider physics~\cite{Addazi:2018nzm}.}, either by direct detection of new particles or due to their impact on precision measurements of many observables, such as masses and couplings in the different SM sectors (scalar, gauge or fermionic). Still, most models have parameter space corners which will remain inaccessible to present and near-future colliders, but whose early universe dynamics results in GWs detectable by LISA. Fig.~\ref{fig:HL-LHC} illustrates this complementarity in a singlet extension, showing the values of $\alpha$ and $\beta/H_*$ for different parameter points and their prospective detection by LISA and HL-LHC.

Another exciting possibility of complementary observations comes from astrophysics and more specifically gamma ray observations. These are connected with intergalactic magnetic fields, as the latter would lead to pair production  $\gamma + {\bf B} \to e^+ e^-$ followed by scattering of the charged particles off the magnetic field: $e^\pm + {\bf B} \to e^\pm + \gamma$, to produce a secondary $\gamma$. 
This process leads to a  dilution of the spectra observed from distant blazars, which recently enabled a lower bound on the magnitude of intergalactic magnetic fields~\cite{Tavecchio:2010mk,Ando:2010rb,Neronov:1900zz,Essey:2010nd,Chen:2014rsa,Biteau:2018tmv}.
Due to the lack of astrophysical processes to account for magnetic fields in intergalactic voids,
explanations through a primordial origin have been put forward.
One possibility is inflationary magnetogenesis~\cite{Turner:1987bw,Ratra:1991bn,Martin:2007ue,Kobayashi:2014sga} although it has been pointed out recently that models where magnetic fields are produced before the EWPT generically lead to problems with baryon isocurvature constraints~\cite{Kamada:2020bmb}.
Another interesting possibility involves a FOPT~\cite{Vachaspati:1991nm,Sigl:1996dm,Tevzadze:2012kk}. While there are many differences between the magnetic fields produced by both scenarios,
notably involving the uncertainty of their evolution 
after the transition~\cite{Durrer:2013pga}, the magnetic field is governed by the same basic parameters as the GW signal.
In particular this means that the transition in question has to be strong enough to produce a magnetic field amplitude above the known lower bounds.
This, of course, leads to a certain correlation between the two possible signals, and it was found that the parameter space predicting a GW signal observable by LISA can also facilitate generation of a magnetic field strong enough to explain the
observed diffusion of the blazar spectra for some models predicting a
first order EWPT~\cite{Ellis:2019tjf}.
Moreover, the presence of helicity in the magnetic fields produced at a FOPT plays a crucial role on
its dynamical evolution, playing a crucial role in the resulting restrictions on the strength of the FOPT~\cite{Brandenburg:2017neh, Brandenburg:2017rnt}, as well as providing circularly polarised SGWB. Such polarisation can also be measured by LISA providing a distinct signal of this scenario~\cite{Ellis:2020uid,Domcke:2019zls}.\footnote{Other
proposed measures for detecting magnetic
helicity in voids involve exploring the twisting of photon paths coming from multiple
blazar sources~\cite{Vachaspati:2020blt, Tashiro:2014gfa, Kahniashvili:2005yp}.}

\newpage
%%%%%%%%%%%%%%%%%%%%%%%%%%%%%%%%%
% Here Sec. 7 starts

\section{Cosmic Strings}
\label{sec:CosmicStrings}

\small \emph{Section coordinators: M.~Lewicki, L.~Sousa. Contributors: P.~Auclair, J.J.~Blanco-Pillado, C-F.~Chang, Y.~Cui, D.G.~Figueroa, A.~Jenkins, S.~Kuroyanagi, M.~Lewicki, M.~Sakellariadou, K.~Schmitz, L.~Sousa, D.~Steer, J.~Wachter.}

\normalsize

\subsection{Introduction}

Cosmic defects are stable energy configurations that can be formed in early universe phase transitions, usually due to a spontaneous symmetry-breaking process driven by some scalar field(s) acquiring a non-zero expectation value within a (topologically non-trivial) vacuum manifold. 
If the symmetry broken is global, all non-constant field configurations produce energy-momentum, and are loosely  referred to as {\it global defects}. When the broken symmetry is gauged, we speak of {\it local defects} instead. Here we will focus on cosmic strings, which correspond to stable one-dimensional defect solutions of field theories~\cite{Nielsen:1973cs}, independently of whether they are global or gauge.  Alternatively, cosmic strings can also be identified with cosmologically stretched fundamental strings from String Theory, formed e.g.~at the end of brane inflation~\cite{Dvali:2003zj,Copeland:2003bj}.

The energy per unit length of a string is $\mu \sim \eta^2$, with $\eta$ a characteristic energy scale. The string tension, at least for the simplest cases, is typically of order $\mu$. In the case of topological strings, $\eta$ represents the energy scale of the phase transition. A network of strings formed in the early universe emits GWs throughout most of the history of the universe, generating a SGWB from the superposition of many uncorrelated emissions. Here we forecast the constraints that LISA may put on the dimensionless combination 
\begin{eqnarray}
G\mu \sim10^{-6} \left(\frac{\eta}{\rm{10^{16} ~GeV}}\right)^2\,,
\end{eqnarray}
where $G = 1/M_{p}^2$ is Newton's constant, and $M_p = 1.22\times 10^{19}$ GeV the Planck mass. We note that various potential observational signatures of cosmic string networks, other than GW emission, have been discussed in the literature. These include anisotropies in the CMB~\cite{Ade:2013xla,Charnock:2016nzm,Lizarraga:2016onn,Ringeval:2010ca}, lensing events \cite{Vilenkin:1984ea,Bloomfield:2013jka}, and cosmic ray emission~ \cite{Brandenberger:1986vj,Srednicki:1986xg,Bhattacharjee:1991zm,Damour:1996pv,Wichoski:1998kh,Peloso:2002rx,Sabancilar:2009sq,Vachaspati:2009kq,Long:2014mxa}. (See \cite{Hindmarsh:1994re,Vilenkin:2000jqa,Sakellariadou:2006qs,Vachaspati:2015cma} for a review.) Currently, CMB data from the Planck Satellite~\cite{Ade:2013xla} imply $G\mu < 10^{-7}$ for NG, Abelian-Higgs, and semi-local strings. The most stringent bounds, however, come from searches for the SGWB, with PTA constraining $G\mu$ for NG strings to be $G\mu \lesssim 10^{-11}$~\cite{Blanco-Pillado:2017rnf,Ringeval:2017eww}, and LIGO-Virgo observations constraining it to be as low as $G\mu < 2\times10^{-14}$, depending on the string network model~\cite{LIGOScientific:2019vic,Abbott:2017mem}. As we will show, LISA will be sensitive to string tensions with $G\mu \gtrsim 10^{-17}$ for NG strings, improving current upper bounds by $\sim 10$ orders of magnitude relative to CMB constraints, by $\sim 6$ orders of magnitude relative to current PTA constraints, and even by $\sim$3 orders of magnitude relative to future constraints from next generation of PTA experiments.~\footnote{The more recent work 
\cite{Boileau:2021gbr} shows that, depending on different assumptions on the astrophysical background and the galactic foreground, the forecast sensitivity of LISA should be reduced this to $G\mu \gtrsim \mathcal O(10^{-16} - 10^{-15})$.}

As the characteristic width $\delta \sim 1/\eta$ of a cosmic string is much smaller than the size of the horizon, in the following we mainly assume that strings are well described by the NG action, which is the leading-order approximation when the curvature scale of the strings is much larger than their thickness.  We refer to such strings as NG strings (we focus on string networks without junctions). Cosmic string networks are expected to reach an attractor solution known as {\it scaling}, for which the energy density of the network remains a fixed fraction of the background energy density in the universe. When strings within the network collide, they intercommute (i.e.~``exchange partners'') and reconnect after the collision (technically speaking this corresponds to an intercommutation probability $\mathcal{P}=1$, which we mainly assume, though we will also comment briefly on the case $\mathcal{P}<1$). As a result, loops are formed whenever a string self-intersects or two curved strings collide. Sub-horizon sized loops decouple from the cosmological evolution and oscillate under their own tension, emitting GWs in this process. The relativistic nature of the strings typically leads them to form {\it cusps}, corresponding to points on the string momentarily moving at the speed of light \cite{Turok:1984cn}. Furthermore, intersections of strings will also generate discontinuities in their tangent vector known as {\it kinks}.  Cusps and kinks generate GW burst emissions~\cite{Damour:2000wa,Damour:2001bk}, which add up all throughout cosmic history to form a SGWB. 

Besides sub-horizon loops, the network also contains long strings that that stretch across a Hubble volume. These are either infinite or super-horizon loops, and they are also expected to emit GWs. However, the dominant contribution is generically produced by the superposition of radiation from many sub-horizon loops along each line of sight. 

Cosmic string networks are expected to create one of the largest SGWB of cosmological origin known. 
In the following we analyse the ability of LISA to probe the SGWB emitted by a network of cosmic strings, considering leading models of the string networks as described in the literature. The prospects for detection or, alternatively, for setting up new stringent constraints on string parameters, are very encouraging. 

\subsection{Network modelling}\label{sec:network-models}

The SGWB generated by cosmic string networks in a given frequency $f$ has, at the present time, contributions from all the loops created throughout the history of the universe that have emitted GWs that have a frequency $f$ today. The number density $\mathsf{n}(l,t)$ of non-self-intersecting, sub-horizon, cosmic string loops of invariant length $l$ at any cosmic time $t$ is then the crucial ingredient in this computation. In this section, we review the models for the loop number density that have been proposed in the literature.

\subsubsection{Model 1}\label{sec:model1}
We start by considering an analytical approach --- originally developed in Ref.~\cite{Kibble:1984hp} and later extended in Refs.~\cite{Caldwell:1991jj,DePies:2007bm,Sanidas:2012ee,Sousa:2013aaa} --- which is based on two assumptions. The first assumption is that the production of loops is the dominant energy-loss mechanism in the evolution of the long string network. In this case, as we shall see, the loop production function  $f(l,t)~dl$ --- which gives us the number density of loops of lengths between $l$ and $l+dl$ produced per unit time and per unit volume --- is determined by the large-scale evolution of the long string network. Here, we shall use the approach introduced in Ref.~\cite{Sousa:2013aaa} and use the Velocity-dependent One-Scale  model~\cite{Martins:1996jp,Martins:2000cs} to describe its evolution. This model provides an analytical description of the evolution of the characteristic length  $L\equiv (\mu/\rho_\infty)^{1/2}$ --- where $\rho_\infty$ is the energy density of the long string network --- and of the root-mean-squared  velocity $\bar{v}$ of the network~\cite{Martins:1996jp,Martins:2000cs}:
\begin{eqnarray}
\frac{d\bar{v}}{dt} & = & \left(1-\bar{v}^ 2\right)\left[\frac{k(\bar{v})}{L}-2H\bar{v}\right]\,,\label{eqn:vosv}\\
\frac{dL}{dt} & = & \left(1+\bar{v}^ 2\right)HL+\frac{\tilde{c}}{2}\bar{v}\,, \label{eqn:vosL}
\end{eqnarray}
with
\begin{equation}
k(\bar{v})=\frac{2\sqrt{2}}{\pi}\left(1-\bar{v}^2\right)\left(1+2\sqrt{2}\bar{v}^3\right)\frac{1-8\bar{v}^6}{1+8\bar{v}^6}\,.
\label{eqn:curavture}
\end{equation} 
The curvature parameter $k(\bar{v})$ accounts (to some extent) for the effects of small-scale structure~\cite{Martins:2000cs}.

Here, $\tilde{c}$ is a phenomenological parameter that quantifies the efficiency of the loop-chopping mechanism, that may be calibrated with simulations (for NG strings, $\tilde{c}=0.23\pm0.04$ fits both radiation and matter era simulations~\cite{Martins:2000cs}). Using Eq.~(\ref{eqn:vosL}), we find that the energy lost as a result of loop production is given by
\begin{equation}
\left.\frac{d\rho_\infty}{dt}\right|_{\rm loops}=\tilde{c} \bar{v} \frac{\rho_{\infty}}{L}=\mu\int_0^{\infty}{lf(l,t)dl}\,,
\label{eqn:cVOS}
\end{equation}
and determines the normalisation of loop production function.

The second assumption of this model (which will be somewhat relaxed later) is that all the loops are created with a length $l$ that is a fixed fraction of the characteristic length of the long string network: $l=\alpha_L L$, where $\alpha_L<1$ is a (constant) free-parameter of the model. We then have 
\begin{equation}
f(l,t) = \left(\frac{\mathcal{F}}{f_r}\right) \frac{\tilde{c}}{l}\frac{\bar{v}}{L^3} \delta\left(l - \alpha_LL\right)\,,
\label{eqn:voslpf}
\end{equation}
where $\bar{v}$ and $L$ are given by Eqs.~(\ref{eqn:vosv}) and (\ref{eqn:vosL}). Here the normalisation of $f(l,t)$ is determined using Eq.~(\ref{eqn:voslpf}), except for the correction factor $f_r$ and $\mathcal{F}$. The first factor, $f_r\sim \sqrt{2}$, is introduced to account for the energy lost as a result of the redshifting of the peculiar velocities of loops~\cite{Vilenkin:2000jqa}. Also, since in general we do not expect all loops to be created with exactly the same size but to follow a distribution of lengths, a second factor $\mathcal{F}$ is included to account for the effect of the spread of the distribution (See Refs.~\cite{Sanidas:2012ee,Sousa:2020sxs} for a detailed analysis of this effect.) For NG strings, $\mathcal{F}$ is estimated to be $\mathcal{O} (0.1)$~\cite{Blanco-Pillado:2013qja}.

Note that Eq.~(\ref{eqn:voslpf}) is valid throughout any cosmological era, even during the radiation-to-matter and matter-to-dark-energy transitions (in which the network is not in a linear scaling regime). As a result, it allows us to compute $f(l,t)$ through cosmic history in a realistic cosmological background~\cite{Sousa:2013aaa}. The number density of loops $\mathsf{n}(l,t)$ at all times may then be found by solving
\begin{equation}
\mathsf{n}(l,t)  = \int_{t_{i}}^{t}{dt' f(l',t') \left(\frac{a(t')}{a(t)}\right)^3}\,,
\end{equation}
while accounting for the decrease in the length of loops caused by the emission of GWs $l = l' + \mathrm{\Gamma}
G\mu\,(t'-t)$ (where $\Gamma$ is a dimensionless constant characterising GW emission efficiency ---  that we define more precisely in Sec.~\ref{ssec:SGWB-from-loops} ---, $l$ is the length of a loop created at a time $t'$ with a length $l'$ at a later time $t>t'$).

The loop-size parameter $\alpha_L$ may either be calibrated with numerical simulations (as we shall do in Sec.~\ref{sec:detection-prospects}), but may be also be treated as a free parameter of the model to explore a wider variety of scenarios (as in Sec.~\ref{super}). Although $\alpha_L$ is the natural parameter of this model, we will express our results in terms of $\alpha=\alpha_L \xi_r$ (where $\xi_r$ is the value of $L/t$ in the radiation era) in order to ease comparison with other loop distribution models.

\subsubsection{Model 2}\label{sec:model2}

The second model we consider is a loop number density distribution, $\mathsf{n}(l,t)$, extracted from large scale numerical simulations of the string networks~\cite{Blanco-Pillado:2013qja}. These number densities were first obtained from the integration of the loop production function $f(l,t)$ of non-self-intersecting loops computed directly from the simulation. Recently, these loop number densities were also determined directly from simulations without the intermediate use of a loop production function~\cite{Blanco-Pillado:2019tbi}. Both approaches produce the same result.

In the following we give the resulting distributions for three types of loops: those created in the radiation era, emitting GWs in the radiation era; those created in the radiation era, emitting GWs in the matter era; and those created in the matter era, emitting GWs in the matter era.

With $\Omega_\mathrm{rad}$ the fraction of the critical density in radiation, and understanding the redshift $z$ to be a function of time, we can write
\begin{subequations}\label{eqn:II-all}\begin{align}
    \mathsf{n}_\mathrm{r,r}(l,t) &= \frac{0.18}{t^{3/2}(1+\Gamma G\mu t)^{5/2}}\Theta(0.1-l/t)\,,\label{eqn:II-rr}\\
    \mathsf{n}_\mathrm{r,m}(l,t) &= \frac{0.18(2\sqrt{\Omega_\mathrm{rad}})^{3/2}(1+z)^3}{(l+\Gamma G\mu t)^{5/2}}\Theta(0.09\, t_\mathrm{eq}/t-\Gamma G\mu-l/t)\,,\label{eqn:II-rm}\\
    \mathsf{n}_\mathrm{m,m}(l,t) &= \frac{0.27-0.45(l/t)^{0.31}}{t^2(l+\Gamma G\mu t)^2}\Theta(0.18-l/t)\,,\label{eqn:II-mm}
\end{align}\end{subequations}
where $\Theta(x)$ is the Heaviside step function, and the subscript ``r,m'' indicates ``loops produced in the (r)adiation era, emitting in the (m)atter era'', and so on. For Eqs.~(\ref{eqn:II-rr}) and (\ref{eqn:II-mm}), the cutoffs are due to the maximum size of loop which can be produced in these eras. For Eq.~(\ref{eqn:II-rm}), there is an additional term in the cutoff which models the decay due to GW emission; that is, a loop formed in the radiation era must have some minimum size in order to survive into the matter era. Eq.~(\ref{eqn:II-all}) contains all the information necessary to compute the SGWB due to decaying loops, as outlined in Sec.~\ref{ssec:SGWB-from-loops}.

Analysis of Eq.~(\ref{eqn:II-mm}), in comparison to Eqs.~(\ref{eqn:II-rr}) and (\ref{eqn:II-rm}), shows it to be negligible for $G\mu\lesssim 10^{-10}$ and thus for cosmic string tensions of interest to LISA. The cosmic string generated SGWB potentially to be observed by LISA will be for radiation-era loops, integrated up to the present. Of particular interest are radiation-era loops emitting in the matter era; these loops contribute to a ``bump'' in the SGWB spectrum which may, depending on $G\mu$, fall within the LISA band (see Sec.~\ref{sec:detection-prospects}). For this particular reason, but also more broadly, it is important to understand how gravitational backreaction affects the evolution of loops. At present, the decay of loops into GWs is accounted for by $\Gamma G\mu$ terms, but this does not take into account potential changes to the loop population due to, e.g.~loop fragmentation as a result of backreacted trajectories. While current theoretical studies~\cite{Blanco-Pillado:2018ael,Blanco-Pillado:2019nto}, and numerical studies of simple models~\cite{Robbins:2019qba}, suggest that the change to loop trajectories will be small, it is important to keep in mind that no current, simulation-inferred network model incorporates gravitational backreaction directly.

\subsubsection{Model 3}

The third model we consider is based on the analytical studies of the small-scale structure and the correlation functions of individual infinite strings \cite{Dubath:2007mf,Polchinski:2007rg,Polchinski:2006ee}.
This analysis predicts a power-law loop production function in terms of scaling units,
whose slope is parametrised by a parameter $\chi$ linked to the fractal dimension of the strings.
The production of cosmic string loops is suppressed by gravitational back-reaction, on scales below
\begin{equation}
    \gamma_\mathrm{c} = \Upsilon (G\mu)^{1+2\chi},
\end{equation}
in which $\Upsilon = \mathcal{O}(20)$.
The characteristics of the loop production function were later inferred from measurements of the loop number density in large-scale simulations of NG strings~\cite{Ringeval:2005kr, Lorenz:2010sm}.\footnote{Note that the relationship between the loop number density and the loop production function is not always one-to-one, see Ref.~\cite{Auclair:2019zoz} for a review.} 
The numerically inferred loop-production function is then extended from large scales down to the gravitational back-reaction scale, resulting in an extra population of small loops.

The loop number density of Model 3 can be approximated by the following expressions
\begin{align}
    t^4 \mathsf{n}_\mathrm{r, r}(l,t) &=
    \begin{cases}
        \dfrac{0.08}{(\gamma + \Gamma G\mu)^{3-2\chi_\mathrm{r}}} & \text{ if } \Gamma G\mu < \gamma\\[4mm]
        \dfrac{0.08 (1/2 - 2\chi_\mathrm{r})}{(2-2\chi_\mathrm{r}) \Gamma G\mu \gamma^{2-2\chi_\mathrm{r}}} & \text{ if } \gamma_{\rm c} < \gamma < \Gamma G\mu\\[4mm]
        \dfrac{0.08(1/2 - 2\chi_\mathrm{r})}{(2-2\chi_\mathrm{r}) \Gamma G\mu \gamma_c^{2-2\chi_\mathrm{r}}} & \text{ if } \gamma < \gamma_{\rm c}\\[6mm]
    \end{cases}
\\
    t^4 \mathsf{n}_\mathrm{m, m}(l, t)  &=
    \begin{cases}
        \dfrac{0.015}{(\gamma + \Gamma G\mu)^{3-2\chi_\mathrm{m}}} & \text{ if } \Gamma G\mu < \gamma\\[4mm]
        \dfrac{0.015 (1 - 2\chi_\mathrm{m})}{(2-2\chi_\mathrm{m}) \Gamma G\mu \gamma^{2-2\chi_\mathrm{m}}} & \text{ if } \gamma_{\rm c} < \gamma < \Gamma_{\rm d} G\mu\\[4mm]
        \dfrac{0.015(1 - 2\chi_\mathrm{m})}{(2-2\chi_\mathrm{m}) \Gamma G\mu \gamma_c^{2-2\chi_\mathrm{m}}} & \text{ if } \gamma < \gamma_{\rm c}\\[6mm]
    \end{cases}
\\
    t^4 \mathsf{n}_\mathrm{r,m}(l, t) &= \left(\frac{t}{t_{\rm eq}}\right)^4 \left(\frac{1+z}{1+z_{\rm eq}}\right)^3 t_{\rm eq}^4
    \mathsf{n}_{\rm rad}^{(3)}\left[\frac{\gamma t + \Gamma_{\rm d} G\mu (t-t_{\rm eq})}{t_{\rm eq}}\right]
\end{align}
in which $(\chi_\mathrm{r}, \chi_\mathrm{m}) = (0.2, 0.295)$.
This extra population of small loops dominates the SBGW in the high frequency spectrum as discussed in Refs.~\cite{Abbott:2017mem,Auclair:2019wcv,Abbott:2021ksc} and hence can lead to very different constraints on $G\mu$ to that of the two above-mentioned models.
The energy density of these small loops is very large and the question of energy balance, in the context of the one-scale model, has been raised in Ref.~\cite{Blanco-Pillado:2019vcs}.

\subsubsection{Field theory}
\label{subsec:fieldTheory}

The above modellings are all based on NG strings, which are infinitely thin.
Cosmic strings appear however naturally as solitonic solutions of classical field theory models~\cite{Nielsen:1973cs}, so in principle they can decay not only by GW radiation but also directly into excitations of their elementary constituents. For example, it has been observed with numerical simulations that global (axionic) strings decay into the massless Goldstone modes present in the vacuum of the theory~\cite{Davis:1989nj}. For local strings with no long-range interactions, the excitations in the vacuum are however massive, and hence are naturally expected to be suppressed for long wavelengths comparable to the length of the strings~\cite{Martins:2003vd,Olum:1999sg,Olum:1998ag}. Furthermore, recent simulations of individual loops in the Abelian-Higgs model ~\cite{Matsunami:2019fss} report that extrapolating their results to large loop sizes, the GW emission should be expected to dominate over particle emission for loops larger than a certain critical length scale. 

In contrast, large-scale field theory simulations of Abelian-Higgs strings~\cite{Vincent:1997cx,Hindmarsh:2008dw,Daverio:2015nva,Hindmarsh:2017qff} observe that the network of long strings reaches a scaling regime, thanks to energy loss into classical radiation of the scalar and gauge fields involved. As a consequence of this, sub-horizon loops formed during the network evolution decay very promptly. This intriguing discrepancy has been under debate for over twenty  years, but the origin of this radiation is not currently understood.

The similarities and differences between field theory and NG simulations of string networks can then be summarised as follows: the infinite strings are rather similar in curvature radius and length density, but loops decay into field modes in the field theory simulations. In field theory simulations, the string energy density goes into radiated modes of the fields, which do not belong to the string network anymore. As a consequence, the string loops decay within a Hubble time, and hence do not continue to contribute as a source of GWs. In the NG picture, this channel does not exist, and instead the energy of the infinite strings goes into loops, which then decay via GWs. Our analysis in the following is mostly based on the NG classical evolution of strings. We assume, as supported by NG simulations, that loops are formed throughout cosmic history, and they decay into GWs. Our discussion about the ability of LISA to measure a GW background from cosmic strings is therefore based on this fundamental assumption. 

\subsection{Computation of the gravitational wave spectrum from loops}\label{ssec:SGWB-from-loops}
The incoherent superposition of GWs emitted from oscillating cosmic string loops leads to a SGWB. The calculation of the SGWB from the cosmic string networks have been widely studied in the literature \cite{Vachaspati:1984gt,Blanco-Pillado:2013qja,Blanco-Pillado:2017oxo,Blanco-Pillado:2017rnf,Ringeval:2017eww,Vilenkin:1981bx,Hogan:1984is,Accetta:1988bg,Bennett:1990ry,Caldwell:1991jj,Siemens:2006yp,DePies:2007bm,Olmez:2010bi,Sanidas:2012ee,Sanidas:2012tf,Binetruy:2012ze,Kuroyanagi:2012wm,Kuroyanagi:2012jf,Sousa:2013aaa,Sousa:2014gka,Cui:2018rwi,Chang:2019mza,Jenkins:2018nty,Gouttenoire:2019kij}. The generic form of a GW spectrum in logarithmic intervals of frequency,  introduced in Eq.~(\ref{Omegagw}), also applies to the case with cosmic strings. 
For the cosmic string application, Eq.~(\ref{Omegagw}) is implemented by integrating over the GW emission from all the loops throughout cosmic history that contribute to a certain frequency.\footnote{We neglect the GW contribution from long (horizon-spanning) strings as being subdominant to the GW contribution from loops for NG strings. See Refs.~\cite{Buchmuller:2013lra,Matsui:2019obe}.} In the following subsections, we review two methods of calculating SGWB from string loops that have been developed in the literature along with recent updates. There are ingredients that are common to both approaches. The first ingredient is  $\mathsf{n}(l,t)$, the number density of non-self-intersecting, sub-horizon, cosmic string loops of invariant length $l$ at cosmic time $t$, As shown in Sec.~\ref{sec:network-models}, $\mathsf{n}(l,t)$ can be estimated by analytical or simulation-based methods. Another characteristic function is the gravitational loop power spectrum $P_{\textrm{GW}}(f,l)$, which may be determined as an average based on simulations, or approximated by focusing on the analytical high-frequency behaviour of particular events on the strings, e.g.~cusps and kinks.

\subsubsection{Method I}

The first method to calculate $\Omega_{\textrm{GW}}(t_0,f)$ \cite{Vachaspati:1984gt,Blanco-Pillado:2013qja,Blanco-Pillado:2017oxo,Blanco-Pillado:2017rnf,Caldwell:1991jj,Siemens:2006yp,DePies:2007bm,Olmez:2010bi,Kuroyanagi:2012wm,Kuroyanagi:2012jf,Sousa:2013aaa,Cui:2018rwi} assumes that $P_{\textrm{GW}}(f,l)$ takes the form of 
\begin{align}
P_{\textrm{GW}} (f,l) = G\mu^2 l P(f l),
\end{align}
where $P(f, l)$ is an averaged function that can be computed from an ensemble of loops of length $l$ (but of different shapes) based on simulations. 

Integrating all emissions throughout the history of the universe and taking into account the redshift effects, the GW energy density for a particular frequency $f$ as measured today is
\begin{align}
\frac{d\rho_{\textrm{GW}}}{df} (t_0,f ) = G \mu^2 \int^{t_0}_{0} dt \left( \frac{a(t)}{a_0}\right)^4 \int^\infty_0 dl\, l\, \mathsf{n}(l,t) P\left(\frac{a_0}{a(t)}f l\right).
\end{align}
The GW radiation power of an isolated loop of length $l$ can be computed using the standard formulae in the weak gravity regime \cite{Vachaspati:1984gt,Vilenkin:1986ku,Garfinkle:1987yw}. As a simple approximation, we assume that the loops evolve in flat space in a periodic manner, and thus emit GW at discrete frequencies
\begin{align}
\label{Eq: Harmonic}
\omega_n = 2\pi n /T,
\end{align}
where $T = l/2$ is the oscillation period, and the harmonic modes $n = 1,2, \dots$. As a consequence, we replace $P(fl)$ by $P_n$, a discrete function of the harmonic mode number. For an individual loop, a simple (monochromatic) power spectrum can be obtained assuming the emission is dominated by specific events (e.g.~cusps, kinks) \cite{Vachaspati:1984gt,Burden:1985md,Garfinkle:1987yw,Binetruy:2009vt}
\begin{align}
\label{eq:Pn}
P_n = \frac{\Gamma}{\zeta(q)} n^{-q},
\end{align} 
where $\zeta(q)$ is Riemann zeta function and $\Gamma = \sum_{n=1}^{\infty} P_n$ is the total power of emission, which is found to be highly peaked around $\Gamma \sim 50$ \cite{Blanco-Pillado:2015ana,Blanco-Pillado:2013qja,Blanco-Pillado:2017oxo,Wachter:2016rwc,Blanco-Pillado:2018ael,Chernoff:2018evo,Blanco-Pillado:2019nto}. The power law parameter $q$ is $4/3$, $5/3$, or $2$ for GW emissions dominated by cusps, kinks, or kink-kink collisions, respectively.\footnote{We should keep in mind that the these simple power laws may not be a good approximation at low $n$. In this case, the entire loop's structure becomes important.}
With this method, the GW energy density today can be calculated by summing over loop harmonic modes \cite{Blanco-Pillado:2017oxo,Blanco-Pillado:2013qja}.
\begin{align}
\frac{d\rho_{\textrm{GW}}}{df} (t_0,f ) =  G\mu^2 \sum_{n=1}^{\infty} C_n(f) P_n,
\end{align}
where 
\begin{align}
\label{eq:Cn}
C_n(f) = \frac{2n}{f^2}\int^\infty_0 \frac{dz}{H(z)(1+z)^4} \mathsf{n}\left( \frac{2n}{(1+z) f}, t(z) \right).
\end{align}
Note that in practice, only a finite number of modes need to be included in calculations, but the necessary number of modes to ensure a good convergence for a reliable result depends on the background cosmology. For standard cosmology $\Omega_{\textrm{GW}}(t_0,f)$ converges by summing over $10^3-10^5$ modes (depending on the value of q)~\cite{Sanidas:2012ee}. However, recent studies \cite{Cui:2019kkd,Gouttenoire:2019kij,Blasi:2020wpy} demonstrated that more than $10^5$ modes may be necessary to ensure a convergence to the correct power law at high frequency. For instance, in the presence of an early matter dominated era, summing over a small number of modes gives $f^{-1}$ at high $f$ while an $f^{-1/3}$ relation emerges with higher modes included.

\subsubsection{Method II}

The second method analytically estimates $P_{\textrm{GW}}(f,l)$ based on unresolved burst events on the strings \cite{Ringeval:2017eww,Abbott:2017mem,Siemens:2006yp,DePies:2007bm,Kuroyanagi:2012wm,Kuroyanagi:2012jf,Jenkins:2018nty,Auclair:2019wcv}. The cosmic string GW emission may be dominated by cusps, kinks, or kink-kink collisions. The contribution to the SGWB from unresolved bursts is given by
\begin{align}
\frac{d\rho_{\textrm{GW}}}{df} (t_0, f) = f^2 \int^\infty_{z_{\rm min}} dz \int^\infty_0 dl \, h^2(l,z,f) \frac{d^2 R(z,l)}{dz dl},
\end{align}
where $\frac{d^2 R(z,l)}{dzdl}$ denotes the burst rate per unit loop length $l$ and per unit redshift $z$ which is proportional to $\mathsf{n}(l,t)$. $z_{\rm min}$ will be defined shortly. The amplitude $h(l,z,f)$ is the Fourier transform of the waveform of the bursts \cite{Siemens:2006yp,Damour:2000wa,Damour:2001bk,Binetruy:2009vt} and reads
\begin{align}
h(l,z,f) = A_q (l,z) f^{-q},
\end{align}
with
\begin{align}
A_q(l,z) = g_1^{(q)} \frac{G \mu H_0 l^{2-q}}{(1+z)^{q-1} \varphi_r(z)}, \;\;\;\; \varphi_r(z) \equiv H_0 \int^z_0 \frac{dz^\prime}{H(z^\prime)} dz.
\end{align}
Again the power $q$ is equal to $4/3$, $5/3$ and $2$ for cusps, kinks and kink-kink collisions, respectively.  The calibration constant $g_1^{(q)}$ accounts for loop geometry effects on different bursts. Since the cusps and kinks radiate non-isotropically, the above waveform is only valid for directions near the cusp or kink direction. A cutoff angle $\theta_{\rm cutoff}$ is thus introduced to account for the geometric beaming effect:
\begin{align}
\theta_{\textrm{cutoff}} (l, z, f) = \left( \frac{1}{g_2 f (1+z) l} \right)^{1/3},
\end{align}
with $g_2 = \frac{\sqrt{3}}{4}$. The burst rate is then given by \cite{Damour:2000wa,Damour:2001bk,Olmez:2010bi}
\begin{align}
\frac{d^2 R(z,l)}{dzdl} = 2 \frac{4 \pi  \varphi_r^2(z)}{(1+z)^3 H(z) H_0^2} \left( \frac{ \mathsf{n}(l,t(z))}{l(1+z)}\right) \Delta(l,z,f)\,,
\end{align}
where the quantity 
\begin{align}
\Delta (l,z,f) \simeq \left( \frac{\theta_{\textrm{cutoff}} (l,z,f) }{2} \right)^{3(2-q)} \Theta\left( 1 - \theta_{\textrm{cutoff}}(l,z,f) \right),
\end{align}
represents the fraction of observable bursts \cite{Olmez:2010bi}. 

Earlier burst events that cannot be resolved at a GW detector thereby contribute to a SGWB. Integrating over $z$ and $l$, the spectrum due to a given type of burst is
\begin{align}\label{Eq: method-II Omega_f}
\Omega_{\textrm{GW}}(f) = \frac{\left( g_1^{(q)} \right)^2 g_2^{q-2}}{2^{4-3q}} \frac{N_q}{3}  \frac{(G \mu)^2}{H_0^2} (2\pi f)^3\,
%\times
%\nonumber\\
\quad\int^\infty_0 dx \int^\infty_{z_{\textrm{min}}(x,f)} dz \frac{(f t(x))^{-2-q}}{(1+z)^{4+q}} x^{1-q} \frac{\mathsf{n}(x)}{H(z)}, 
\end{align}
where $x\equiv l/t$, $z_{\textrm{min}}$ is the solution to $\theta_{\textrm{cutoff}} (l, z_{\textrm{min}},f) = 1$, and $N_q \sim \mathcal{O}(1)$ is the average number of bursts per oscillation in a loop \cite{Ringeval:2017eww,Jenkins:2018nty,Abbott:2017mem}. An alternative calculation of $\Omega_{\textrm{GW}}(f)$ by integrating over $z, h$ is shown in Ref.~\cite{Cui:2019kkd}.

More recent bursts of large amplitude could be resolved individually and thus detected as transient events by GW detectors, which is a different type of signal relative to a SGWB. If a burst is to be resolved in a given frequency band $f$, it must produce a strain greater than the experimental sensitivity $h > h_{\rm exp}$ with rate less than $f$. The rate of such events is~\cite{Siemens:2006vk,Auclair:2019wcv}
\begin{equation}
R_{\rm exp}(f) = \int_{0}^{z_*} \!dz \int_{\max(h_{\rm min},h_{\rm exp})}^{h_{\rm max}} 
\!\!\!dh \; \frac{d^2R}{dz\, dh}(h,z,f)\,,
\end{equation}
where $z_*$ enforces the rate condition and is given by
\begin{equation} \label{eq:astar}
f= \int_{0}^{z_*}\!dz \int_{h_{\rm min}}^{h_{\rm max}}\!dh \;  
\frac{d^2R}{dz\, dh}(h,z,f) \, .
\end{equation}
For a given cosmic string model, the SGWB channel is usually more sensitive than individual bursts, providing stronger constraints on parameters such as $G\mu$ \cite{Abbott:2017mem}. Nevertheless in certain motivated scenarios, such as string formation before or during inflation, a SGWB could be suppressed and burst events would become the leading channel for discovery \cite{Cui:2019kkd}.

In addition to the Gaussian SGWB and the resolvable burst events as discussed above, an unresolved string signal from a set of infrequent bursts at low redshift $z\ll 1$ emits a distinct non-Gaussian and non-continuous popcorn-like signal. 
This effect is negligible on the current pulsar timing limit $G \mu < 10^{-11}$ 
\cite{Regimbau:2011bm,Olmez:2010bi}, and has no effect on the SGWB that is emitted from a large loop size $\alpha \sim 0.1$ \cite{Blanco-Pillado:2017oxo,Siemens:2006yp,Olmez:2010bi,Binetruy:2012ze}. However, if the burst occurs in our neighbourhood with strong amplitude, then  GW detectors may be able to identify the signal \cite{Chernoff:2017fll,Helfer:2018qgv}. Such a coherent signal may provide a complementary detection mechanism: cosmic strings may generate a non-negligible amount of GW memory~\cite{Aurrekoetxea:2020tuw} that would not be captured by
SGWB probes~\cite{Jenkins:2021kcj}.

\subsubsection{Templates for the stochastic gravitational wave background generated by cosmic string loops\label{temp1}}

\paragraph{Models 1 and 2}

An analytical approximation for the contribution of each of three loop populations --- loops that decay in the radiation era, radiation-era loops that survive into the matter era, and loops created in the matter era --- to the SGWB was derived in Ref.~\cite{Sousa:2020sxs}. Therein, they found that radiation-era loops give rise to a SGWB of the form
\begin{equation}
\mathrm{\Omega}_{\rm GW}^r(f)=\frac{128}{9}\pi A_r \Omega_r\frac{G\mu}{\epsilon_r}\left[\left(\frac{f(1+\epsilon_r)}{B_r\Omega_m/\Omega_r+f}\right)^{3/2}-1\right]\label{Ogwr2}\,,
\end{equation}
where we have defined $\epsilon_r=\alpha/(\Gamma G\mu)$ and
\begin{equation}
A_i=\frac{{\tilde c}}{\sqrt{2}}\mathcal{F}\frac{v_i}{\xi_i^3}\quad\mbox{and}\quad B_i=\frac{2 H_0 \Omega_i^{1/2}}{\nu_i\Gamma G\mu}\,,
\end{equation}
and where the labels $i=r,m$ are used to refer to the values of the corresponding variables in the radiation and matter eras, respectively. In the radiation era, we have $\nu_r=1/2$, $\xi_r=0.271$, $v_r=0.662$ and $A_r=5.4\mathcal{F}$. This contribution dominates the high-frequency region of the SGWB spectrum (for $f\gg B_r\Omega_m/\Omega_r$), giving rise to a plateau of amplitude
\begin{equation}
\label{eqn:plateaualpha}
\mathrm{\Omega}^{\rm plateau}_{\rm GW}\, h^2=\frac{128}{9}\pi A_r \mathrm{\Omega}_{\rm rad}\,h^2 \frac{G\mu}{\epsilon_r}\left[\left(\epsilon_r+1\right)^{3/2}-1\right]\simeq 1.02\times 10^{-2} \frac{G\mu}{\epsilon_r}\left[\left(\epsilon_r+1\right)^{3/2}-1\right]\,.
\end{equation}
The contribution of the loops that are created in the radiation era but survive into the matter era is of the form
\begin{eqnarray}
\mathrm{\Omega}_{\rm GW}^{rm}(f)=32\sqrt{3}\pi \left(\Omega_m\Omega_r\right)^{3/4}H_0\frac{A_r}{\Gamma}\frac{(\epsilon_r+1)^{3/2}}{f^{1/2}\epsilon_r}\left\{\frac{\left(\frac{\Omega_m}{\Omega_r}\right)^{1/4}}{\left(B_m\left(\frac{\Omega_m}{\Omega_r}\right)^{1/2}+f\right)^{1/2}}\left[2+\frac{f}{B_m\left(\frac{\Omega_m}{\Omega_r}\right)^{1/2}+f}\right]-\right.\label{Ogwrm}\\
\left.-\frac{1}{\left(B_m+f\right)^{1/2}}\left[2+\frac{f}{B_m+f}\right]\right\}\nonumber\,,
\end{eqnarray}
and gives rise to a peak in the low-frequency portion of the spectrum (for $f\ll B_r\Omega_m/\Omega_r$).

Matter-era loops also give rise to a peak-like contribution in the same frequency range, albeit with a different shape:
\begin{equation}
\mathrm{\Omega}_{\rm GW}^m(f)=54\pi H_0\Omega_m^{3/2}\frac{A_m}{\Gamma}\frac{\epsilon_m+1}{\epsilon_m}\frac{B_m}{f}\left\{\frac{2B_m+f}{B_m(B_m+f)}-\frac{1}{f}\frac{2\epsilon_m+1}{\epsilon_m(\epsilon_m+1)}+\frac{2}{f}\log{\left(\frac{\epsilon_m+1}{\epsilon_m}\frac{B_m}{B_m+f}\right)}\right\}\,,
\label{Ogwm}
\end{equation}
where $\epsilon_m=\epsilon_r\xi_m/\xi_r$ and we have $\nu_m=2/3$, $\xi_m=0.625$, $v_m=0.583$ and $A_m=0.39\mathcal{F}$. The shape of the SGWB generated by cosmic string loops in this frequency range is then determined by the interplay between these two contributions. For larger values of $\alpha$, $\mathrm{\Omega}_{\rm GW}^{rm}$ dominates, although its relative importance decreases as $\alpha$ decreases (since $\mathrm{\Omega}_{\rm GW}^{rm}\sim\alpha^{1/2}$). So, for small enough $\alpha$, the contribution of matter-era loops dominates the SGWB spectrum in the low-frequency range.

It was demonstrated in Ref.~\cite{Sousa:2020sxs} that the SGWB is well described by an approximation of the form
\begin{equation}
\mathrm{\Omega}_{\rm GW}(f)=\mathrm{\Omega}_{\rm GW}^r(f)+\mathrm{\Omega}_{\rm GW}^{rm}(f)+\mathrm{\Omega}_{\rm GW}^m(f)
\label{approx}
\end{equation}
for $\alpha\gtrsim\Gamma G\mu$ and $f<3.5\times 10^{10}/(1+\epsilon_r)\,\,{\rm Hz}$ (larger frequencies  are outside the sensitivity windows of the current major GW experiments; see however Ref.~\cite{Sousa:2020sxs} for a description of how to extend these results for larger frequencies). For $\alpha\lesssim\Gamma G\mu$, loops survive significantly less than a Hubble time and decay effectively immediately on cosmological times scales. In this case, no loops produced in the radiation era are expected to survive into the matter era and, as a result, the contribution $\mathrm{\Omega}_{\rm GW}^{rm}$ should be switched off. However, for small enough $\alpha$ (i.e.~$\alpha\lesssim  0.1\Gamma G\mu$), this approximation further simplifies to~\cite{Sousa:2014gka,Sousa:2020sxs}
\begin{equation}
\mathrm{\Omega}_{\rm GW}(f)=\frac{64\pi}{3} G\mu \Omega_r A_r+54\pi\frac{H_0\Omega_m^{3/2}}{\epsilon_m\Gamma}\frac{A_m}{f}\left[1-\frac{B_m}{\epsilon_m}\frac{1}{f}\right]\,.
\end{equation}
Note that, since Model 1 may be calibrated to describe Model 2~\cite{Auclair:2019wcv},  Eq.~(\ref{approx}) can also be used as a template for this model by setting $\mathcal{F}=0.1$ and $\alpha=0.1$.\footnote{As discussed in Sec.~\ref{sec:model2}, for $G\mu\lesssim 10^{-10}$, the contribution from matter-era loops $\mathrm{\Omega}_{\rm GW}^m$ is negligible and does not need to be included.} Note also that, although these approximations only provide a description of the SGWB generated by the fundamental mode of emission, one may use it to construct an analytical approximation up to an arbitrary number of modes of emission $n_*$ for any $q$ using
\begin{equation}
\mathrm{\Omega}_{\rm GW}(f,q,n_*)=\sum_{j=1}^{n_*}\frac{j^{-q}}{\mathcal{E}}\mathrm{\Omega}_{\rm GW}(f/j)\,,\label{eq:summation}
\end{equation}
with $\mathcal{E}=\sum_m ^{n_*} m^{-q}$.

\begin{figure}
\centering
\includegraphics[width=0.6\textwidth]{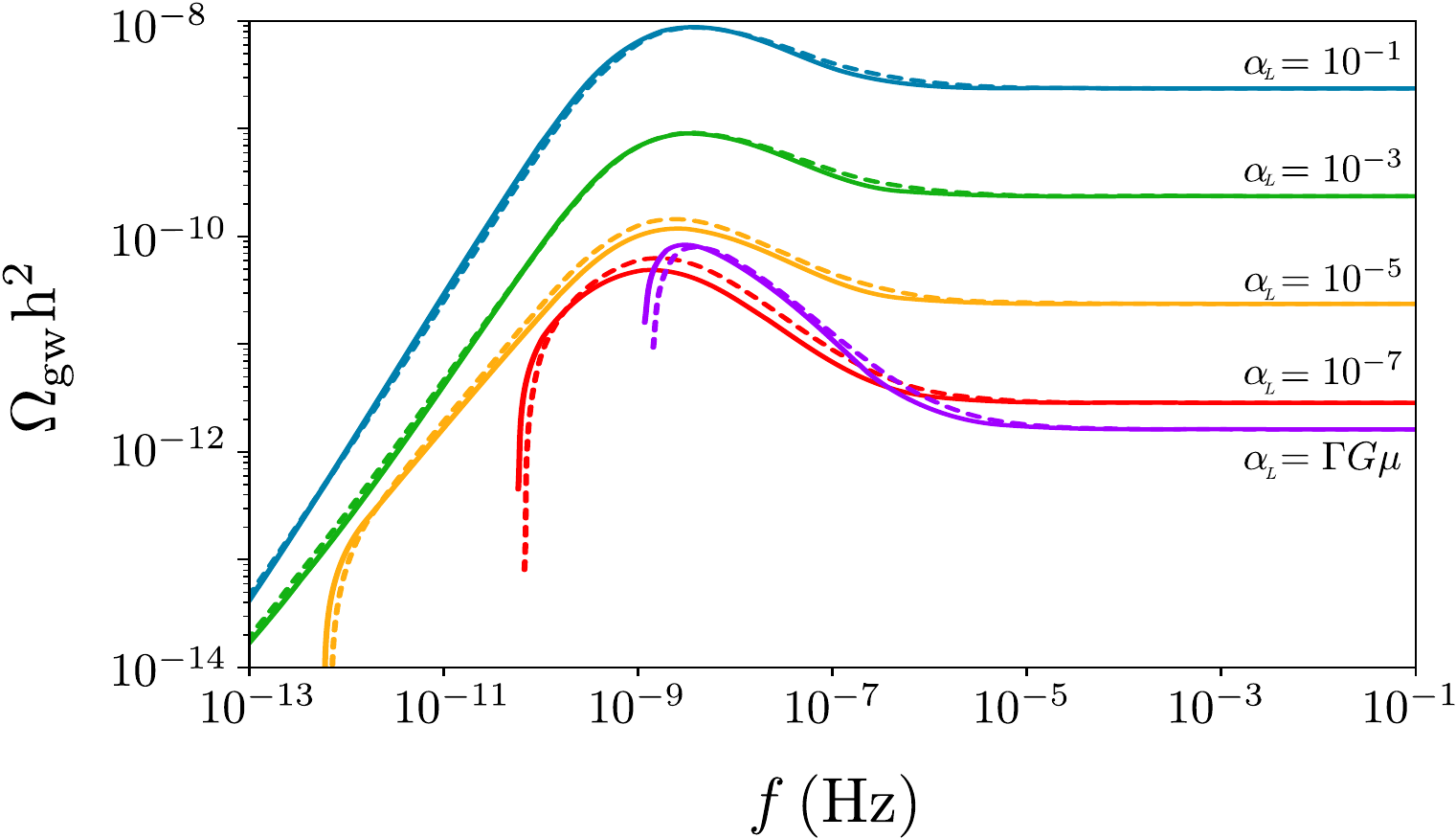}
\caption{\small Analytical approximation to the SGWB generated by cosmic string networks with $G\mu=10^{-10}$ and different values of $\alpha$ for Model 1. The solid lines represent the approximation to the SGWB in Eq.~(\ref{approx}), while the dashed lines correspond to the SGWB obtained numerically. Figure taken from Ref.~\cite{Sousa:2020sxs}.}
\label{comp-large}
\end{figure}

\paragraph{Model 3}
A first review of the SGWB of Model 3 was conducted in Ref.~\cite{Ringeval:2017eww}, and we present below analytical approximations derived in Ref.~\cite{Auclair:2019wcv}.
The loop distribution of Model 3 is similar on large scales to the two above-mentioned models, but is also characterised by an extra population of small loops with invariant lengths smaller than $\Gamma G \mu t$. This extra population is treated separated from the rest of the loop distribution resulting in five contributions, i.e.~the extra populations of small loops emitting GWs either during the radiation era or the matter era, and the three populations mentioned in the previous section.

The contribution from radiation-era loops is given by
\begin{equation}
    \Omega_\mathrm{GW}^{r} = \dfrac{64\pi C_r \Omega_r}{3\Gamma (2-2\chi_r)}(\Gamma G\mu)^{2\chi_r} \left(1+\dfrac{4H_r(1+z_\mathrm{eq})}{f\Gamma G\mu} \right)^{2\chi_r-2},
\end{equation}
where $C_r = 0.08$ and $\chi_r = 0.2$.
This formula is characterised by a plateau at high frequency.
The loops produced during radiation-era that survive during matter era also contribute to the SGWB
\begin{equation}
    \Omega_\mathrm{GW}^{rm} = \frac{54 \pi C_r H_m \Omega_m}{\Gamma f (\Gamma G\mu)^{1- 2\chi_r}}(1+z_\mathrm{eq})^{3(2\chi_r-1)/2} \left[\frac{x^{2-6\chi_r}}{2-6\chi_r} {}_2F_{1}({3-2\chi_r},{2-6\chi_r};{3-6\chi_r};{-\frac{3 H_m x}{f\Gamma G\mu}})\right]^{\sqrt{1+z_\mathrm{eq}}}_1.
\end{equation}
Matter-era loops give rise to a SGWB given by
\begin{equation}
    \Omega_\mathrm{GW}^{m} = \frac{2 \times 3^{2\chi_m} \pi C_m \Omega_m}{H_m^{2-2\chi_m} \Gamma f^{2\chi_m - 2}} (\Gamma G\mu)^{2} \left[\frac{x^{2\chi_m-4}}{2\chi_m-4} {}_2F_{1}({3-2\chi_m},{4-2\chi_m};{5-2\chi_m};{-\frac{f\Gamma G\mu}{3H_m x}})\right]^{\sqrt{1+z_\mathrm{eq}}}_{1},
\end{equation}
where $C_m = 0.015$ and $\chi_m = 0.295$, and ${}_2F_{1}$ is the Gauss hypergeometric function.

The extra population of small loops plays an important role in determining the SGWB, particularly at high frequencies.
Due to their small size ($< \Gamma G \mu t$) their contribution is cutoff at low frequencies and their lifetime is too short for them to survive from the radiation-era to the matter-era.
The contribution of the extra population of small loops emitting GWs during the radiation era is given by the piece-wise formula
\begin{align}
    \Omega_\mathrm{GW}^{r, \mathrm{epsl}} &= \frac{64 \pi C_r \Omega_r(1/2 - 2\chi_r)}{3 (1-2\chi_r)(2-2\chi_r)} G\mu \gamma_c^{2\chi_r-1} \\
    & \nonumber \times \begin{cases}
        0 & \text{if } f < 4 (1+z_\mathrm{eq}) H_r (\Gamma G\mu)^{-1}\\
        \left[4(1+z_\mathrm{eq})H_r /(\gamma_c f)\right]^{2\chi_r-1} - \left(\Gamma G\mu / \gamma_c\right)^{2\chi_r - 1} & \text{if } f < 4(1+z_\mathrm{eq}) H_r \gamma_c^{-1}\\
        (2-2\chi_r) - 4(1+z_\mathrm{eq})H_r (1-2\chi_r) / (\gamma_c f) - \left(\Gamma G\mu /\gamma_c\right)^{2\chi_r-1} &  \text{if } f > 4(1+z_\mathrm{eq}) H_r \gamma_c^{-1}
    \end{cases}
\end{align}
and the contribution of the extra population of small loops emitting GWs during the matter era is given by the piece-wise formula
\begin{align}
    \Omega_\mathrm{GW}^{m, \mathrm{epsl}} &= \frac{54 \pi C_m H_m \Omega_m (1 - 2\chi_m)}{(3-2\chi_m)(2-2\chi_m) f} G\mu \gamma_c^{2\chi_m-2} \left(\frac{3 H_m}{\gamma_c f}\right)\\
    & \nonumber \times \begin{cases}
        0 & \text{if } f < 4 H_m (\Gamma G\mu)^{-1}\\
        \left[\frac{3 H_m}{\gamma_c f}\right]^{2\chi_m - 3}\left\{1 - \left[3H_m/ (\Gamma G\mu f)\right]^{3-2\chi_m}\right\} & \text{if } f < 4 H_m \sqrt{1+z_\mathrm{eq}}(\Gamma G\mu)^{-1}\\
        \left[\frac{3 H_m}{\gamma_c f}\right]^{2\chi_m - 3}\left[1 - (1+z_\mathrm{eq})^{-(3-2\chi_m)/2}\right] & \text{if } f < 4 H_m \gamma_c^{-1}\\
        (3-2\chi_m)f\gamma_c / (3 H_m) + (2\chi_m - 2) - \left(\frac{f \gamma_c}{3H_m \sqrt{1+z_\mathrm{eq}}}\right)^{3-2\chi_m} & \text{if } f < 4 H_m \sqrt{1+z_\mathrm{eq}} \gamma_c^{-1}\\
        \left[\frac{3 H_m}{\gamma_c f}\right]^{-1}(3-2\chi_m) \left[1 - (1+z_\mathrm{eq})^{-1/2} \right]& \text{if } f > 4 H_m \sqrt{1+z_\mathrm{eq}} \gamma_c^{-1}
    \end{cases}.
\end{align}
Similarly to the previous section, these approximations assume that the GW are emitted by the fundamental mode, ie. by the oscillations of the loop.
One can use Eq.~\eqref{eq:summation} to construct predictions with an arbitrary number of modes.

\begin{figure}
    \centering
    \includegraphics[width=0.6\textwidth]{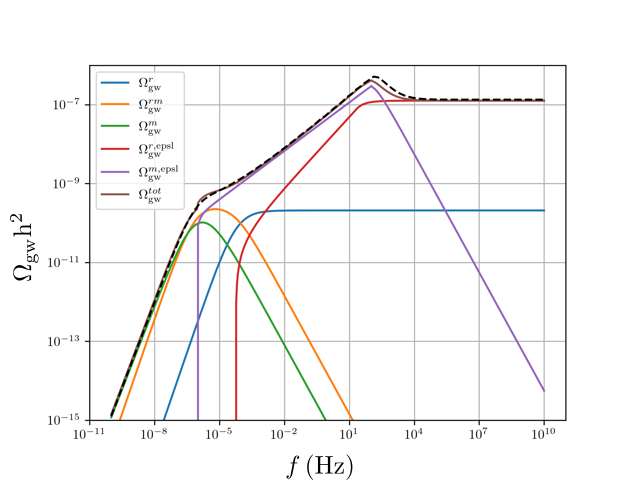}
    \caption{\small Figure to demonstrate the validity of the templates for Model III. The dark dashed line is the result from numerical integration with only the fundamental mode and $G(z) = 1$.}
\end{figure}

\subsection{LISA detection prospects}\label{sec:detection-prospects}

Searches by current GW experiments (by LIGO/VIRGO and PTAs) on power spectra of the form $\Omega_{\rm GW}(f) = A f^n$ have provided upper bounds on the amplitude $A$ for different fixed values of the spectral index $n$~\cite{LIGOScientific:2019vic,Lentati:2015qwp,Arzoumanian:2018saf}. Very recently, the NANOGrav and EPTA collaborations have even reported an excess that might be due to a SGWB signal~\cite{Arzoumanian:2020vkk, Chen:2021rqp}, although the origin is still uncertain. As no statistical significant evidence of the expected quadrupolar spatial correlations has been found, the detected signal can still be ascribed to some unknown systematics. More data is needed in order to discern whether this signal is a real SGWB or not, and if it is, whether it is due to BSM  scenarios~\cite{Ellis:2020ena, DeLuca:2020agl, Neronov:2020qrl} or to astrophysics~\cite{Sesana:2016ljz}.

The PLS curve~\cite{Thrane:2013oya} is a useful construction that  graphically quantifies the ability of a detector to measure a SGWB with a spectrum characterised by a power-law in frequency. Recently, the LISA Cosmology Working Group has presented a new technique for a systematic reconstruction of a SGWB signal without assuming a power-law spectrum~\cite{Caprini:2019pxz,Flauger:2020qyi}. This consists of separating the entire LISA band into smaller frequency bins, and then to reconstruct the signal within each bin, where it can be locally well-approximated in terms of a power law. This method can be used to reconstruct signals with arbitrary spectral shapes, taking into account instrumental noise at each frequency bin. The use of this technique for assessing the detectability of the SGWB from a cosmic string network is particularly needed, as the spectral shape of the signal is not a simple plateau for the lowest $G\mu$ values that LISA can probe. Furthermore, the SGWB spectrum from string networks can also exhibit scale-dependent features within the LISA frequency band, such as changes in the number of relativistic degrees of freedom or the early universe EoS, see Sec.~\ref{sec:nonStandard}.

While a multi-bin analysis technique for the detection of the SGWB from string networks is currently a work in progress, here we content ourselves with using the PLS as a criterion for detection; namely the SGWB spectrum must be equal or above the PLS curve. We will use the LISA PLS as introduced by Ref.~\cite{Thrane:2013oya}, but using the most updated LISA sensitivity curves based on the final configuration of LISA and new knowledge of its noise. (See Ref.~\cite{LISA_docs} for all current LISA documentation.)  

Claiming detection of a given SGWB from cosmic string loops (say for a given tension and other fixed string network parameters), can be roughly interpreted as the detection of the signal after 3 years of accumulated data (corresponding to 4 years of LISA operation), with a SNR\,$\geq 10$. Since the shape can be more complicated than a simple power law, a more elaborated analysis following Refs.~\cite{Caprini:2019pxz,Flauger:2020qyi} is required to assess the SNR for a given detection, see also Ref.~\cite{Karnesis:2019mph}. Here, for the moment, we simply quantify the parameter space compatible with a detection, without quantifying the SNR associated to such detection. We do not reconstruct such parameter space with appropriate statistical techniques, though such work is already in the pipeline of coming work from the LISA collaboration. 

\begin{figure}
    \centering
    \includegraphics[height=9cm]{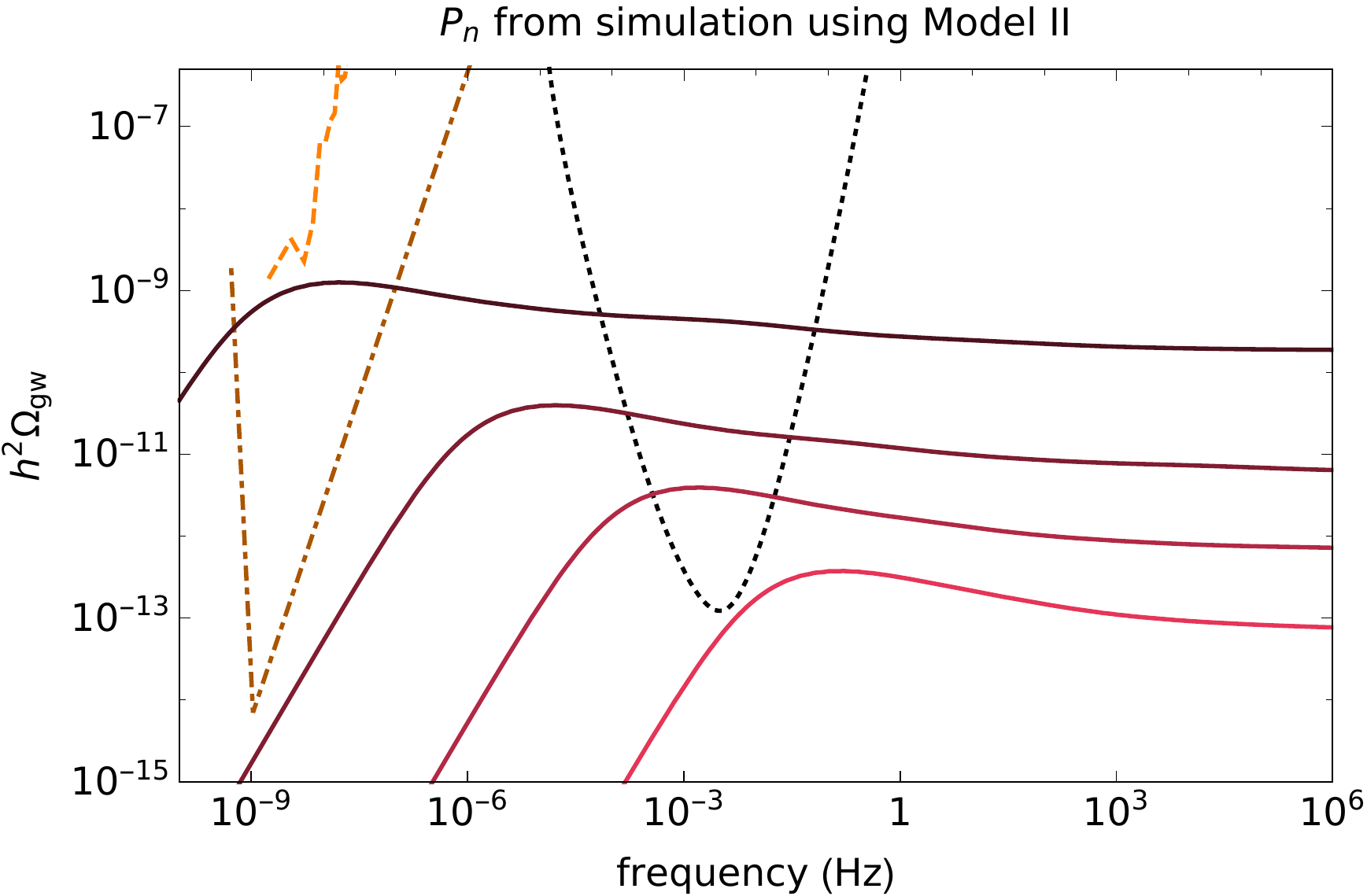}
    \caption{\small 
    Solid red curves show cosmic string SGWB curves for a range of $G\mu$ values. From the darkest most high up line to the lightest lowest one these read: $G\mu=10^{-10}$, $G\mu=10^{-13}$, $G\mu=10^{-15}$ and $G\mu=10^{-17}$. The $P_n$ used in computation of these spectra was inferred from simulations~\cite{Blanco-Pillado:2017oxo}, and the loop number density is from Model 2.
    The dashed orange curve shows the sensitivity of EPTA. The  dark orange dash-dotted line shows the projected SKA sensitivity. The dotted black line shows the LISA PLS of SNR\,$=10$.}\label{fig:fiducialSGWB}
\end{figure}

\begin{figure}
    \centering
    \includegraphics[height=9cm]{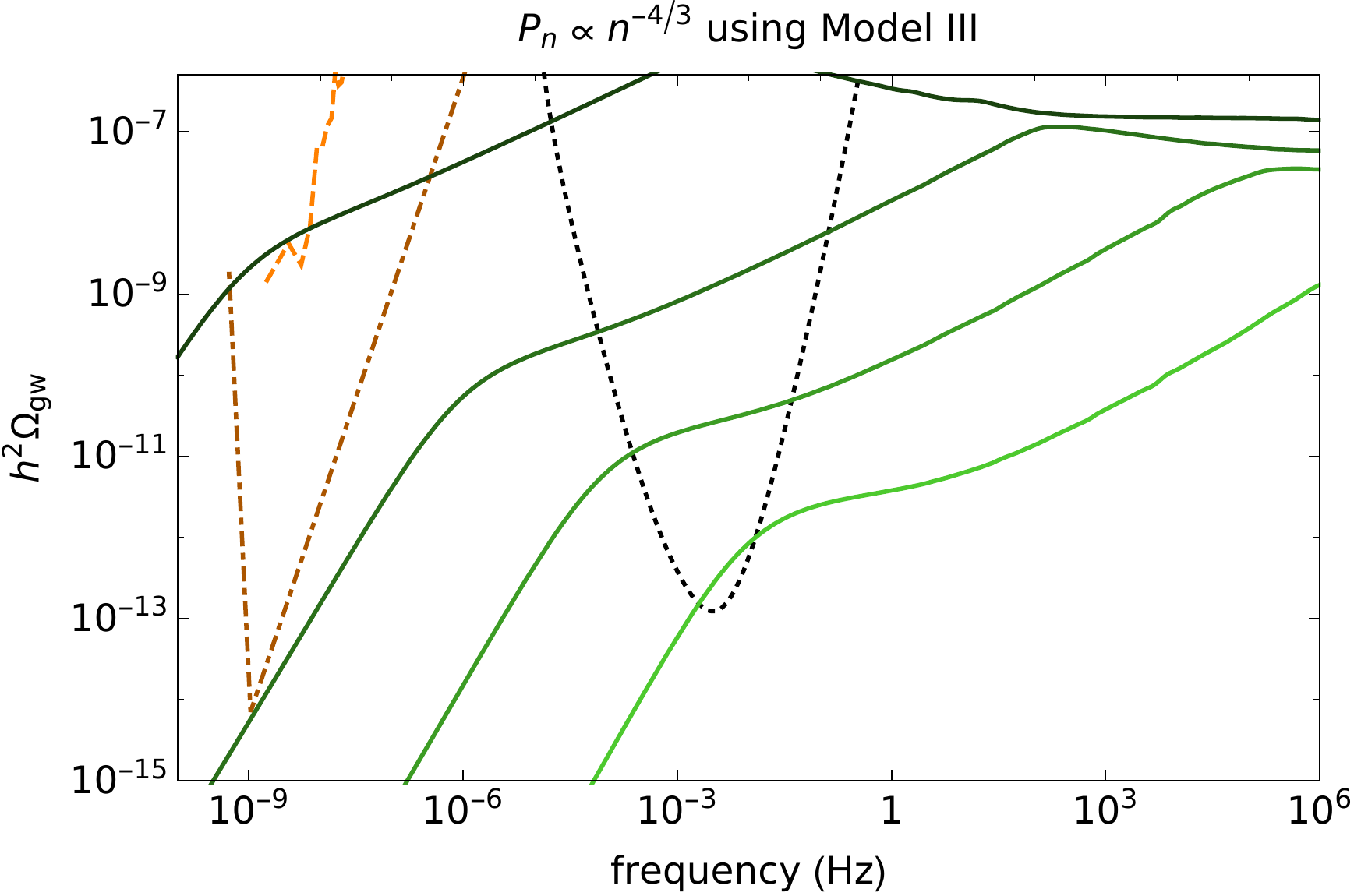}
     \caption{\small Identical to Fig.~\ref{fig:fiducialSGWB}, however, $P_n\propto n^{-4/3}$ and using the loop number density from Model 3~\cite{Lorenz:2010sm}.}
   \label{fig:combined-LRS-large}
   \end{figure}

In Figs.~\ref{fig:fiducialSGWB} and \ref{fig:combined-LRS-large}, we present  numerical results for the SGWB generated by cosmic string loops, in a standard cosmological background. The LISA band is well suited to set strong constraints on the string tension, thanks to the natural shift of the ``bump'' in the SGWB to larger frequencies as $G\mu$ decreases. This effect can be clearly seen in these figures, where we show a sequence of SGWB spectra over the LISA PLS, as we vary the string tension. We can find in this way the lowest $G\mu$ for which an intersection between the SGWB spectrum and the PLS still takes place. While the exact bound depends on our choice of model and $P_n$, in the regime LISA will probe, all three models --- Model 2 and Model 3, as well as Model 1 (when calibrated to describe NG simulations by setting $\alpha=0.1$ and $\mathcal{F}=0.1$) --- predict a string tension bound of $\mathcal{O}(10^{-17})$. 
We note that the trailing edge of the bump of the SGWB (that scales as $\Omega_\text{GW}\propto f^{3/2}$)  will be the last part of the spectrum to pass through the LISA sensitivity band.

Note also that for string tensions $G\mu > 10^{-16}$, LISA will probe the high-frequency side of the SGWB bump, the particular shape of which depends on how the number of particle degrees of freedom change across the universe history. This is relevant because while the three models predict roughly equal lower bounds for the LISA window, Models 1 and 2 disagree with Model 3 at high frequencies. These discrepant regions will pass through the LISA band and hence, in the event of a detection, we could discern among the different models.

\subsubsection{Anisotropies in the stochastic gravitational wave background}

GW sources with an inhomogeneous spatial distribution would lead to anisotropies in the SGWB, in addition to the anisotropies induced by the nature of spacetime along the line of propagation of GWs. The formalism \cite{Jenkins:2018nty} to study anisotropies induced by the distribution of string loops was consequently applied for the string distribution of Models 2 and 3.
It has been shown  that the angular spectrum of the 2-point correlator is relatively insensitive to the particular choice of  the string loop model: regardless of the model, the anisotropies are driven by local Poisson fluctuations in the number of loops, and the resulting angular power spectrum is spectrally white (i.e.~$C_\ell=\text{constant}$ with respect to $\ell$) \cite{Jenkins:2018nty}.

\begin{figure}[t]
\centering
\includegraphics[width=0.7\textwidth]{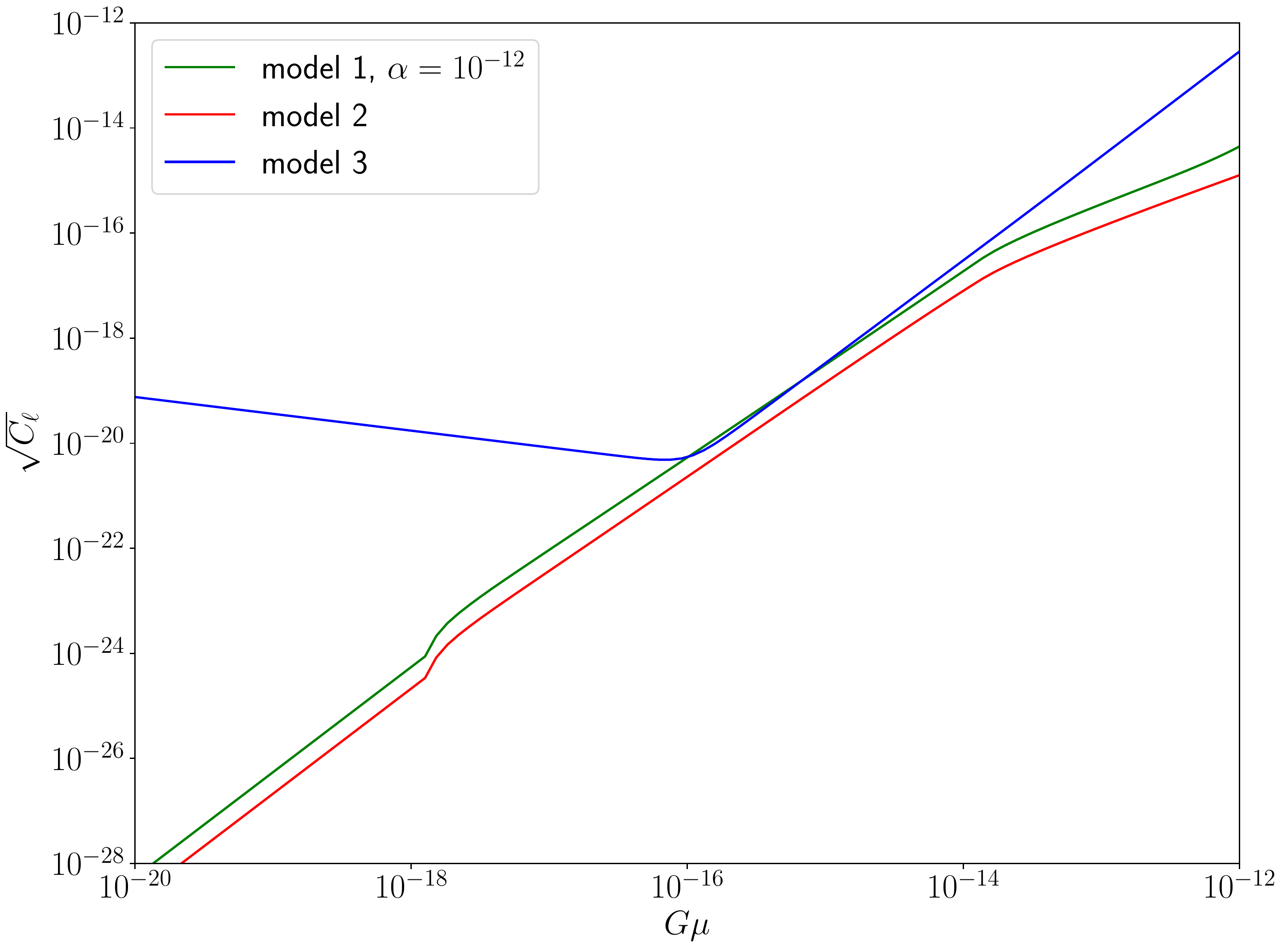} 
\caption{\small \label{fig:C_ells}
Amplitude of the SGWB anisotropies for different cosmic string network models, as a function of the string tension. We use a representative LISA-band GW frequency of $1~\mathrm{mHz}$. Note that the spectra here are not normalised with respect to the monopole, so $\sqrt{C_\ell}$ is proportional to $\Omega_\mathrm{GW}$. We note that the angular spectrum is constant with respect to $\ell$, since the anisotropies are driven by local Poisson fluctuations in the number of loops.}
\end{figure}

In Fig.~\ref{fig:C_ells} we show the amplitude of the SGWB angular power spectrum for models 2 and 3 as a function of $G\mu$.
We also include Model 1 with a small value of the initial loop size $\alpha$, as this has been studied in the literature as a source of significant anisotropies in the PTA frequency band~\cite{Kuroyanagi:2016ugi}.
We find that, regardless of the loop model and the string tension, the predicted $C_\ell$ spectrum is unfortunately far too small to be detected with LISA.

\subsection{Extended and alternative considerations}
\label{sec:extended}

In the previous sections we focused on GW signals from cosmic string loops in most conventional scenarios, i.e.~NG strings evolving in the standard cosmology background with particle content of the SM. Nevertheless there are well motivated variations encoding new physics that can lead to subtle or dramatic deviations to the standard prediction of string SGWB spectra that we have demonstrated. In this section we review a few representative cases. We will start with the fact that long strings also radiate a SGWB that add up to the previously discussed SGWB from loops. We will also consider cosmic string scenarios with different loop sizes and forecast the parameter space available for exploration with LISA. We will then move on to the possible signals from networks of cosmic strings formed during a global symmetry breaking which as we will see are quite distinct as then particle production and not GW production becomes the dominant emission mode. Next we will turn to cosmic strings emerging from superstring theory and in particular the possibility of smaller reconnection probability they can feature. We finally consider the case of metastable cosmic strings.

Another key aspect is that the general broad band nature of cosmic strings SGWB enables \textit{cosmic archaeology} which would allow us to probe the expansion history prior to BBN as well as new particles beyond the SM that are difficult to access by other means. We will not cover this method here, instead referring the interested reader to Sec.~\ref{sec:nonStandard} where methods of probing the expansion history are discussed and specifically to its Sec.~\ref{subsec:nonStandardCosmicStrings} dealing with cosmic strings.

\subsubsection{Long strings}\label{ssec:InfiniteStrings}

We have focused so far on the GW emission by sub-horizon size string loops. Long strings, however, either infinite or super-horizon loops, emit also GWs. One contribution to this signal originates due to the relativistic motion of the long strings, as the network energy-momentum tensor adapts itself to maintain scaling during cosmic evolution. As a result, an emission of GWs  sourced by the anisotropic stress of the network, takes place around the horizon scale at every moment~\cite{Krauss:1991qu,JonesSmith:2007ne,Fenu:2009qf,Figueroa:2012kw}. This SGWB background is actually expected to be emitted by any scaling network of cosmic defects, independently of the topology and origin of the defects~\cite{Figueroa:2012kw}. It represents an {\it irreducible background} generated by any defect network (that has reached scaling). For NG cosmic string networks, such irreducible background represents however a sub-dominant signal compared to the background emitted from the loops discussed before. In the case of field-theory string networks (for which simulations to date indicate the absence of ``stable'' loops~\cite{Vincent:1997cx,Hindmarsh:2008dw,Daverio:2015nva,Hindmarsh:2017qff,Hindmarsh:2021mnl}), it is instead the dominant GW signal emitted by the network.

The energy density spectrum of this irreducible SGWB from long strings is predicted to be scale-invariant for the modes emitted during radiation domination (RD)~\cite{Figueroa:2012kw}. The amplitude of the background depends on the fine details of the so called \textit{unequal-time-correlator} of the network's energy-momentum tensor, which can be only accurately obtained from very large scale lattice simulations of defect networks. In the case of global defects, the scale-invariant GW power spectrum has been analytically estimated in Refs.~\cite{JonesSmith:2007ne,Fenu:2009qf}. The amplitude of the spectrum plateau for global strings has been recently calibrated in lattice field theory simulations~\cite{Figueroa:2020lvo}, obtaining
\begin{eqnarray}\label{eq:PlateauAmplitudeToday}
h_0^2\Omega_{\rm GW}^{(0)} 
\simeq 3.13\cdot 10^{-13} \left(\frac{G\mu}{10^{-6}}\right)^2\,.
\end{eqnarray}
Using the latest Planck CMB constraints~\cite{Lopez-Eiguren:2017dmc}, Ref.~\cite{Figueroa:2020lvo} finds that the amplitude of the plateau satisfies
\begin{eqnarray}\label{eq:GWtodayAmpl}
h^2\Omega_{\rm GW}^{(0)} < 9.7\cdot 10^{-15}\,,
\end{eqnarray}
with the number in the right hand side of the inequality corresponding to the amplitude when the CMB bound $10^{12}(G\mu)^2 \leq 0.031$ is saturated. The amplitude in Eq.~(\ref{eq:GWtodayAmpl}) is larger, for instance, than the maximum amplitude expected (as bounded by current CMB constraints~\cite{BICEP:2021xfz}) for the quasi-scale invariant GW background in slow-roll inflation~\cite{Caprini:2018mtu}, $h_0^2\Omega_{\rm GW}^{\rm (inf)} \lesssim 10^{-16}$. It is however, still too small to be observed by LISA, and it is clearly subdominant when compared to the amplitude of the dominant GW signal from the long lived NG loops, which scales as $(G\mu)^{1/2}$. 

One can also consider the contribution to the GW spectrum coming from the accumulation of small-scale structure on long strings. These kinks are the product of the multiple intercommutations that infinite strings suffer over the course of their cosmological evolution, and were noticed early on in numerical simulation of cosmic networks~\cite{Bennett:1987vf,Sakellariadou:1990nd}. The emission of GW from individual infinite strings modulated by kinks has been calculated in Refs.~\cite{Sakellariadou:1990ne,Hindmarsh:1990xi}. Using these results, one can also compute the spectrum produced by these kinks on a network assuming the simple model in which their characteristic scale is given by $\alpha t$. At high frequencies one can then estimate that the radiation-era plateau of this contribution should be~\cite{Vilenkin:2000jqa}
\begin{eqnarray}
    h^2\mathrm{\Omega}_{\rm GW} \simeq  \frac{128 \pi^2}{3 \xi^2 \alpha}  \, h^2\mathrm{\Omega}_{\rm rad}(G\mu)^2\,,
\end{eqnarray}
which, for $\alpha\approx 0.1$ and $\xi_{\rm r}=0.271$, shows a rough agreement with the value obtained from field theory simulations. On the other hand, recently, Ref.~\cite{Matsui:2019obe} has calculated the GW spectrum produced by kink-kink collisions on long strings, and found that the amplitude is larger than in previous estimates. This is because the characteristic scale $\alpha$ turns out to be much smaller than $0.1$ according to their semi-analytic estimation of the kink number distribution. 

\subsubsection{Agnostic approach to loop size}

In Sec.~\ref{sec:detection-prospects}, we have analyzed the detection prospects for NG strings using either Model 1 (calibrated to NG simulations) or simulation-inferred Models 2 and 3. Here, we extend the analysis a bit further, by using Model 1 to study scenarios with different loop sizes. As we have seen in Sec.~\ref{temp1}, although the typical shape of the SGWB spectrum generated by cosmic strings is roughly independent of $\alpha$, the amplitude of the radiation-era plateau and the characteristics of the peak  --- its shape, height and broadness --- are highly dependent on the size of loops. As a matter of fact, the amplitude of the spectrum generally decreases with decreasing $\alpha$ and, therefore, LISA should, in general, be less sensitive to scenarios in which loops are created with a smaller size.

\begin{figure}
    \centering
    \includegraphics[width=0.85\textwidth]{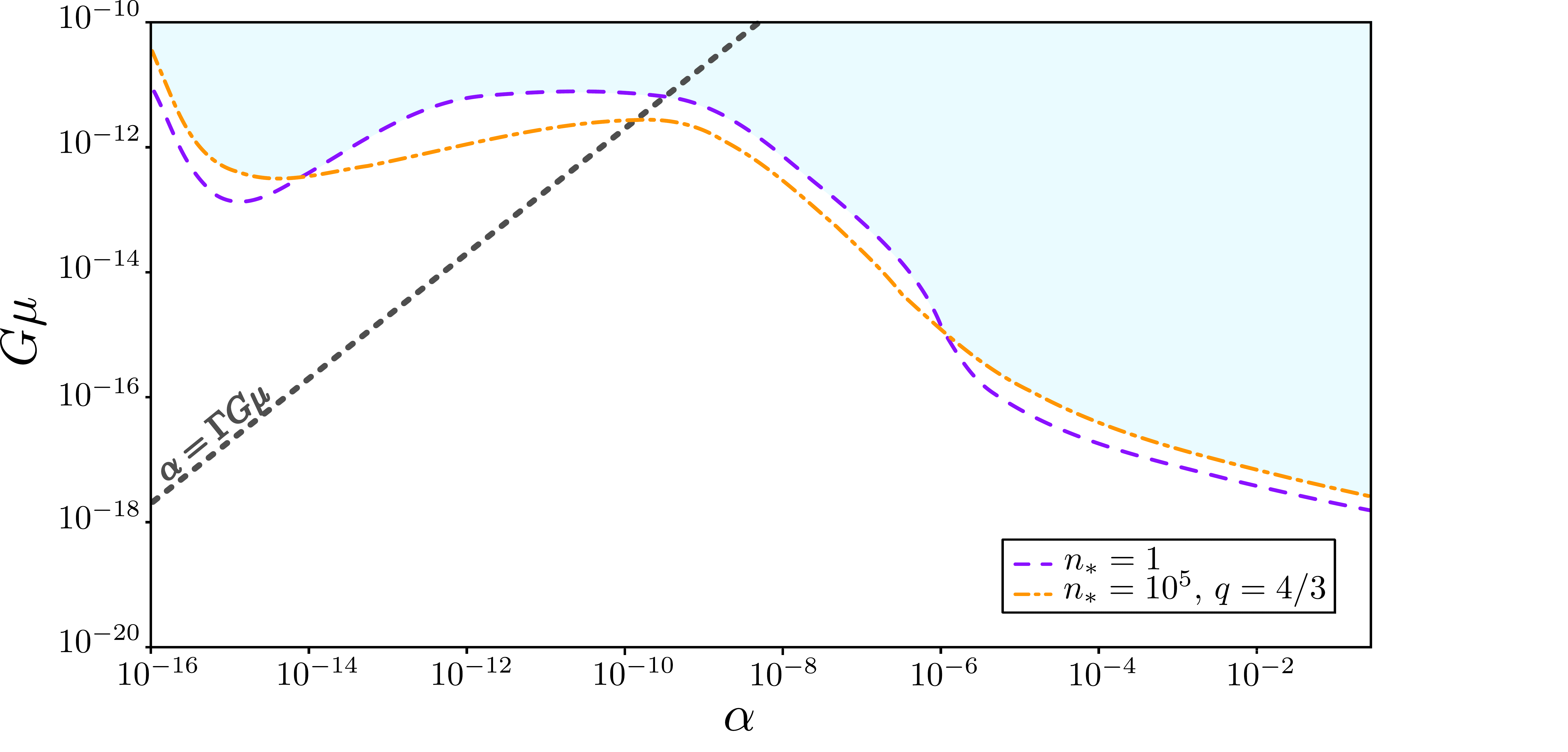}
    \caption{\small Projected constraints on $G\mu$ of the LISA mission for cosmic string scenarios characterised by different loop-size parameter $\alpha$ for $n_*=1$ (dashed line) and $n_*=10^5$, with $q=4/3$ (dash-dotted line). The shaded area corresponds to the region of the $(\alpha,G\mu)$ parameter space that will be fully available for exploration with LISA. The dotted line corresponds to scenarios for which $\alpha=\Gamma G\mu$, so that the region above this line corresponds to cosmic string models in which loops are small, while the region bellow corresponds to the large loop regime. Figure taken from  Ref.~\cite{Auclair:2019wcv}.}
    \label{fig:agn}
\end{figure}

The $(\alpha,G\mu)$ parameter space available for exploration with LISA --- which we plot in Fig.~\ref{fig:agn} --- is characterized in Ref.~\cite{Auclair:2019wcv}. This analysis finds that LISA should be able to probe about $16$ orders of magnitude in loop size and, thus, it will have a good capability to detect string models that deviate from the standard NG scenario. Note, however, that the projected constraints on cosmic string tension are also less stringent as the loop size decreases. Nevertheless, LISA shall be able (conservatively) to probe cosmic string scenarios in which loop production is significant up to tensions 
\begin{equation}
G\mu<8\times 10^{-12}\,,
\end{equation}
independently of loop size~\cite{Auclair:2019wcv}. This corresponds to an improvement of almost $5$ order of magnitude over current equivalent constraints~\cite{Sanidas:2012ee} and of more than $2$ orders of magnitude over the projected SKA constraints~\cite{Sanidas:2012tf}.

%%%%%%%%%%%%%%%%%%%%%%%%%%%%%%%%%%%%%%%%%%%%%%%%%%%%%%%%%%%%%%%%%%%%%%%%%%%%%%%%%%%%%%%%%%%%%%%
%%%%%%%%%%%%%%%%%%%%%%%%%%%%%%%%%%%%%%%%%%%%%%%%%%%%%%%%%%%%%%%%%%%%%%%%%%%%%%%%%%%%%%%%%%%%%%%

\subsubsection{Global strings}

Global cosmic strings are generically predicted in BSM frameworks with e.g.~a post-inflationary global U(1) symmetry breaking which may be associated to axion-like DM. The GW emission from global strings have been less actively studied than that from NG strings, with only few papers having addressed an explicit computation of the SGWB from global/axion strings, see Refs.~\cite{JonesSmith:2007ne,Fenu:2009qf,Giblin:2011yh,Figueroa:2012kw,Chang:2019mza,Gouttenoire:2019kij,Figueroa:2020lvo,Gorghetto:2021fsn}. Studying this topic is however particularly timely given the increasing interest in axion DM. In the following we review the semi-analytical approach taken in the recent work Ref.~\cite{Chang:2019mza}. 

The approach taken in Ref.~\cite{Chang:2019mza} systematically follows a similar procedure as for NG strings which is reviewed in Sec.~\ref{sec:model1} but there are a few key differences for global strings.
First, global strings have a time-dependent string tension \cite{Copeland:1990qu, Dabholkar:1989ju, Vilenkin:2000jqa}
\begin{equation}
\label{Eq: Global string tension}
\mu(t) = 2 \pi \eta^2 \hbox{ln}\left(L/\delta\right) \equiv 2 \pi \eta^2 N, 
\end{equation}
where $L\simeq H^{-1}\xi^{-1}$ is the string correlation length, $\xi$ is the number of long strings per horizon volume, $\delta\simeq1/\eta$ is the string thickness, and $N\equiv\hbox{ln}\left(L/\delta\right) \simeq \hbox{ln}(\eta\xi^{-1}t)$ is time-dependent. Once reaching the scaling regime, the long string energy density evolves as
\begin{equation}
\rho_\infty = \xi(t) \frac{\mu(t)}{t^2},
\end{equation}
where $\xi(t)$ quickly approaches a constant for NG strings, yet needs to be determined for global strings. In addition, global string loops have an additional significant decay channel, through Goldstone emission. Taking these into account, the evolution equations of a global string network are as follows based on a Velocity-dependent One-Scale model \cite{Martins:2018dqg,Vilenkin:2000jqa}:
\begin{align}
\left(2 - \frac{1}{N} \right) \frac{dL}{dt} = 2 H L \left( 1 + v^2 \right) + c v + \sigma \frac{v}{N},\;\;\;\; \frac{dv}{dt} = \left( 1 - v^2\right) \left[ \frac{k}{L} - 2 H v \right],
\end{align}
where $k\sim 0.28$ is a momentum parameter, $v \sim 0.57$ is the averaged long string velocity, $c\sim 0.5$ is the loop chopping parameter and $\sigma \sim 5.83$ is the Golstone radiation parameter. These values are obtained \cite{Chang:2019mza} from recent simulation results \cite{Gorghetto:2018myk,Kawasaki:2018bzv,Hindmarsh:2019csc,Klaer:2017ond}.

As presented in Eq.(\ref{Eq: Global string tension}), reliable numerical simulations of global string networks require a huge scale separation in string length $L \propto 1/H(t)$ and core width $\delta$. Thus the global string simulation is much more challenging. Nevertheless, it has seen rapid development in recent years \cite{Gorghetto:2018myk,Kawasaki:2018bzv,Hindmarsh:2019csc,Klaer:2017ond,Buschmann:2019icd,Figueroa:2020lvo}, while uncertainties remain to be resolved with future higher resolution simulations.
Due to the lack of dedicated simulation results for global string loop distribution at formation time, a reasonable benchmark inspired by Model 2 for the NG case is considered while the effects of alternative possibilities of loop distribution are discussed in Ref.~\cite{Chang:2019mza}. 

\begin{figure}[t]
\centering
\includegraphics[width=0.85\textwidth]{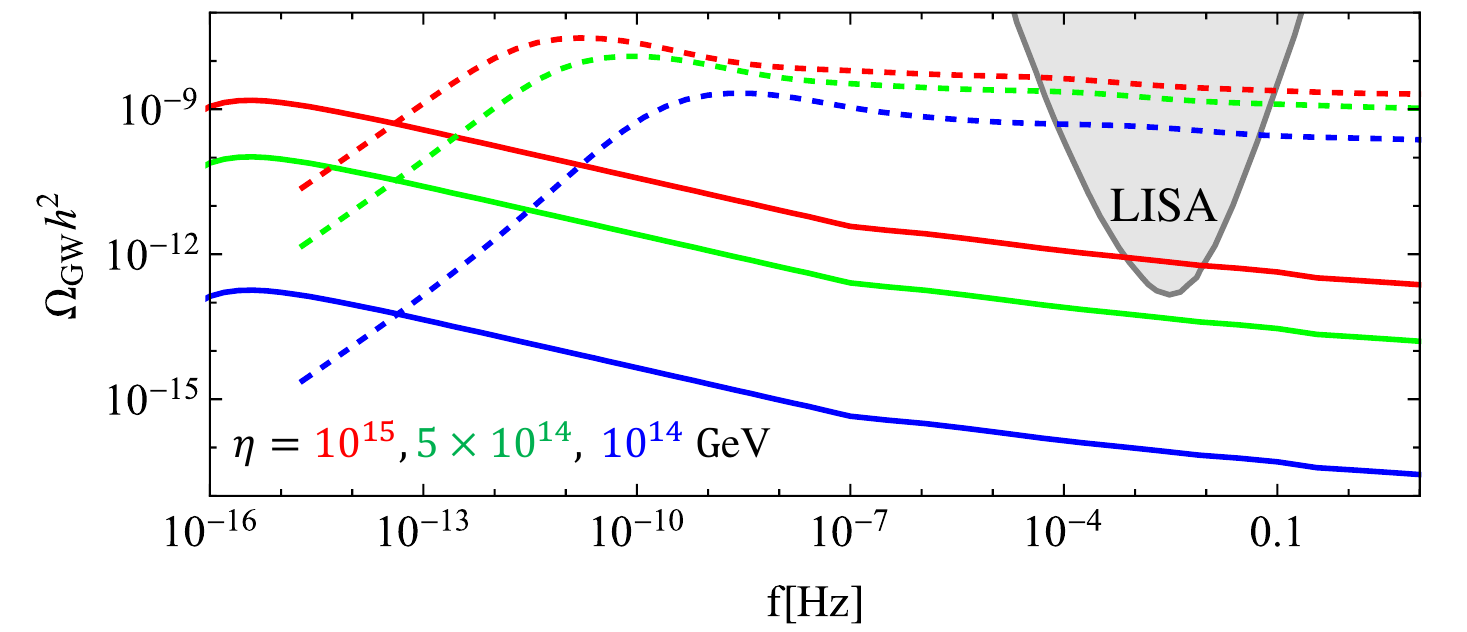} 
\caption{\small \label{Fig: Global_String_Spectrum} SGWB spectrum from a global (solid) and gauge (dashed) string network in standard cosmology with loop length parameter $\alpha = 0.1$, and symmetry breaking scale $\eta =  10^{15}, 5 \times 10^{14}, 10^{14}\,$GeV for red, green, blue, respectively. The grey region is the LISA sensitivity.}
\end{figure}

Once formed, a loop oscillates and loses energy by the rate \cite{Vilenkin:1986ku, Battye:1995hw, Battye:1993jv, Vilenkin:2000jqa}
\begin{equation}
dE/dt=-\Gamma G\mu^2-\Gamma_a\eta^2, \label{eq: powers}
\end{equation} where the right hand side represents GW and Goldstone radiation in order. Studies show that $\Gamma\simeq 50$ \cite{Vilenkin:1981bx,BlancoPillado:2011dq,Blanco-Pillado:2013qja,Blanco-Pillado:2017oxo}, $\Gamma_a\simeq 65$ \cite{Vilenkin:2000jqa}. 
Consequently the length of a loop after its formation time $t_i$ would evolve as
\begin{equation}
\ell(t) \simeq \alpha t_i - \Gamma G \mu (t-t_i) - \kappa (t-t_i), \label{eq: ell_evol}
\end{equation}
where $\kappa \equiv \Gamma_a/(2\pi N)$. The SGWB from global strings can be computed following the similar procedure of computing the SGWB for NG strings while taking into account these distinctions. Fig.~\ref{Fig: Global_String_Spectrum} illustrates the results with varying symmetry breaking scale $\eta$, assuming standard cosmic history, with 
the NG string GW spectra shown in contrast. As can be seen, the global string amplitudes are more sensitive to scale $\eta$, i.e.~$\Omega_{\textrm{GW}}^{\textrm{global}} \propto \eta^{4}$ and the NG string $\Omega_{\textrm{GW}}^{\textrm{NG}} \propto \eta$. The spectrum falls off at $f\sim \frac{1}{t_0} \sim 10^{-17}\,$Hz due to the last GW emissions, today. Meanwhile, in the late matter domination period, it follows the frequency dependence $\Omega_{\textrm{GW}}^{\textrm{global}}(f) \propto f^{-1/3}$ for summation of higher oscillation normal modes $n \gg 10^{5}$ as demonstrated in very recent studies \cite{Cui:2019kkd, Blasi:2020wpy}. Then, at $f \sim 10^{-8}\,$Hz, the logarithmic time dependence of $\mu$ made a gradually, logarithmically declining plateau towards high $f$ on the GW spectrum, instead of a large plateau in NG strings in RD period.

In very recent simulation studies \cite{Saurabh:2020pqe}, they have concluded that the global string loop lifetime is of the order the loop initial length $l \sim \alpha t$ which is in a good agreement with the prediction of Eq.~(\ref{eq: ell_evol}) by energy conservation. However, the amplitude of the SGWB from large-scale simulations of global string networks from Refs.~\cite{Figueroa:2020lvo,Figueroa:2012kw}, c.f.~Eqs.~(\ref{eq:PlateauAmplitudeToday}) and  (\ref{eq:GWtodayAmpl}), obtain an amplitude of the SGWB notably smaller than the above semi-analytical prediction. More work is therefore needed, with further investigation and improvements on both global string simulation and analytical analysis, in order to understand the origin of this discrepancy.

\subsubsection{Cosmic superstrings}\label{super}
Several brane-inflationary scenarios predict the copious production of cosmic superstrings (see e.g.~Refs.~\cite{Copeland:2003bj,Dvali:2001fw,Jones:2003da}): fundamental strings -- or F-strings -- and 1-dimensional Dirichlet branes -- or D-strings -- that grow to macroscopic sizes and play the cosmological role of cosmic strings. These cosmic superstrings, as a result of their quantum nature, do not always exchange partners when they collide: strings may simply pass through each other without intercommutation. In other words, the probability of intercommutation $\mathcal{P}$ is -- unlike that of ordinary strings~\cite{Shellard:1987bv,Verbiest:2011kv} -- smaller than $1$. In fact, it was shown  that $10^{-3} \lesssim \mathcal{P} \lesssim 1$ in collisions between F-strings and $10^{-1} \lesssim \mathcal{P} \lesssim 1$ for D-string collisions~\cite{Jackson:2004zg}.

Cosmic superstring networks are then expected to lose energy less efficiently and, consequently, to be significantly denser than networks of ordinary cosmic strings and to generate a SGWB with a higher amplitude as a result. As a matter of fact, the loop-chopping parameter -- which quantifies the energy that is lost in the form of loops -- is expected to be such that $\tilde{c} \propto \mathcal{P}^\gamma$, where $\gamma>0$.\footnote{The exact value of the exponent $\gamma$ is still a matter of debate -- $\gamma=1/2$ was observed in NG simulations in Minkowski space~\cite{Sakellariadou:2004wq} and $\gamma=1/3$ in both radiation- and matter-era simulations~\cite{Avgoustidis:2005nv} -- and, thus, here we discuss the effects of $\mathcal{P}$ (mostly) qualitatively.} For such weakly interacting networks, with $\tilde{c}\ll 1$, the amplitude of the SGWB is expected to roughly scale as~\cite{Avelino:2012qy,Sousa:2016ggw}
\begin{equation}
 \mathrm{\Omega}_\text{\rm GW}\propto {\tilde{c}}^{-2}\propto \mathcal{P}^{-2\gamma}\,.
\end{equation}
The constraints on $G\mu$ derived on the previous sections for $\mathcal{P}=1$ are then conservative for cosmic superstrings: for $\mathcal{P}<1$ the bounds should necessarily be tighter.

Note, however, that the length of loops produced by cosmic superstring networks is not known, since the precise number density of loops was not yet measured in simulations.\footnote{There is some evidence that the reduction of the intercommuting probability is more efficient in suppressing the production of large loops than that of small loops~\cite{Avgoustidis:2005nv}, which seems to indicate that smaller $\alpha$ ($\sim \Gamma G\mu$) may be favoured for these networks.} Nevertheless, conservative ``$\alpha$-independent'' constraints on cosmic superstring tension -- obtained using the radiation-era plateau of the SGWB generated by small loops -- can be derived. In fact, LISA will be able to probe cosmic strings up to a tension of $G\mu\sim 10^{-12}, 10^{-13}, 10^{-14}$, for $\mathcal{P}=10^{-1}, 10^{-2}, 10^{-3}$ respectively~\cite{Auclair:2019wcv}.\footnote{For larger loop sizes, the strength of the constraints may increase by up to $6$ orders of magnitude~\cite{Auclair:2019wcv}.}

Note also that there are relevant aspects of cosmic superstring dynamics that were not taken into account when deriving these constraints. In particular, when superstrings of different types collide, they are expected to bind together to create a (heavier) third type of string. This is expected to lead to networks with junctions and a hierarchy of tensions -- whose dynamics differ from that of ordinary string networks~\cite{Copeland:2003bj,Copeland:2006eh,Copeland:2006if,Copeland:2007nv,Avgoustidis:2007aa,Rajantie:2007hp,Sakellariadou:2008ay,Avgoustidis:2009ke,Avgoustidis:2014rqa} -- and to have an impact on the shape and amplitude of the SGWB~\cite{Pourtsidou:2010gu,Sousa:2016ggw}. Moreover, there are several other important aspects regarding the GW emission by cosmic superstrings that need to be clarified -- most notably the number and strength of the cusps \cite{Elghozi:2014kya} as well as the possible coupling of superstrings to other fields -- before a detailed study of the parameter space available to LISA can be performed.

\subsubsection{Metastable strings}

Up to this point, the discussion in this section focused on stable cosmic strings, i.e.~cosmic superstrings or strings whose stability is protected by the nontrivial vacuum topology in the underlying field theory. However, in certain scenarios, cosmic strings can become metastable, with important implications for the expected signal in GWs. In the following, we will discuss metastable strings in models with an enlarged gauge group at higher energies, as they often arise in the context of grand unified theories. For other models resulting in cosmic strings with a finite lifetime, see e.g.~Refs.~\cite{Kamada:2014qja,Kamada:2015iga,Bettoni:2018pbl}.

\begin{figure}
    \centering
    \includegraphics[width=0.495\textwidth]{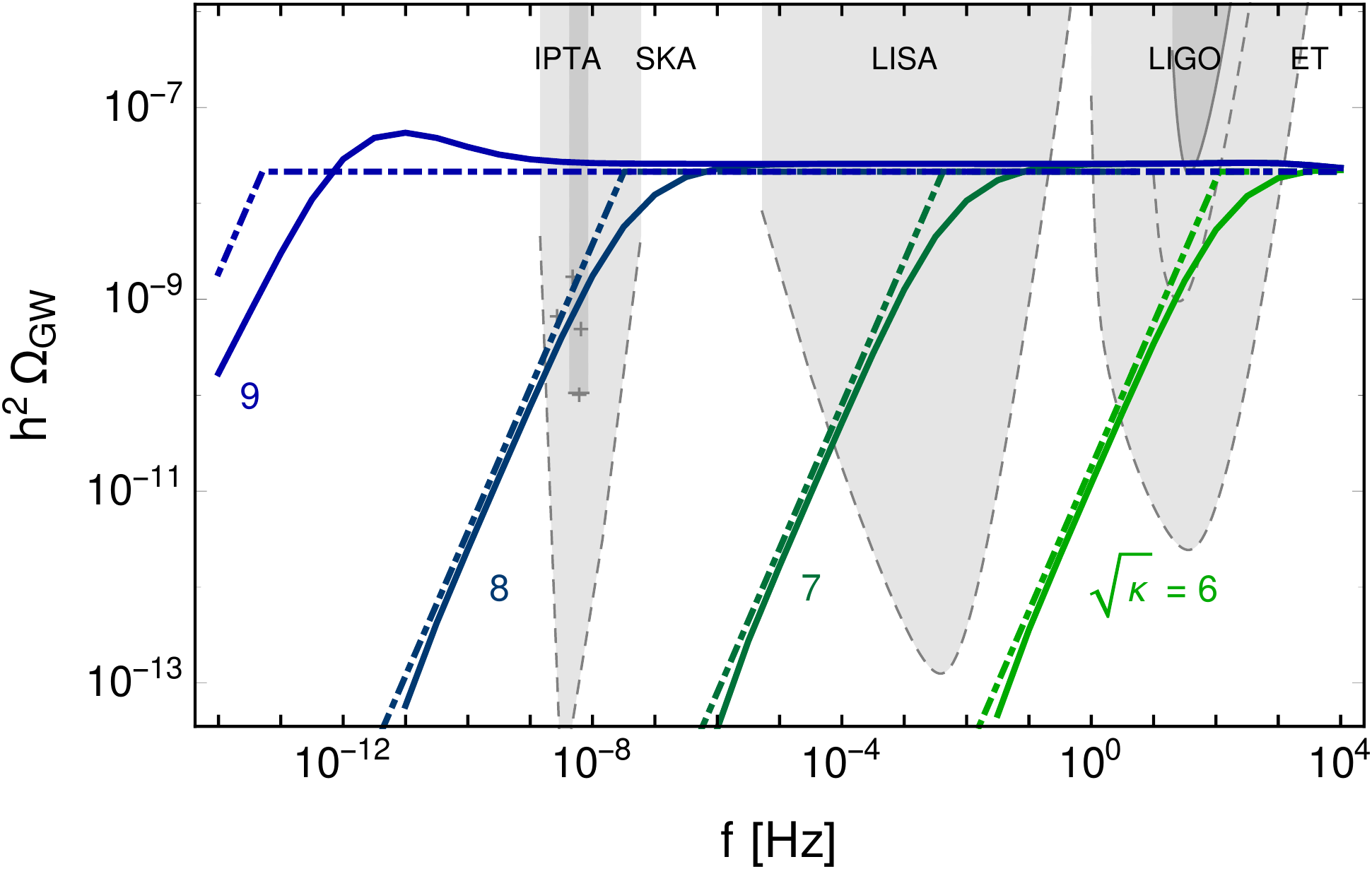}
    \includegraphics[width=0.48\textwidth]{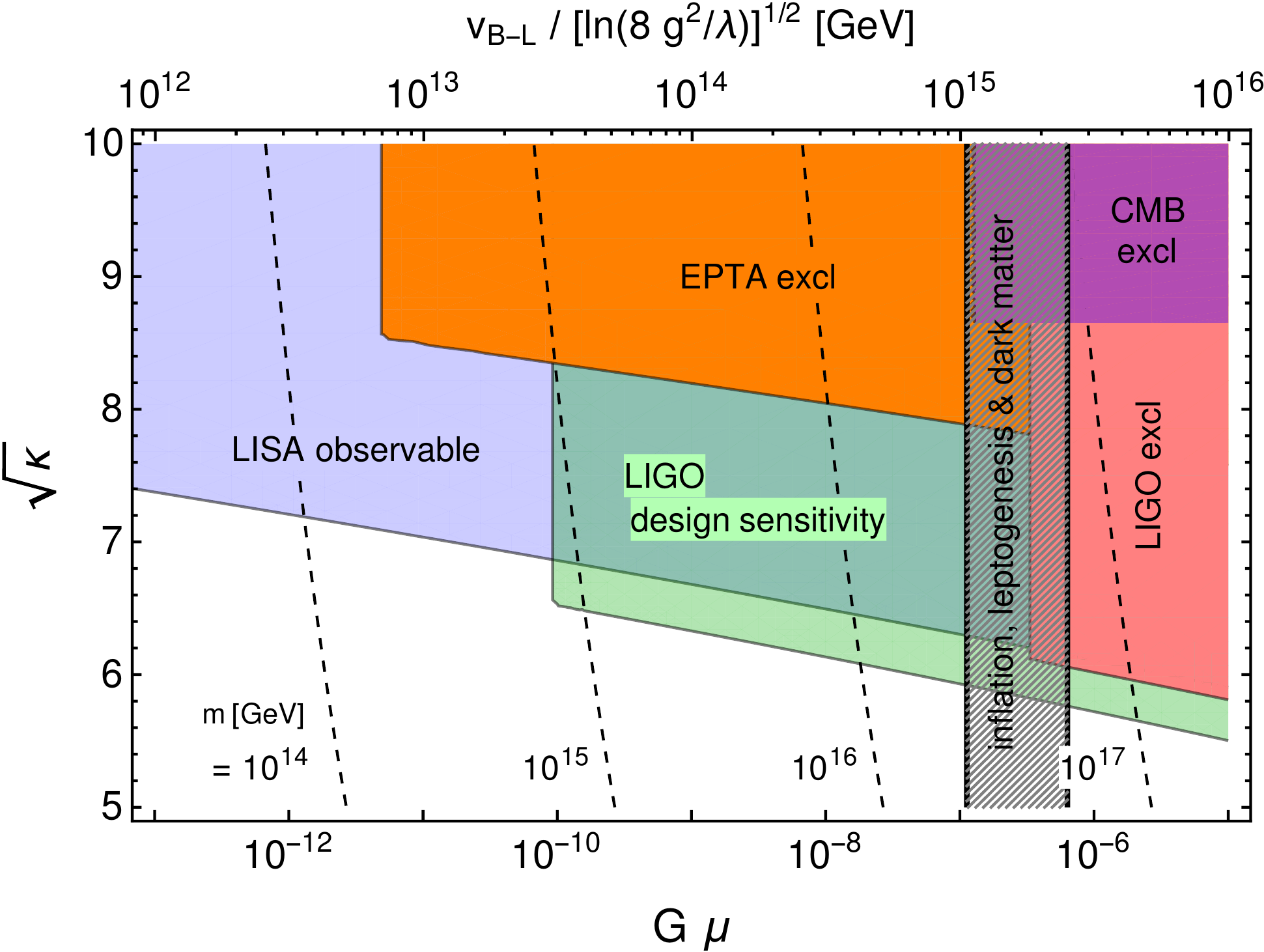}
    \caption{\small Left panel: SGWB signals from a metastable string network for $G\mu = 2 \times 10^{-7}$ and $\alpha = 0.1$, for different values of $\sqrt{\kappa}$; see Eq.~\eqref{eq:Gammad}. The solid lines represent the numerical result, the dashed lines show the analytical approximation in terms of a flat plateau at high frequencies and an $f^{3/2}$ rise at low frequencies. Right panel: Existing and future constraints on the metastable string parameter space. The hashed vertical band marks the prediction of the $U(1)_{B-L}$ model in Ref.~\cite{Buchmuller:2019gfy}, where both plots appeared for the first time.}
    \label{fig:metastable}
\end{figure}

Consider a $U(1)$ model giving rise to cosmic strings and its embedding in a gauge group $G$ with symmetry breaking pattern $G \rightarrow H' \times U(1) \rightarrow H$. The second symmetry breaking step will still produce cosmic strings as before. However, if $G$ and $H$ are such that the vacuum manifold of the broken subgroup $G/H$ is simply connected, these cosmic strings are metastable, i.e.~classically stable, but quantum mechanically unstable against the nonperturbative Schwinger production of monopole--antimonopole pairs~\cite{Vilenkin:1982hm,Preskill:1992ck,Martin:1996ea,Martin:1996cp}. These monopoles are the topological defects associated with the first symmetry breaking step, $G \rightarrow H' \times U(1)$. They nucleate along the strings, such that the network breaks apart into individual string segments with a monopole on the one end and an antimonopole on the other end. These dumbbell-like objects~\cite{Nambu:1977ag} keep losing energy via the emission of GWs, until the defect network has fully disappeared. This late-time contribution to the GW spectrum requires further investigation. In the following, we will neglect it and focus on the emission of GWs from string loops prior to their decay.

Models based on the symmetry breaking pattern $G \rightarrow H' \times U(1) \rightarrow H$ have recently been investigated in Refs.~\cite{Buchmuller:2019gfy,Buchmuller:2020lbh,Buchmuller:2021dtt}.
While Refs.~\cite{Buchmuller:2019gfy,Buchmuller:2020lbh} consider the breaking chain $SO(10) \rightarrow G_{\rm SM} \times U(1)_{B-L} \rightarrow G_{\rm SM}$, with $G_{\rm SM}$ denoting the SM gauge group, Ref.~\cite{Buchmuller:2021dtt} studies the breaking of $G = SU(2)_R \times U(1)_{B-L}$ in an extended electroweak sector down to the hypercharge group $H = U(1)_Y$ via $H' = U(1)_R \subset SU(2)_R$. In both cases, $U(1)_{B-L}$ is the Abelian gauge symmetry associated with baryon-minus-lepton number, $B\!-\!L$, and both models thus predict the production of a network of metastable cosmic $B\!-\!L$ strings.%
\footnote{Both models also predict the production of a monopole gas during the first symmetry breaking step. These monopoles are, however, diluted by a stage of inflation preceding the production of the metastable string network and are hence irrelevant.}

The lifetime of the collapsing string network is controlled by the monopole--antimonopole nucleation rate per unit length~\cite{Leblond:2009fq} (see also Ref.~\cite{Monin:2008mp}),
\begin{equation}
\label{eq:Gammad}
\Gamma_d = \frac{\mu}{2\pi}\,e^{-\pi\kappa} \,,\quad \kappa = \frac{m^2}{\mu} \,,
\end{equation}
where $\mu$ denotes the cosmic string tension and $m$ is the mass of the monopoles nucleating on the string. The parameter $\kappa$ can, at least in principle, be calculated in terms of the parameters of the grand unified theory model; see Ref.~\cite{Buchmuller:2021dtt} for an example. The string network then decays around a redshift $z_d$, which can be estimated by equating the decay rate of an average-sized string with the expansion rate~\cite{Buchmuller:2019gfy,Leblond:2009fq},
\begin{equation}
z_d \simeq 8.4\left(\Gamma G\mu\,\frac{\Gamma_d}{H_0^2}\right)^{1/4} \,.
\end{equation}
Here, $\Gamma \sim 50$ is defined in Eq.~\eqref{eq:Pn} and $H_0$ is the present-day Hubble parameter. In order to compute the SGWB signal from the collapsing string network, the lower boundary in the redshift integral in Eq.~\eqref{eq:Cn} needs to be replaced by $z_d$. Fixing $G\mu = 2 \times 10^{-7}$, this results in the GW spectra shown in the left panel of Fig.~\ref{fig:metastable}. These spectra saturate the CMB bound on $G\mu$ and are close to the sensitivity reach of existing ground-based interferometers; see the right panel of Fig.~\ref{fig:metastable}. At high frequencies, the SGWB signal exhibits the usual plateau, while going to lower frequencies, it decays in proportion to $f^{3/2}$. For $\sqrt{\kappa} \lesssim 6$, the string network is only short-lived, such that the signal dies off at frequencies above the LISA band, whereas for $\sqrt{\kappa} \gtrsim 8$, LISA will not be able to distinguish the signal from a standard string signal. In this case LISA will be able to access $G\mu$ values that would otherwise be excluded by PTA measurements. A LISA detection of a standard string signal at $G\mu$ in excess of current PTA bounds would therefore point to metastable strings with $\sqrt{\kappa} \lesssim 8$. In fact, values of $\sqrt{\kappa} \simeq 8$ are probed by existing PTA observations~\cite{Buchmuller:2020lbh}. Even larger values of $\sqrt{\kappa}$ can leave an imprint in CMB spectral distortions~\cite{Kite:2020uix}. The most interesting value from the perspective of LISA is $\sqrt{\kappa} \sim 7$, which can be easily achieved in realistic models~\cite{Buchmuller:2021dtt}. For $\sqrt{\kappa} \sim 7$, LISA will see the turnover in the spectrum caused by the collapse of the string network and hence uncover valuable information on new physics close to the energy scale of grand unification.

\newpage
%%%%%%%%%%%%%%%%%%%%%%%%%%%%%%%%%
% Here Sec. 8 starts

\section{Inflation}
\label{sec:Inflation}

\small \emph{Section coordinators: E.~Dimastrogiovanni. Contributors:  
D.~Comelli, E.~Dimastrogiovanni, M.~Fasiello, D.G.~Figueroa, J.~Fumagalli, L.~Iacconi, A.~Malhotra, S.~Matarrese, A.~Mazumdar, L.~Pilo, S.~Renaux-Petel, A.~Ricciardone, R.~Rollo, G.~Tasinato, 
V.~Vennin, D.~Wands, L.~Witkowski. 
}
\normalsize

\subsection{Introduction}

Can LISA detect GWs  produced at the time of the  big bang? This section aims to investigate such a possibility focusing on cosmic inflation. 

Inflation is the leading paradigm for describing the very early phases of cosmic expansion. Furthermore, it offers a  mechanism for GW production that can be probed with GW experiments. The inflationary era is a short phase of accelerated expansion believed to have occurred within the first instants of the history of our universe and leading to the hot big bang phase, which characterises the standard cosmological evolution. This framework was  initially proposed and developed   to solve  basic problems within the standard big bang cosmology~\cite{Guth:1980zm,Starobinsky:1980te,Linde:1981mu,Albrecht:1982wi}. After the original proposal, it was soon realised that inflation comes equipped with a mechanism of particle production, an inevitable consequence of quantum mechanics applied to an accelerating cosmological spacetime. The same mechanism is responsible for generating both the  
primordial scalar anisotropies \cite{Mukhanov:1981xt,Hawking:1982cz,Starobinsky:1982ee,Guth:1982ec,Bardeen:1983qw} that source
the evolution of cosmic structures at large scales, and a SGWB of primordial GWs   \cite{Grishchuk:1974ny,Starobinsky:1979ty}.
 
The basic setup of GW production during inflation is as follows.  Consider the dynamics  of spin-2, transverse-traceless
fluctuations $h_{ij}(t,\,\vec x)$ around a flat FLRW line element
 \begin{equation}
 d s^2\,=\,-d t^2+a^2(t)  \left[ \delta_{ij}+h_{ij}(t,\,\vec x)\right]\,d x^i d x^j 
  \end{equation}
 with $a(t)$ being the scale factor. During inflation one typically assumes that the Hubble parameter $H\,=\,\dot a/a$ is nearly constant, leading to a quasi-exponential expansion. The dynamics of spin-2 fluctuations is obtained by expanding the Einstein-Hilbert action at quadratic order in $h_{ij}$ around the homogeneous FLRW solution above. Cosmological  perturbations can be quantized according to the basic rules of quantum field theory (see e.g.~Ref.~\cite{Birrell:1982ix}). Quantum mechanics  converts the large gradients characterising the spacetime  geometry, associated
 with the rapid cosmological expansion, into the production of spin-2 quanta whose amplitude  freezes for wavelengths larger
 than the Hubble horizon $H^{-1}$.

After inflation ends, the Hubble scale starts to increase, reaching the size of the wavelength of the fluctuations produced during inflation. The latter re-enter the horizon with a very large occupation number, behaving as classical stochastic variables, and forming a  primordial SGWB (see e.g.~Refs.~\cite{Caprini:2018mtu,Maggiore:2018sht} for reviews). A detection of the (as of today undetected) inflationary SGWB  would provide key information on the physics of the early universe. Moreover, it would be the first direct experimental evidence of the quantization (see \cite{Riotto:2002yw} for a review and \cite{Polarski:1995jg} for a discussion on the transition to a semi-classical behaviour) of spin-2 gravitational interactions:  the production of GWs is based on quantum mechanical notions applied to cosmology.

The simplest class of inflationary models  relies on a single scalar field, the inflaton, characterised by a standard kinetic term and driving the acceleration. In this configuration, the inflaton
homogeneous and time-dependent profile  rolls slowly down an almost flat potential throughout the inflationary phase~\cite{Liddle:2000cg,Baumann:2009ds}. Such scenarios are associated to rather specific predictions in terms of the inflationary SGWB properties.
They consist in an almost scale-invariant spectrum, whose amplitude is too small for detection by direct GW experiments such as LISA and whose slope is slightly red-tilted (the amplitude decreases as frequency increases). At present the most promising path to detection of an inflationary SGWB from standard single-field slow-roll models is through its footprints on the B-mode polarisation of the CMB radiation (see e.g.~Ref.~\cite{Kamionkowski:2015yta} for a  review). 

There are nevertheless plenty of good reasons to go beyond vanilla models of inflation. Such a step is motivated for example by the need to embed inflationary physics  in the broader context of particle physics and especially within quantum  theories of gravity. Upon exploring a richer inflationary dynamics, one must also revisit the corresponding predictions of primordial GW properties.

Particularly relevant for this work is the fact that in scenarios beyond single-field slow-roll, the amplitude of the primordial SGWB can be enhanced, and this may well occur in the LISA frequency band. As a result, LISA has the potential and the opportunity to probe our understanding of the early universe and, in particular, of the inflationary phase. In this section we expand on such a possibility. We refer the reader to Ref.~\cite{Bartolo:2016ami} for previous work on the topic by the LISA Cosmology Working Group.
 
 We shall consider here the following scenarios:
\begin{itemize}

\item When embedding inflation in particle physics and string theory setups~\cite{Lyth:1998xn,Baumann:2014nda}, several additional fields may enter the game thereby changing the basic predictions of inflation. As we shall see in Sec.~\ref{sec-GWaddf}, the presence of extra degrees of freedom brings about the possibility of enhanced cosmological correlators and associated signatures. Of particular interest for us will be the case where such dynamics affects the tensor sector and leads to an amplification of the primordial GW signal.  Additional properties of the SGWB that may reveal a multi-field mechanism include chirality, non-Gaussianity, and specific frequency profiles that can help
distinguish the primordial signal from its astrophysical counterpart.
 
\item A SGWB with sufficiently large amplitude to allow detection by LISA can be induced at second order in perturbations by a boosted scalar power spectrum. The mechanism leading to such enhancement typically gives rise also to oscillatory features. In Sec.~\ref{sec:features} we discuss the characteristic oscillatory frequency profiles of the SGWB that can be probed in the LISA frequency ranges. In the process, we discuss novel methods to identify the different signatures associated to a number of inflationary models.
 
\item The inflationary mechanism   can be realised  in scenarios with alternative spacetime symmetry breaking patterns, in which the background of the fields driving inflation depends also on space coordinates. As we discuss in Sec.~\ref{sec:supersolid}, the dynamics of the corresponding tensor sector are different than in standard scenarios, and can lead to a primordial SGWB signal within reach of LISA. 
 
\item Explicit UV completions of inflation motivated by non-local versions of Starobinsky's $R+\alpha\,R^2$ model  are discussed in Sec.~\ref{AM}. There we also show how considering such approaches reflects on the properties of the primordial GW spectrum.
  
\item  Inflation is an early period of accelerated expansion that must come to end so as to give way to   cosmological evolution within the framework of standard  big bang cosmology.  In the transition phase the inflaton field couples with or decays into SM particles. This process may lead to GW production and to a SGWB spectrum whose properties are investigated in Sec.~\ref{sec:preheating}.

\item  All the scenarios mentioned above point to the fact that the primordial SGWB from inflation can be produced by qualitatively distinct mechanisms, all of which are markedly different from astrophysical processes. Sec.~\ref{sec:disob} aims to put together and investigate  the features of the inflationary SGWB -- chirality, frequency profile, anisotropies --  that can be  probed by LISA.
\end{itemize}

\subsection{Gravitational waves sourced by additional fields during inflation
}\label{sec-GWaddf}
\subsubsection{Axion-gauge field inflation}
\label{a-gf}
Ubiquitous in particle physics, the existence of axion-like particles driving inflation has received considerable attention (see Ref.~\cite{Pajer:2013fsa} for a comprehensive review) starting with the well-known natural inflation proposal \cite{Freese:1990rb,Adams:1992bn}. The appeal of such a setup relies in part on their approximately shift-symmetric potential  protecting the inflaton mass from large corrections. 
As a result of non-perturbative contributions from gauge field configurations (instantons), the potential acquires the typical cosine profile 
  \begin{figure}
 \centering
\includegraphics[scale=0.37]{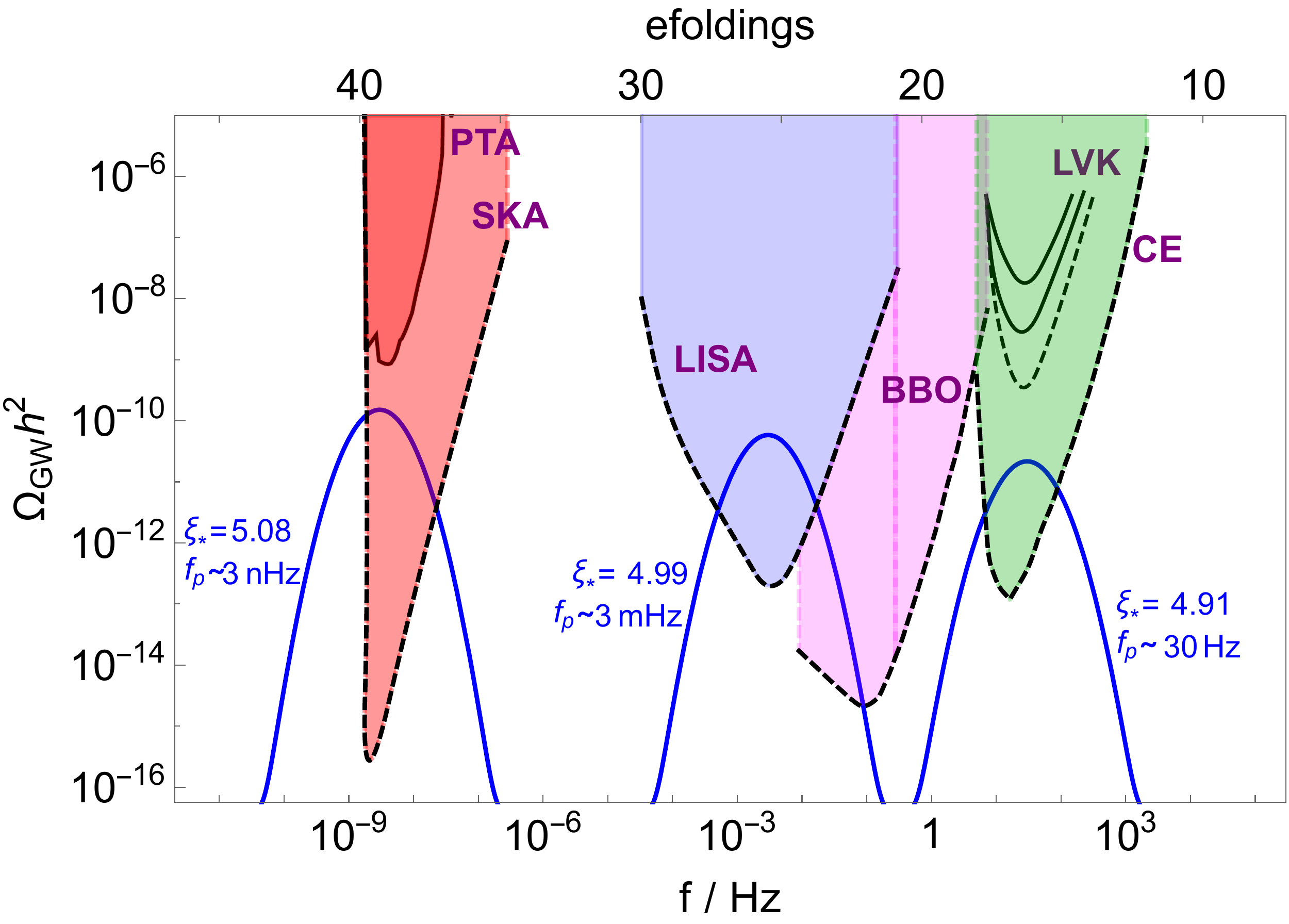}
\caption{\small Axion-Abelian gauge field coupling can produce a large amplitude and varying width GW background depending on the evolution of fields during inflation. One can have a bump/Gaussian peak at scales where the axion rolls fast. The  figure shows three examples that are relevant for current and forthcoming/future interferometers, namely PTA (frequencies  around $10^{-10}-10^{-7}$Hz), LISA (frequencies $10^{-5}-10^{-1}$Hz) and Advanced LIGO-Advanced Virgo-KAGRA (around 1-100 Hz). In all those examples, the upper limit of the GW background is set by constraints from enhanced curvature perturbations and PBH bounds. Interestingly, density perturbations peaked at PTA and LISA scales can also generate PBHs which can be phenomenologically relevant and possibly constitute some fraction of DM. Figure adapted from Refs.~\cite{Garcia-Bellido:2017aan,Unal:2018yaa}.}
\label{Caner}
 \end{figure}

 \begin{equation}
  \label{5_eq_axion_p}
  V(\chi)=\Lambda^4 \Bigg[1+\cos\left(\frac{\chi}{f_\chi}\right) \Bigg].
 \end{equation}
 The axion-like field $\chi$ is driven by a potential at a scale $\Lambda$ and with decay constant $f_\chi$. In the absence of other fields, a nearly scale-invariant scalar spectral index, as required by CMB observations, leads to a large axion decay parameter  $f_\chi \sim M_{P}$. On the other hand, a number of reasons favour inflationary realisations with a sub-Planckian $f_\chi$: (i) quantum gravity effects are expected to break the shift symmetry, as well as any global symmetry, at the Planck scale through the formation  of a (virtual) BH; (ii) in typical string theory constructions one finds $f_\chi < M_P$ \cite{Banks:2003sx}. It follows that a field content allowing for a smaller decay constant whilst preserving the naturalness of the potential, is of interest in this context. Coupling the axion-inflaton with gauge fields is a simple and intriguing possibility. Perhaps the most studied example is that of a Chern-Simons-type coupling
 \begin{equation}
 \label{fftilde}
     \lambda\, \frac{\chi}{4 \, f_\chi}\, F\tilde{F},
 \end{equation}  
 which preserves the shift symmetry and ``dissipates'' some of the inflaton kinetic energy into the gauge sector, in so flattening its effective potential without the need to resort to a large $f_\chi$. Several specific axion-gauge fields realisations are found in the literature, employing both Abelian \cite{Anber:2009ua} and non-Abelian \cite{Adshead:2012kp} gauge modes. Their phenomenology is a rich one, but we shall focus here mostly on the gravity (i.e.~tensor) sector. The Abelian and non Abelian scenarios share some key features:\\
 
 \noindent $\bullet$  Inspection of the equations of motion for the gauge sector in Eqs.~(\ref{Ab}) and (\ref{nonAb}), reveals that the effect of the coupling to the axion-like inflaton depends also on the popularisation of the gauge field, and it is  controlled by the parameter $\xi \equiv \lambda \dot{\chi}/ (2 f_\chi H$). More specifically, denoting by $A_\pm$ the mode functions of the two circular polarisations of the gauge field, one has~\cite{Garretson:1992vt,Anber:2009ua,Maleknejad:2011sq,Adshead:2012kp}
\begin{equation}
\label{Ab}
{\rm Abelian \, case} %\,\cite{Garretson:1992vt,Anber:2009ua}
:\qquad\quad \; A^{\prime \prime}_{\pm} + \left( k^2 \pm \frac{2 k}{\tau} \xi  \right) A_{\pm}=0 \; ; \quad\qquad\qquad\quad\quad\;
 \end{equation}
\begin{equation}
\label{nonAb}
{\rm non \, Abelian \, case} \, %\cite{Maleknejad:2011sq,Adshead:2012kp} 
:\quad t^{\prime \prime}_{\pm} + \Bigg[ k^2 + \frac{2}{\tau^2} \left( m_Q \xi \pm k \tau (m_Q+\xi)  \right)  \Bigg]t^{}_{\pm}\simeq 0\; ;
 \end{equation}
where $\tau$ is conformal time. The quantity $t$ in Eq.~(\ref{nonAb}) is the wave-function associated with the tensor degrees of freedom $t_{ia}$, stemming from choosing SU(2) as the gauge group \cite{Maleknejad:2011jw};  the parameter $m_Q$ is defined as $m_Q\equiv g Q/H$ where $g$ is the gauge coupling, $H$ the Hubble rate, and $Q$ the background of one of the (three) gauge sector scalar modes. We note the slow roll relation $\xi \simeq m_Q + 1/ m_Q$, so that the two parameters that control the production in the Abelian and non Abelian case coincide in the large $\xi$ limit. 

It is intuitively clear then that the effect of the Chern-Simons  coupling, driven by $\xi$, is stronger as one approaches the end of inflation where a large kinetic term $\dot{\chi}$ breaks the slow-roll condition.
 \\
 
\begin{figure}[h!]
 \centering
\includegraphics[scale=0.9]{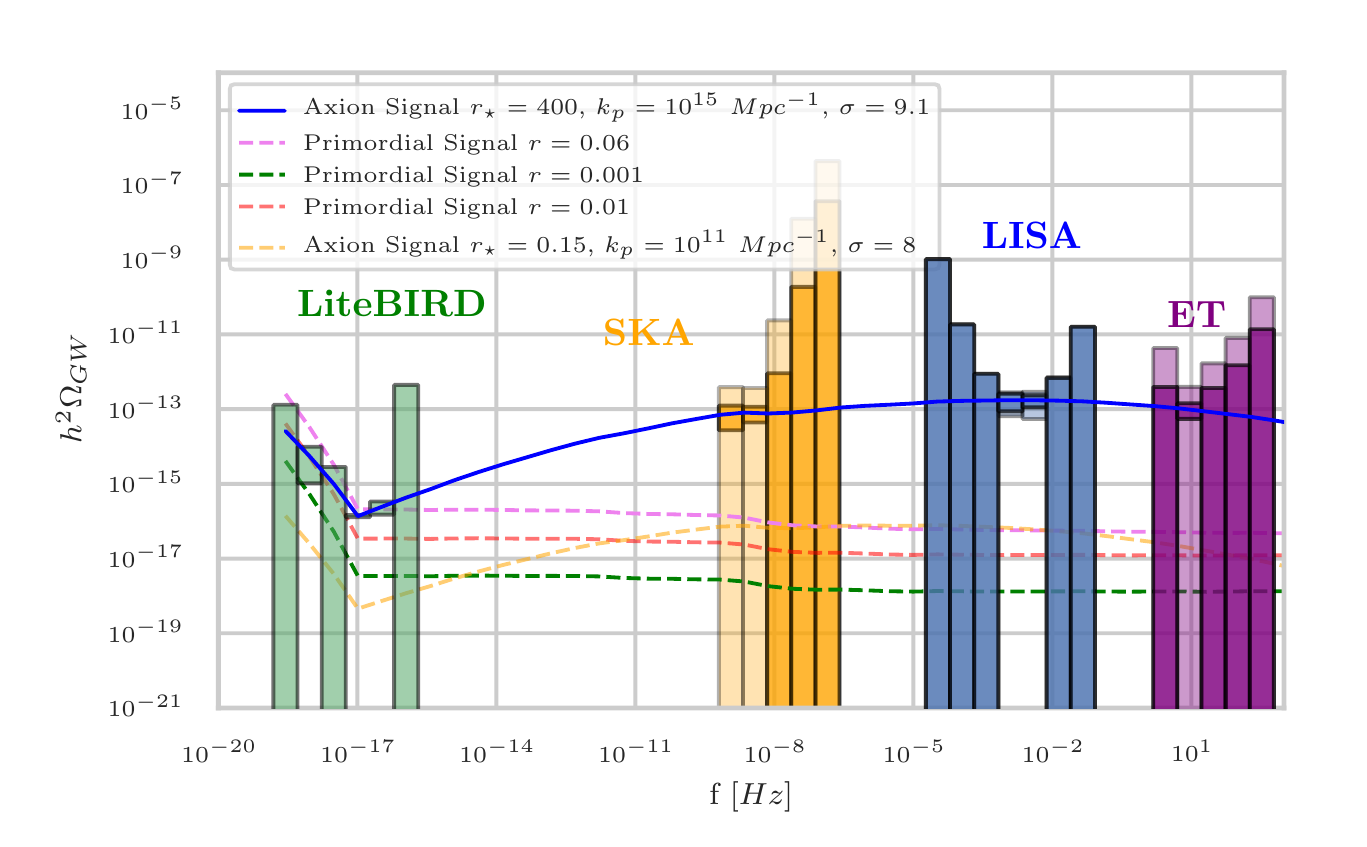}
\caption{\small 
The GW energy density $\Omega_{\rm GW}$ as a function of frequency sourced by SU(2) axion-gauge field setups (solid blue line and yellow dashed line). A signal outside the reach of certain probes may still be accessed via LISA  at small scales (and e.g.~LiteBIRD at very large scales). The vertical bars are the expected 1$\sigma$ error bars estimated  with and without the presence of astrophysical foregrounds (light and dark shared areas, respectively). Different colours are used for LiteBIRD (green), SKA (orange), LISA (blue), and ET (purple).
 Logarithmic binning in wavenumber is employed with $\Delta \ln k=1.2$. For the sake of comparison, the tensor spectrum predicted in some illustrative single-field slow-roll scenario (dashed lines) are displayed, including their BICEP2/Keck/Planck upper bound $r = 0.06$. The quantity $k_p$ is the pivot scale and the quantity $\sigma$ in the upper left box is a parameter of the SU(2) model, typically of order $1-10$. For details see Ref.~\cite{Campeti:2020xwn} from which the figure is taken.}
\label{Omegakyk}
 \end{figure}

 \noindent $\bullet$ Given that the amplitude of the relevant gauge degrees of freedom is enhanced by the coupling, and  that these modes source GWs, the tensor power spectrum is typically blue in these scenarios. This makes them of immediate interest for GW detectors, such as LISA (see e.g.~Figs.~\ref{Caner} and \ref{Omegakyk}, respectively for the Abelian and for the non-Abelian cases).

The fact that minimal single-field slow-roll scenarios predict instead a slightly red-tilted GW power spectrum, well below the sensitivity of LISA and possibly of the proposed BBO, is very much relevant here. A detection of a \textit{primordial} GW signal at small scales will strongly point to a multi-field inflation mechanism such as the ones discussed in this and the next subsection.   \\
  
   \begin{figure}[]
\hspace{1.7cm}	\includegraphics[scale=0.19]{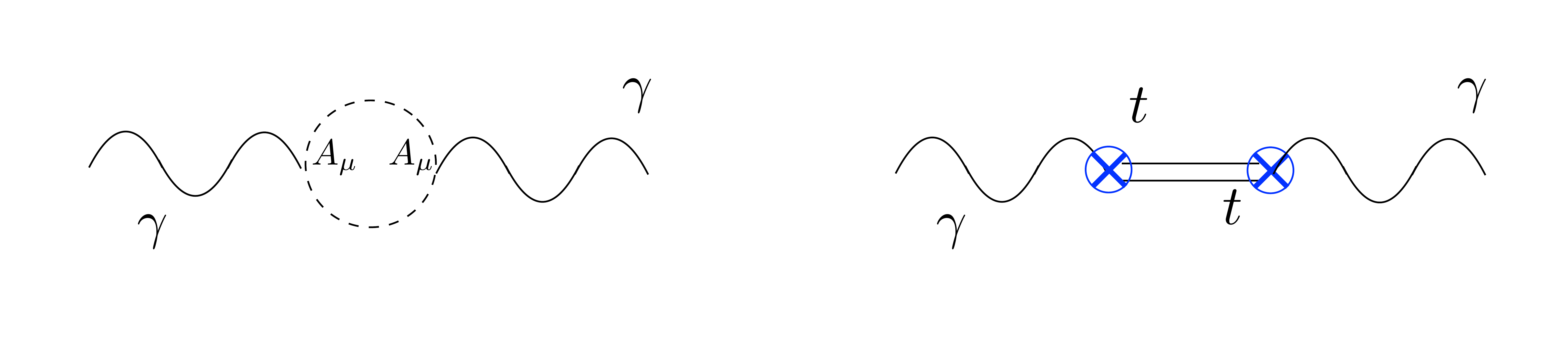}
	\caption{\small Feynman diagrams for the GW power spectrum.  Left panel: The gauge fields $A_{\mu}$ source GW non-linearly in the Abelian case. Right panel: The $SU(2)$ scenario, where gauge tensor degrees of freedom $t_{ij}$  source GWs already at tree level.}
	   	\label{GWsources}
\end{figure}
  
\noindent  $\bullet$ Polarisation-dependent equations of motion and solutions for gauge fields feed directly into a polarisation dependent GW spectrum. If the sourced contribution is comparable or larger than the ever-present vacuum tensor fluctuations the GW spectrum will be a \textit{chiral} one. This is particularly interesting a signature because (i) it is testable with LISA 
by means of the kinematically induced dipole  \cite{Domcke:2019zls}  
and (ii) is rather uniquely associated  with a Chern-Simons-type coupling, at least in terms of chirality of primordial origin.

Further characterisation of these inflationary scenarios  necessarily includes the study of their scalar sector, especially vis a vis CMB constraints as well as scalar, tensor and mixed higher-point functions. The possibility of large tensor non-Gaussianity (i.e.~the GW bispectrum) \cite{Anber:2012du,Namba:2015gja,Agrawal:2017awz} is worth mentioning here, with the reminder that its shape function is typically very similar to the so-called equilateral template. As a result, it is best tested at CMB scales (namely by studying the impact on the CMB of modes with wavelength not much smaller than the present horizon), rather than at higher frequencies where propagation effects strongly suppress the signal \cite{Bartolo:2018evs}. Moreover, models with axions and gauge fields can also produce a SGWB peaked at scales relevant for various GW detectors, \cite{Domcke:2016bkh,Garcia-Bellido:2016dkw} including aLIGO/AVirgo/Kagra, LISA and PTAs. They can also produce peaked density perturbations that can seed an abundance of PBHs compatible with that of DM~\cite{Bugaev:2013fya,Garcia-Bellido:2016dkw,Domcke:2017fix,Garcia-Bellido:2017aan}, as we briefly touch upon in Sec.~\ref{Pzeta}.  We shall also point to the intriguing possibility of similar scenarios being responsible for the matter-antimatter asymmetry of the universe \cite{Maleknejad:2014wsa,Caldwell:2017chz}.

For the sake of completeness we should also mention some caveats one may extract from recent literature on the subject as well as several emerging new research paths. 
Coupling the axion with the gauge sector allows for a sub-Planckian dynamics by essentially feeding the axion kinetic energy into the gauge sector. This effect is regulated by the coupling $\lambda$. The ensuing flattening of the inflaton effective potential makes for a viable scalar spectral index $n_s$ at large scales. At the same time, the (enhanced) gauge sector also sources GWs. It turns out that the parameter space granting the appropriate  $n_s$ may correspond to an overproduction of GWs \cite{Dimastrogiovanni:2012ew}. 

In addition, the amplified gauge fields also source scalar density perturbations, and it challenging to generate observable GW without overproducing scalar modes \cite{Barnaby:2010vf,Namba:2015gja,Papageorgiou:2018rfx,Papageorgiou:2019ecb}. Several routes have been explored %\cite{Adshead:2016omu} 
to preserve the intriguing GW phenomenology of these models whilst also satisfying CMB and PBH bounds. One of the most economical choices is to introduce a spectator scalar field \cite{Namba:2015gja,Dimastrogiovanni:2016fuu,Obata:2016tmo}. This weakens the production of density perturbations and the corresponding scalar sector constraints on the axion-like particle and delivers viable scenarios both in the Abelian and non-Abelian cases. The back-reaction dynamics \cite{Peloso:2016gqs,Maleknejad:2018nxz} as well as the requirements stemming from perturbativity bounds \cite{Ferreira:2015omg,Peloso:2016gqs,Dimastrogiovanni:2018xnn,Papageorgiou:2018rfx,Papageorgiou:2019ecb} have been recently studied for these setups.

The possibility of  Schwinger pair production has been explored in the same context  as well as the coupling with fermionic fields \cite{Domcke:2018eki,Mirzagholi:2019jeb}. In the same spirit, it is natural to investigate couplings with SM particles \cite{Ferreira:2017lnd,Domcke:2019qmm}. Also of note are several successful embeddings of axion gauge fields models within supergravity and string theory constructions \cite{DallAgata:2018ybl,McDonough:2018xzh,Holland:2020jdh}.

\subsubsection{Non-minimally coupled (spinning) fields}
As a useful approach that is complementary to the one in the previous section, we will briefly expand here on a recently introduced EFT of non-minimally coupled (spinning) fields \cite{Bordin:2018pca}. The presence of multiple fields during inflation is completely natural from the top-down perspective \cite{Baumann:2014nda}. The benefit of an effective approach to such dynamics lies in the fact it provides a general, unified, description  of the inflationary field content and the ensuing cosmological signatures. It is well-known that the presence of spinning fields makes for a richer phenomenology\footnote{For example, scalar and mixed non-Gaussianities display a characteristic extra angular behaviour in the squeezed limit \cite{Arkani-Hamed:2015bza}. Even more relevant in this context is the fact that extra tensor modes, such as those of spin-2 particles, can directly source GWs.}, one that is certainly worth exploring. 

The presence of increasingly demanding unitary constraints as one goes up the spin ladder, typically implies that higher $(s\geq 2)$ spin fields decay within of a few $e$-folds during inflation, to the detriment of their imprints on early universe observables. This is remedied by coupling the extra spinning content directly to the inflaton: the background breaks de Sitter isometries and weakens the related unitarity constraints, thus allowing for a lighter (more long-lived) particle content. As a result, extra spin-2 fields can for example directly source the GW spectrum providing the leading contribution to the signal \cite{Bordin:2018pca}. 

\begin{figure}[h!]
   	\centering
	\includegraphics[scale=0.69]{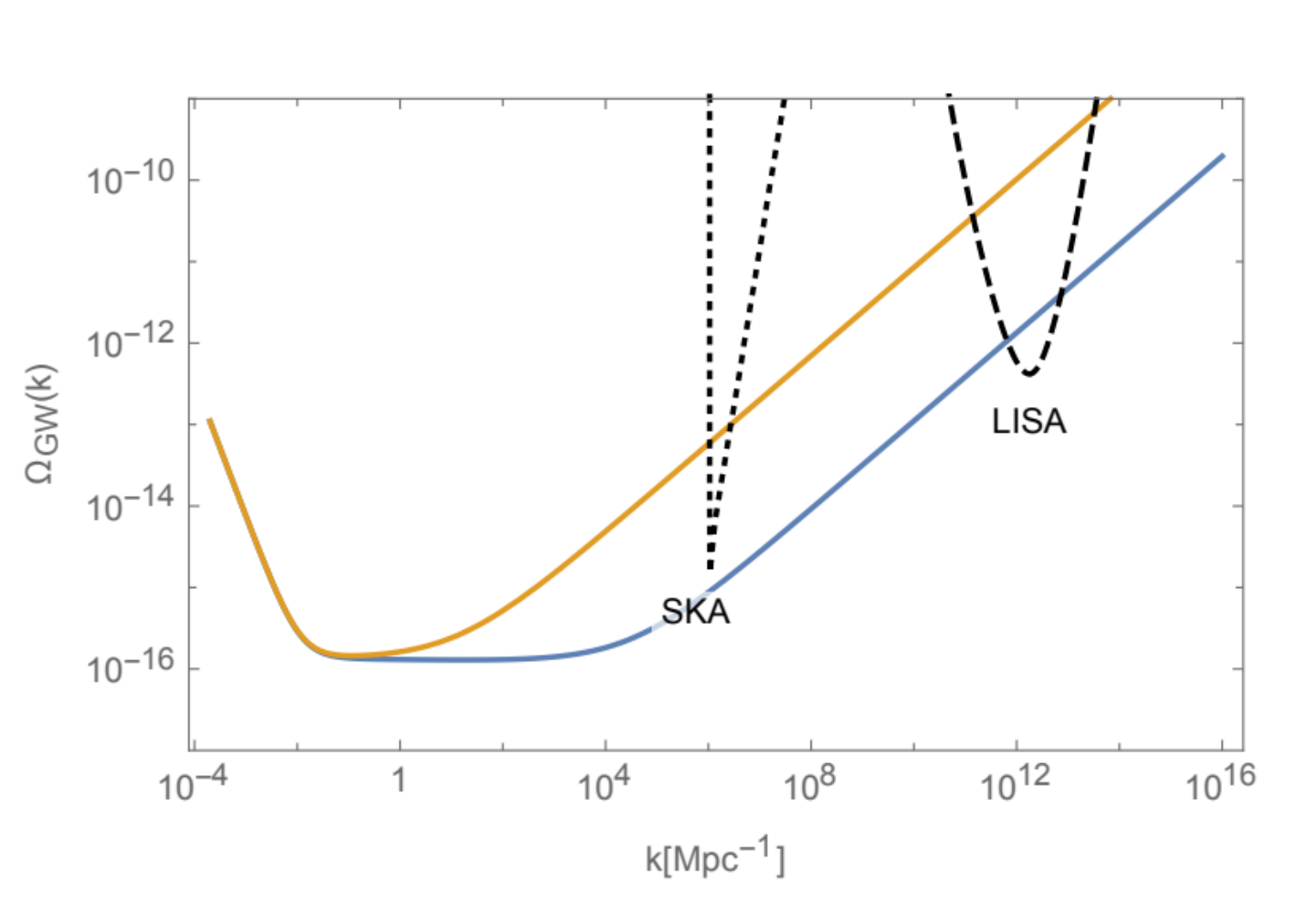}
	\caption{\small The GW energy density of the EFT when including the contribution of an extra spin-2 field whose helicity-2 mode has a time-dependent sound speed. The two colours correspond to different values of $\rho/H$, respectively $(3 \times 10^{-3}, 4 \times 10^{-4})$, supporting a signal within reach for both SKA and LISA, or for LISA only. In generating both lines, the time dependence of the sound speed has been kept constant to $s_2\equiv\dot{c}_2/(H c_2)\simeq -0.2$. Figure taken from  Ref.~\cite{Iacconi:2020yxn}.}
	   	\label{c2}
\end{figure}

It is instructive to report a few of the leading operators in the Lagrangian coupling an extra spin-2 field $\sigma$ with the  standard tensor modes $\gamma_{ij}$ and the Goldstone boson $\pi$, related to the  scalar curvature  $\zeta$ via  $\zeta\sim -H \pi$:
\begin{eqnarray}
\mathcal{S} \supseteq
\mathcal{S}_{\text{free}}^{(2)}+\mathcal{S}_{\text{int}}^{(2)} +\mathcal{S}_{\text{int}}^{(3)}&=&\int dt\,  d^3x\, a^3 \left(L_{\text{free}}^{(2)}+L_{\text{int}}^{(2)} +L_{\text{int}}^{(3)}\right) \label{33} \\
L_{\text{free}}^{(2)} &=& \frac{1}{4} \Big( (\dot{\sigma}^{ij})^2 -c_2^2a^{-2}(\partial_i \sigma^{jk})^2 -\frac{3}{2}(c_0^2-c_2^2)a^{-2}(\partial_i \sigma^{ij})^2 -m^2(\sigma^{ij})^2 \Big) \cr
L_{\text{int}}^{(2)} &=& - \frac{\rho}{\sqrt{2 \epsilon} H} a^{-2} \partial_i \partial_j \pi_c \sigma^{ij} +\frac{1}{2} \rho \dot{\gamma_c\,}_{ij} \sigma^{ij}  \cr
L_{\text{int}}^{(3)} &=&  -\frac{\rho}{2\epsilon_1 H^2 M_{\rm Pl}}a^{-2} (\partial_i\pi_c \partial_j\pi_c \dot{\sigma}^{ij}+2H \partial_i\pi_c \partial_j\pi_c {\sigma}^{ij} ) -\mu(\sigma^{ij})^3 +\dots \nonumber \,,
\end{eqnarray}
where $c_i$ is the sound speed for the helicity-i mode of the spin-2 particle $\sigma$. The first line of Eq.~(\ref{33}) contains the free quadratic Lagrangian for $\sigma$ whilst the second and third lines  respectively describe the quadratic and cubic mixing of $\sigma$ with standard scalar and tensor modes. In addition to vacuum fluctuations, the GW power spectrum receives a contribution proportional to the quadratic coupling $\rho$, which may well be the leading one. Upon allowing a time dependent $\rho$ or a time dependent helicity-2 sound speed $c_2$, it is possible to obtain a blue GW spectrum \cite{Iacconi:2019vgc,Iacconi:2020yxn} and, more in general, one with a non-trivial scale dependence (see Fig.~\ref{c2}). There is a substantial portion of parameter space in this EFT that delivers a GW power spectrum detectable by LISA. 

Further, even within bounds ensuring  perturbativity control, lack of gradient instabilities, and compliance with CMB bounds, this setup supports large squeezed non-Gaussianities, including the purely tensorial three-point function. It follows that the EFT field content may also be tested by probing GW anisotropies (see Sec.~\ref{Anisotropies-inflation}) as well as via cross-correlating the SGWB with the CMB (see Sec.~\ref{cross-corr} below).

\subsubsection{Models that produce peaks in the power spectrum of scalar perturbations}
\label{Pzeta}

Primordial density waves in the early radiation-dominated universe interact gravitationally with GWs and therefore generate a SGWB~\cite{Mollerach:2003nq,Ananda:2006af,Baumann:2007zm}. As discussed in Sec.~\ref{subsec:second-order-SGWB}, LISA is sensitive to GWs with frequency $f$ generated from primordial density perturbations with power spectrum ${\cal P}_\zeta(f) \approx 10^{-4}$ which re-entered the Hubble-horizon at a temperature $T\approx 10^5(f/mHz)$~GeV. 
These density waves can be generated from quantum fluctuations that are swept up by the accelerated expansion during a preceding period of inflation and exit the Hubble-scale $N$ $e$-folds before the end of inflation. By assuming instantaneous reheating (a finite period of reheating can reduce this number), one obtains
\begin{equation}
    N % \approx 31 + \ln \left( \frac{V_*^{1/4}}{M_{\rm Pl}} \right) - \ln \left( \frac{f}{{\rm mHz}} \right) 
     \approx 22 + \ln \left( \frac{V_*^{1/4}}{10^{15}\, {\rm GeV}} \right) - \ln \left( \frac{f}{{\rm mHz}} \right) 
    \,,
\end{equation}
where $N=\int H\, dt$ is the integrated expansion, and  $V_*$ is the energy density during inflation. 

It is often assumed that the power spectrum of primordial density perturbations from inflation can be described by a power law, ${\cal P}_\zeta\propto f^{n-1}$, whose amplitude and spectral index are fixed by observations on CMB scales, where ${\cal P}_\zeta\approx 10^{-9}$ and $n<1$. This is not  general~\cite{Chluba:2015bqa}. However, within slow roll, the deviation from a simple power law (running of the spectral index) is limited (second order in slow-roll parameters) and thus a significant enhancement in the primordial power spectrum on the scales probed by LISA requires the breakdown of the usual slow-roll approximations. 

Quantum fluctuations in the fields driving the accelerated expansion naturally generate primordial density fluctuations in the subsequent radiation era, and there may be additional fluctuations from fields that, although subdominant during inflation (so-called spectator fields), affect the subsequent evolution between the end of inflation and the epoch when modes re-enter the Hubble-horizon. 
The dimensionless density contrast, $\zeta$, can be identified with the perturbation in the local integrated expansion, $\zeta=\delta N$. In the large-scale limit (neglecting spatial gradients) this can be related to the field fluctuations at first order $\zeta=({\partial N}/{\partial \vec\Phi})\cdot \delta\vec\Phi$, where $N(\vec\Phi)$ corresponds to the background expansion as a function of the local field values, $\vec\Phi$, during inflation.

For a single inflaton field (or more generally for adiabatic field fluctuations in a multi-field setting~\cite{Liddle:1998jc,Gordon:2000hv}) ${\partial N}/{\partial \varphi}=H/\dot\varphi$ while the amplitude of quantum field fluctuations at Hubble exit are given by ${\cal P}_{\delta\varphi}\simeq H/2\pi c_s$, where $c_s$ is the sound speed. Given that the Hubble rate at Hubble-exit ($k=aH$) necessarily decreases during inflation, there are essentially two ways to boost the primordial density perturbations in single field models. Either one can decrease the speed of the inflaton, $\dot\varphi$ faster than the slow-roll evolution, known as ultra-slow roll~\cite{Inoue:2001zt,Kinney:2005vj}, or one can rapidly decrease the sound speed, $c_s$, e.g.~due to non-decoupling of massive fields~\cite{Achucarro:2010jv}. Ultra-slow roll describes a regime where the potential gradient becomes negligible and the field becomes friction dominated, $\dot\varphi\propto e^{-3N}$, which can occur in models with a near-inflection point in the potential. Typically the ultra-slow roll phase is transient, leading to a localised step \cite{Starobinsky:1992ts,Leach:2001zf} or broad peak \cite{Garcia-Bellido:2017mdw} in the power spectrum. In some cases the departure from slow-roll may interrupt inflation, before inflation restarts in an ultra-slow roll phase~\cite{Roberts:1994ap,Leach:2000yw,Ragavendra:2020sop}.

If the energy scale of inflation is as low as $V_*^{1/4}\sim10^6$~GeV then LISA could be sensitive to physical processes that end inflation and reheat the universe. The coherent oscillation of the inflaton (and, possibly, other field) that characterised this stage may lead to resonant amplification of scalar metric perturbations near the Hubble scale at the end of inflation, a phenomenon known as metric preheating~\cite{Nambu:1996gf,Finelli:1998bu,Jedamzik:2010dq,Easther:2010mr}. Density perturbations produced close to the Hubble scale at this stage may become large in models where inflation ends due to an instability, triggering a phase transition. This may be second-order, as in hybrid inflation models~\cite{Linde:1993cn,Copeland:1994vg}, where a tachyonic instability in a second ``waterfall'' field leads to the rapid growth of quantum fluctuations, or in the geometrical destabilisation of inflation, due to a geometrical instability in a negatively-curved field-space~\cite{Renaux-Petel:2015mga}. A FOPT at the end of inflation requires the rapid nucleation of bubbles of a broken symmetry phase, whose collisions both reheat the universe and can generate GWs~\cite{Caprini:2015zlo}.

More generally fluctuations in fields orthogonal to the instantaneous trajectory in field-space, and hence independent of the adiabatic field fluctuations, can enhance the primordial scalar power spectrum, ${\cal P}_\zeta$, after Hubble exit~\cite{Wands:2002bn}. Any non-geodesic ``turn'' in field space during inflation converts isocurvature field perturbations at Hubble-exit into an additional source of curvature perturbations on super-Hubble scales~\cite{Gordon:2000hv,GrootNibbelink:2001qt,Palma:2020ejf,Fumagalli:2020adf}. In a different mechanism, spectator axion fields might also exhibit a brief phase of fast roll during inflation, generating a localised peak in the power of density perturbations \cite{Namba:2015gja}. Alternatively, the effect of spectator fields may be completely subdominant during inflation but boost the primordial power spectrum after the end of inflation as in the curvaton~\cite{Enqvist:2001zp,Lyth:2001nq,Moroi:2001ct} or other modulated reheating~\cite{Dvali:2003em,Kofman:2003nx} scenarios. Quite generally we expect otherwise light scalar fields present during inflation to acquire effective masses of order the Hubble scale during inflation~\cite{Dine:1995uk,Baumann:2011nk}, suppressing their fluctuations on large scales, leading to a steep blue tilt, dominating the power spectrum on small scales, as in the axion-like curvaton model~\cite{Kawasaki:2012wr}.

\subsection{Small-scale primordial features}
\label{sec:features}

Embeddings of inflation in high energy theory motivate the exploration of inflationary mechanisms beyond the single-field slow-roll framework. For example, UV completions of inflation in string compactifications typically introduce many new degrees of freedom that ultimately contribute to the inflationary dynamics. Thus, from the UV point of view, single-field slow roll models have the semblance of toy-models that capture the essence of inflation without being fully realistic. 

Avenues for going beyond single-field slow-roll are plentiful and, as a result, a large number of models have been constructed. However, instead of proceeding model by model, there is also a more systematic way for going beyond the simplest version of inflation. The idea is to characterise departures from single-field slow-roll in terms of their effect on the scalar power spectrum.  

These signatures of a departure from single-field slow-roll have been termed ``features'' after corresponding characteristic properties of the scalar power spectrum. See Refs.~\cite{Chen:2010xka, Chluba:2015bqa, Slosar:2019gvt} for reviews on this topic. Using the language of features, departures from single-field slow-roll inflation can be assigned to one of the two following broad classes:\footnote{Oscillations of heavy fields act as ``primordial standard clocks'', leading to feature signals resembling that of a sharp feature at larger scales and of a resonant feature at smaller scales. }
\begin{enumerate}
    \item A so-called ``sharp feature'' is characterised by an oscillation in the scalar power spectrum that is periodic in the wavenumber $k$.
    This arises whenever there is some sudden transition during inflation, like e.g.~a step in the inflation potential or a sharp turn in the trajectory. 
    \item A so-called ``resonant feature'' denotes an oscillation in the scalar power spectrum in $\log(k)$. It arises when some components of the background oscillate with a frequency larger than the Hubble scale, inducing a resonance with the oscillations of the quantum modes of the density perturbations, a typical example being axion monodromy inflation \cite{Silverstein:2008sg,Flauger:2009ab}.
\end{enumerate}
Over the relevant range of scales, the scalar power spectrum corresponding to each class can be written as
\begin{align}
    \label{eq:P_of_k_sharp}
    \textrm{Sharp:} \quad & \mathcal{P}_\zeta(k)=\mathcal{P}_0(k) \Big[1 + A_{\textrm{lin}} \cos \big(\omega_{\textrm{lin}} k + \vartheta_{\textrm{lin}} \big) \Big] \, , \\
    \label{eq:P_of_k_resonant}
    \textrm{Resonant:} \quad & \mathcal{P}_\zeta(k)=\mathcal{P}_0(k) \Big[1 + A_{\textrm{log}} \cos \big(\omega_{\textrm{log}} \log(k/k_\star) + \vartheta_{\textrm{log}} \big) \Big] \, ,
\end{align}
with $k_\star$ being some arbitrary reference scale introduced for dimensional reasons. That is, a sharp or resonant feature is described by an oscillation with amplitude $A_{\textrm{log}/\textrm{lin}}$ about an envelope $\mathcal{P}_0(k)$. The precise form on the envelope will depend on the model, but can be taken as sufficiently smooth over the period of oscillations.\footnote{In realistic models the amplitude and frequency of the oscillation as well as phase offset can run with the scale $k$, but if this running is sufficiently ``slow'', the templates \eqref{eq:P_of_k_sharp} and \eqref{eq:P_of_k_resonant} will still be applicable over a suitable range of scales.} 
For sharp features, the frequency $\omega_{\textrm{lin}} \sim 1/k_{\textrm{f}}$, where $k_{\textrm{f}}$ corresponds to the scale that crosses the Hubble radius at the time of the feature. As for resonant ones, one has
 $\omega_{\textrm{log}} \sim M/H$, where $M$ is the frequency of the background oscillations.

At large scales ($k \lesssim 1 \textrm{ Mpc}^{-1}$) features are severely constrained by CMB and LSS data, which mandate a nearly scale-invariant power spectrum with amplitude $\mathcal{P}_\zeta \sim 10^{-9}$. In contrast, at small scales ($k \gg 1 \textrm{ Mpc}^{-1}$), CMB and LSS constraints do not apply and the scalar power spectrum can depart significantly from scale-invariance. This opens the possibility that the feature constitutes the dominant contribution to the scalar power spectrum at small scales.

For such contribution to be sufficiently large  to make  the SGWB detectable by current or forthcoming interferometers, the scalar power spectrum at the scale of the feature needs to be significantly enhanced compared to its value at CMB scales. Specifically, $\mathcal{P}_\zeta$ should increase from $\sim\! 10^{-9}$ to $\sim\! 10^{-4}$ by moving $k$ from the CMB to the LISA momentum scale.
 GW observatories such as LISA are thus sensitive to feature models where the scalar power spectrum exhibits a peak at small scales, where the CMB and LSS constraints do not apply, as previously explained in Sec.~\ref{Pzeta}. This is not an unrealistic expectation: such localised enhancements of the power spectrum occur frequently in inflation models with features. In fact, the mechanism of amplification of scalar fluctuations is often also responsible for producing the oscillatory feature and vice versa, as observed e.g.~in Refs.~\cite{Ballesteros:2018wlw,Palma:2020ejf,Fumagalli:2020adf,Braglia:2020eai,Fumagalli:2020nvq,Tasinato:2020vdk,Braglia:2020taf}. GW observatories such as LISA can thus test models of inflation at scales inaccessible to CMB and LSS surveys.

What is interesting for GW astronomy is that features in the scalar power spectrum lead to corresponding features in the frequency profile of the corresponding scalar-induced GWs \cite{Fumagalli:2020nvq,Braglia:2020taf,Fumagalli:2021cel}. Here we focus on the contribution to the SGWB sourced by scalar fluctuations when they re-enter the horizon during RD after inflation, see Sec.~\ref{subsec:second-order-SGWB}.\footnote{In addition, there will also be a scalar-induced contribution to the SGWB sourced during inflation. The latter is slow-roll suppressed compared to the post-inflationary contribution and will be ignored in the following, even though it can dominate in selected models, see e.g.~Ref.~\cite{Zhou:2020kkf}.}
Remarkably, this inherently nonlinear effect does not lead to oscillations being washed out. One rather finds that a sharp feature leads to a periodic modulation in the spectral shape of the SGWB contribution, while a resonant feature produces a corresponding log-periodic modulation. As a result, over some range of scales the spectral shape of $\Omega_\textrm{GW}(k)$ can be matched by the following templates \cite{Fumagalli:2020nvq,Braglia:2020taf,Fumagalli:2021cel}:
    \begin{align}
    \label{eq:sharp_template} 
    \textrm{Sharp:} \quad & \Omega_{\textrm{GW}}(k) = \overline{\Omega}_{\textrm{GW}}(k) \Big[1+ \mathcal{A}_\textrm{lin} \cos \big(\omega_\textrm{lin}^\textsc{gw} k + \phi_\textrm{lin}  \big) \Big] \, , \\
    \label{eq:resonant_template}
    \textrm{Resonant:} \quad & \Omega_{\textrm{GW}}(k) = \overline{\Omega}_{\textrm{GW}}(k) \Big[1+ \mathcal{A}_{\textrm{log},1} \cos \big(\omega_\textrm{log} \log (k/k_\star) + \phi_{\textrm{log},1} \big) \\
    \nonumber & \hphantom{\Omega_{\textrm{GW}}(f) = \overline{\Omega}_{\textrm{GW}}(k) \Big[1} + \mathcal{A}_{\textrm{log},2} \cos \big(2 \omega_\textrm{log} \log (k/k_\star) + \phi_{\textrm{log},2} \big) \Big] \, ,
\end{align}
with $\omega_\textrm{lin}^\textsc{gw} = \sqrt{3} \omega_\textrm{lin}$. Here, $\overline{\Omega}_{\textrm{GW}}(k)$ refers to the GW fraction with the oscillatory component averaged out. 

\begin{figure}[h!]
 \centering
\includegraphics[width=0.9\textwidth]{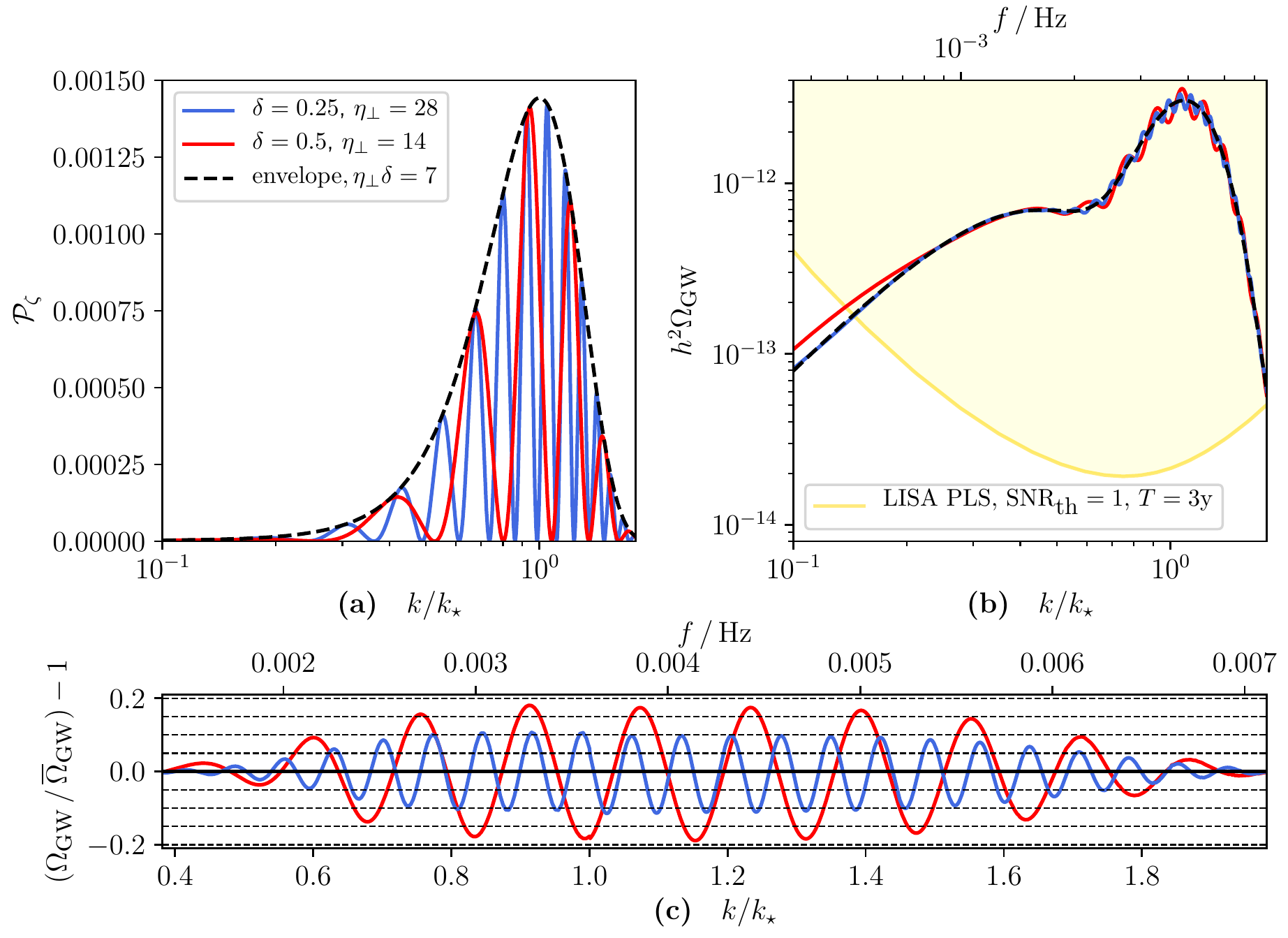}
\caption{\small Scalar power spectrum $\mathcal{P}_{\zeta}$ \textbf{(a)} and the scalar-induced GW energy density $\Omega_{\textrm{GW}}$ \textbf{(b)} for two inflation models (see text for details) with a sharp feature and enhanced fluctuations from a strong turn in the inflationary trajectory, together with the PLS for SNR threshold $\textrm{SNR}_\textrm{th}=1$ and total observation time $T=3$ years assumed in the analysis of ~\cite{Fumagalli:2020nvq}. The black dashed line in \textbf{(a)} shows the envelope of $\mathcal{P}_\zeta$ and in  \textbf{(b)} the corresponding GW energy density $\overline{\Omega}_{\textrm{GW}}$. The $\mathcal{O}(1)$-oscillations in $\mathcal{P}_\zeta$ are processed into $\mathcal{O}(10 \%)$ modulations on the principal peak of $\Omega_{\textrm{GW}}$. Over the frequency range of the principal peak, the modulations in $\Omega_{\textrm{GW}}$ can be modelled as cosine-oscillations about $\overline{\Omega}_{\textrm{GW}}$, as can be seen in \textbf{(c)}. To translate $k/k_\star$ into $f$ / Hz, we considered $N_\star=31.5$, i.e.~the peak in $\mathcal{P}_\zeta$ occurs at wavenumber $k=k_\star$, which is 31.5 $e$-folds larger than the CMB value. Figures adapted from Ref.~\cite{Fumagalli:2020nvq}.}
\label{fig:P_Omega_ratios_SmallScaleFeatures}
 \end{figure}

The above is best illustrated by an example. As has been observed in Refs.~\cite{Palma:2020ejf,Fumagalli:2020adf}, an amplification of scalar fluctuations can be achieved in multi-field inflation when the inflationary trajectory exhibits a strong turn. If the turn is also sufficiently sharp, i.e.~it is executed in a short time interval, this leads to a sharp feature. In simple cases the scalar power spectrum can then be computed analytically as \cite{Palma:2020ejf,Fumagalli:2020nvq}
\begin{equation}
\label{eq:P-analytic-xi}
\frac{\mathcal{P}_\zeta(k)}{{\cal P}_0} =   \, \frac{e^{2 \sqrt{(2-\kappa)\kappa} \, \eta_\perp \delta} }{2 (2-\kappa) \kappa}
\times \sin^2\left(e^{-\delta/2}\kappa \eta_\perp+\arctan\left(\frac{\kappa}{\sqrt{(2-\kappa)\kappa}}\right) \right)\, ,
\end{equation}
which is valid for $\kappa \leq 2$ and where $\kappa \equiv k / k_\star=k/(k_\textrm{f} \eta_\perp)$.
Here $\eta_\perp$ quantifies the strength of the turn in terms of the departure from a geodesic, $\delta$ is the duration of the turn in units of $e$-folds and $k_\textrm{f}$ is the scale crossing the Hubble radius at the time of the sharp turn (see Ref.~\cite{Fumagalli:2020nvq} for generalisations). As can easily be checked, Eq.~(\ref{eq:P-analytic-xi}) is indeed of the form of Eq.~(\ref{eq:P_of_k_sharp}) characteristic of a sharp feature.  

In Fig.~\ref{fig:P_Omega_ratios_SmallScaleFeatures} (a) we plot $\mathcal{P}_\zeta(k)$ in Eq.~(\ref{eq:P-analytic-xi}) for two example models and in Fig.~\ref{fig:P_Omega_ratios_SmallScaleFeatures} (b) the corresponding GW energy density $\Omega_{\textrm{GW}}(k)$. We can then make the following observations, which will not only hold for the model at hand, but will be generic for a sharp feature and can be also generalised to the case of a resonant feature:
\begin{itemize}
    \item The overall spectral shape $\overline{\Omega}_{\textrm{GW}}(k)$ is determined by the envelope $\mathcal{P}_0(k)$. If this is sufficiently narrowly-peaked, the spectral shape of $\overline{\Omega}_{\textrm{GW}}(k)$ consists of a broad lower peak at lower frequencies and a narrow principal peak at higher frequencies. This is what is observed in Fig.~\ref{fig:P_Omega_ratios_SmallScaleFeatures} (b). The principal peak occurs at $k=2/\sqrt{3} \, k_\star$, where $k_\star$ is the frequency where the envelope $\mathcal{P}_0$ has its maximum, and this can be understood to arise from resonant amplification \cite{Kohri:2018awv}.  
    \item     The oscillatory running of the SGWB induced by the feature is visible as a modulation of $\Omega_\textrm{GW}(k)$ on the principal peak, see again Fig.~\ref{fig:P_Omega_ratios_SmallScaleFeatures} (b). In the vicinity of the maximum the modulation is well-matched by the template in  Eq.~(\ref{eq:sharp_template}) for a sharp feature, as can be seen in Fig.~\ref{fig:P_Omega_ratios_SmallScaleFeatures} (c) where we plot the ratio $\Omega_\textrm{GW}(k)/ \overline{\Omega}_{\textrm{GW}}(k)$: Especially near the centre of the peak this exhibits a sinusoidal oscillation with near-constant amplitude. 
    \item The modulation in $\Omega_\textrm{GW}(k)$ can be understood as a superposition of resonance peaks sourced by the individual maxima of $\mathcal{P}_\zeta(k)$. Labelling the maxima in $\mathcal{P}_\zeta(k)$ by $k_{i}$, the resonance analysis predicts a series of peaks in $\Omega_\textrm{GW}(k)$ at \cite{Cai:2019amo}:
    \begin{align}
    \label{eq:kmaxij-def}
    k_{\textrm{max},ij} = \frac{1}{\sqrt{3}}(k_{i}+ k_{j}) \, , \quad \textrm{with} \quad k_{\textrm{max},ij} > |k_{i}- k_{j}| \, .
    \end{align}
    Applying this to the sharp and resonant feature case one can show that this predicts the periodic structure recorded in the templates Eq.~(\ref{eq:sharp_template}) and Eq.~(\ref{eq:resonant_template})~\cite{Fumagalli:2020nvq,Fumagalli:2021cel}.
    \item The amplitude of oscillation in $\Omega_\textrm{GW}(k)$ is typically attenuated compared to the amplitude of oscillation in $\mathcal{P}_\zeta(k)$. For example, a sharp feature with $A_\textrm{lin}=1$ leads to a modulated GW spectrum with $\mathcal{A}_\textrm{lin} \sim \mathcal{O}(10 \%)$~\cite{Fumagalli:2020nvq,Braglia:2020taf}.
    \item Consider a feature in the scalar power spectrum with an associated peak at $k_\star = k_\textsc{cmb} e^{N_\star}$, with $N_\star$ the number of $e$-folds after horizon-crossing of the CMB modes. This bump in $\mathcal{P}_{\zeta}(k)$ will produce a peak in $\Omega_\textrm{GW}$ whose frequency $f_\textrm{peak}$ is related to $N_\star$ as 
    \begin{align}
    N_\star \approx \ln \bigg( \frac{f_\textrm{peak}}{\textrm{Hz}} \bigg) +37 \, .
    \end{align}
    For the peak to fall into the frequency range of maximal sensitivity of LISA, $f=10^{-3}$ - $10^{-2}$ Hz, the enhancement and the feature in $\mathcal{P}_{\zeta}(k)$ has to occur $N_\star \sim 30$ - $32$ $e$-folds after the generation of the CMB modes, making LISA sensitive to a wide range of models producing features during the later stages of inflation.
\end{itemize}
We thus find that features in the scalar power spectrum, encoding departures from single-field slow-roll inflation, lead to corresponding features in the scalar-induced contribution to the SGWB. If the feature is associated with a sufficient enhancement of the relevant scalar fluctuations, the resulting contribution to the SGWB is in principle detectable by GW observatories like LISA. As features encode departures from single-field slow-roll without reference to explicit models, this leads to the exciting prospect that SGWB measurements can be used to learn about inflation in a model-independent fashion. 

\subsection{Effective Field Theory of broken space-reparametrisations}
\label{sec:supersolid}
LISA can shed light also on the symmetries characterising the early universe, and in particular the inflationary period. Standard models of inflation are based on 
the assumption that the inflaton background field is only time dependent, and this is well described within the EFT of inflation framework~\cite{Creminelli:2017sry}. However within the standard EFT framework, it is hard to produce GWs which have an amplitude large enough to be detected by interferometers. On the other hand, if space-diffeomorphisms are broken during inflation, and so the background field can have a space-dependent vacuum expectation value, then GWs are amplified at small scales becoming a potential target for GW detectors, like LISA. In particular, the breaking of space-diffeomorphism allows the graviton to be massive, since there are no symmetries preventing tensor fluctuations from acquiring a mass during inflation. If this is the case, the graviton is characterised by an action which is the most general one for tensor fluctuations:
\begin{equation}
    S_{T}^{(2)}=\frac{M_{P}^2}{8}\int d^4 x a^{2}(\eta)\left[h'_{ij}{h'}^{ij} - (k^2 + a^2(\eta) m_{g}^2)h_{ij}h^{ij}\right]\,,
\end{equation}
where $a(\eta)$ is the scale factor and $m_{g}$ is the graviton mass, which is the distinctive feature of a setup that breaks space-diffeomorphism. Whereas such action can be derived within an EFT framework, there are models which realise such a possibility, like Solid~\cite{Endlich:2013jia} and Supersolid~\cite{Koh:2013msa,Cannone:2014uqa,Cannone:2015rra,Bartolo:2015qvr} inflation. They are described by Lagrangians which contain three (or four) scalar fields, which respect some internal symmetries ensuring  the homogeneity and isotropy of the background. Such models can lead  to a blue spectrum for tensor modes, as well as to other distinctive properties
in the tensor spectrum~\cite{Endlich:2013jia,Ricciardone:2016lym,Bartolo:2015qvr,Cannone:2014uqa, Akhshik:2014gja,Akhshik:2014bla}.
Such an action leads, in the small graviton mass limit ($|m_h/H|\ll1$), to a tensor primordial power spectrum like
\begin{equation}\label{eq:PS}
{\mathcal P}_h\,=\,\frac{2\,H^2}{\pi^2\,M_{P}^2}\,\left( \frac{k}{k_*}\right)^{n_T}\,,
\end{equation}
with the tensor spectral index equal to
\begin{equation}\label{eq:tensortilt}
n_T\,=\,\frac23\,\frac{m_h^2}{H^2}\,.
\end{equation}
Notice that a blue spectrum, $n_T>0$, requires a positive $m_h^2$. This is the case of interest, 
since it enhances the tensor spectrum at small scales, and can lead to a signal detectable by LISA. A model-independent analysis of such kind of models, and the possibility to detect the GW signal in the range of frequencies and energy densities probed by LISA has been carried out in Ref.~\cite{Bartolo:2016ami}.

\be
\varphi^0= \bar \varphi(t), \, \qquad \varphi^a= x^a   \, , \quad a
=1,2,3 \, ;
\ee
this time 4D diffeomorphisms are completely broken, leaving only a global symmetry as a leftover, where an SO(2) rotation is performed both to the fields and to the spatial coordinates. To allow for a FLRW background solution, the action
for the scalar fields dynamics needs to be symmetric under internal rotations and shift transformations
\be
\varphi^a \to {\cal R}^a_b \, \varphi^b \, , \qquad    \pmb{{\cal R}}^t
\pmb{{\cal R}}=\mathbf{1}\, ,   \qquad \varphi^A \to  \varphi^A + c^A\,,
\qquad  \qquad A=0,1,2,3 \, .
\label{back}
\ee
In general, the fluctuations around  Eq.~(\ref{back}) can be interpreted as the phonons of a supersolid~\cite{Son:2005ak,Celoria:2017bbh}, while the special case where only the ``spatial'' Stueckelberg fields $\varphi^a$ (with $a=1,2,3$) are present corresponds to a solid~\cite{Endlich:2012pz}. Similar models have been proposed as massive deformation of gravity~\cite{Ballesteros:2016gwc,Celoria:2017hfd}. In the scalar sector there are two propagating degrees of freedom that mix non-trivially both at early and late times and, after exiting the horizon, give rise to non-trivial cross-correlations with distinctive features for primordial non-Gaussianity. One key feature is that
the production of GWs during inflation can be enhanced by the cubic interaction of the graviton with phonons~\cite{Celoria:2020diz};  the spectral index of the  secondary produced GWs is blue-tilted with the amplitude within the LISA sensitivity.

\subsection{UV complete models of $R+\alpha R^2$ model}
\label{AM}

Starobinsky's $R+\alpha R^2$ model of inflation matches the latest CMB data extremely well~\cite{Akrami:2018odb,Starobinsky:1983zz}. Since massless gravity contains only derivative interactions, it invites higher and infinite covariant derivative contributions as well. One particularly attractive model of higher 
derivative theory of gravity which contains infinite covariant 
derivatives and generalizes the quadratic 
curvature theory of gravity 
has been presented in Refs.~\cite{Biswas:2005qr,Modesto:2011kw,Biswas:2011ar}.
It has several cosmological implications, from providing the initial conditions from a non-singular cosmology to modifying the power spectrum for the primordial GWs~\cite{Koshelev:2017tvv,Koshelev:2020foq}. One particular subset of the action has been studied deeply, which generalizes Starobinsky's model and recovers it in the IR limit:
\begin{equation}\label{NC-action}
S=\int d^{4}x\sqrt{-g}\left(\frac{M_{p}^{2}}{2}R+	\frac{\lambda}{2}
\bigg[R{\cal F}_{1}(\Box_{s})R+W_{\mu\nu\rho\sigma}{\cal F}_{2}(\Box_{s})W^{\mu\nu\rho\sigma}\bigg]\right)\,.
\end{equation}
Here $\Box_{s}={\Box}/{M_{s}^{2}}$ with $M_{s}$ being the
scale of non-locality, and $\lambda$ is a dimensionless parameter useful to control the effect of higher curvature contributions. In the limit $M_s \rightarrow \infty$, one recovers the local Starobinsky's model of inflation. The ghost-free condition around an inflationary, i.e.~approximate de Sitter background, constrains the gravitational form factors ${\cal F}_1$ and ${\cal F}_2$~\cite{Biswas:2016etb,Biswas:2016egy}. Noticeably, the scalar power spectrum does not get any modification compared to the local Starobinsky's model of $R+\alpha R^2$ inflation~\cite{Starobinsky:1983zz}. However, the gravitational power spectrum, the tilt and the tensor to scalar ratio $r$ all get modified in an interesting manner~\cite{Koshelev:2017tvv,Koshelev:2020foq}:
\begin{eqnarray}
	{\cal P}_{T} &=& \frac{1}{12\pi^{2}{\cal F}_{1}(m^2/M_s^2)}(1-3\epsilon) 
	e^{-2\gamma_T(-{\bar{R}}/{2 M_s^{2}})}  |_{k=aH}\,, \cr
		n_{T} &\equiv &  \frac{d\ln{\cal P}_{T}}{d\ln k} |_{k=aH} 	
		\approx -\frac{3}{2N^{2}}-(\frac{2}{N}+\frac{3}{2N^{2}})\frac{\bar{R}}{2 M_s^{2}}
		\gamma_T(-\frac{\bar{R}}{2 M_s^{2}}) |_{k=aH}\,, \cr
		r &=&\frac{12}{N^{2}}e^{-2\gamma_T(-\bar{R}/{2M_s^{2}})}|_{k=aH}\,.
\end{eqnarray}
The crucial difference here in comparison with the local Starobinsky model is that the tensor power spectrum is scaled by an exponential factor of $\gamma_T$ evaluated at the pole of the tensor mode $\bar R/ (6 M_s^2)$, where $\bar R= 3.7\times 10^{-8}(55/N_\ast)^3M_p^2$ denotes the Ricci scalar evaluated at the pivot scale $k_\ast=aH$, and $N_\ast$ denotes the number of $e$-foldings of inflation corresponding to the pivot scale.
Accordingly, also the tensor tilt gets modified. The ghost-free condition demands that $\gamma_T$ is an entire function (namely, that it can be represented as a power series that converges everywhere in the complex plane) ~\cite{Biswas:2016etb,Biswas:2016egy}. In the local $R^2$ model one obtains  $ r={12}/{N^{2}}=3(1-n_s)^2$ as it follows from the original computation of scalar and tensor power spectra generated during inflation~\cite{Starobinsky:1983zz}.  From the CMB data \cite{BICEP:2021xfz} we can infer that the constraint $r< 0.036$ implies $\gamma_T > -1.05$ at $N_\ast=55$, while there are no constraints on the tilt in the tensor power spectrum, due to lack of data. LISA could be able to test it, although there are indications that at LISA frequencies the nonlocal model is indistinguishable from local Starobinsky inflation~\cite{Koshelev:2017tvv,Calcagni:2020tvw} (section 
\ref{sec:sgwbqg}). Additional mechanisms~\cite{Bartolo:2016ami} or a 
modification in the theory (e.g., in the form factor) could reopen this opportunity.

\subsection{Preheating}\label{sec:preheating}

Preheating is characterised by non-perturbative particle production mechanisms~\cite{Traschen:1990sw,Kofman:1994rk,Shtanov:1994ce,Kaiser:1995fb,Khlebnikov:1996mc,Prokopec:1996rr,Kaiser:1997mp,Kofman:1997yn,Greene:1997fu,Kaiser:1997hg}, which typically take place after inflation in many models of particle physics (see Refs.~\cite{Allahverdi:2010xz,Amin:2014eta,Lozanov:2019jxc} for reviews). Following the end of inflation, interactions between the inflaton and some other field species -- the {\it preheat field(s)} -- can induce an exponential growth of the modes of the preheat field(s) within certain bands of momenta. A paradigmatic example of this is {\it parametric resonance}~\cite{Kofman:1994rk,Kofman:1997yn,Greene:1997fu,Figueroa:2016wxr}, though there are other mechanisms (see below). The field gradients created during this stage can generate a sizeable anisotropic stress to source GWs, with the specific details of the resulting GW spectrum depending strongly on the considered scenario~\cite{Khlebnikov:1997di,Easther:2006gt,Easther:2006vd,GarciaBellido:2007dg,GarciaBellido:2007af,Dufaux:2007pt,Dufaux:2008dn,Kusenko:2008zm,Kusenko:2009cv,Fenu:2009qf,Figueroa:2011ye}. In general, preheating mechanisms are very efficient in transferring a significant fraction of the total energy available from the inflationary sector into the preheat field(s), and as a result, a large amount of GWs are radiated in the process. Furthermore, if the inflaton-preheat field coupling is tuned to certain values, the resulting SGWBs may develop large anisotropies at cosmological scales~\cite{Bethke:2013aba, Bethke:2013vca}.

For illustrative purposes, let us consider a standard scenario where the inflaton oscillates around the minimum of its potential after the end of inflation. For example we can consider a power-law potential $V(\phi) = \frac{1}{p} \lambda \mu^{4-p} \phi^p$, with $\lambda$ a dimensionless coefficient, $\mu$ some mass scale, and $p$ an integer index $p \geq 2$. Denoting as $t_{\star}$ the end of inflation, the inflaton oscillates for $t \gtrsim t_{\star}$ with a time-dependent frequency $\Omega_{\rm osc} \equiv \omega_{\star} (t /t_{\star})^{1- 2/p}$, $\omega_{\star} \equiv \sqrt{\lambda} \mu^{(2 - p/2)} \phi_{\star}^{(p/2 - 1)}$, where $\phi_{\star} \equiv \phi (t_{\star})$~\cite{Figueroa:2016wxr}. To be specific, let us also consider a quadratic interaction $g^2 \phi^2 \chi^2$ between the inflaton $\phi$ and a the preheat field $\chi$, with $g$ a dimensionless coupling constant. If the  \textit{resonance parameter} $q_{\star} \equiv g^2 \phi_{\star}^2 /\omega_{\star}^2$ is much larger than unity $q_{\star} \gg 1$, the preheat field is excited through a process of {\it broad resonance} during the inflaton oscillations, with the amplitude of the resonant modes growing exponentially inside a Bose-sphere of radius $k \lesssim k_{\star} \sim q_*^{1/4} \omega_{\star}$. This radiates GWs efficiently within a similar band of momenta. At the end of the process the resulting energy density spectrum of GWs exhibits a peak with amplitude and location (redshifted to frequencies today) given by~\cite{Figueroa:2017vfa}
\begin{subequations}
    \begin{gather}
        f \simeq  8 \cdot 10^{9} \left( \frac{\omega_{\star}}{{\rho}_{\star}^{1/4}}\right)\, \epsilon_{\star}^{\frac{1}{4}} q_{\star}^{\frac{1}{4} + \eta}~{\rm Hz} \label{eq:preheating-1} \,,\\
        \Omega_{\text{GW}}^{0}(f) \simeq \mathcal{O}(10^{-9}) \times \epsilon_{\star} \, \mathcal{C} \frac{\omega_{\star}^{6}}{\rho_{\star} M_p^{2}}\,q_{\star}^{-\frac{1}{2}+\delta} \label{eq:preheating-2} \,,
    \end{gather}
\end{subequations}
where $\rho_{\star}$ is the total energy density at $t=t_{\star}$, $\eta$ and $\delta$ are parameters that account for non-linearities of the system, and $\mathcal{C}$ is a constant that characterises the strength of the resonance. The factor $\epsilon_{\star} \equiv (a_{\star} /a_{\rm RD})^{1 - 3 w}$ parametrises the expansion history between the end of inflation and the onset of RD, assuming an averaged EoS $w \neq 1/3$ during this period (if $w = 1/3$ then $\epsilon_* = 1$). The values for $\mathcal{C}$, $\eta$, and $\delta$, can only be determined with classical lattice simulations in a model by model basis. 

Leaving aside non-linear effects for simplicity, we observe that the frequency and amplitude of the peak, roughly scale as $f \sim q_{\star}^{1/4}$ and $\Omega_{\text{GW},0} \sim q_{\star}^{-1/2}$. This means that in order to shift the peak to observable frequencies, we need to decrease $q_\star$; but in doing so, we further decrease the amplitude of the signal. Using this linear approximation we see that decreasing the peak frequency say by an order of magnitude, implies the reduction of the amplitude of the background by two orders of magnitude. Furthermore, we see that these backgrounds can only be expected down to a minimum frequency, as efficient GW production requires $q_* > 1$ to sustain broad resonance during the field instability, and hence $q_* = 1$ marks a minimum frequency we can think of. Even though such scaling behaviours are modified when considering corrections due to non-linearities, the logic persists~\cite{Figueroa:2017vfa}, and as a result these backgrounds are forced to be peaked at high-frequencies way above those accessible by LISA. To be concrete, considering for instance a quartic potential around the minimum, we obtain $f \simeq (10^7 - 10^8)  \,  {\rm Hz}$  and $\Omega_{\text{GW}, 0} \simeq (10^{-13} - 10^{-11}) $, when assuming the range $q_{\star} \in (1,10^4)$.\footnote{The GW spectrum in the quartic potential case also features additional peaks, see Ref.~\cite{Figueroa:2017vfa} for more details.} For a quadratic potential around the minimum of the potential, we find instead (assuming $\epsilon_{\star} = 1$) a frequency $f \simeq (10^8 - 10^9) \, {\rm Hz}$ and amplitude $\Omega_{\text{GW},0} \simeq (10^{-12} - 10^{-11})$, when considering a resonance parameter $q_{\star} \in (10^4,10^6)$. The conclusion is clear, these backgrounds are completely out of reach of current and (so far) planned direct detection GW experiments.

If the scalar field interactions induce a tachyonic effective mass in the preheat species, GWs can also be strongly produced, e.g.~during hybrid preheating~\cite{GarciaBellido:2007dg,GarciaBellido:2007af,Dufaux:2008dn}. In this case, contrary to parametric resonance, the present day frequency and amplitude of the generated SGWB might be tuned to peak at small frequencies, while retaining a large amplitude. This depends on the model parameters, but in general a strong fine-tuning is required for this to happen. A similar circumstance arises in the case of the GWs produced from oscillons produced during hilltop preheating~\cite{Antusch:2016con,Antusch:2017vga,Amin:2018xfe}, which upon similar fine tuning of the potential parameters, can also lead to observable GW backgrounds at low frequencies. 

If the field species involved in preheating are of a different nature than just scalar fields, then new channels of GW production open up. For example, GWs can be produced during the out-of-equilibrium excitation of fermions after inflation, both for spin-1/2 \cite{Enqvist:2012im,Figueroa:2013vif,Figueroa:2014aya} and spin-3/2 \cite{Benakli:2018xdi} fields. In this case a SGWB with large amplitude is also forced to peak at high frequencies. GWs can also be generated when the preheat fields are (Abelian and non-Abelian) gauge fields. For example gauge fields could be coupled to a charged scalar field via standard gauge covariant derivatives like in Refs.~\cite{Dufaux:2010cf,Figueroa:2016ojl,Tranberg:2017lrx}, or to a pseudo-scalar field through a derivative axial coupling as in Refs.~\cite{Adshead:2018doq,Adshead:2019igv,Adshead:2019lbr}. Preheating can be remarkably efficient in the second case~\cite{Adshead:2015pva,Cuissa:2018oiw}, with the energy density produced in GWs possibly reaching up to $\sim 1\%$
of the total energy in the system for the strongest coupling strengths~\cite{Adshead:2019igv,Adshead:2019lbr}. The peak frequency is however also very large in these scenarios as well. 

In conclusion, preheating mechanisms are capable of creating very large SGWBs, but these are naturally peaked at very high frequencies, which are typically beyond the LISA window. Some particular models can sustain a sizeable amplitude at LISA frequencies, but only at the expense of a strong fine-tuning of their parameters.

\subsection{Summary of distinctive gravitational wave observables from inflation}
\label{sec:disob}

\noindent In what follows we describe the relevant observables that can be used to break possible degeneracies between the SGWB from inflation, and the one due to other cosmological sources, as well as to astrophysical ones.

\subsubsection{Chirality}
\label{inflation-chirality}

Although parity violation has so far been observed only in weak interactions, it is important to investigate whether the same phenomenon may occur also in the very early universe, during inflation. In the primordial context, we shall refer to chirality whenever the two polarisations of GWs have different solutions. For convenience, one may introduce the parameter  $\chi \equiv |P_{\gamma}^{L}-P_{\gamma}^{R} | / \sum_{\lambda} P^{\lambda}$, whose range extends up to $\chi=1$.
Several mechanisms for generating chiral GWs are found in the literature. The common  trait to such realisations is the presence of a Chern-Simons-type interaction. In the case of the EM field strength such term reads
\begin{equation}
\label{CStypeEM}
g(\chi) F_{\mu\nu}\tilde{F}^{\mu\nu}\;,
\end{equation}
where $\tilde{F}^{\mu\nu}$ is the dual EM field. The reader will recognise the similarities between Eq.~(\ref{fftilde}) and (\ref{CStypeEM}). The comparison underscores the fact that the field strength $F$ need not be that of electromagnetism nor is it limited to Abelian gauge fields. In  modified gravity theories, Chern-Simons gravity is obtained by promoting the three dimensional gravitational Chern-Simons term to 4D \cite{Jackiw:2003pm,Alexander:2009tp}:
\begin{equation}
\label{CStypeGR}
f(\phi) R^{\sigma}_{\;\mu\nu\rho}\tilde{R}^{\;\mu\nu\rho}_{\sigma}\;,
\end{equation}
where $\tilde{R}$ is the dual Riemann tensor. The matter field $\phi$ is conveniently identified as the inflaton in models of the early phase acceleration. The parity violation ensuing from Chern-Simons-type coupling in the early universe can be tested across a vast range of scales, from CMB \cite{Lue:1998mq,Thorne:2017jft} to interferometers \cite{Crowder:2012ik,Smith:2016jqs,Domcke:2019zls}, including LISA. Both types of interactions in Eqs.~(\ref{CStypeEM}) and (\ref{CStypeGR}) have been extensively studied~\cite{Anber:2009ua,Satoh:2010ep,Crisostomi:2017ugk,Bartolo:2017szm,Bartolo:2018elp,Qiao:2019hkz,Bartolo:2020gsh,Bordin:2020eui}), sometimes considered together \cite{Lue:1998mq,Mirzagholi:2020irt}, and investigated as emerging from quantum gravity (in the sense of Ho\v{r}ava-Lifshitz \cite{Horava:2009uw} theories)~\cite{Takahashi:2009wc}.

\subsubsection{Frequency profile}

The single-field slow-roll inflationary paradigm predicts a slightly red-tilted GW spectrum. The tilt of the tensor power spectrum is indeed proportional to the slow-roll parameter, $n_T\simeq -2\epsilon\ll 1$. Such a GW signal is well below the sensitivity of LISA and could even elude detection from more  sensitive instruments, including the proposed BBO. This is the key notion behind the claim that a detection of primordial GWs by LISA would be a smoking gun for an inflationary mechanism that goes beyond the single-field slow-roll scenario. With some interesting exceptions \cite{Mylova:2018yap} one may go further and state that a primordial signal at LISA scales (and sensitivity) is strongly suggestive of multi-field (or multi-clock) dynamics.

A typical case in point is that of GW sourced by extra fields during inflation. Whenever the sourced contribution goes beyond vacuum fluctuations, the richer multi-field dynamics can give rise to a non-trivial frequency dependence, from a blue GW spectrum to a ``bump-like'' structure and to oscillatory features mimicking the ones of the scalar power spectrum.
For all these examples there exist explicit realisations, as detailed in Secs.~\ref{sec-GWaddf} and \ref{sec:features}.

In studying the SGWB frequency profile one ought to be aware of the bounds at CMB scales, those set at intermediate scales by PTAs and, perhaps most importantly for LISA, those set at the relatively close frequencies accessed by LIGO/Virgo.

\subsubsection{Anisotropies}
\label{Anisotropies-inflation}

The angular resolution of the LISA detector might enable a detection of another peculiar feature useful in the SGWB characterisation process:
the anisotropy (direction dependence) in the energy density. Such
anisotropies contain information about the generation process of GWs and
their propagation across cosmic inhomogeneities.
Using a Boltzmann equation
approach, the contribution coming from the generation mechanism retains a
frequency dependence, which is peculiar of the SGWB, and, due to the
non-thermal nature of the graviton distribution function at their
decoupling time~\cite{Bartolo:2019oiq, Bartolo:2019yeu}.
The contribution arising from the propagation of GWs through
large-scale cosmological scalar (and tensor) background perturbations,
happens to be larger compared to the same effect for CMB photons. So, in a similar way to CMB photons, the SGWB from e.g.~inflation is affected by the Sachs-Wolfe and the Integrated
Sachs-Wolfe effects. The former is a gravitational redshift, due to the difference of the gravitational potential at the moment of the SGWB production and today. The latter is due to the variation of the gravitational potential along the line of sight from the SGWB production to its the detection, integrated in time~\cite{Bartolo:2019oiq, Bartolo:2019yeu}.

It turns out that LISA is sensitive to these effects in the angular power spectrum if the SGWB
isotropic energy density is $\Omega_{\rm GW} \gtrsim 10^{-12}$~\cite{Bartolo:2019oiq,
Bartolo:2019yeu, DallArmi:2020dar}.
For a representative inflationary model, namely axion
inflation \cite{Cook:2011hg, Garcia-Bellido:2016dkw}, it  was shown that 
the predicted SGWB can be within the reach of LISA and exhibits anisotropies with a large frequency dependence, a possible distinctive target for the SGWB detection and characterisation in LISA.
%a primordial GW signal visible at interferometer scales can indeed lead to anisotropies in the SGWB, with a large frequency dependence, becoming a possible distinctive target for LISA.

In the remaining part of this subsection we will discuss GW anisotropies that result from squeezed primordial non-Gaussianity. This is yet another case in point for the use of GW anisotropies as a probe of the production mechanism for GWs. Let us briefly describe the genesis of these non-Gaussianity-sourced anisotropies. A squeezed primordial bispectrum encodes a coupling between two short-wavelength modes and one long-wavelength mode. The effect of the very same coupling is also manifest at the level of the power spectrum of the short-wavelength modes, in the form of a modulation by the long-wavelength mode \cite{Jeong:2012df,Dai:2013kra,Brahma:2013rua,Dimastrogiovanni:2014ina,Dimastrogiovanni:2015pla}. The magnitude and form of this modulation depends on the specific type of interactions (as dictated by the inflationary Lagrangian) and on the nature (e.g scalar vs. tensor) of the long-wavelength mode. Let us consider the case of a tensor bispectrum:
\begin{equation}
\langle h_{\textbf{k}_{1}}^{}h_{\textbf{k}_{2}}^{}h_{\textbf{q}_{}}^{} \rangle  =(2\pi)^{3}\delta^{(3)}(\textbf{k}_{1}+\textbf{k}_{2}+\textbf{q}_{})B^{\text{ttt}}(\textbf{k}_{1},\textbf{k}_{2},\textbf{q}_{}) \,.
\end{equation}
In the squeezed limit, $q\ll k_{1}\simeq k_{2}$, and for models preserving statistical isotropy and parity, the long wavelength tensor mode imprints a quadrupolar modulation in the primordial power spectrum of the short-wavelength modes \cite{Dimastrogiovanni:2019bfl}:
\begin{equation}\label{ps-anisotropy}
P_{h}^{\text{mod}}(\textbf{k},\textbf{x})=P_{h}(k)\left[1+\mathcal{Q}_{\ell m}(\textbf{k},\textbf{x})\,\hat{n}_{\ell}\hat{n}_{m} \right]\,, 
\end{equation}
where $\textbf{k}= k\hat{n}$ and we have defined
\begin{equation}
\mathcal{Q}_{\ell m}(\textbf{k},\textbf{x})\equiv\int \frac{d^{3}q}{(2\pi)^{3}}e^{i\textbf{q}\cdot\textbf{x}}\sum_{\lambda_{3}}h_{\ell m}^{\lambda_{3}}(\textbf{q})F_{\rm NL}^{\text{ttt}}(\textbf{k},\textbf{q})\,.
\end{equation}
Here $F_{\rm NL}^{\text{ttt}}(\textbf{k},\textbf{q})$ is the amplitude of the tensor bispectrum in the squeezed limit, normalised by the product of the power spectra, $P_{h}(k)\cdot P_{h}(q)$.  The anisotropy in Eq.~(\ref{ps-anisotropy}) determines a contribution to the energy density contrast (see Eq.~(\ref{dcontrast}) for the definition of $\delta_{\text{GW}}$) given by
\begin{equation}
\delta_{\text{GW}}(k,\hat{n})=\mathcal{Q}_{\ell m}(\textbf{k},\textbf{d})\,\hat{n}_{\ell}\hat{n}_{m}\,,
\end{equation}
where $\textbf{d}=-\hat{n}d$, $d\equiv\eta_{0}-\eta_{\text{in}}$ being the separation in conformal time between horizon re-entry of the k-mode and the present time. 

In a similar way, for inflationary models predicting a long-short mode coupling between scalars and tensors, a long-wavelength scalar fluctuation will modulate the power spectrum of GWs on small scales:
\begin{equation}\label{ps-anisotropy2}
P_{h}^{\text{mod}}(\textbf{k},\textbf{x})=P_{h}(k)\left[1+ \int \frac{d^{3}q}{(2\pi)^{3}}e^{i\textbf{q}\cdot\textbf{x}}\,\zeta(\textbf{q})F_{\rm NL}^{\text{stt}}(\textbf{k},\textbf{q}) \right]\,.
\end{equation}
This generates an anisotropic component for the GW energy density measured at present time, which leads to
\begin{equation}\label{zeta-anisotr}
\delta_{\text{GW}}(k,\hat{n})= \int \frac{d^{3}q}{(2\pi)^{3}}e^{-i d\hat{n}\cdot\textbf{q}}\zeta(\textbf{q})F_{\rm NL}^{\text{stt}}(\textbf{k},\textbf{q}) \,,
\end{equation}
$with F_{\rm NL}^{\text{stt}}(\textbf{k},\textbf{q})$ being the amplitude of the scalar-tensor-tensor bispectrum in the squeezed limit, normalised by $P_{\zeta}(q)\cdot P_{h}(k)$. 

\begin{figure}
	\centering
	\includegraphics[width=0.47\textwidth]{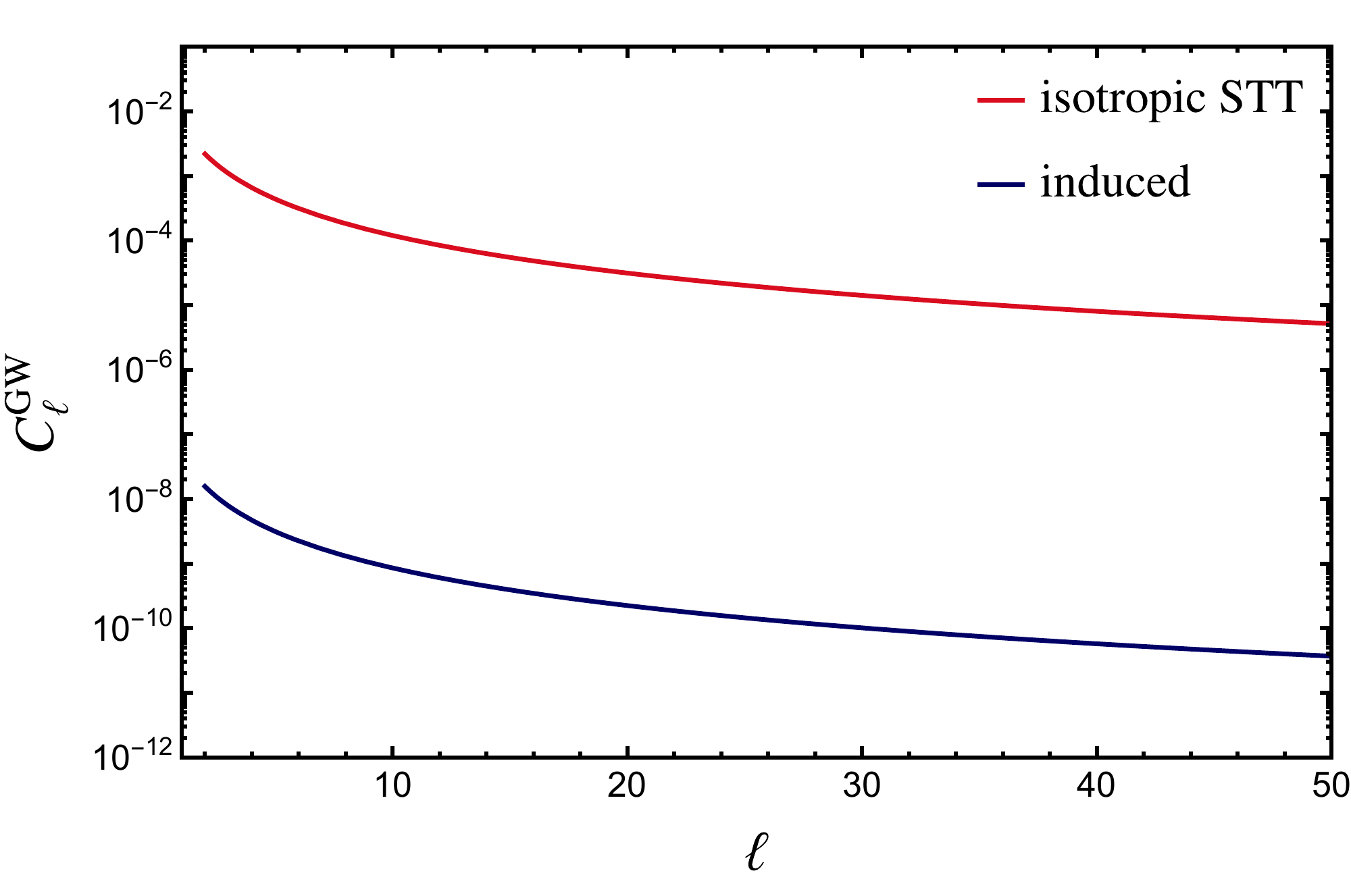} \hfill
\includegraphics[width=0.47\textwidth]{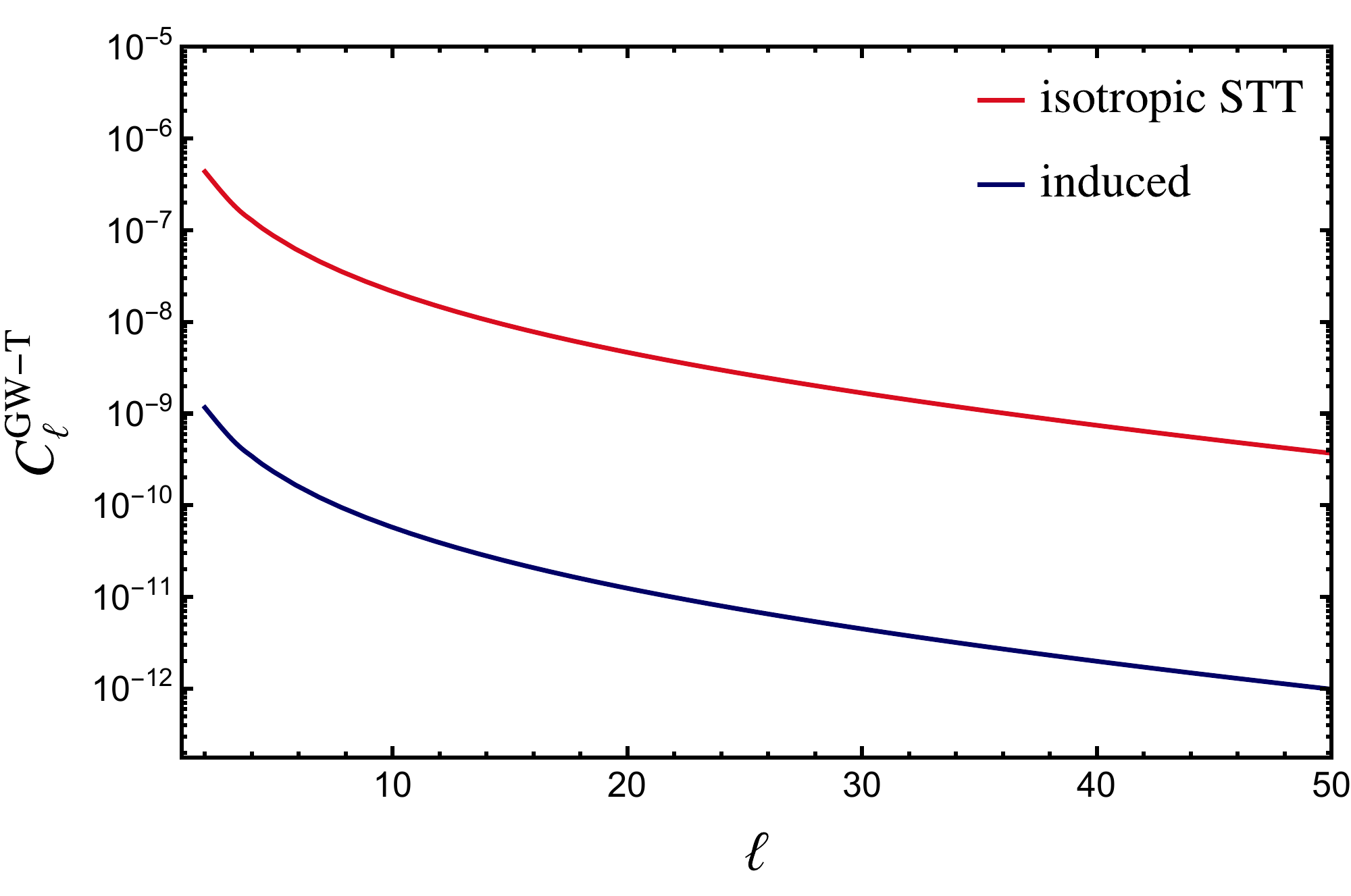}
	\caption{\small Left panel: Auto-correlation of GW anisotropies induced by a scalar-tensor-tensor squeezed bispectrum with $F_{\text{NL}}^{\text{stt}}\sim \mathcal{O}(10^{3})$ (red line), and of GW anisotropies arising from propagation through the perturbed background (blue line). Right panel: Cross-correlation of GW anisotropies and CMB temperature anisotropies in the same two cases.}
		\label{auto_cross_normalised}
\end{figure}

Primordial non-Gaussianity is one of the most informative probes of field interactions during inflation. Already at the level of the bispectrum (the lowest-order non-Gaussian correlator), a variety of constraints can be placed on inflationary models based on its momentum dependence (a.k.a. shape) and overall amplitude. In particular, bispectra that receive substantial contribution in the \textsl{squeezed} momentum configuration are a smoking gun for inflationary dynamics that go beyond the single-field slow-roll realisation. This is due to the fact that, whenever the inflationary dynamics is guided by a single effective degree of freedom, the leading order contribution to the squeezed bispectrum can be removed by a gauge transformation \cite{Maldacena:2002vr,Tanaka:2011aj,Pajer:2013ana}. Single-clock models therefore predict a suppressed bispectrum. On the other hand, typical classes of models that can lead to sizeable soft limits for cosmological correlators include multiple fields, excited initial states, or an inflationary background that breaks space diffeomorphism invariance. 

While at large (e.g.~CMB) scales primordial bispectra can be directly constrained, this is not the case at small scales due to loss of coherence from propagation in the perturbed universe (see e.g.~discussion in Sec.~\ref{subsec:second-order-SGWB}). Anisotropies are therefore a key observable  for constraining tensor and mixed non-Gaussianity at interferometer scales.

As an example, in the left panel of Fig.~\ref{auto_cross_normalised} we display the auto-correlation $\langle \delta_{\text{GW},\ell_{1}m_{1}}\delta_{\text{GW},\ell_{2}m_{2}} \rangle=\delta_{\ell_{1}\ell_{2}}\delta_{m_{1} m_{2}}\mathcal{C}_{\ell_{1}}^{\text{GW}}$ arising from anisotropies of the kind described in Eq.~(\ref{zeta-anisotr}) (red line). For the sake of comparison, in the same plot we show the auto-correlation for anisotropies imprinted as gravitons travel in the perturbed background after horizon re-entry (blue line) \cite{Alba:2015cms,Contaldi:2016koz,Bartolo:2019oiq,Bartolo:2019yeu}. 

The recent works \cite{Contaldi:2020rht,Bartolo:2022pez} study the capability of LISA to access the SGWB anisotropies. See Sec.~\ref{sec:pipesgwb} for more details.

\subsubsection{Cross-correlations}
\label{cross-corr}

\begin{figure}[!h]
	\centering
	\includegraphics[width=0.6\textwidth]{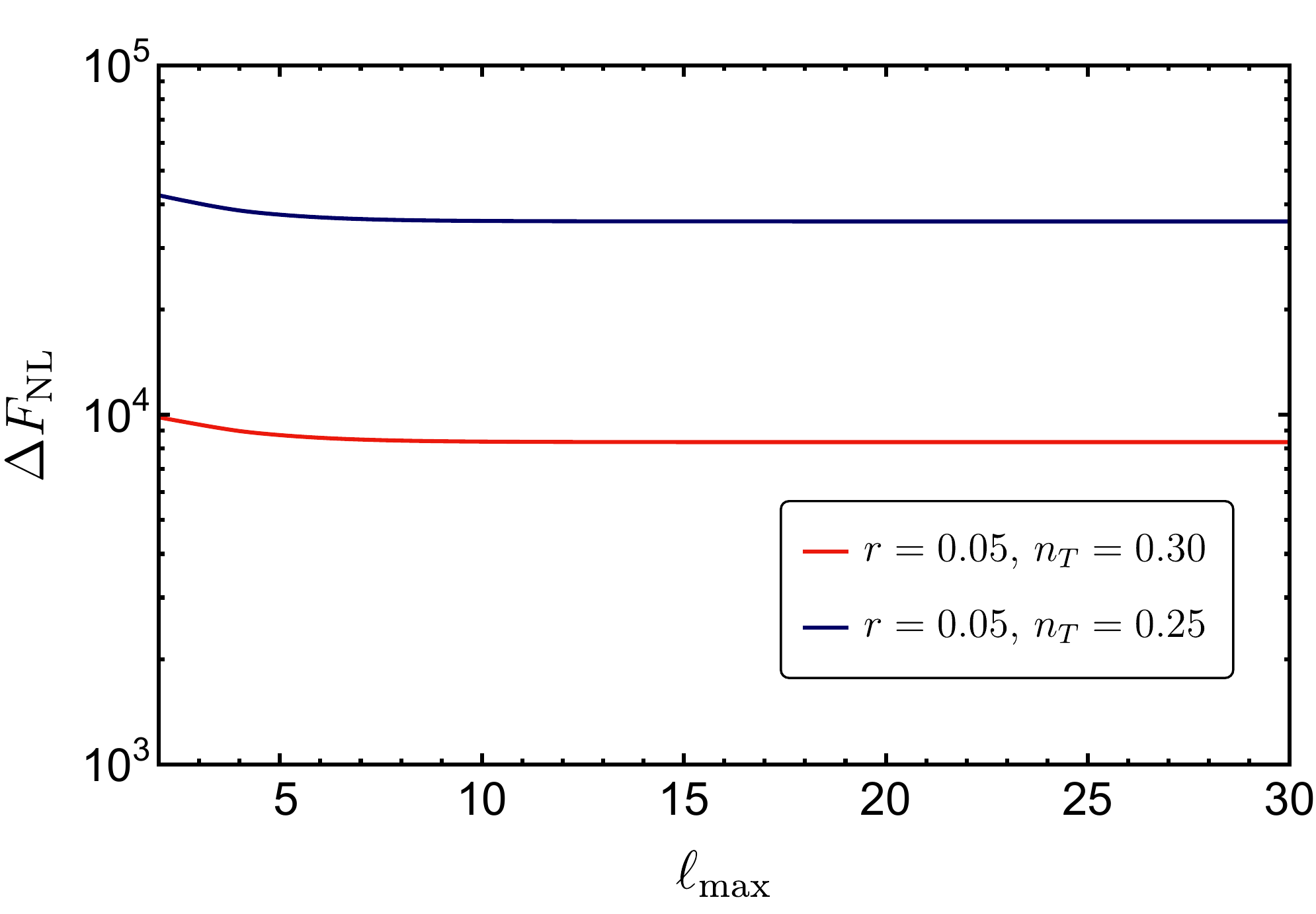}
	\caption{\small The expected 1$\sigma$ error in determing $F_{\rm NL}^{\text{stt}}$ with cross-correlations of CMB temperature anisotropies and GW anisotropies, as a function of $\ell_{\rm max}$.}
	\label{Ameek2}
\end{figure}

Anisotropies such as those just discussed in Sec.~\ref{Anisotropies-inflation} are produced by the effect of long-wavelength (scalar or tensor) perturbations and are therefore correlated with anisotropies in the CMB \cite{Adshead:2020bji,Malhotra:2020ket}. These cross-correlations provide an additional observable for constraining primordial non-Gaussianity (see right panel of Fig.~\ref{auto_cross_normalised} for an application). As an example, we report in Fig.~\ref{Ameek2} the error in determining $F_{\rm NL}^{\text{stt}}$ using a measurement of these cross-correlations. The calculation of the error follows that of Ref.~\cite{Malhotra:2020ket} and it is adapted here, under that assumption that Taiji and LISA should happen to fly together \cite{Ruan:2020smc,Orlando:2020oko}. We see that for the combination $\{{r=0.05,n_T=0.30}\}$ one would be able to detect $F_{\rm NL}^{\text{stt}}$ of order $10^4$ or larger.  

For any given inflationary model generating non-trivial scalar-tensor-tensor or tensor-tensor-tensor bispectra in the squeezed limit, the overall level of the GW anisotropies is determined by the specific form of the angular dependence of the bispectra and by the magnitude of the $F_{\rm NL}$ parameters. Typically, there exist minimum values of  $F_{\rm NL}$ below which the GW anisotropy map is dominated by noise.  Cross-correlations can be particularly helpful in these cases as they have the potential to constrain smaller levels of non-Gaussianity~\cite{Malhotra:2020ket}.

\clearpage
%%%%%%%%%%%%%%%%%%%%%%%%%%%%%%%%%
% Here Sec. 9 starts

\section{Tests of non-standard pre-Big-Bang nucleosynthesis cosmology via the SGWB}
\label{sec:nonStandard}

\small \emph{Section coordinator: G.~Calcagni. Contributors: G.~Calcagni, C-F.~Chang, Y.~Cui, D.G.~Figueroa, S.~Kuroyanagi, M.~Lewicki, A.~Mazumdar, G.~Servant, P.~Simakachorn.}
\normalsize

\subsection{Introduction}

In the standard model of cosmology, the universe begins with a inflationary epoch that first induces an exponential growth of the universe and then reheats it with a very hot plasma. In the standard picture, it is the energy density of such a plasma that dominates the expansion of the universe until the matter and DE domination final stages. It is however worth remembering that this picture is based on several theoretical assumptions and has only been tested up to temperatures $T\approx 5\, {\rm MeV}$ thanks to data from CMB~\cite{Aghanim:2018eyx} and successful predictions of BBN~\cite{Cyburt:2015mya}. The period between that temperature and inflation still holds many unknowns~\cite{Allahverdi:2020bys}. In this section, we will discuss how detection of a SGWB by LISA would allow us to probe the evolution of the universe during that time. We will begin with the possibility of probing the expansion rate including the number of relativistic degrees of freedom thanks to the SGWB sources discussed in the previous sections and move on to possible departures from GR. 

According to the standard inflationary paradigm in GR, scalar and tensor perturbations are generated during an early era of accelerated expansion induced by a slow-rolling phase. In the vanilla inflationary scenario, the primordial tensor spectrum is red-tilted at CMB scales, i.e.~its amplitude decreases with frequency. If inflation is  followed by the long-lasting RD period, then also the produced SGWB, late-time remnant of the primordial tensor spectrum, turns out to be red-tilted and its amplitude at the mHz -- Hz frequencies remains too small to be detected by any present or planned near-future GW detector. However, both deviations from the vanilla inflationary scenario (see Sec.~\ref{sec:Inflation}) and the usual RD epoch of the standard model of cosmology can enhance the SGWB to amplitudes large enough to reach the LISA sensitivity. In the latter case, for instance, if the expansion rate of the universe between the end of inflation and the onset of RD, is dictated by an effective EoS parameter $w \neq 1/3$, the inflationary SGWB is naturally tilted at high frequencies, for the modes that re-entered the Hubble radius during such non-standard epoch. If the EoS is stiff, i.e.~$w > 1/3$, then the tilt is blue, thus enhancing the power of the GW spectrum at large frequencies, and hence making this branch of the spectrum potentially observable with LISA. If, on the contrary, the EoS is $w < 1/3$, then the tilt at high frequencies is red, and hence the spectrum goes below the standard plateau amplitude due to modes crossing during RD.

Any modification of the expansion rate occurring after a FOPT also leaves its imprint on the SGWB produced by that phase transition. The modified redshift results in an overall shift of the entire signal but, more interestingly, characteristic features might also be produced in the signal. This is possible for modes with lengths larger than the horizon size at the time of the transition. If these modes enter the horizon during a time of modified expansion, the typical $f^3$ slope at low frequencies will be modified allowing us to decode the EoS at that time. However, as this feature is always present significantly below the main amplitude peak of the FOPT SGWB (see Sec.~\ref{sec:PTs}), measuring the expansion using this method will always be much more challenging than the initial detection of the spectrum.   

Cosmic strings are another possible GW source which would also allow us to probe the expansion history. As a network of strings would continuously produce GWs throughout most of cosmological history (see Sec.~\ref{sec:CosmicStrings}), also the expansion history is encoded in the resulting spectral shape. This, in fact, includes not only drastic modifications such as periods of non-standard expansion, but also subtler modifications including the number of degrees of freedom contributing to the radiation dominating at early times. As we will see, an accurate observation of a SGWB from cosmic strings will always give us some information on the expansion rate of the very early universe.

Finally, we will also explore the possibility to test  quantum gravity cosmological models via SGWB observations at LISA. After excluding a plethora of models whose SGWB does not reach the instrument's sensitivity curve, we single out two pre-big-bang scenarios as candidates that could generate a signal relevant for LISA science.

%%%%%%%%%%%%%%%%%%%%%%%%%%%%%%%%%%%%%%%%%%%%%%%%%%%%%%%%%%%%%%%%%%%%%%%%%%%%%%%%%%%%%%%%%%%%%%%%%%%%%%%%%%%%%%
\subsection{Non-standard expansion histories}

%%%%%%%%%%%%%%%%%%%%%%%%%%%%%%%%%%%%%%%%%%%%%%%%%%%%%%%%%%%%%%%%%%%%%%%%%%%%%%%%%%%%%%%%%%%%%%%%%%%%%%%%%%%%%%
\subsubsection{Inflation}

During inflation, tensor perturbations are expected to be excited due to quantum fluctuations at sub-Hubble scales. Due to the exponential expansion, the initially small wavelengths of these tensors are stretched to scales exponentially larger than the inflationary Hubble radius. As a result, when inflation ends the universe is filled with a stochastic tensor background with a (quasi-)scale invariant power spectrum at super-horizon scales~\cite{Grishchuk:1974ny,Starobinsky:1979ty, Rubakov:1982df,Fabbri:1983us}. The spectrum is usually parametrised as
\be
\left<h_{ij}(t,\mathbf{x})h^{ij}(t,\mathbf{x})\right>\equiv\int \frac{dk}{k} P_h(t,k) \quad \Longleftrightarrow \quad P_h(k)\simeq \frac{2}{\pi^2}\left(\frac{H_{\text{inf}}}{ M_{\text{Pl}}}\right)^2\left({k\over k_*}\right)^{n_T} %\label{inf}
\,;\hspace{0.5cm}n_T \simeq -2\epsilon \simeq -{r\over8} \,,\label{sec9eq1}
\ee
with $k_{*}$ a pivot scale, $H_{\rm inf}$ is the inflationary Hubble scale when $k_*$ exited the Hubble radius during inflation and $P_h$ is the dimensionless tensor power spectrum (here $M_{\text{Pl}}$ is the reduced Planck mass). In an exact {\it de Sitter} background, the tensor spectrum would be exactly scale invariant, i.e.~with $n_T = 0$. In slow-roll inflation, however, a small (``slow-roll suppressed") red tilt is developed, with an amplitude smaller than (but of the order of amplitude of) the tensor-to-scalar ratio $r$ evaluated at $k_*$. Since the latter is constrained (at $k_*/a_0 = 0.05~{\rm Mpc}^{-1}$) by the B-mode in the CMB as $r \lesssim 0.036$~\cite{BICEP:2021xfz}, we can immediately infer an upper bound on the scale of inflation as $H_{\rm inf} \lesssim H_{\rm max} \simeq 5.1\times 10^{13}\,{\rm GeV}$. This also implies naturally that the red tilt can only be very small, $-n_T \leq 0.005 \ll 1$. 

During the expansion history following inflation, the tensor modes re-enter successively back inside the Hubble radius, turning into a proper classical (yet stochastic) background of GWs. Once the modes become sub-Hubble, they start oscillating with a decaying amplitude $h_{ij} \propto 1/a$ (this is independent of the expansion rate~\cite{Caprini:2018mtu}), propagating as relativistic degrees of freedom. Since modes with different wavelengths re-enter the Hubble radius at different moments of cosmic evolution, different modes may propagate through different periods of expansion. As a result, modes with very different wavelengths may sustain very different amplitudes with respect to each other, depending on the rate of expansion of the universe at their time of Hubble-crossing. To characterise the spectrum of the SGWB today, we can write the energy density spectrum normalised to the critical density $\rho_{\text{crit}}=3M_{\rm Pl}^2H^2$, as~\cite{Caprini:2018mtu}
\be
\begin{aligned} \label{transferfunc}
&\Omega_{\text{GW}}(t, k) \equiv\dfrac{1}{\rho_{\text{crit}}}\dfrac{d\rho_{\text{GW}}(t,k)}{d\ln k} = \frac{k^2}{12a^2(t)H^2(t)}\Delta_h^2(t,k)\,,\\
&\Delta_h^2(t,k)\equiv T_h(t,k) \Delta_{h,\text{inf}}^2(k)\,,~~~ T_h(t,k) \simeq {1\over2}\left[{a_k\over a(t)}\right]^2\,,
\end{aligned}
\ee

where $T_h(t,k)$ is the transfer function characterising the expansion history between the moment $t = t_k$ of horizon re-entry of a given mode $k$, and a later moment $t > t_k$. Here $t_k$ is implicitly defined from the condition $a_kH_k \equiv k$, with $a_k \equiv a(t_k)$, and $H_k\equiv H(t_k)$. The factor $1/2$ is due to the time-average of the rapidly-oscillating sub-horizon wave.

The transfer function today can be evaluated as
\begin{eqnarray}
 T_h(k) \simeq {1\over2}\left({a_k\over a_0}\right)^2 \simeq {1\over2}\mathcal{G}_k\Omega_{\rm rad}^{(0)}\left({a_0H_0\over a_k H_k}\right)^2\,,~~~%\label{eq:Gk}
\mathcal{G}_k \equiv \left(g_{*,k}\over g_{*,0}\right)\left(g_{s,0}\over g_{s,k}\right)^{4/3}\,,
\end{eqnarray}
where $g_s$ and $g_*$ are the relativistic number density of species contributing to the total entropy and energy densities, respectively. Using $\Omega_{\rm rad}^{(0)} \simeq 9\cdot 10^{-5}$, $g_{s,0} \simeq 3.91$, $g_{*,0} = 2$, and $g_{s,k} \simeq g_{*,k} \simeq 106.75$, so that $\mathcal{G}_k \simeq 0.65$, we obtain that tensor modes crossing during the RD epoch lead to an energy density spectrum with amplitude
\be \label{eq:InfGWtodayRD}
 \Omega_{\rm GW}^{(0)}{\Big |}_{\rm inf}  \simeq \mathcal{G}_k { \Omega_{\rm rad}^{(0)}\over12\pi^2}\left(H_{\rm inf}\over m_{\rm Pl}\right)^2\left(H_{\rm inf}\over H_{\rm max}\right)^2\left(f\over f_{\rm Pl}\right)^{n_T}  \simeq 5\times 10^{-16}\left(H_{\rm inf}\over H_{\rm max}\right)^2\left(f\over f_{\rm Pl}\right)^{n_T}\,,
\ee
with $n_T$ given by the original inflationary tilt. In other words, the RD energy spectrum retains the (quasi-) scale invariant spectral shape of the original inflationary tensors. We note that in evaluating Eq.~(\ref{eq:InfGWtodayRD}) we have considered for concreteness that $g_{s,k}, g_{*,k}$ equal the SM degrees of freedom  before electroweak symmetry breaking, and hence they are independent of $k$. In reality the number of SM relativistic degrees of freedom change with the scale, but for simplicity in Eq.~(\ref{eq:InfGWtodayRD}) we considered an identical suppression for all the modes as $\mathcal{G}_k \simeq 0.65$, so that we can provide a single number for the amplitude of this plateau. 

If between the end of inflation and the onset of RD there is a period of expansion characterised by an EoS different than that of radiation $w \neq 1/3$, the inflationary GW energy density spectrum develops a tilt within the range of scales corresponding to the modes crossing the Hubble radius during such period. As a result, the (quasi-)scale invariance of the original tensor spectrum is lost. This feature is actually quite interesting from an observational point of view, as we might detect or constrain in this way the post-inflationary expansion history, and hence the properties of the fields driving the expansion between inflation and RD~\cite{Giovannini:1998bp,Giovannini:1999bh,Boyle:2005se, Watanabe:2006qe, Boyle:2007zx,Kuroyanagi:2008ye,Kuroyanagi:2010mm,Kuroyanagi:2014qza,Figueroa:2018twl,Figueroa:2019paj,Gouttenoire:2021jhk}. 

In order to understand better the above discussion, let us consider that between the end of inflation and the onset of RD, the expansion rate is dictated by an effective EoS parameter $w \neq 1/3$. In scenarios where the inflaton potential is a monomial, $V(\phi) \propto \phi^p$, the inflaton oscillates around the minimum of such a potential after inflation, so that an effective EoS, averaged over inflaton oscillations, emerges as $w \simeq (p-2)/(p+2)$~\cite{Turner:1983he,Lozanov:2016hid}. For $p < 4$ the EoS is in the range $0 < w < 1/3$. For $p = 2$, the oscillations of a massive free field lead to an energy density redshifting on average (over one oscillation cycle) as  $\left\langle\rho_\phi \right\rangle_{\rm osc} \propto 1/a^3$, analogous to that of non-relativistic particle species. Thus, an EoS of $w \simeq 0$ emerges in that case. Nothing prevents however the possibility of a stiff dominated  stage with EoS $w = w_s$ for {\small$1/3 < w_s < 1$}. Such a case can be actually achieved quite naturally if the kinetic energy of the inflaton dominates after inflation, either through inflaton oscillations~\cite{Turner:1983he} under a steep potential (e.g.~$V(\phi) \propto \phi^p$ with $p > 4$), or simply by an abrupt drop of the inflationary potential at at moment that triggers itself the end of inflation.

\begin{figure}[t]
\centering
\includegraphics[width=7cm]{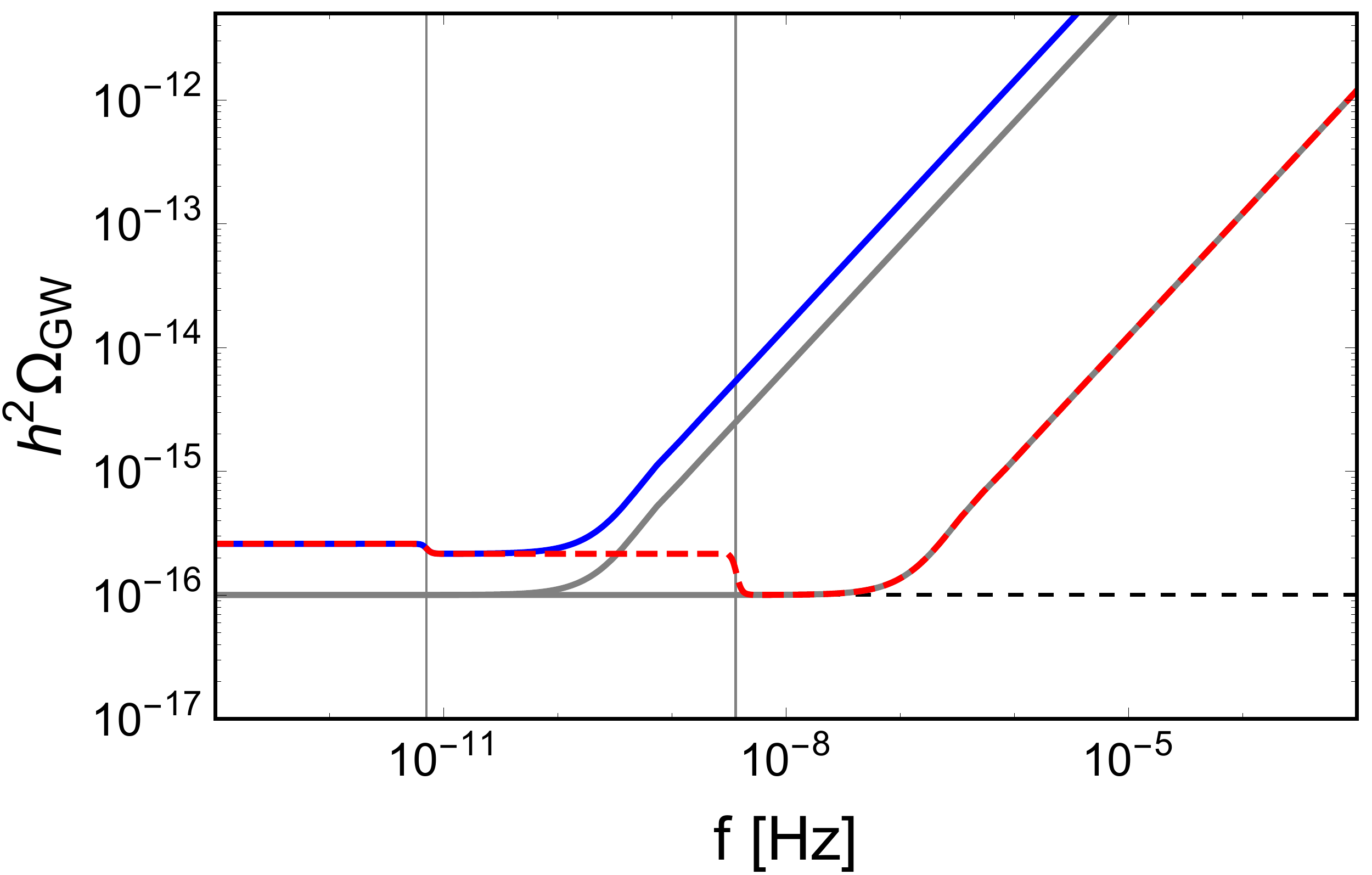}~~~~
\includegraphics[width=7cm]{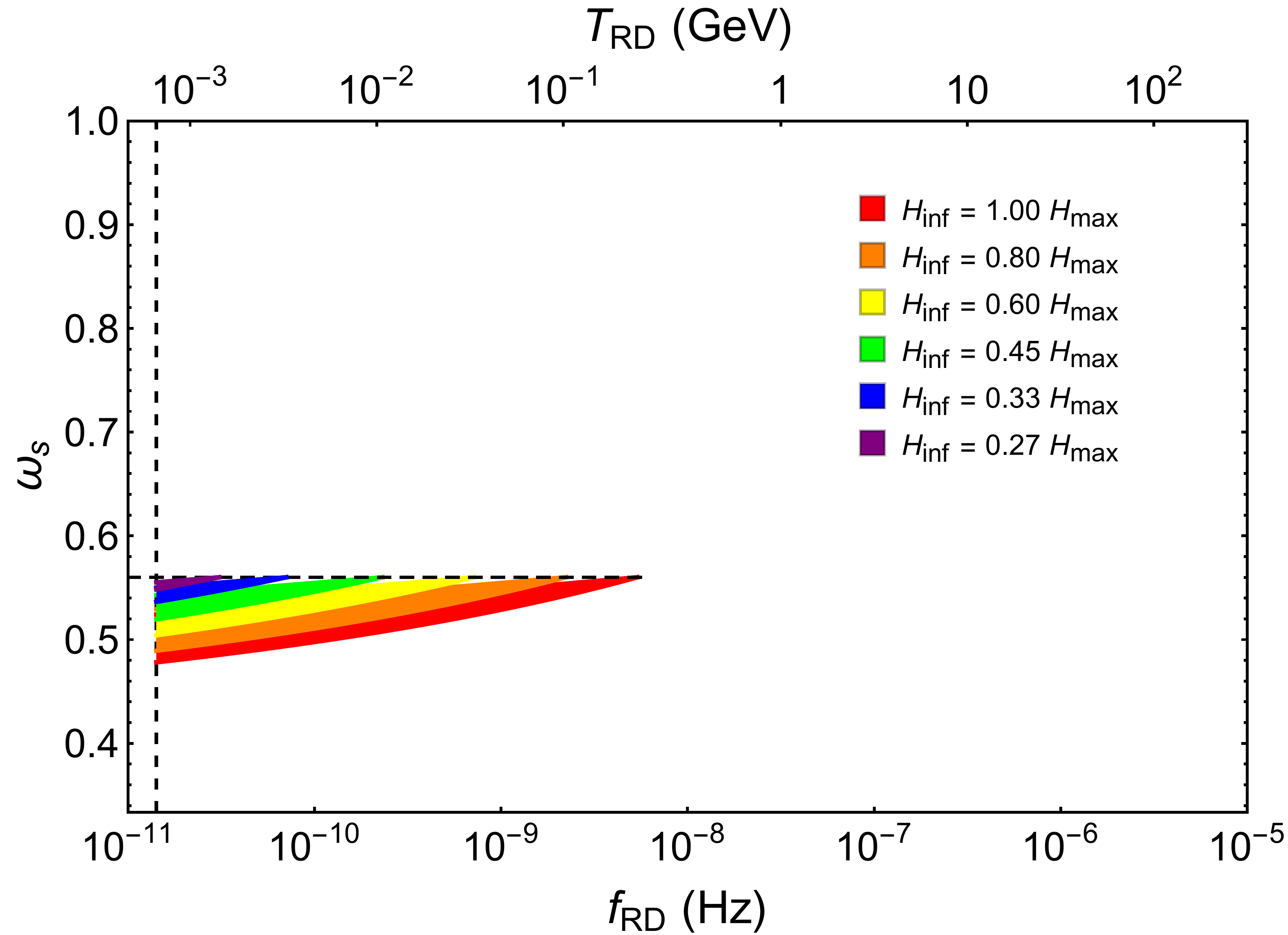}
\caption{\small Left panel: GW energy spectra including changes in the number of relativistic degrees of freedom, depending on whether the  transition from stiff domination  to RD  takes place before (red dashed line) or after (blue solid line) than the QCD phase transition. For comparison we show the corresponding spectra (grey solid line) without correcting for changes in the number of degrees of freedom. Figure taken from Ref.~\cite{Figueroa:2019paj}. Right panel: Parameter space region that LISA can probe after removing the region (above the dashed horizontal line) incompatible with the BBN bound on GW backgrounds. Here $w_s$ refers to the EoS for a {\it stiff} background.}
\label{fig:GWstiffLISA} 
\end{figure}

Propagating the inflationary tensor modes through the epoch starting immediately after the end of inflation, leads to a GW energy density spectrum today, expressed as a function of present-day frequencies $f=k/(2\pi a_0)$, like~\cite{Figueroa:2019paj}
\begin{eqnarray}\label{eq:GWfullSpectrumInstant}
\Omega_{\text{GW}}^{(0)}(f) = \Omega_{\rm GW}^{(0)}{\Big |}_{\rm inf} \times \mathcal{W}(f/f_{\rm RD}) \times \mathcal{C}_{w}\,\left({f\over f_{\rm RD}}\right)^{2\left({3w-1\over 3w+1}\right)} \simeq \Omega_{\rm GW}^{(0)}{\Big |}_{\rm inf}
\times\left\lbrace
\begin{array}{crl}
1 & \hspace*{-0.3cm}, & f \ll f_{\rm RD} \vspace*{0.3cm}\\
\mathcal{C}_{w}\,\left({f\over f_{\rm RD}}\right)^{2\left({3w-1\over 3w+1}\right)} & \hspace*{-0.3cm}, & f \gg f_{\rm RD} \\
\end{array}
\right.
\end{eqnarray}
with $\mathcal{C}_{w}$ a numerical prefactor in the range $1 < \mathcal{C}_{w} < 2^{5/2}/\pi \simeq 1.80$ for $1/3 < w < 1$, and where $f_{\rm RD} \equiv k_{\rm RD}/(2\pi a_0)$ is the frequency corresponding to the horizon scale at the onset of RD, $k_{\rm RD} = a_{\rm RD}H_{\rm RD}$, and $\mathcal{W}(x)$ is a window function connecting the two sides of the spectrum: the high-frequency branch on the one hand, corresponding to modes crossing during the non-standard period ($f \gg f_{\rm RD}$), and the low frequency branch corresponding to modes crossing during RD ($f \ll f_{\rm RD}$). 
The tilt in Eq.~(\ref{eq:GWfullSpectrumInstant}) indicates that models with a stiff epoch between the end of inflation and the onset of RD are actually very appealing observationally, as in these scenarios a blue tilt is developed in the GW energy density spectrum at large frequencies, with $0 < n_T < 1$ ~\cite{Giovannini:1998bp,Giovannini:1999bh,Riazuelo:2000fc,Sahni:2001qp,Tashiro:2003qp,Boyle:2007zx,Giovannini:2008zg,Giovannini:2008tm,Caprini:2018mtu,Bernal:2019lpc,Figueroa:2019paj}. This potentially opens up the possibility of detection by upcoming direct detection GW experiments, including LISA. The presence of a stiff period is actually well-motivated theoretically in scenarios like Quintessential inflation~\cite{Peebles:1998qn,Peloso:1999dm,Huey:2001ae,Majumdar:2001mm,Dimopoulos:2001ix,Wetterich:2013jsa,Wetterich:2014gaa,Hossain:2014xha,Rubio:2017gty}, gravitational reheating~\cite{Ford:1986sy,Spokoiny:1993kt} (with the caveats explained in Ref.~\cite{Figueroa:2018twl}), the Higgs-reheating scenario~\cite{Figueroa:2016dsc} and  generalisations~\cite{Dimopoulos:2018wfg,Opferkuch:2019zbd}. The GW spectrum in these scenarios is controlled by $w$, $f_{\rm RD}$, and $H_{\rm inf}$, and the parameter space compatible with a detection by various experiments has been recently analysed in Refs.~\cite{Bernal:2019lpc,Figueroa:2019paj}. In Fig.~\ref{fig:GWstiffLISA} we can see the form of the spectrum in the left panel (including scale-dependent changes in $\mathcal{G}_k$ due to changes in the number of relativistic species). In the right panel of the same figure we plot the parameter space that LISA can probe.\footnote{A short kination era disconnected from the inflation sector may also occur inside the radiation era due to e.g.~axion dynamics. This would largely impact the inflationary SGWB as well as other primordial backgrounds, with striking enhanced prospects at LISA \cite{
Co:2021lkc, Gouttenoire:2021jhk, Gouttenoire:2021wzu}.}

Consistency with upper bounds on SGWBs, like the BBN and CMB constraints, rules out a significant fraction of the observable parameter space. (See discussion at the end of Sec.~\ref{subsec:GWprobeIntro}.) This renders for instance this signal as unobservable by Advanced LIGO, independently of the parameter space $\lbrace w, f_{\rm RD}, H_{\rm inf}\rbrace$~\cite{Figueroa:2019paj}. The GW background remains detectable in LISA, but only in a small island of parameter space, see right panel of Fig.~\ref{fig:GWstiffLISA} corresponding contrived scenarios having a low EoS $0.46 \lesssim w \lesssim 0.56$, a high inflationary scale $H_{\rm inf} \gtrsim 10^{13}$ GeV at the same, and a low transition frequency $10^{-11}~{\rm Hz} \lesssim f_{\rm RD} \lesssim 3.6\times10^{-9}~{\rm Hz}$ (or equivalently a low reheating temperature $1~{\rm MeV} \lesssim T_{\rm RD} \lesssim 150$ MeV).

If there is an intermediate phase with EoS in the range $0 \leq w < 1/3$, the transfer function of the inflationary GW spectrum in Eq.~(\ref{eq:GWfullSpectrumInstant}), develops instead a red-tilted high-frequency branch, corresponding to the modes propagating through that phase. If we could detect the (quasi-)scale invariant GW plateau part of the spectrum, the transition to the high-frequency branch due to the non-standard era, could be used to determine the reheating temperature $T_{\rm RD}$ of the universe. This is because the end of the non-standard era also corresponds to the onset of RD, and hence the pivot frequency $f_{\rm RD}$ separating the two branches in the energy density spectrum informs directly about the energy scale at the onset of RD~\cite{Nakayama:2008wy,Kuroyanagi:2008ye, Kuroyanagi:2010mm,DEramo:2019tit}. Unfortunately, given the suppression of the amplitude of the plateau, c.f.~Eq.~(\ref{eq:InfGWtodayRD}), the determination of $T_{\rm RD}$ by this method cannot be probed with LISA, and only a futuristic experiment like BBO or DECIGO~\cite{Crowder:2005nr,Seto:2001qf,Sato:2017dkf} might have a chance. 

Another interesting opportunity is to look for a signal of particles beyond the SM. When particles becomes non-relativistic as the temperature drops below the mass and decouple from thermal equilibrium, they no longer contribute to the radiation energy density. This can be seen as a change in the values of $g_{*,k}$ and $g_{s,k}$, resulting in a temporary speed-up of the Hubble expansion rate. SM particles induce changes in $g_{*,k}$ and $g_{s,k}$ in the frequency range of $\sim 10^{-12}$ to $10^{-5}$~Hz \cite{Watanabe:2006qe}, which is beyond the range of LISA sensitivity curve. However, any change due to BSM particles with mass near $100$~TeV leave an imprint on the SGWB spectrum at the LISA sensitivity. This effect is not detectable in the case of standard slow-roll inflation, since the amplitude of the SGWB is far below the LISA sensitivity as in Eq.~(\ref{eq:InfGWtodayRD}), while a strongly blue-tilted primordial spectrum would provide an opportunity to test BSM physics by LISA \cite{Caldwell:2018giq}.

%%%%%%%%%%%%%%%%%%%%%%%%%%%%%%%%%%%%%%%%%%%%%%%%%%%%%%%%%%%%%%%%%%%%%%%%%%%%%%%%%%%%%%%%%%%%%%%%%%%%%%%%%%%%%%
\subsubsection{Phase transitions}
We discussed the production of GWs in FOPTs in detail in Sec.~\ref{sec:PTs}. We will now discuss the impact a period of non-standard evolution would have on these backgrounds. The first most obvious change will come in through modified red-shifting of the spectra from the time of their production until today~\cite{Allahverdi:2020bys, Gouttenoire:2021jhk}. Simply parametrising the non standard evolution through a barotropic parameter $w$ we find for the abundance
\begin{equation}
\begin{aligned} \label{eq:PTOmegaredshift}
\Omega_{{\rm GW},0} &= \left(\frac{a_*}{a_0}\right)^4  \left(\frac{H_*}{H_0}\right)^2 \Omega_{{\rm GW},*}
=  1.67\times 10^{-5} h^{-2} 
\left[\frac{100}{g_s(T_{\rm RD})} \right]^\frac{4}{3}
\left[\frac{g_*(T_{\rm RD})}{100} \right] 
\left(\frac{H_*}{H_{\rm RD}}\right)^{2\frac{3w-1}{3w+3}} \Omega_{{\rm GW},*} \,, 
\end{aligned}
\end{equation}
and for the frequency
\begin{equation}
\begin{aligned} \label{eq:PTfredshift}
f_0 & = \frac{a_*}{a_0} f_* =\frac{a_{\rm RD} H_{\rm RD}}{a_0} \frac{a_* H_*}{a_{\rm RD} H_{\rm RD}} \frac{f_*}{ H_*} \nonumber\\
&= 1.65\times 10^{-5} \,{\rm Hz}\, \left( \frac{T_{\rm RD}}{100\,{\rm GeV}} \right) 
\left(\frac{100}{g_s(T_{\rm RD})} \right)^\frac{1}{3}
\left(\frac{g_*(T_{\rm RD})}{100} \right)^\frac{1}{2} \left(\frac{f_*}{H_*}\right)\left(\frac{H_*}{H_{\rm RD}}\right)^{\frac{3w+1}{3w+3}} \, ,
\end{aligned}
\end{equation}
where the asterisk ($*$) denotes quantities at the time of the transition, while ``RD" refers to the moment when the universe becomes radiation dominated. In a decelerating universe $H_*/H_{\rm RD}>1$ and the direction of these modifications depends on the barotropic parameter. In radiation domination, $w=1/3$, the amplitude does not change as GWs redshift at the same rate as the background. For an expansion dominated by energy density redshifting slower than radiation $w<1/3$ the amplitude will keep decreasing while in the opposite case of an extra component redshifting faster $w>1/3$ the abundance would increase looking at only the effect of redshift. 

Another crucial modification comes from the conditions at the time of the transition. While typically the GW production is not crucially modified~\cite{Barenboim:2016mjm,Guo:2020grp}, the transition is linked to the radiation component and if that is subdominant the GW abundance also  needs to be corrected. In terms of the strength of the transition this leads to a simple modification by 
\begin{equation}\label{eq:PTWGabundanediminish}
\alpha\propto \rho_R^{-1}  \rightarrow (\rho_{\rm tot}-\rho_V)^{-1} \;, 
\end{equation}
where $\rho_R$ is radiation density while $\rho_V$ is the difference between the initial and final state energy densities. Finally $\rho_{\rm tot}$ is the total energy density which in domination of any additional component is of course the dominant contribution.

This modification makes observation of scenarios where the transition occurs during the domination of an extra component rather unlikely. Even in the optimistic case of kination (with $w=1$) where the redshift of the signal would increase the amplitude from Eq.~\eqref{eq:PTWGabundanediminish} we get $(\rho_V/\rho_{\rm tot})^2\approx (H_*/H_{\rm RD})^{-4\frac{3w-1}{3w+3}}$ which results in an overall suppression. Due to this, to get any hope of observation we would need the modified expansion to end almost immediately after the transition~\cite{Barenboim:2016mjm} despite the fact that the two are governed by completely separate mechanisms. 
Thus, the most promising models from the observational point of view are the ones in which the transition occurs in standard circumstance, but it is strong enough to itself modify the expansion history, as in this case Eq.~\eqref{eq:PTWGabundanediminish} does not apply. One example here would be a very strong transition in which the field dominates the total energy density and after the transition as the field oscillates around its minimum it is causing a matter dominated period which will last until the field decays~\cite{Ellis:2020nnr}.

All the modifications of the spectrum we discussed up to now, essentially come down to shifts of the entire spectrum that could also be mimicked by simple changes in the parameters of the transition. We now switch to a modification changing the spectral shape which could provide a smoking gun signal for a modified expansion period. 

\begin{figure}[t]
\centering
\includegraphics[width=0.635\textwidth]{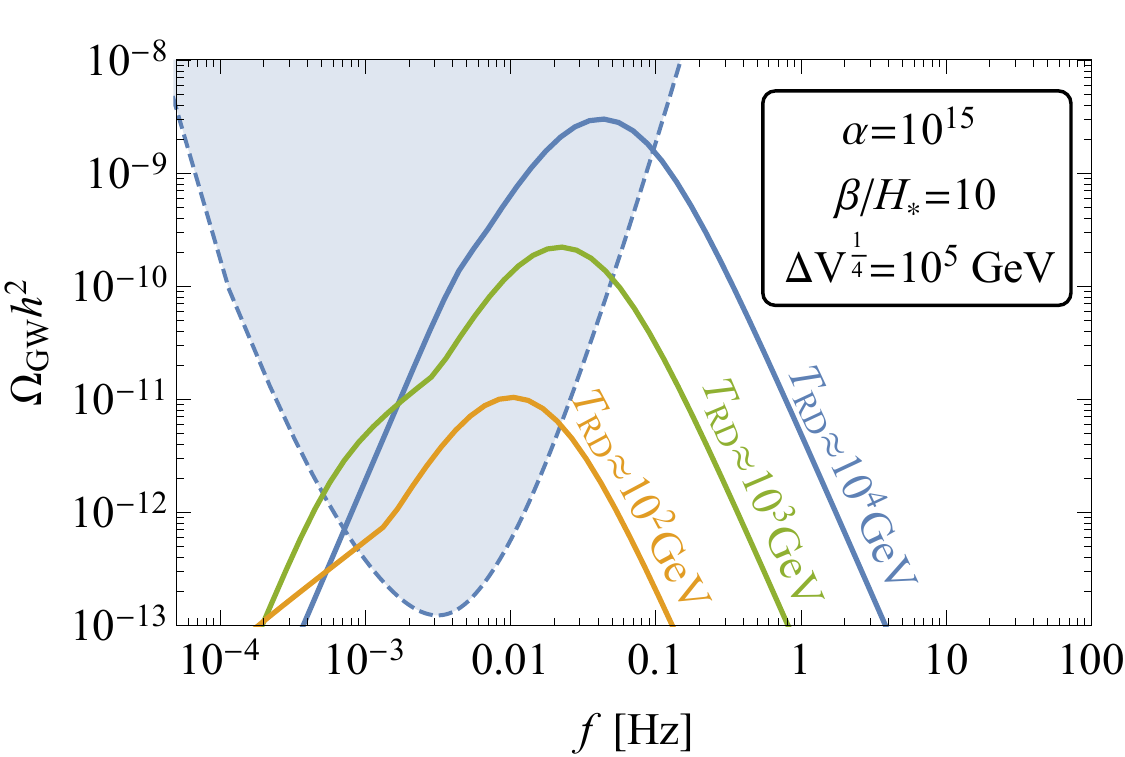} 
\caption{\small \label{Fig:CosmicStringTemperatureReach}The lines show the GW spectrum produced in FOPT by bubble collisions for the indicated transition parameters. From top to bottom, the decay width of the scalar field decreases leading to a lengthening period of effectively matter-dominated expansion as the field oscillates around its minimum before finally decaying. The resulting reheating temperatures read $T_{\rm RD }=10^4$ GeV, $T_{\rm RD}=10^3$ GeV and $T_{\rm RD}=10^2$ GeV, respectively.}
\end{figure}

This modification has to do with the fact that at scales larger than the horizon size at the time of the transition the features of our source become irrelevant. The source at this large scale is effectively just white noise and GWs entering the horizon at later time during RD always predict a spectrum proportional to $f^3$~\cite{Caprini:2009fx,Unal:2018yaa,Cai:2019cdl} at frequencies corresponding to superhorizon scales $f<f_{H_*}=a_* H(a_*)/2\pi$. This behaviour, however, depends on the expansion rate at the time when GWs enter the horizon~\cite{Hook:2020phx}. In our relevant example of matter domination, it would create an $f^1$ plateau in the spectrum for $f_{H_{\rm RD}}<f<f_{H_*}$. The main issue with observation of this feature comes from the fact that the peak of the signal has to do with the characteristic scale of the transition, which is typically much smaller than horizon size such that $f_*/f_{H_*}\propto \beta/H$, which is typically much bigger than one. As a result, this feature will typically appear significantly below the peak and, given that the abundance decreases quickly off peak, it will also have a much smaller abundance. Concerning LISA scales, in Fig.~\ref{Fig:CosmicStringTemperatureReach}, we show several examples of this modification on bubble collision spectra calculated according to Ref.~\cite{Lewicki:2020azd} using parameter examples not difficult to realise in classically conformal models~\cite{Ellis:2020nnr}. From top to bottom the lines show the decay width of the scalar field $\Gamma_\phi/H=1$, $\Gamma_\phi/H=10^{-2}$  and $\Gamma_\phi/H=10^{-4}$ which results in longer matter-dominated periods and reheating temperatures of $T_{\rm RD}=10^4$ GeV, $T_{\rm RD}=10^3$ GeV and $T_{\rm RD}=10^2$ GeV, respectively. We also stress that departure from standard cosmology that we have just explained, is not the only option leading to relevant chances of the SGWB frequency profile. For instance, the occurrence of  an intermediate matter and kination era  inside the radiation epoch as motivated by e.g.~early axion dynamics also impacts the  low-frequency slope of the SGWB signal~\cite{Gouttenoire:2021wzu}.

%%%%%%%%%%%%%%%%%%%%%%%%%%%%%%%%%%%%%%%%%%%%%%%%%%%%%%%%%%%%%%%%%%%%%%%%%%%%%%%%%%%%%%%%%%%%%%%%%%%%%%%%%%%%%%
\subsubsection{Cosmic strings} \label{subsec:nonStandardCosmicStrings}
We now turn to the case of GW spectra produced by cosmic strings that we first discussed in Sec.~\ref{sec:CosmicStrings}. We focus on analytical modelling described in Sec.~\ref{sec:model1} with numerical factors set to agree with Model 2 described in Sec.~\ref{sec:model2}.

\begin{figure}[t]
\centering
\includegraphics[width=0.735\textwidth]{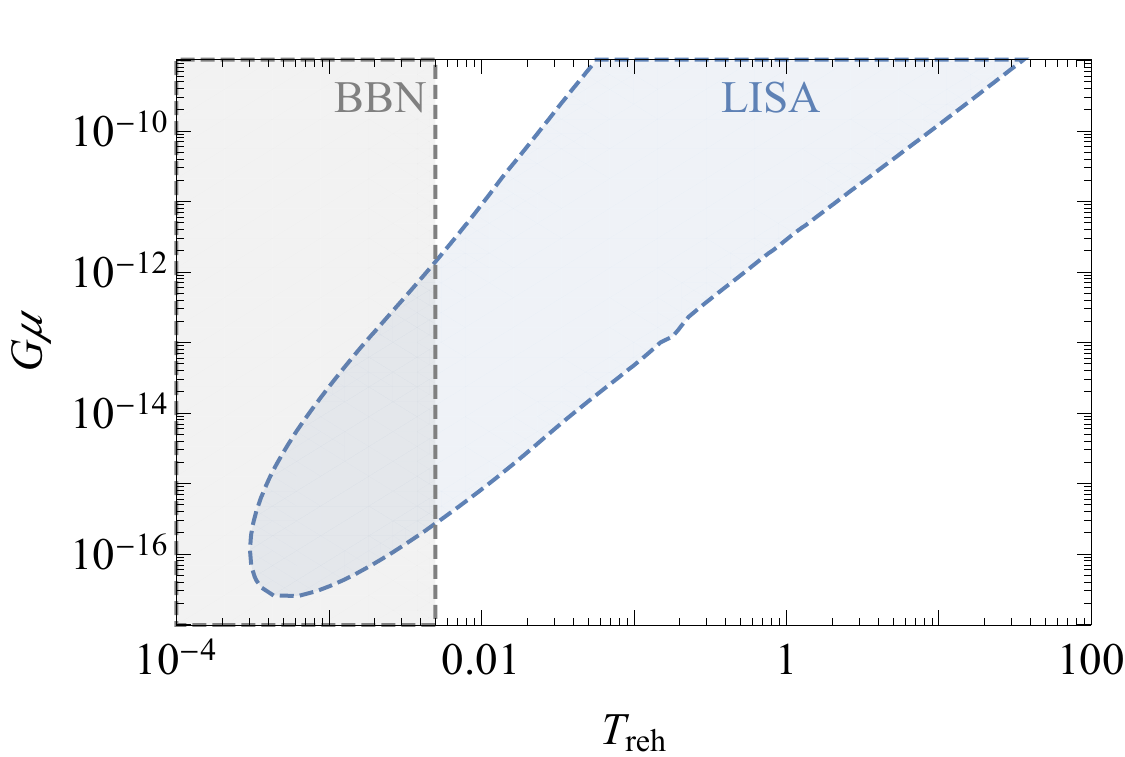} 
\caption{\small \label{Fig:CosmicStringTemperatureReach2} Range of temperature $T_{\rm RD}$ to which LISA can probe the cosmological expansion rate using a spectrum from a cosmic string network with the indicated string tension $G\mu$. The grey region indicates temperatures where modifications of the expansion rate would already be in tension with BBN.}
\end{figure}

These spectra are very convenient laboratories to study the expansion history of the early universe, since in RD they simply produce a flat plateau. Any features beyond that can be linked to modifications of the expansion rate from minor modifications caused by variations in the number of degrees of freedom to simply a different power-law caused by domination of energy density red-shifting differently than radiation~\cite{Cui:2017ufi,Cui:2018rwi,Gouttenoire:2019kij,Gouttenoire:2019rtn}. The feature will appear in the spectrum above at the characteristic frequency $f_{\rm RD}$ corresponding to the temperature $T_{\rm RD}$ at which the expansion begins to follow the standard radiation-dominated picture (with SM number of degrees of freedom)~\cite{Cui:2018rwi}:
\begin{equation} \label{eqn:fdeltaforlargealpha}
f_{\rm RD}=
  (8.67\times 10^{-3} \, {\rm Hz})\,
\left(\frac{T_{\rm RD}}{\rm GeV} \right)
\left(\frac{ 10^{-11}}{G\mu}\right)^{1/2}
  \left[\frac{g_*^{\rm SM}(T_{\rm RD})}{g_*^{\rm SM}(T_0)}\right]^\frac{8}{6} \left[\frac{g_s^{\rm SM}(T_0)}{g_s^{\rm SM}(T_{\rm RD})}\right]^{\frac{7}{6}}\, ,
\end{equation}
where $G\mu$ is the tension characterising the string network. We stress that, for local strings, the relevant energy scale is not the size of the horizon, in contrast with phase transitions for instance. This is because loops do not suddenly decay after production, in contrast with other cosmological sources of GW. So, for LISA, the cosmic-strings GW spectrum is sensitive to MeV scale, rather than the 100 GeV$-$TeV scale \cite{Gouttenoire:2019kij}. In Fig.~\ref{Fig:CosmicStringTemperatureReach2}, we show the part of the $(G\mu,T_{\rm RD})$ parameter space in which LISA can probe the spectra at $f_{\rm RD}$ testing the standard cosmological expansion rate.

For global strings, the turning-point frequency turns out to be~\cite{Gouttenoire:2019kij}
\begin{equation}
\label{eqn:globalstrings}
f_{\rm RD}=\left(\frac{T_{\rm RD}}{\rm GeV} \right)\left(\frac{0.1}{\alpha} \right) \left[\frac{g_*^{\rm SM}(T_{\rm RD})}{g_*^{\rm SM}(T_0)}\right]^\frac{1}{4}\times  \begin{cases}
     8.9\times 10^{-7}\, {\rm Hz} \quad \hbox{for} \ \ 10\,\%, \\
   7.0\times 10^{-8}\, {\rm Hz} \quad \hbox{for} \ \ 1\,\%,
    \end{cases}
\end{equation}
where $\alpha$ is the loop size with respect to the Hubble horizon and the percentage indicates the relative deviation $|(\Omega_{\rm st}-\Omega_{\rm nonst})/\Omega_{\rm st}|$ of the non-standard spectrum with respect to the standard one. 
Thus, in contrast to local strings,the turning-point frequency is independent of the string tension  (cf.~ Eqs.~(\ref{eqn:fdeltaforlargealpha}) and (\ref{eqn:globalstrings})).

\begin{figure}[t]
\centering
\includegraphics[width=0.735\textwidth]{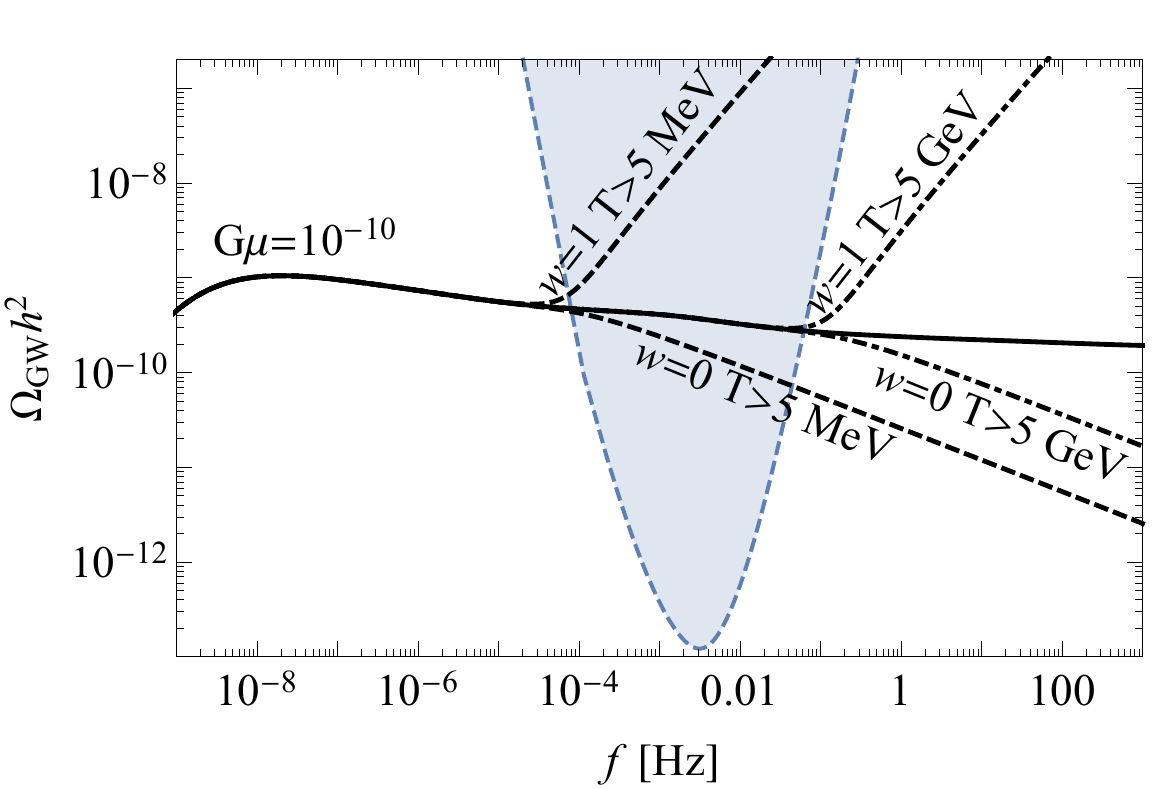} 
\caption{\small \label{Fig: CosmicStringModCosmoPlot} GW spectrum produced by a cosmic string network with $G\mu=10^{-10}$ together with spectra produced if that network evolved in an early matter domination period (with $w=0$) or kination (with $w=1$) lasting up until $T_{\rm RD}=5$ MeV and $T_{\rm RD}=5$ GeV. We note that the blue tilted branches in the figure are only shown for representative purposes, as they violate BBN and CMB constraints [recall discussion at the end of Sec.~\ref{subsec:GWprobeIntro}], so for a realistic effect compatible with those bounds, a much smaller stiff EoS is required, somehow larger than (but very close to) $w \approx 1/3$.}
\end{figure}

Coming back to the local-string case, assuming that at earlier times the expansion is governed by a simple EoS with $\rho\propto a^{3(1+w)}$, we can approximate the slope of the spectrum at $f>f_{\rm RD}$ as
\begin{equation}
    \Omega_{\rm GW}(f>f_{\rm RD})\propto
    \begin{cases}
     f^{2\frac{3w-1}{3w+1}} \quad \hbox{for} \ \ w \geq 5/21, \\
   f^{-1/3} \quad \ \ \hbox{for} \ \  w < 5/21.
    \end{cases}
\end{equation}
We show examples of modified spectra for spectrum from a cosmic string network $G\mu=10^{-10}$ including early matter domination (with $w=0$) and kination (with $w=1$) lasting up until $T_{\rm RD}=5$ MeV and $T_{\rm RD}=5$ GeV in Fig.~\ref{Fig: CosmicStringModCosmoPlot}. 
A change in the number of degrees of freedom would instead create a smooth step in the spectrum with the total change in abundance given by \cite{Cui:2018rwi},
\begin{equation}
\Omega_{\rm GW}(f\gg f_{\rm RD})\simeq \Omega_{\rm GW}^{\rm SM}(f)   \left( \frac{g_*^{\rm SM}}{g_*^{\rm SM} + \Delta g_*} \right)^{\frac13},
\end{equation}
where the SM index denotes quantities computed assuming the number of degrees of freedom as in the SM while $\Delta g_*$ is the number of new degrees of freedom decoupling at $T_{\rm RD}$. According to this simple formula probing the abundance of the plateau with $\mathcal{O}(1\%)$ accuracy would allow us to discover inclusion of even several new degrees of freedom with larger amounts requiring less accuracy. The range of temperatures we can survey is again given by Fig.~\ref{Fig:CosmicStringTemperatureReach2} although ascertaining the exact accuracy concerning our possible estimation of the number of degrees of freedom in this range would require further scrutiny. It was also shown that measurements of the turning-point frequency in the SGWB from cosmic strings due to a temporary matter era induced by a massive unstable particle can enable to probe unstable particles with lifetime in the range $10^{-8}$  to  0.1 second, extending significantly the well-known BBN bound \cite{Gouttenoire:2019rtn}.

In addition to the potential of probing a new EoS and new relativistic degrees of freedom with a SGWB from cosmic strings, (pre-)inflationary dynamics may leave distinctive imprints in GW signals. 
Although a cosmic string network formed during or before inflation is diluted by inflation, it may come back into horizon at later times and lead to detectable GW signals. In this case, the SGWB typically gets suppressed especially at higher frequencies~\cite{Guedes:2018afo}. Nevertheless, well-motivated BSM scenario of an intermediate short inflationary period inside the radiation era can lead to detectable effects. With LISA,  intermediate inflation scales in the range  $0.1- 10^{12}$ GeV can be probed \cite{Gouttenoire:2019kij}. Besides, a GW burst signal could be significant and be the leading signal~\cite{Cui:2019kkd}, in contrast to the conventional cases where the SGWB is the more sensitive means to probe cosmic strings. The dedicated study of the correlated burst and SGWB signals could reveal information about when the strings come back into the horizon as well as related inflationary dynamics.

Non-standard pre-BBN cosmology can significantly distort the vanilla shape of the SGWB from cosmic strings. It can turn it into a peaked shape \cite{Gouttenoire:2019kij}. Note that a peaked shape can also be generated in standard cosmology in the presence of a particle production cut-off. However, this works only for small string tension and lies outside of LISA. So, a peaked GW spectrum from comic strings is a smoking-gun signature of a non-standard era (less stiff than radiation, i.e.~matter or inflationary era). Note that this peak is much broader than the peak from the FOPT SGWB signal and can be distinguished.

\subsection{SGWB in quantum gravity}\label{sec:sgwbqg}

In Sec.~\ref{sec:models} we saw some modifications to the propagation of GWs of astrophysical origin that could happen in quantum gravity. Here we will concentrate on the potential quantum-gravity effects on primordial SGWBs. (See Ref.~\cite{Addazi:2021xuf} for an overview of the experimental and theoretical constraints on quantum gravity.)

Early-universe models embedded in or inspired by theories of quantum gravity have the potential to leave an observable imprint in the SGWB. In fact, these cosmological scenarios usually predict modifications in the shape of the primordial scalar and tensor spectra, which are generated either from the quantum fluctuations of an inflationary field or by an alternative mechanism. If these modifications include a blue tilt at high frequencies, they could overcome the CMB bounds and still give rise to a primordial SGWB $\Omega_{\rm GW}(f)$ reaching the sensitivity curves of the present or future interferometers.

We can classify early universe models related to quantum gravity according to the tilt of the tensor spectrum, which we call ${\cal P}_{\rm t}(f)$ to distinguish the full spectrum from the exact power-law parametrization $\Delta_{h,{\rm inf}}^2$ in Eq.~(\ref{sec9eq1}). 

This classification is based on the overall trend of the full spectrum. 
We should note that in some models a positive tensor index $n_T>0$ at CMB scales does
not imply a blue tilt at higher frequencies. The models are the following (see Ref.~\cite{Calcagni:2020tvw} for a full list of references):
\begin{itemize}
    \item \emph{Red-tilted tensor spectrum}: the large class of string-cosmology models with flux compactification (not to be confused with the cosmic-strings models discussed above); one semi-classical solution in Wheeler--DeWitt canonical quantum cosmology;  loop quantum cosmology, where gravity is quantized canonically using the Ashtekar--Barbero connection. There are three main approaches to describe perturbations in loop quantum cosmology: with a dressed metric, in hybrid quantization and with effective constraints, the latter in turn being characterised by the presence of inverse-volume corrections, holonomy corrections or both. The first two and the third in the presence of inverse-volume corrections predict a red-tilted tensor spectrum.
    \item \emph{Blue-tilted tensor spectrum}: another semi-classical solution in Wheeler--DeWitt canonical quantum cosmology; loop quantum cosmology in the effective-constraints approach in the presence of holonomy corrections;  non-local Starobinsky inflation, a model embedded in non-local quantum gravity where early-universe acceleration is driven by curvature;  string-gas cosmology (again, unrelated to cosmic-strings models), where primordial spectra are generated by a thermal bath of strings most of which are wrapped around compact extra dimensions;  new ekpyrotic scenario, where the spectra are generated by the collision of branes;  Brandenberger--Ho non-commutative inflation, an inflationary stage realised in a geometry which is fuzzy at microscopic scales;  multi-fractional inflation, an inflationary era realised in a geometry whose dimension changes with the probed scale; pre-big-bang scenario, where the dualities of string theory suggest the existence of a phase prior to the big bang.
\end{itemize}
Models with red-tilted tensor spectrum do not generate a observable SGWB detectable at ground- or spaced-based interferometers. Regarding models with a blue-tilted spectrum, quantum modifications in Wheeler--DeWitt canonical cosmology are too small to be observable by any of the present or planned interferometers \cite{Calcagni:2020tvw}. Loop quantum cosmology in the effective-constraints approach in the presence of holonomy corrections is already ruled out observationally \cite{Bolliet:2015raa}. The tensor spectrum of non-local Starobinsky inflation (see Sec.~\ref{AM}) tends asymptotically to the spectrum of local Starobinsky inflation, hence it is unobservable even if the amplitude at interferometer scales is larger than at CMB scales \cite{Calcagni:2020tvw}. String-gas cosmology, the new ekpyrotic scenario, Brandenberger--Ho non-commutative inflation and multi-fractional inflation can all reach the sensitivity of DECIGO but, unfortunately, not of LISA \cite{Calcagni:2020tvw}.

A model worth further investigation is the \emph{pre-big-bang scenario} \cite{Gasperini:1992em}, where CMB constraints are respected and, at the same time, the amplitude of the SGWB at high frequencies can increase to touch the sensitivity curves of GW experiments, possibly including LISA \cite{Gasperini:2016gre}. In Ref.~\cite{Gasperini:2016gre}, the sensitivity threshold of a six-link, 5M-km-arm-length configuration was assumed \cite{Caprini:2015zlo}. An updated analysis could determine the parameter space of the model opening an observational window.

While the above model has a singular hypersurface at which perturbations are matched, another non-standard pre-BBN expansion history is realised by bouncing models where the universe slowly contracts \cite{Ben-Dayan:2016iks,Ben-Dayan:2018ksd,Artymowski:2020pci}, and then expands without a big-bang singularity or inflation (e.g.~\cite{Battefeld:2014uga,Lehners:2010fy} for reviews). In conformal time, before the bounce the scale factor goes as $a(\eta) \sim (-\eta)^b$, where $0<b \ll 1$ for slow contraction. Considering vacuum fluctuations on top of this background metric and using the standard power-law parametrization $P_h \sim f^{n_T}$, the expected SGWB spectral tilt from vacuum fluctuations is $n_{T,{\rm vacuum}}\simeq 3-|2b-1|\simeq 2+2b$, resulting in a blue spectrum \cite{Starobinsky:1979ty, Boyle:2003km, Artymowski:2020pci} possibly slashing through the LISA sensitivity band. Hence, the SGWB signal is directly probing the geometry of the universe $a(\eta)$ making LISA the natural arena to test alternatives to inflation. As is well known, bouncing models with only one scalar field typically predict vacuum fluctuations generating spectra with $n_{T,{\rm vacuum}}=n_{s,{\rm vacuum}}-1\sim 2$. One still has to make sure that the slow-contracting bouncing models actually generate a viable scalar spectrum in accord with CMB observations, for instance by a curvaton mechanism or some other alternative. Specifically, coupling to gauge fields has been considered with an interaction term $\mathcal{L} \propto I^2(\phi)(F^2-\gamma F \tilde F)$, where $I$ is a function of the scalar. As seen in Sec.~\ref{inflation-chirality}, gauge fields can provide an additional source of chiral GWs, whose tilt we denote by $n_{T,{\rm sourced}}$ and is such that $n_{T,{\rm sourced}}=n_{s,{\rm  sourced}}-1\approx-0.04$ in accord with CMB observations, however $r\sim 1/9$ which is above current bounds. Hence, the scalar spectrum needs to be generated by some curvaton or entropic mechanism from a second scalar field. Contrary to the inflationary case in Sec.~\ref{a-gf}, the spectral tilt $n_{T,{\rm sourced}}$ is constant in the bouncing case \cite{Ben-Dayan:2016iks,Ben-Dayan:2018ksd}. Once CMB observations are determined by a curvaton mechanism, then depending on the parameters of the model the sourced tensor spectrum may be detected by LISA with $0.15 < n_{T,{\rm sourced}} < 0.31$ or $0.85 < n_{T,{\rm sourced}}<1.1$. If one wishes to match CMB observations solely with the sourced spectrum, then one open question is carrying out the calculation of the spectra across the bounce and check that the model consistently generates sourced spectra observed at CMB scales and vacuum spectra observed by LISA. This is the opposite to what happens in the inflationary models discussed in Sec.~\ref{sec-GWaddf}, where the tensor spectrum observed on CMB scales must be the vacuum one while LISA can only observe the sourced spectrum.

Another model worth exploring in the LISA context might be the non-perturbative gravity and bouncing universe of Refs.~\cite{Biswas:2005qr,Biswas:2010zk,Biswas:2012bp}, where it is possible to explore a pre-big-bang phase by modelling it by a string-gas dominated Hagedorn phase~\cite{Biswas:2006bs,Biswas:2014kva}. In this non-perturbative extension of Einstein's gravity, there exists a bouncing phase given by the scale factor $a(t)=\cosh{\lambda t}$ in the presence of radiative matter and a non-zero cosmological constant. Such a solution would also permit a stiff fluid in a Hagedorn phase where the universe is primarily dominated by string winding modes \cite{Biswas:2006bs,Biswas:2014kva}. Their eventual decay into radiation, and exit from the bounce would yield the standard big-bang cosmology. Primordial perturbations are created during the thermal phase with thermal statistical initial conditions which lead, specifically, to a blue tilt in the tensor power spectrum, while keeping the amplitude of the matter spectrum within the \textsc{Planck} observation. The blue tilt in the power spectrum, along with the matter power spectrum, constrains the scale entering in the gravitational sector to be around $10^{-4}M_{\rm Pl}$~\cite{Biswas:2014kva}.

In parallel with the observability of these and any other model related to quantum gravity with a blue-tilted SGWB (e.g.~Ref.~\cite{Dapor:2020jvc}), it will be important to assess their theoretical robustness, which we have not discussed here. In general, models directly derived from quantum gravity should yield more rigid predictions and be better falsifiable than those only inspired by quantum-gravity phenomena. A faithful and realistic description of a high-energy, high-curvature generation mechanism would move this sector of GW astronomy beyond the level of \emph{ad hoc} model building towards a mini-program contributing to LISA science, according to the following algorithm:
\begin{enumerate}
    \item Selection or construction of early-universe models embedded into or inspired by quantum gravity, under the criterion of giving a blue-tilted spectrum ${\cal P}_{\rm t}(f)$ of primordial tensor fluctuations.
    \item Control of the underlying theoretical steps leading from the fundamental theory to the cosmological model: assumptions, approximations, parameter space, fine tunings.
    \item Generation of the SGWB spectrum $\Omega_{\rm GW}(f)$ from the primordial spectrum ${\cal P}_{\rm t}(f)$ via transfer functions.
    \item Comparison of the theoretical SGWB with the LISA sensitivity curve and constraints on the parameter space of the model. Non-detection of a SGWB can be used to rule out theories or to constrain their parameter space. If it was detected, one could further investigate characteristics of the SGWB such as anisotropies or the local spectral tilt at LISA frequencies to extract information on the underlying physics.
\end{enumerate}
In order to achieve the last goal, we will capitalise on the LISA SGWB search and use some of the pipelines described in Sec.~\ref{sec:pipesgwb}.

%%%%%%%%%%%%%%%%%%%%%%%%%%%%

\newpage
%%%%%%%%%%%%%%%%%%%%%%%%%%%%%%%
% Here Sec. 10 starts

\section{Primordial black holes}
\label{sec:PBH}

\small \emph{Section coordinators: S.~Clesse, J.~Garcia-Bellido. Contributors: S.~Clesse, V.~De Luca,  J.M.~Ezquiaga, G.~Franciolini, J.~Garcia-Bellido, R.~Kumar Jain, S.~Kuroyanagi, I.~Musco, T.~Papanikolaou, M.~Peloso, S.~Renaux-Petel, A.~Riotto, E.~Ruiz Morales, M.~Scalisi, O.~Sergijenko, C.~Unal, C.~Joana, V.~Vennin, D.~Wands.}
\normalsize

\subsection{Introduction}

\begin{figure}[!b]
     \centering
	\includegraphics[width = 0.9\textwidth]{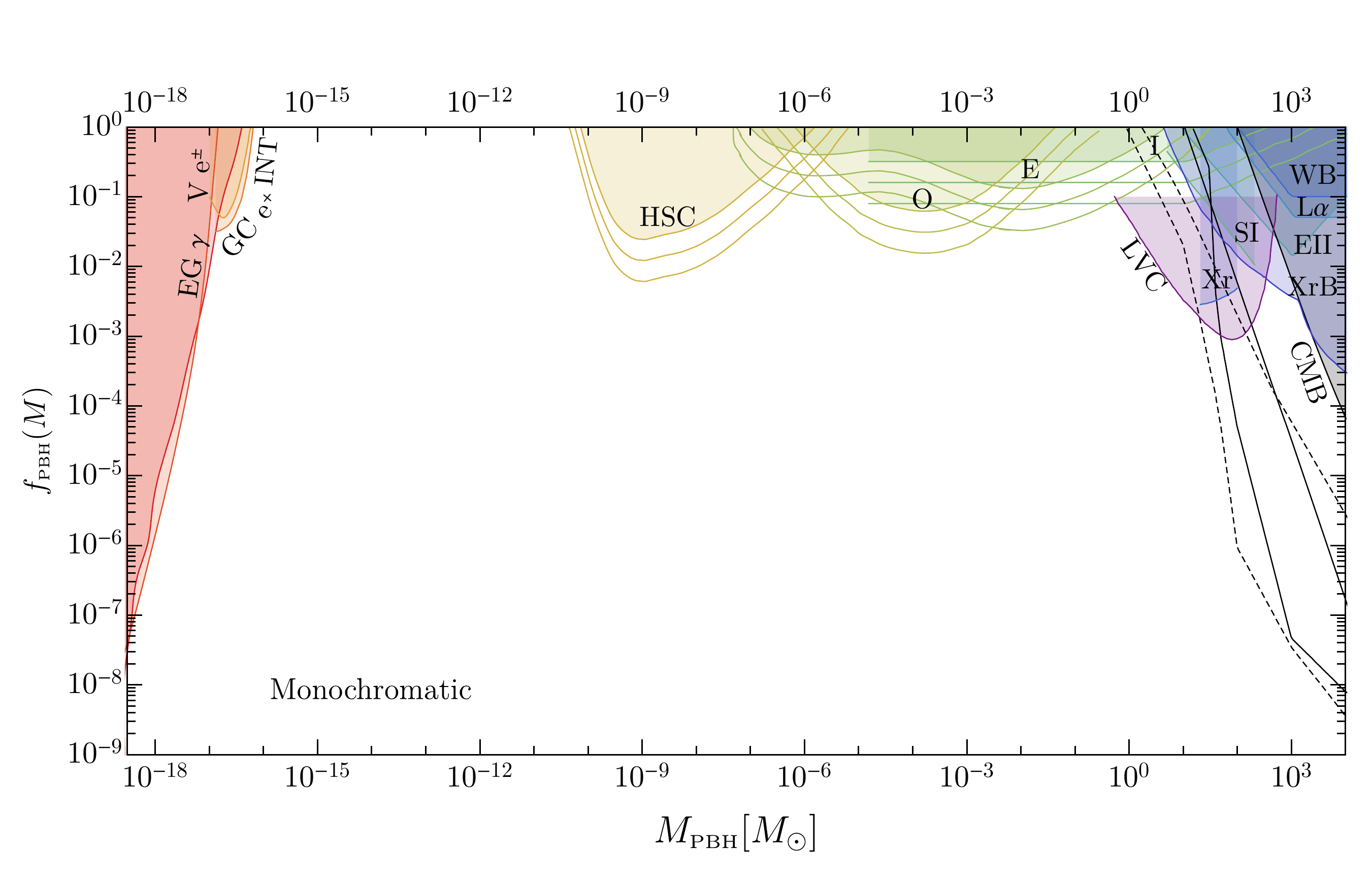}
	\caption{\small Most stringent limits on the DM fraction made of PBHs, $f_{\rm PBH}$, coming from the Hawking evaporation producing extragalactic gamma-ray (EG$\gamma$), e$^\pm$ observations by Voyager 1 (Ve$^\pm$), positron annihilations in the Galactic Center (GCe$^+$), gamma-ray observations by INTEGRAL (INT), microlensing searches by Subaru HSC (HSC), MACHO/EROS (E), OGLE (O) and Icarus (I), from CMB limits (CMB), GW observations by LIGO/Virgo (LVC), wide binaires in the galactic halo (WB), the ultra-faint dwarf galaxies Eridanus II (EII) and Segue 1 (S1), X-rays towards the galactic center (XrB) and Lyman-$\alpha$ limits (L$\alpha$).  For microlensing and CMB limits, the different lines indicate some degree of uncertainties, respectively due to PBH clustering and disk vs. spherical accretion.  Microlensing limits only apply to the fraction of PBHs uniformly distributed in galactic halos and are less stringent if a  non-negligible fraction $f_{\rm \rm clust}$ of PBHs are in clusters.  We show the limits for $f_{\rm \rm clust}=0, 0.4$ and $0.8$.  For LVC, the rate suppression of early binaries still allows $f_{\rm PBH} \gtrsim 0.1$.  \emph{All these limits apply to monochromatic models} and can be model dependent.  Recasting them to realistic PBH models with arbitrary mass functions requires a careful analysis. Figure adapted from Ref.~\cite{DeLuca:2020agl}.}
	\label{fig:fPBHlimits}
\end{figure}

The idea that BHs may have formed in the early universe comes back to the late 1960's with the precursor work of Zel'dovich and Novikov~\cite{1967SvA....10..602Z} and to the 1970's with the works of Hawking and Carr \cite{Hawking:1971ei,Carr:1974nx,Carr:1975qj} and of Chapline \cite{Chapline:1975}.  Already in Refs.~\cite{Carr:1974nx,Chapline:1975} it was mentioned that such PBHs  could contribute to the suspected DM in the universe or to the seeds of MBHs.  The first formation scenarios in the context of inflation were proposed in the 1990's \cite{Dolgov:1992pu,Carr:1994ar,GarciaBellido:1996qt} but these usually led to (evaporating) PBHs of very small mass.  In the late 1990's stellar-mass PBHs have been seriously considered as a DM candidate, following the possible detection (e.g.~in the MACHO survey) of several microlensing events towards the Magellanic clouds \cite{Aubourg:1993wb,Alcock:1996yv}. However, the EROS~\cite{Tisserand:2006zx} and OGLE~\cite{Wyrzykowski:2010bh,Wyrzykowski:2010mh,Wyrzykowski:2011tr,Novati:2013fxa} surveys later set more stringent limits on the PBH abundance, and at the same time, very stringent constraints from CMB observations were claimed in Ref.~\cite{Ricotti:2007au}.
Fig.~\ref{fig:fPBHlimits} summarizes the current constraints on the PBH abundance in the idealized limit that all PBHs have the same mass.

Since 2016, the real game-changer that has rekindled the idea that PBHs may exist and constitute from a significant fraction to the totality of the DM~\cite{Bird:2016dcv,Clesse:2016vqa,Sasaki:2016jop} has been the first GW detection from a BH merger by Advanced LIGO/Virgo~\cite{Abbott:2016blz}.  Nowadays, the importance of the different PBH binary formation channels, the possible abundance of PBHs, and their viable mass function, are subject to an intense activity and are under discussion (for recent reviews, see e.g.~Refs.~\cite{Carr:2020xqk,Carr:2020gox}). Furthermore, since PBHs are formed by the collapse of large density perturbations, PBHs are accompanied by a SGWB sourced by these perturbations at second-order. It has been calculated  that if BHs detected by LIGO/Virgo have primordial origin, there is an inevitable  accompanying SGWB  peaking around PTA frequencies~\cite{Ando:2017veq,Garcia-Bellido:2017aan}.\footnote{This SGWB is ineludible in the sense that it does not require any further assumption other than GR and large density perturbations. It is a standard SGWB formed by anisotropic stress which is quadratic order in scalar perturbations.}  In late 2020 NANOGrav has claimed the possible detection of a SGWB at nHz frequencies~\cite{Arzoumanian:2020vkk}, which could have been sourced by the density perturbations at the origin of  stellar-mass PBH formation~\cite{DeLuca:2020agl,Kohri:2020qqd,Kohri:2020qqd}.  
In this context, LISA will search for the GW signatures of PBHs~\cite{Garcia-Bellido:2017aan,Cai:2018dig,Bartolo:2018evs,Unal:2018yaa} and will be complementary to ground-based GW detectors~\cite{Maggiore:2019uih,Reitze:2019iox} and EM probes~\cite{Ali-Haimoud:2019khd}, in order to prove or exclude the existence of PBHs, to evaluate the possible contribution to the DM, to the seeds of MBHs at high redshift, and to distinguish PBHs from SOBHs, on a wide range of mass scales.  Any firm detection would open a new window on the physics at play in the very early universe and a possible way to solve various long-standing astrophysical and cosmological puzzles~\cite{Clesse:2017bsw,Carr:2019kxo}.

This section is organized as follows:  after reviewing the principal formation scenarios (Sec.~\ref{sec:PBHform}), we will consider the principal sources of GWs related to PBHs and their detectability with LISA, which are the SGWB generated at second-order by the non-linear cosmological density fluctuations at the origin of PBH formation (Sec.~\ref{subsec:second-order-SGWB}), the PBH binaries and hyperbolic encounter (Sec.~\ref{sec:PBHsGWsources}) which may unveil a possible primordial origin of MBH seeds if they are detected at high redshifts (Sec~\ref{sec:PBHhighz}).

\subsection{Formation scenarios}  \label{sec:PBHform}

Hereafter  we first provide a rapid overview of the principal mechanisms that can lead to large curvature fluctuations and PBH formation, which can be related to the early-universe  phenomena discussed in Secs.~\ref{sec:PTs}, \ref{sec:CosmicStrings} and \ref{sec:Inflation}.  We then review the general theory of PBH formation from large curvature fluctuations.  Finally we discuss some recent developments  related to PBH formation, such as non-linear and non-Gaussian effects, thermal history, that are all relevant for the estimation of the curvature threshold to lead to PBH formation, as well as their possible abundance and mass distribution.

\subsubsection{Origin of curvature fluctuations}

There exists a broad variety of scenarios leading to large curvature fluctuations and PBH collapse, which can be classified as below.  Some scenarios are detailed in the corresponding sections of this white paper.

\begin{enumerate}
\item \textbf{Single-field inflation:}  Slow-roll single field inflation models can produce a power spectrum amplitude of curvature fluctuations on smaller scales larger than the ones probed by CMB and large scale structures observations. The simplest example is to have two subsequent inflationary phases, the second one lasting less than 50 $e$-folds (roughly).  Potentials with inflection points~\cite{Garcia-Bellido:2017mdw,Germani:2017bcs}, like critical Higgs inflation~\cite{Ezquiaga:2017fvi}, can also lead to a transient enhancement of the primordial power spectrum. More generically, in the slow-roll approximation, any potential leading to a transient reduction of the speed of the inflaton will lead to large curvature fluctuations, eventually leading to PBH formation. Another possibility is to invoke a variation of the sound speed during inflation, see e.g.~Ref.~\cite{Ozsoy:2018flq}.

\item \textbf{Multi-field inflation:}   Large curvature fluctuations may also arise in multi-field models, e.g.~during the waterfall phase of {\it hybrid inflation}~\cite{Clesse:2015wea} or due to {\it turning trajectories} in the inflationary landscape~\cite{Fumagalli:2020adf}.  Eventually, the power spectrum will not only exhibit a broad or sharp peak on scales that are relevant for PBHs, but for a sufficiently sharp turn, this is accompanied by oscillatory features that may lead to specific signatures in the PBH population and to oscillatory patterns in the scalar-induced SGWB~\cite{Fumagalli:2020nvq,Braglia:2020taf,Palma:2020ejf}. It is also possible that curvature fluctuations are generated by the tunneling of the inflaton towards a local minimum of the field space~\cite{Garriga:2015fdk,Deng:2017uwc,Kusenko:2020pcg} and the subsequent {\it bubble collapse}.  Another intriguing possibility has been proposed by employing axions and their interaction with the {\it gauge fields} to enhance primordial density perturbations and producing PBHs \cite{Bugaev:2013fya,Linde:2012bt,Domcke:2017fix,Garcia-Bellido:2016dkw,Garcia-Bellido:2017aan,Ozsoy:2020kat,Ozsoy:2020ccy}.  In Natural Inflation \cite{Freese:1990rb,Adams:1992bn}, the inflaton is a pseudo-scalar particle protected from quantum corrections for super-Planckian excursions via the shift symmetry.  Also, theories with UV completion predict a large number of pseudo-scalar particles that could be the inflaton or spectator fields.  In this scenario, curvature fluctuations are sourced by enhanced vector modes and can have a non-Gaussian distribution.

\item \textbf{Quantum diffusion:}
The backreaction of quantum fluctuations during inflation makes the dynamics of the fields stochastic, and allow them to explore wider regions of the inflationary potential. This makes the tails of the distribution functions of primordial density fluctuations much heavier \cite{Vennin:2015hra, Pattison:2017mbe, Ezquiaga:2019ftu,Figueroa:2020jkf}, since they decay exponentially instead of in a Gaussian way, which boosts the production of PBHs afterwards. 

\item \textbf{Curvaton and stochastic spectator fields:}
PBHs could have formed due the existence of a curvaton field, e.g.~with a simple modification of the original curvaton scenario \cite{Lyth:2001nq}.  The primordial curvature perturbations on CMB scales are produced by the inflaton, which acts very similarly to the standard single-field scenario, while the curvaton field becomes responsible for perturbations on smaller scales, at the origin of PBH formation.  In another scenario~\cite{Carr:2019hud}, a stochastic spectator field experiences quantum fluctuations during inflation making it exploring a wide range of the potential, including its slow-roll part but without having any impact on the inflationary dynamics.  In regions where the field acquires a value allowing slow-roll, after inflation but when these are still super-horizon, additional expansion is produced locally, which generates curvature fluctuations that later collapse into PBHs.  This model has the advantage that the primordial curvature power spectrum remains at level observed at CMB scales almost everywhere, except in PBH-forming regions.

\item \textbf{Preheating:}  If inflation is followed by a  preheating, when the inflaton oscillates coherently at its ground state and decays to other degrees of freedom,  resonant amplification of the quantum field fluctuations take place ~\cite{Kofman:1994rk,Kofman:1997yn}. These are accompanied with a resonant amplification of curvature fluctuations~\cite{Finelli:1998bu,Bassett:1999mt,Jedamzik:1999um,Bassett:1999cg}, which may collapse and form PBHs ~\cite{Martin:2019nuw, Auclair:2020csm}.  The PBHs that form are typically very light and only
if reheating completes at
very low energy (below the electroweak scale) the formed PBHs would have a relevant mass for the LISA frequency range (for GW radiation from PBH binaries).  It is however also possible that these curvature fluctuations source a detectable SGWB at second order~\cite{Papanikolaou:2020qtd}. 

\item \textbf{Phase transitions:}  The formation of PBHs may have been facilitated in FOPTs~\cite{Jedamzik:1999am}, in non-equilibrium second order phase transitions~\cite{Rubin:2000dq} and in specific realizations of the QCD transition~\cite{Davoudiasl:2019ugw}.  

\item \textbf{Early matter era:}  The curvature threshold for the PBH collapse depends on the EoS of the universe.  It goes to zero in a matter dominated era, and so it is possible that PBH have formed from standard $\mathcal O(10^{-5})$ inflationary curvature fluctuation if the early universes has undergone a transient matter-dominated era.  Several mechanisms have been proposed to be at the origin of such early matter era, see e.g.~Refs.~\cite{Nayak:2009pk,Ballesteros:2019hus,Green:1997pr}.  

\item \textbf{Cosmic strings, domain walls:} Topological defects may have lead to the production of PBHs.  For instance, cosmic string cusps can collapse gravitationally into PBHs with a mass function that could extend up to stellar masses~\cite{Jenkins:2020ctp}.  The collapse of domain walls, e.g.~produced by tunnelling during inflation~\cite{Deng:2016vzb,Liu:2019lul} or in QCD axion models~\cite{Ferrer:2018uiu,Ge:2019ihf} is another class of PBH formation scenarios.  

\item \textbf{Primordial magnetic fields:}  It is suspected that the required seeds for extragalactic magnetic fields have an origin in the early universe.  These primordial magnetic fields induce an anisotropic stress that can act as a source of large super-Hubble curvature fluctuations, leading to PBH formation with a broad possible range of masses~\cite{Saga:2020ics}.

\end{enumerate}

\subsubsection{Standard theory of primordial black holes formation from Gaussian curvature fluctuations}  \label{sec:PBHformstd}

If PBHs have been formed due to the collapse of large primordial curvature perturbations, the fraction of the density of the universe made of PBHs at the time of their formation, $\beta_{\rm form}$ (usually defined per unit of logarithmic mass interval), is determined by the probability that the amplitude of a primordial curvature fluctuation, measured in terms of the mass excess of the density contrast $\delta$ 
is above a certain threshold $\delta_{\rm c}$ when the perturbation re-enter the cosmological horizon, corresponding to the Hubble length $1/H$, where $H$ denotes the Hubble parameter. The corresponding PBH mass $M$ is therefore linked to the Hubble horizon mass of the collapsing fluctuation, and so also to its size and its corresponding wavenumber $k_M = a H$ (and $a$ is the scale factor), 
\be
M = \gamma M_{\rm H} = \left( \frac{3 H^2}{8 \pi G} \right) \frac{c^3}{H^3}~,
\ee
where $\gamma$ is the ratio between the final PBH mass and the collapsing Horizon mass, which depends on the formation model details (a realistic range is $0.1\div1$). This should take into account that, when the perturbation amplitude $\delta$ is close enough to the threshold $\delta_{\rm c}$ (i.e.~$(\delta - \delta_{\rm c}) \lesssim 10^{-2}$), the mass follows the scaling law of critical collapse:
\be
M = \mathcal{K} (\delta - \delta_{\rm c})^\Gamma M_{\rm H}\,,
\ee
where $\Gamma$ depends only on the EoS ($\Gamma \simeq 0.36$ in the radiation dominated universe) and $\mathcal{K}$ depends on the shape of the perturbation (typically $1\div10$).

If the curvature fluctuations originate from inflation, the PBH mass can be related to the time of Hubble exit of the corresponding scale during inflation, expressed in terms of the number of $e$-folds before the end of inflation $N_*$ (see e.g.~Ref.~\cite{GarciaBellido:1996qt}), which gives, in the case of instantaneous reheating,
\be \label{eq:masspbh}
M \sim 4\pi \frac{M_{\textrm{Pl}}^2}{H_{\textrm{\rm inf}}}e^{2N_*}\,,
\ee
where $M_{\textrm{Pl}}$ is the reduced Planck mass and $H_{\textrm{inf}}$ is the (almost constant) Hubble rate during inflation.

In most inflationary scenarios, the distribution of primordial curvature fluctuations is Gaussian, or almost Gaussian. Some recent scenarios have proposed more complex distributions that can be useful to alleviate or reduce the need of a large power spectrum amplitude. These are shortly discussed later and assuming that the curvature fluctuations are described by a Gaussian distribution the fraction of PBHs at formation is then usually given by
\be
\beta_{\rm form}(M) = 2 \int_{\delta_{\rm c}}^{\infty} \frac{1} {\sqrt{2\pi}\sigma} e^{-\frac{\delta^2}{2\sigma^2}} {\rm d} \delta =  
1 - {\rm erf} \left( \frac{\delta_{\rm c}}{\sqrt 2 \sigma}\right) = {\rm erfc}\left( \frac{\delta_{\rm c}}{\sqrt 2 \sigma}\right)~.
\ee
%where the factor $2$ outside the integral is normalizing the total probability to $1$. 
This expression however does not take into account that when $\delta$ is larger then a certain value $\delta_{\rm max}$ (originally estimated $\delta_{\rm max}\sim 1$ by Bernard Carr in 1975 \cite{Carr:1975qj}) the perturbation forms a separate closed universe, topologically disconnected. A more accurate version can nevertheless be obtained re-normalizing the previous expression as
\be \label{eq:beta_pbh_Gaussian}
\beta_{\rm form}(M)
= \frac{\displaystyle{\int_{\delta_{\rm c}}^{\delta_{\rm max}} \frac{1} {\sqrt{2\pi}\sigma} e^{-\frac{\delta^2}{2\sigma^2}} {\rm d} \delta }} {\displaystyle{\int_0^{\delta_{\rm max}} \frac{1}{\sqrt{2\pi}\sigma} e^{-\frac{\delta^2}{2\sigma^2}} {\rm d} \delta }} =
1 - \frac{{\rm erf} \lp \displaystyle{\frac{\delta_{\rm c}}{\sqrt{2}\,\sigma}}\rp}
{{\rm erf}\lp\displaystyle{\frac{\delta_{\rm max}}{\sqrt{2}\,\sigma}}\rp} = 
\frac{ {\rm erfc}\left( \displaystyle{\frac{\delta_{\rm c}}{\sqrt 2 \sigma}} \right) - {\rm erfc}\left( \displaystyle{\frac{\delta_{\rm max}}{\sqrt 2 \sigma}} \right) }{ 1 - {\rm erfc}\left( \displaystyle{\frac{\delta_{\rm max}}{\sqrt 2 \sigma}} \right)}\,.
\ee
The variance of the field of density perturbations $\sigma$, according to the Gaussian distribution of $\delta$ used before, is given by
\be \label{eq:beta_pbh_Gaussian2}
\sigma^2 = \langle \delta^2 \rangle = \int\limits_0^\infty \frac{\mathrm{d}k}{k} \mathcal{P}_{\delta}(k,r) = \frac{16}{81}\int\limits_0^\infty \frac{\mathrm{d}k}{k}(kr)^4 \tilde{W}^2(k,r) T^2 (k,r) \mathcal{P}_\zeta(k)\,,
\ee
where ${\cal P}_\delta(k,r)$ and $\mathcal{P}_\zeta(k)$ are the density and the curvature power spectrum, while $\tilde{W}(k,r)= 3 (\sin k r - k r \,\cos k r)/(k r)^3$ is the Fourier transform of the top-hat smoothing function and $T(k,r)=\tilde{W}(k,r/\sqrt3)$ is the linear transfer function, both computed at the horizon crossing scale.
 All this shows that a larger power spectrum could increase significantly the fraction of PBHs because the abundance of PBHs is exponentially sensitive to the value of the amplitude.   

One can then calculate the contribution of PBHs to the  density of the universe today, in terms of the fraction of  DM that they represent, given by
\be
f_{\rm PBH} (M)  \sim 2.4 \times \beta_{\rm form}(M) \sqrt{\frac {M_{\rm eq}}{M}} \,,
\ee
where $M_{\rm eq} \simeq 2.8 \times 10^{17} M_\odot $ is  the  horizon  mass  at matter-radiation equality, and the numerical factor corresponds to $ 2 \times (1+ \Omega_b / \Omega_{\rm CDM})$.
In this standard scenario, where PBHs form from Gaussian curvature pertubations, one does not expect spatial clustering at formation larger than the one predicted by the Poisson distribution \cite{Ali-Haimoud:2018dau,Desjacques:2018wuu,Ballesteros:2018swv, MoradinezhadDizgah:2019wjf}.

\subsubsection{Beyond Gaussianity}  \label{sec:non-gaussianity}

The above formalism applies to Gaussian perturbations but it can be generalized to non-Gaussian distributions, by replacing the integrand in the left-hand side of Eq.~(\ref{eq:beta_pbh_Gaussian}) with the appropriate distribution function.

At the perturbative level, even a small amount of local non-Gaussianity~\cite{Lyth:2012yp,Byrnes:2012yx}, or the inevitable non-linear (and hence non-Gaussian) relation between the primordial curvature perturbations and density perturbations can have important effects~\cite{Young:2015kda,Franciolini:2018vbk,DeLuca:2019qsy,Young:2019yug}. In the regime of large non-Gaussianity \cite{Linde:2012bt,Garcia-Bellido:2016dkw,Unal:2020mts}, the amplitude of the power spectrum can produce same abundance of PBHs even if it is many orders of magnitude smaller with respect to Gaussian case. The existence of non-Gaussianity also influences the amount of SGWB accompanying the PBHs \cite{Nakama:2016gzw,Garcia-Bellido:2017aan,Cai:2018dig,Unal:2018yaa}. The fact that the power spectrum needs to deviate from the quasi scale invariance observed on CMB scales to reach fluctuations allowing PBHs to form also implies that it is natural to expect the non-Gaussianity to be strongly scale dependent too.
Moreover, the merger rate of PBHs is strongly impacted by non-Gaussianity~\cite{Young:2019gfc,Atal:2020igj}, because non-Gaussianity couples the long and short wavelength primordial density perturbations, leading to initial spatial
clustering of the PBHs. 

Since PBHs require large fluctuations to form, a perturbative description may not be sufficient. Quantum diffusion of the inflationary fields leads to large deviations from Gaussian statistics on the tails of the distribution functions of primordial density fluctuations~\cite{Pattison:2017mbe, Biagetti:2018pjj, Ezquiaga:2018gbw,Figueroa:2020jkf,Figueroa:2021zah}, which acquire an exponential (rather than Gaussian) profile~\cite{Vennin:2015hra, Pattison:2017mbe, Ezquiaga:2019ftu,Figueroa:2020jkf,
Figueroa:2021zah}. It implies that PBHs can be formed with a much smaller amplitude of the power spectrum than what would be inferred using Gaussian statistics. This cannot be properly taken into account with the usual, perturbative parametrisation of non-Gaussian statistics (such as those based on computing the few first moments of the distribution and the non-linearity parameters $f_{\mathrm{NL}}$, $g_{\mathrm{NL}}$, etc.), which can only account for polynomial modulations of Gaussian tails, and needs to be described with a non-perturbative approach such as the stochastic-$\delta N$ formalism~\cite{ Fujita:2013cna, Vennin:2015hra}.

\subsubsection{The threshold for primordial black holes and the impact of thermal history}

A crucial parameter of the formalism presented above is the critical value $\delta_c$, distinguishing between cosmological perturbations collapsing into PBHs ($\delta > \delta_{\rm c}$) and those ones bouncing back into the surrounding medium ($\delta < \delta_{\rm c}$). This is a fundamental parameter because the resulting PBH abundance is exponentially sensitive to its value. The analysis of the gravitational collapse of curvature perturbations to form PBHs and the appropriate threshold condition has been an active line of research in the past years \cite{Kopp:2010sh,Harada:2013epa,Young:2014ana}. It has been estimated using analytical methods~\cite{Harada:2013epa} but the best approach is to use fully relativistic simulations of PBH formation in spherical symmetry~\cite{Musco:2004ak}. Important results having emerged from recent studies are that its exact value is impacted by non-linear and non-Gaussian effects, and that it depends on the radial profile of the overdensity~\cite{Musco:2018rwt,Young:2019yug,DeLuca:2019qsy,Kehagias:2019eil} as well as on the shape of the primordial power spectrum~\cite{Yoo:2018kvb,Germani:2018jgr}. 

Very recently, a new semi-analytical method tested against simulations in numerical relativity has been proposed in Ref.~\cite{Musco:2020jjb}, computing $\delta_c$ from the shape of the power spectrum, applied to a few particular cases (power-law spectra, log-normal or Gaussian peak, \dots).  In the radiation dominated universe the typical range of the threshold lies within $ 0.4 < \delta_{\rm c} < 0.6 $, with the larger values corresponding to a more peaked shape of the peak of the power spectrum.

On super horizon scales the non-linear amplitude of the curvature profile $\zeta$ is important for the formation of PBHs, and the energy density contrast $\delta\rho/\rho_b$, when the universe is radiation dominated, is expressed in terms of $\zeta$ as
\be
\frac{\delta\rho}{\rho_b} \equiv \frac{\rho(r,t) - \rho_b(t)}{\rho_b(t)} =  - \frac{1}{a^2H^2} \frac{8}{9} e^{-5\zeta(r)/2}\nabla^2 e^{\zeta(r)/2} \,.
\ee
It can be shown that the amplitude $\delta$ is a quadratic function of the curvature profile (see for example Ref.~\cite{Musco:2018rwt} for more details)
\be 
\label{eq:quadratic}
\delta = - \frac{2}{3} \zeta'(r_m) \left[ 2+r_m\zeta'(r_m) \right] = \delta_{\rm \small G} \lp 1 -\frac{3}{8} \delta_{\rm \small G} \rp \,,
\ee
where $\delta_{\rm \small G}\equiv - \frac{4}{3}r_m\zeta^\prime(r_m)$ is the linear component of the amplitude $\delta$. The value of $r_m$ is defined by the location of the peak of the compaction function $\mathcal{C}\equiv2\Delta M/R$ (where $\Delta M = M-M_b$ is the excess of mass with respect the background) and is given by \mbox{$\zeta(r_m) + r_m\zeta^\prime(r_m)=0$}. Given the value of $\delta_{\rm c}$ the threshold value of the linear component $\delta_{\rm \small G}$ is included within $0.5 \lesssim \delta_{\rm c, \small G} \lesssim 0.9$, and from Eq.~\eqref{eq:quadratic} we see that $\delta_{\rm max} = 8/3$, corresponding to the maximum value of $\delta_{\rm \small G}$, above which $\delta$ becomes negative, and does not describe a cosmological perturbation of our universe. 

The threshold $\delta_{\rm c}$ is also sensitive to the EoS at the time of formation. For example, the QCD phase transition makes the EoS to drop, increasing the production of PBHs of mass ${\cal O}(M_\odot)$ \cite{Jedamzik:1996mr,Byrnes:2018clq}. 
The reheating at the end of inflation should have filled the universe with radiation. In the absence of extensions beyond the SM, the universe remains dominated by relativistic particles with an energy density decreasing as the fourth power of the temperature as the universe expands. The number of relativistic degrees of freedom remains constant ($g_{*} = 106.75$) until around $200$ GeV, when the temperature of the universe falls to the mass thresholds of SM particles.

\begin{figure}
     \centering
	\includegraphics[width = 0.44\textwidth]{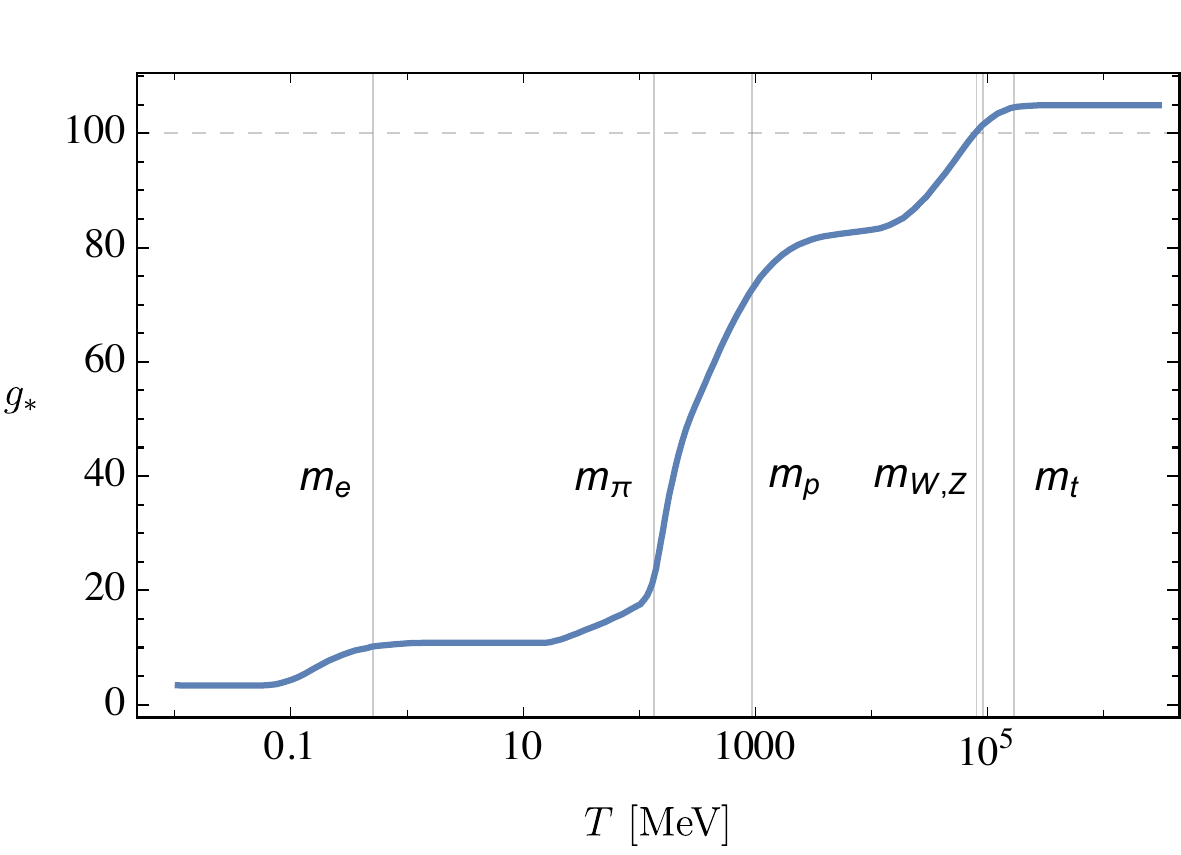}
	\includegraphics[width = 0.50\textwidth]{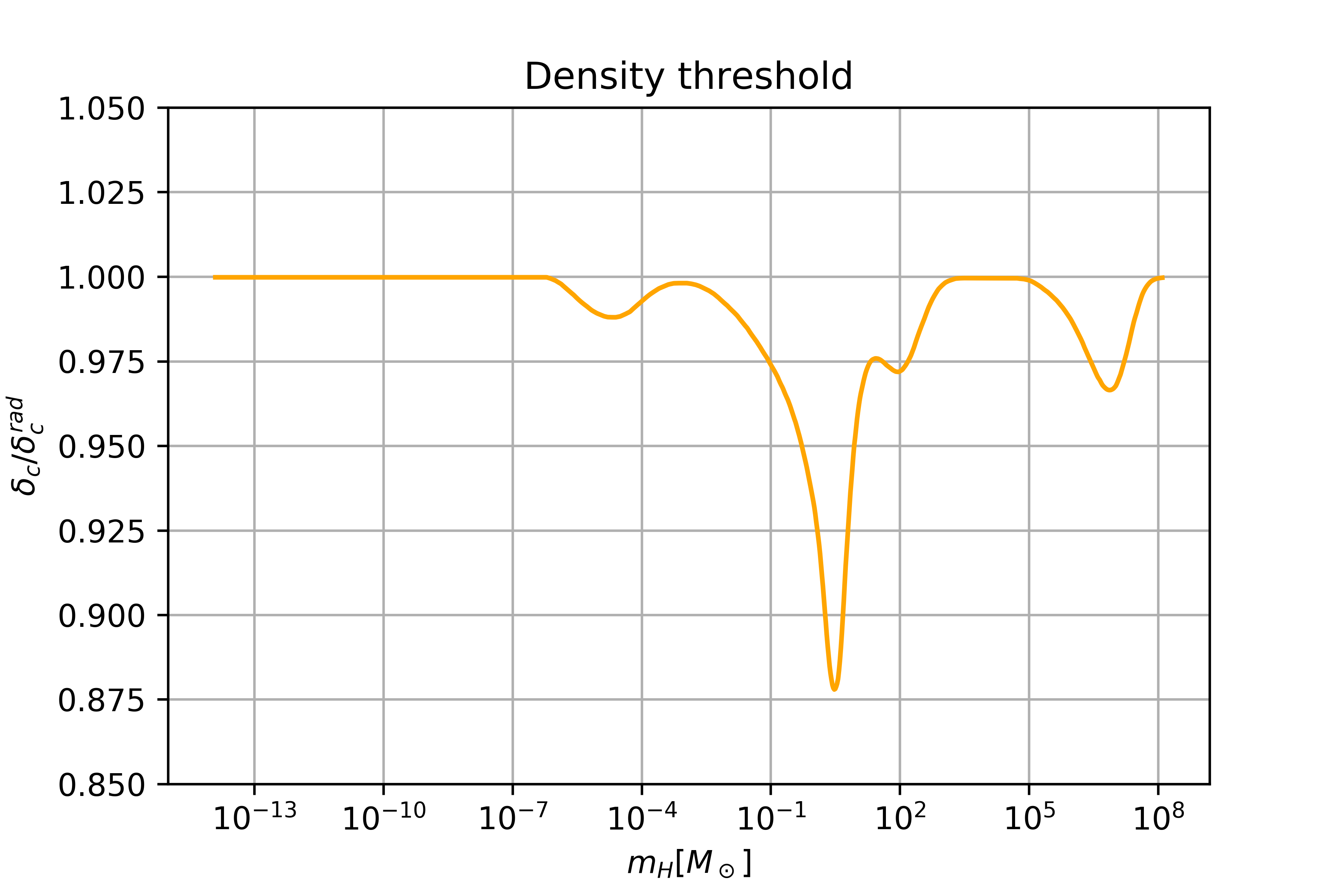}
	\caption{\small Left panel:  Evolution of the relativistic degrees of freedom $g_{*}$ as a function of the temperature. The grey vertical lines correspond to the masses of the electron, 	pion, proton/neutron, $W$, $Z$ bosons and top quark, respectively.  Right panel:  Effect of the evolution of $g_*$ on the critical overdensity $\delta_{\rm c}$ leading to PBH formation, as a function of the Hubble horizon mass (related to the PBH mass by $M = \gamma m_{\rm H}$). }
	\label{fig:gstardeltac}
\end{figure}

The first particle to become non-relativistic is the top quark at $T \simeq m_{\rm t} = 172$ GeV, followed by the Higgs boson at $125$ GeV, and the $Z$ and $W$ bosons at $92$ and $81$ GeV, respectively. These particles become non-relativistic at nearly the same time and this induces a significant drop in the number of relativistic degrees of freedom down to $g_{*} = 86.75$. There are further changes at the $b$ and $c$ quark and $\tau$-lepton thresholds but these are too small to appear in Fig.~\ref{fig:gstardeltac}. Thereafter $g_{*}$ remains approximately constant until the QCD transition at around $200$ MeV, when protons and neutrons condense out of the free light quarks and gluons. The number of relativistic degrees of freedom then falls abruptly to $g_{*} = 17.25$. A little later the pions become non-relativistic and then the muons, giving $g_{*} = 10.75$. Thereafter $g_{*}$ remains constant until $e^{+}e^{-}$ annihilation and neutrino decoupling at around $1$ MeV, when it drops to $g_{*} = 3.36$. 

Whenever the number of relativistic degrees of freedom suddenly drops, it changes the effective EoS parameter $w$.  There are thus four periods in the thermal history of the universe when $w$ decreases. After each of these, $w$ resumes its relativistic value of $1 / 3$ but each sudden drop modifies the probability of gravitational collapse of any large curvature fluctuations present at that time, as shown in Fig.~\ref{fig:gstardeltac}.

As illustrative examples, we have computed the PBH mass functions for two models with an (almost) scale-invariant power spectrum and two different values of the spectral index, $n_{\rm s} = 0.97$ (Model 1) and $n_{\rm s} = 1$ (Model 2).  We assumed $\gamma =0.8$ in both cases.  The imprints of the thermal history on the PBH mass function are clearly visible.  It is worth noticing that these features rely on known physics and are therefore unavoidable for any PBH model with a wide mass function.  The former case corresponds to the scenario proposed in Refs.~\cite{Carr:2019kxo,Clesse:2020ghq} and the latter in Refs.~\cite{Byrnes:2018clq,DeLuca:2020agl}.  They can both account for the totality of the DM and somehow explain some LIGO/Virgo GW events, but produce different abundances in the stellar mass range:  $f_{\rm PBH}(M_\odot) \approx 0.8 $ in the first case, $f_{\rm PBH} (M_\odot)\approx 10^{-4} $ in the second case where the peak lies in the sub-lunar range. We stress that the second example avoids the bounds in the LIGO/Virgo range.

 \begin{figure}
      \centering
  	\includegraphics[width = 0.46\textwidth]{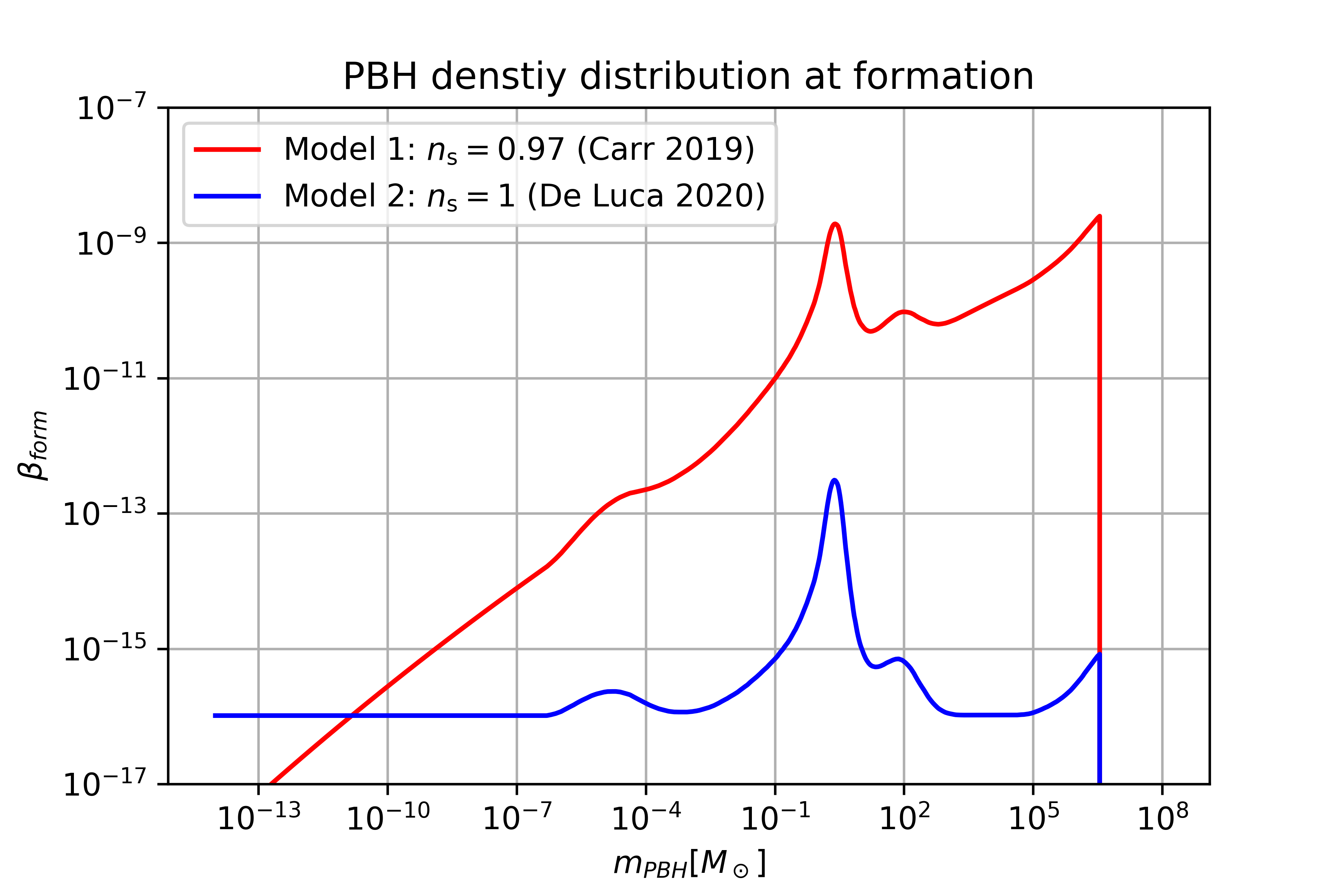}
  	\includegraphics[width = 0.46\textwidth]{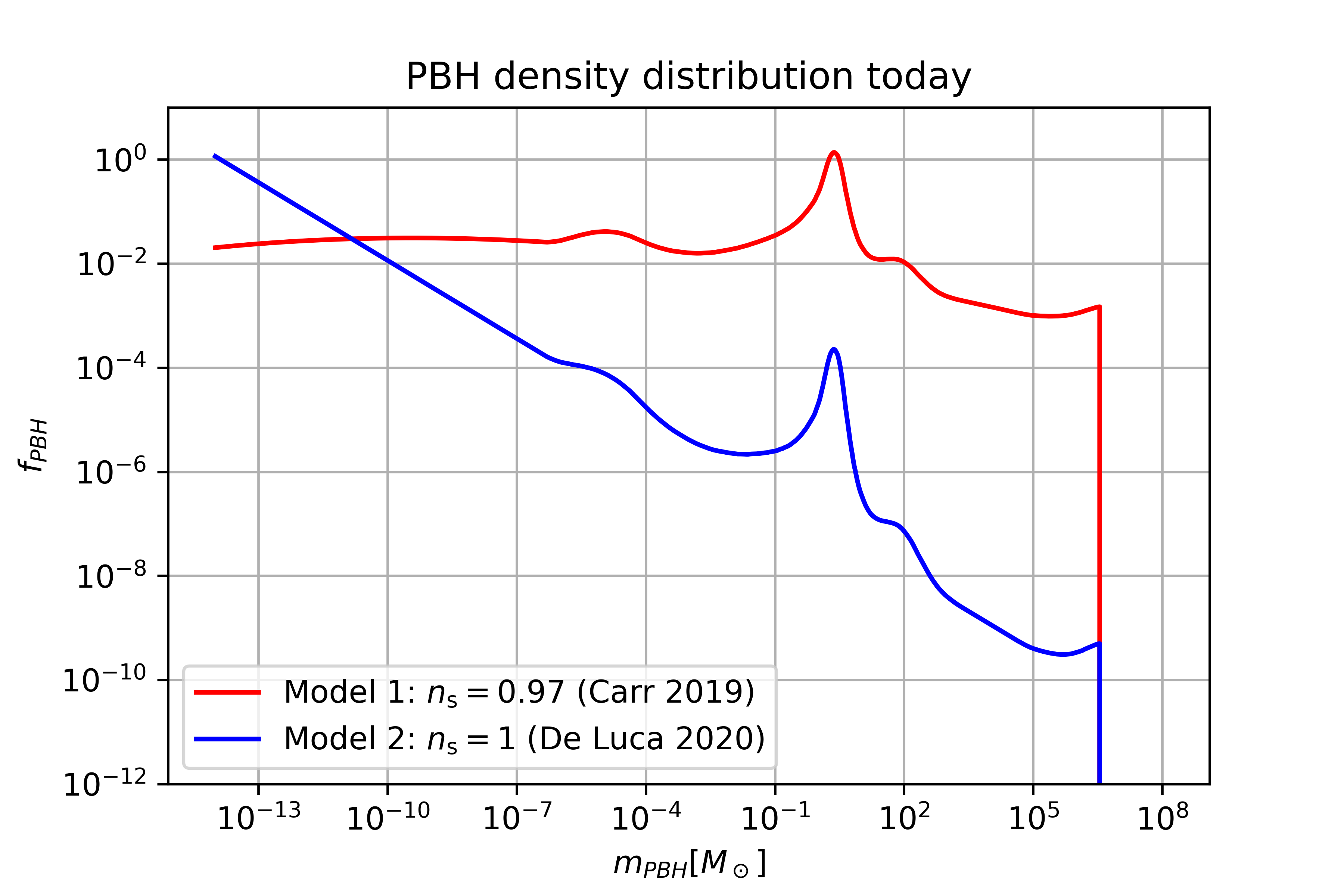}
      \caption{\small PBH density fraction at formation $\beta_{\rm form}$ (left panel) and the corresponding PBH mass function $f_{\rm PBH}$ today (right panel), neglecting the effect of PBH growth by accretion and hierarchical mergers, for two models with a power-law power spectrum and including the effects of thermal history:  Model 1 from Refs.~\cite{Carr:2019kxo,Clesse:2020ghq} with spectral index $n_{\rm s} = 0.97$; Model 2 from Refs.~\cite{DeLuca:2020agl,Byrnes:2018clq} with $n_{\rm s} = 1.$ and a cut-off mass of $10^{-14} M_\odot$.  The transition between the large-scale and small-scale power spectrum is fixed at $k=10^3 {\rm Mpc}^{-1}$.  The power spectrum amplitude is normalized such that both models produce an integrated PBH fraction $f_{\rm PBH} =1$, i.e.~PBH constitute the totality of DM.  A value of $\gamma = 0.8$ was assumed.  }
      \label{fig:fPBH}
  \end{figure}

\subsection{Stochastic gravitational wave background sourced at second order by curvature fluctuations}
\label{subsec:second-order-SGWB}

If PBHs are generated by the collapse of large density perturbations, they are unavoidably associated to the emission of induced GWs at second order by the same scalar perturbations due to the intrinsic nonlinear nature of gravity \cite{Acquaviva:2002ud,Mollerach:2003nq}. 
The phenomenological implications have been investigated in various contexts also associated to PBHs 
\cite{Ananda:2006af,Baumann:2007zm,Bugaev:2009zh,Saito:2009jt,Garcia-Bellido:2017aan,Ando:2017veq,Bartolo:2018qqn,Bartolo:2018rku,Bartolo:2018evs,Clesse:2018ogk,Unal:2018yaa,Chatterjee:2017hru,Wang:2019kaf,Domenech:2019quo,Domenech:2020kqm,Pi:2020otn,Ragavendra:2020sop,Fumagalli:2020nvq}.
If the enhancement of the scalar power spectrum responsible for the generation of PBHs occurs around characteristic scales associated with frequencies between $10^{-7}$ and $10^{-2}$ Hz, this SGWB becomes detectable by GW experiments like LISA.  It is worth emphasizing that contrary to the PBH abundance that is exponentially sensitive to the power spectrum, this SGWB depends on the power spectrum amplitude to the second power.  This way, LISA will even be able to exclude the existence of an extremely tiny fraction of DM made of PBHs (even a single PBH in our universe)~\cite{Clesse:2018ogk}, within a wide mass range.

\begin{figure}
    \centering
	\includegraphics[width = 0.9\textwidth]{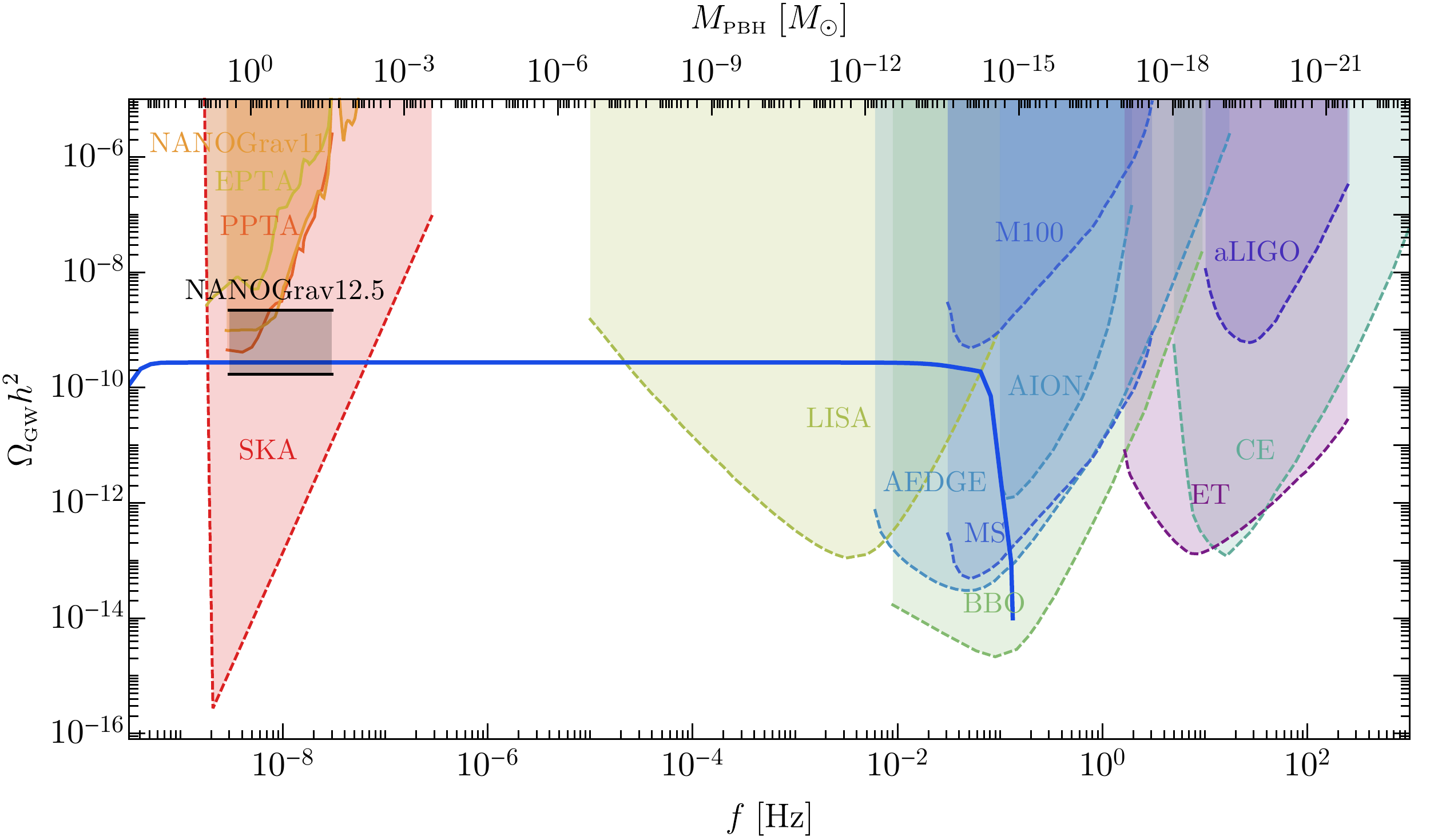}    \caption{\small SGWB sourced at second order by the density perturbations at the origin of PBH formation, for Model 2 of Fig.~\ref{fig:fPBH}. 
	On top of the plot, we show the PBH mass associated to a given GW frequency as in Eq.~\eqref{PBHmassfreq}.
	The LISA sensitivity \cite{Audley:2017drz} and the hint for a detection by NANOGrav 12.5 yr \cite{Arzoumanian:2020vkk} are represented, as well as the constraints coming EPTA \cite{Lentati:2015qwp}, PPTA \cite{Shannon:2015ect}, NANOGrav 11 yrs \cite{Arzoumanian:2018saf, Aggarwal:2018mgp} and future sensitivity curves for planned experiments like SKA \cite{Zhao:2013bba}, DECIGO/BBO \cite{Yagi:2011wg}, CE \cite{Evans:2016mbw}, ET \cite{Hild:2010id}, Advanced LIGO + Virgo collaboration \cite{TheLIGOScientific:2016dpb}, Magis-space (MS) and Magis-100 (M100) \cite{Coleman:2018ozp}, AEDGE \cite{Bertoldi:2019tck} and AION \cite{Badurina:2019hst}.  Figure taken from Ref.~\cite{DeLuca:2020agl}.
}
    \label{fig:PBH-SGWB}
\end{figure}

Figure \ref{fig:fPBH} presents the PBH density fraction at formation $\beta_{\rm form}$ (left panel) and the corresponding PBH mass function $f_{\rm PBH}$ today (right panel) for two models with a power-law power spectrum
(see the caption of the Figure for details). The SGWB associated with one of these two models is shown in Figure 
\ref{fig:PBH-SGWB}, where it is confronted with several current or forecasted experimental limits. The SGWB covers a wide frequency range.  In the ultra-low frequency range, around nHz, PTA experiments like PPTA \cite{Shannon:2015ect}, NANOGrav \cite{Arzoumanian:2018saf} and EPTA \cite{Lentati:2015qwp} give the most stringent constraints on the GWs abundance.  Future experiments like SKA \cite{2009IEEEP..97.1482D} (see also Ref.~\cite{Moore:2014lga}) will greatly improve the sensitivity.   In the LIGO/Virgo frequency range, an additional constraint has been set by the non-observation of a SGWB after O1-O2~\cite{LIGOScientific:2019vic} and O3 runs~\cite{Abbott:2021xxi}.   All these searches can be translated into a constraint on the amplitude of the comoving curvature perturbation at the corresponding scales \cite{Bugaev:2009zh,Byrnes:2018txb,Inomata:2018epa,Unal:2020mts}.  Those bounds are also affecting the maximum allowed PBHs fraction of DM  with the hypothesis that they originate from the collapse of density perturbations.  Detailed studies with the LIGO/Virgo data affecting the mass range $M \in  \left[ 10^{-20},10^{-18} \right] M_\odot $ are reported in Ref.~\cite{Kapadia:2020pir}, while very tight bounds in the mass range $M \in   \left[  10^{-3},1 \right] M_\odot $ are obtained in Ref.~\cite{Chen:2019xse} using the latest NANOGrav data; see also Ref.~\cite{Cai:2019elf} where the dependence of the result to non-Gaussianities is also investigated, finding that local non-Gaussianity can for example alleviate the bounds (see Sec.~\ref{sec:non-gaussianity} for details).    Finally, the next generation multimessenger experiments, CMB distortion (PIXIE) and PTA-SKA, can test the PBH scenario over solar mass robustly, namely they can conclusively detect or rule out the PBHs over solar-mass and the intriguing proposal that the seeds of the MBHs are formed by PBHs \cite{Unal:2020mts} independent of i) statistical properties of perturbations, ii) accretion and merger history and iii) clustering effects. 
 
LISA will be able to provide insights in the intermediate frequencies, and corresponding masses.  Since the emission mostly comes when the corresponding scales cross the horizon, one can relate the GWs frequency to the PBHs mass $M$ as (see for example Refs.~\cite{Saito:2009jt,Garcia-Bellido:2017aan})
\begin{equation}\label{PBHmassfreq}
	f \simeq 3  \, \text{mHz} \left( \frac{\gamma}{0.2}\right)^{1/2}\left( \frac{M}{ 10^{-12}M_\odot}\right) ^{-1/2},
\end{equation}
where the factor $\gamma$ is capturing the relation between the horizon mass at formation and the PBH mass after the collapse.
%The mass-frequency relation can be seen on Fig.~\ref{fig:my_label}.
Notice that the peak frequencies fall within the LISA sensitivity band for PBH masses around $M \sim {\cal O} \lp 10^{-15} \div 10^{-8} \rp M_\odot$ and for this mass range, the PBHs can constitute the totality of the DM. Hence, Ref.~\cite{Garcia-Bellido:2017aan} proposed PBHs in this mass range as DM  and further found that density perturbations forming PBHs lead to GWs detectable by LISA. This proposal has been studied in more detail in Refs.~\cite{Cai:2018dig,Bartolo:2018evs,Bartolo:2018rku,Unal:2018yaa}.

The computation of the resulting SGWB spectrum was originally performed in Ref.~\cite{Ananda:2006af}.  We provide here the main result, assuming a generic form for the power spectrum of curvature fluctuation.   The current GW abundance can then be obtained as
\begin{equation}
	\Omega_\text{\tiny GW}(\eta_0,k) =  \frac{a_f^4\rho_\text{\tiny GW}(\eta_f,k)}{\rho_r(\eta_0)}\Omega_{r,0} =
\frac{g_*(\eta_f)}{g_*(\eta_0)} \left(\frac{g_{*S}(\eta_0)}{g_{*S}(\eta_f)}\right)^{4/3}
\Omega_{r,0} 	\Omega_\text{\tiny GW}(\eta_f,k),
	\label{eq: Omega GW}
\end{equation}
in terms of the present radiation energy density fraction $\Omega_{r,0}$ if the neutrinos were massless.  The crucial quantity is $ \Omega_\text{\tiny GW}(\eta_f,k)$, that is the fractional GW energy density for log interval at the emission epoch $\eta_f$, related to the critical energy density of a spatially flat universe $\rho_c = 3 H^2 M_p^2$.  Assuming that the scalar perturbations $\zeta$ are Gaussian, it can be calculated as 
\begin{align} 
&\left\langle \rho_\text{\tiny GW} \left( \eta ,\, \vec{x} \right) \right\rangle
\equiv \rho_c ( \eta) \, \int d \ln k \; \Omega_\text{\tiny GW} \left( \eta ,\, k \right) \nonumber\\
&=  
  \frac{2 \pi^4 M_p^2}{81  \eta^2 a^2}  \, \int \frac{d^3 k_1 d^3 p_1}{\left( 2 \pi \right)^{6} } 
\frac{1}{k_1^4}\, 
\frac{\left[ p_1^2 -  ( \vec{k}_1 \cdot \vec{p}_1)^2/k_1^2 \right]^2}{p_1^3 \, \left\vert \vec{k_1} - \vec{p}_1 \right\vert^3} \, 
 {\cal P}_\zeta ( p_1) 
{\cal P}_\zeta( |\vec{k_1} - \vec{p}_1|)  
\left( {\cal I}_c^2( \vec{k}_1 ,\, \vec{p}_1) + {\cal I}_s^2( \vec{k}_1 ,\, \vec{p}_1)  \right),
\label{rho1-par}
\end{align} 
where the functions ${\cal{I}}_{c,s}$ are found in Refs.~\cite{Espinosa:2018eve,Kohri:2018awv}.
The integrals need to be done numerically for general power spectra (see Refs.~\cite{Saito:2009jt,Bugaev:2009zh} for analytical calculations in the specific case of a monochromatic or Gaussian curvature spectrum).  

For the frequencies of interest, using 
$f \simeq 8\,  {\rm mHz}  \lp{g_*}/{10} \rp ^{1/4} \lp T/{\rm 10^6 GeV} \rp $,
one can show that the emission of GWs takes place at $\eta_f$ well before the time at which top quarks start annihilating, above which we can assume a RD universe with constant effective degrees of freedom.

The non-linear coupling with the curvature perturbation naturally leads to an intrinsically non-Gaussian GWs signal imprinted in phase correlations. However, the coherence is washed out by the propagation of the waves in the perturbed universe mainly due to time delay effects originated from the presence of large scale variations of gravitational potential~\cite{Bartolo:2018evs,Bartolo:2018rku,Bartolo:2018qqn,Margalit:2020sxp}. This is simply a consequence of the central limit theorem applied to a number $N \sim \lp k_\star \eta_0 \rp^2 \ggg 1$ of independent lines of sight \cite{Bartolo:2018evs}, $k_\star$ being the characteristic perturbation wave-number roughly proportional to the the inverse horizon size at GW emission. Possible small deformations smearing the GWs spectrum can also arise from similar effects \cite{Domcke:2020xmn}.

An interesting primordial signal that is potentially observable is related to scenarios where the scalar power spectrum presents  oscillations of sufficiently large amplitude, characteristic of large particle production mechanisms, leading to oscillatory ${\cal O} (10\%)$ modulations in the frequency profile  of the scalar-induced SGWB \cite{Fumagalli:2020nvq,Braglia:2020taf}, see Sec.~\ref{sec:features} for details.

\subsection{Resolved sources and stochastic gravitational wave backgrounds from primordial black holes binaries and hyperbolic encounters} \label{sec:PBHsGWsources}

PBHs can source GWs in several ways, which will be probed as individual and resolved sources, as a SGWB, or continuous wave signals with LISA.  One can distinguish:
\begin{itemize}
    \item GW chirps from the merging of PBH binaries
    \item SGWB from PBH binaries
    \item Continuous GWs from PBH binaries far from the merging time.
    \item Bursts from close BH-BH interactions
\end{itemize}

The amplitude of these signals depends on the PBH mass function and the resulting merger rates, which themselves depend on the PBH binary formation channel.  Two main channels have been identified:  at formation in the early universe and by tidal capture in dense PBH clusters.  For each we provide an
estimation of the corresponding merger rates.  We then compute the expected signals for a few example models.
For the purpose of calculating the PBH merging/encounter rates and the resulting GW signatures, we use a general mass distribution $f(m_{\rm PBH})$ that would be specified by the underlying formation model.

\subsubsection{Binary formation channels and merger rates}

\textbf{Primordial binaries} may have been created before the epoch of matter-radiation equality, when two PBHs formed sufficiently close to each other for their dynamics to be independent of the expansion the universe.
If this binary takes a time of the order of the age of the universe to merge, the resulting GWs could be detected.   One can estimate the merger rate of such binaries~\cite{Kocsis:2017yty,Raidal:2018bbj} in the simplest scenario where PBHs are not clustered at formation and only contribute to less than about 10\% of the DM as
\begin{align} 
\frac{\d R^{\rm prim} }{\d (\ln m_1) \d (\ln m_2) \d z}
& = 
\frac{1.6 \times 10^6}{{\rm Gpc^3 \, yr}} 
f_\PBH^{\frac{53}{37}} \,
S(m_1,m_2,f_\PBH)
\left[ \frac{t \,m_1 m_2}{t_0(m_1+m_2)^2} \right]^{-\frac{34}{37}} 
 \lp \frac{m_1 + m_2}{M_\odot} \rp^{-\frac{32}{37}}  
 f(m_1) f(m_2) \label{eq:Rprimbinaries}
\end{align}
where $m_1$ and $m_2$ are the BH masses and $S$ accounts for the suppression factor of the merger rate coming from binary disruption in early universe substructures \cite{Raidal:2018bbj}. 
We use the definition of the PBH  mass function $f(m)$ normalised to unity as $\int f(m) \d \ln m  =1$.

Notice that accretion onto PBHs in binaries can be effective for masses above ${\cal O}(10) M_\odot$ with an impact on the merger rate \cite{DeLuca:2020bjf,DeLuca:2020qqa}. When $f_{\rm PBH} \gtrsim 0.1 $, N-body simulations~\cite{Raidal:2018bbj} have shown that PBH clusters can rapidly form and change the lifetime of PBH binaries, due to their tidal force. However, the fraction of unperturbed binaries at the present time is at least of the order of $10^{-2}$ for   $f_{\rm PBH} = 1$ \cite{Vaskonen:2019jpv, DeLuca:2020jug, Jedamzikinprep}. Furthermore, one should also recall that 
not all the binaries end up inside halos, about $10^{-3}$ to $10^{-2}$ remain isolated. For instance, this gives a primordial  merger rate still above the inferred LIGO/Virgo rate, even for large PBH abundances. 

It is worth mentioning that, as investigated in  Ref.~\cite{DeLuca:2020sae}, an overall PBH abundance of the order $f_\PBH \sim 10^{-4}$ would be enough to explain the rate of events in the pair-instability mass gap as inferred by the LIGO/Virgo observations \cite{Abbott:2020tfl}.  
This result depends on the rate of PBH accretion which is needed to avoid CMB constraints in that relevant mass window \cite{DeLuca:2020fpg}.  The redshift dependence of the GW source merger rate at high redshift is a key feature which can be explored to distinguish between the astrophysical BHs and PBHs from LIGO/Virgo \cite{Mukherjee:2021ags} and CE and ET~\cite{Mukherjee:2021itf}.   
LISA will be able to observe events in that mass window \cite{Kaiser:2020tlg} and thus potentially help in constraining the population properties of events falling in the pair-instability mass gap (see  Ref.~\cite{Abbott:2020mjq} and references therein).

\vspace{5mm}

\noindent \textbf{Capture in PBH clusters:}  The second most important PBH binary formation channel is through dynamical capture in dense PBH clusters.  Therefore the PBH clustering properties, which are still uncertain and highly model-dependent, mostly determine their overall merger rate.  In the absence of additional clustering compared to $\Lambda$CDM expectations, one has a rate  $R\thickapprox \mathcal O(1) \, {\rm yr^{-1} \, Gpc}^{-3}$ for equal-mass binaries~\cite{Bird:2016dcv}.  But in the case of wide mass functions, clustering may be  boosted as well as the merger rates, up to the values inferred by LIGO/Virgo if stellar-mass PBHs constitute a significant fraction of the DM.   One can encode the clustering  uncertainties in a single, rescaling parameter $R^{\rm clust}$~\cite{Vaskonen:2019jpv,DeLuca:2020jug}.  Then, the rate distribution can be approximated as~\cite{Bird:2016dcv,Clesse:2016vqa,Sasaki:2018dmp}
\begin{eqnarray}
    \frac{{\rm d} R^{\rm capt}}{{\rm d} \ln m_1 \d \ln m_2 } \approx {\rm Gpc^{-3}yr^{-1} }
    f_\PBH%^{53/21}
    R^{\rm clust}
    \frac{(m_1 + m_2)^{10/7}}{(m_1 m_2)^{5/7}} f(m_1) f(m_2). 
     \label{eq:ratescatpure2}
\end{eqnarray}
One has to notice though that both in the LIGO/Virgo and LISA mass ranges there are various severe constraints on the PBH abundance \cite{Carr:2020gox}, which are not sensitive to uncertainties on the PBH clustering, e.g.~those coming from the CMB anisotropies~\cite{Ali-Haimoud:2016mbv,Poulin:2017bwe,Inman:2019wvr,Serpico:2020ehh}.

\begin{figure}[ht]
    \centering
    Model 1 ($n_{\rm s}=0.97$)\\
	\includegraphics[width = 0.48\textwidth]{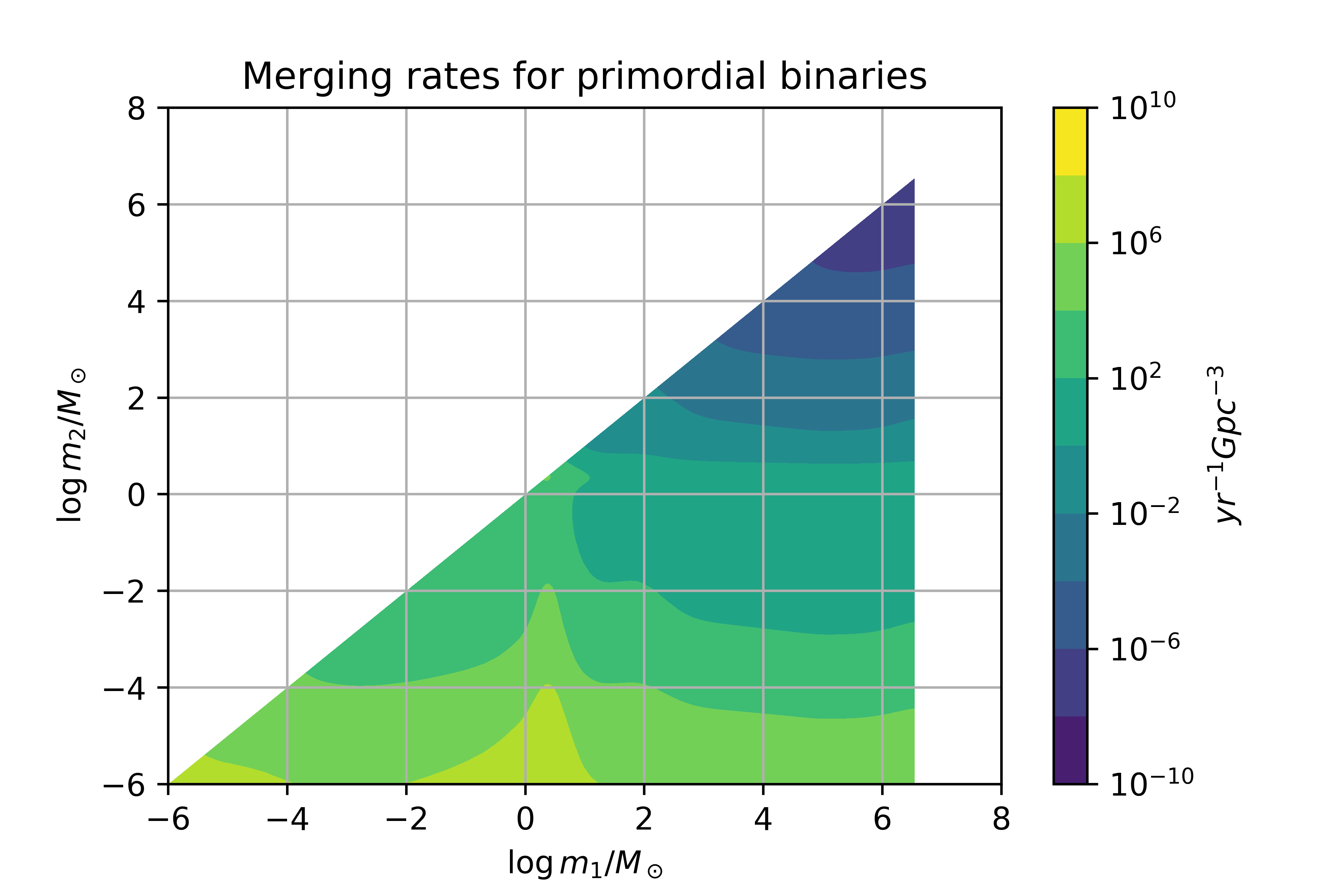}	\includegraphics[width = 0.48\textwidth]{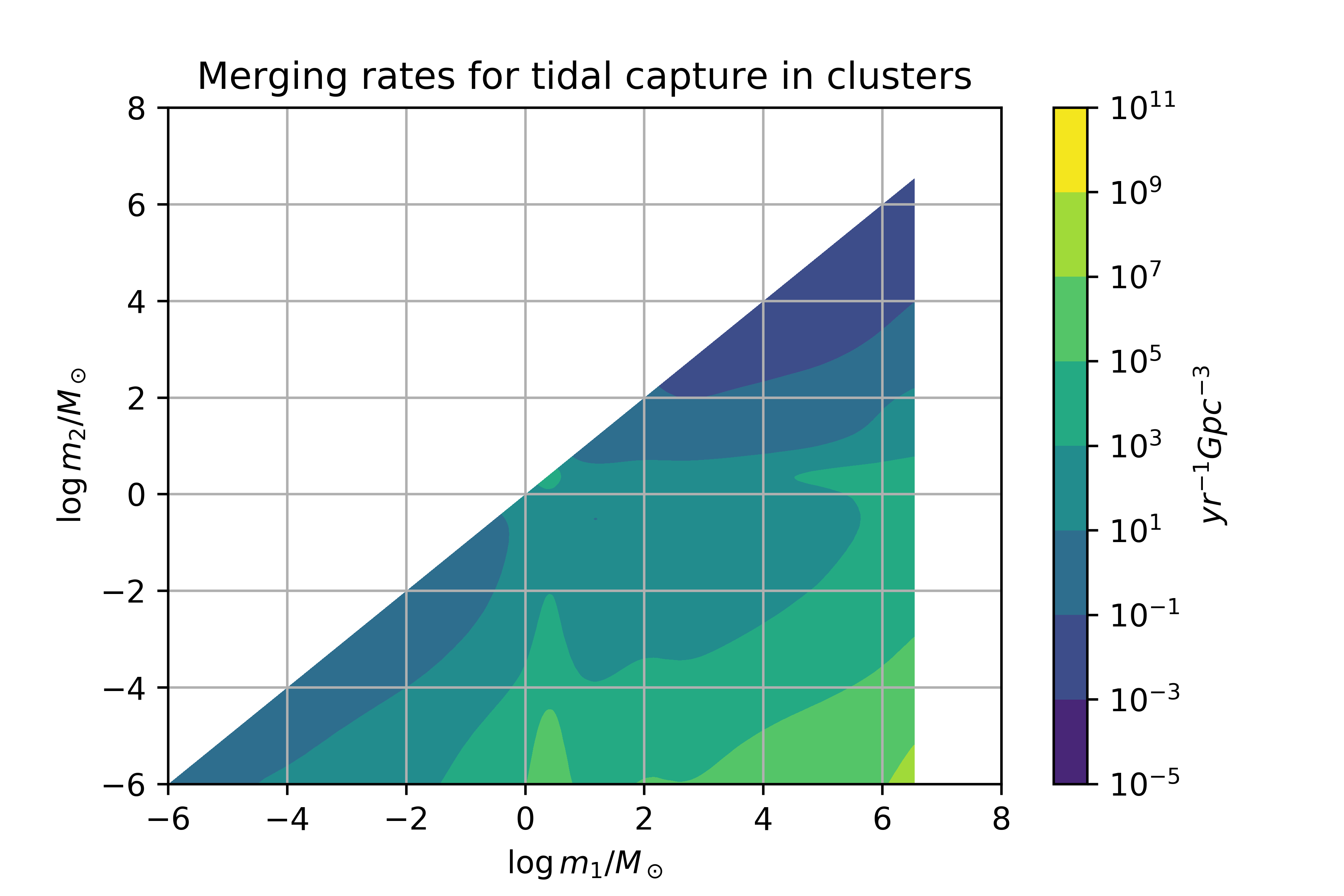}\\
	Model 2 ($n_{\rm s}=1$)\\
		\includegraphics[width = 0.48\textwidth]{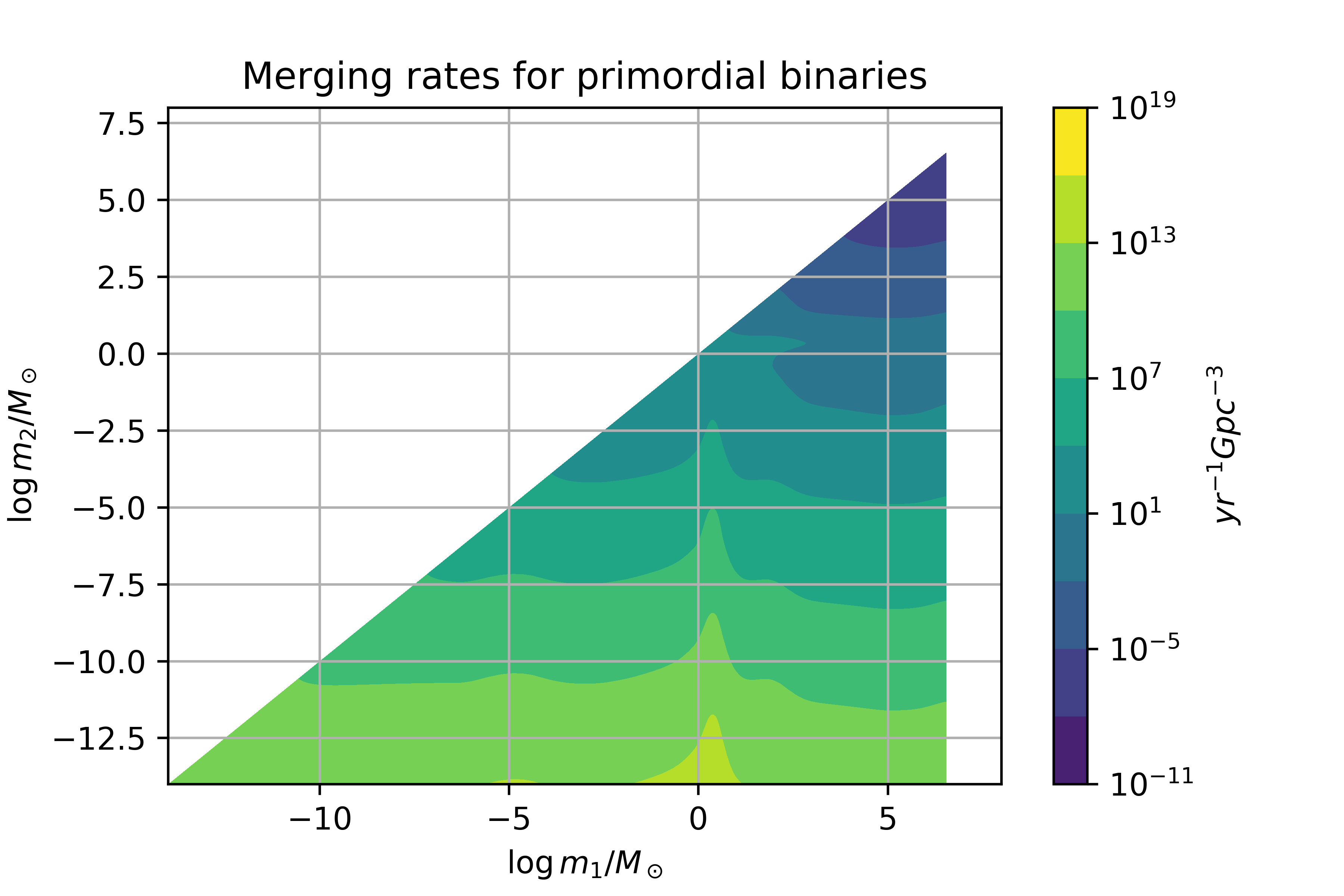}	\includegraphics[width = 0.48\textwidth]{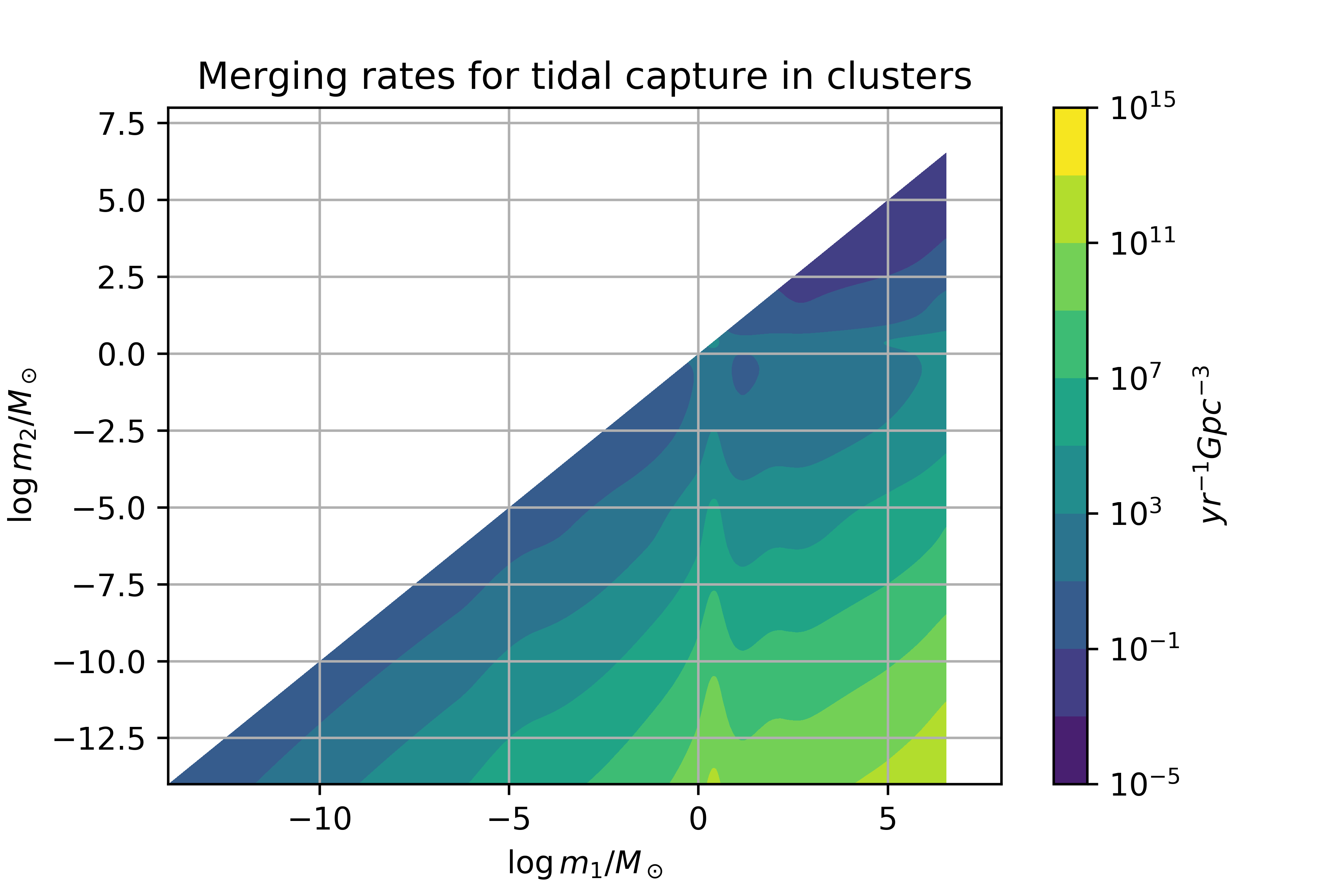}
	\caption{\small Expected merger rates of PBH of mass $m_1$ and $m_2$ for the  Model 1 (top panels) and Model 2 (bottom panels) mass distributions displayed in Fig.~\ref{fig:fPBH} due to the binary formation channels ``primordial binaries" (left panels) and ``tidal capture in halos" (right panels) coming from  Eq.~(\ref{eq:Rprimbinaries}) and Eq.~(\ref{eq:ratescatpure2}) respectively.}
    \label{fig:rates}
\end{figure}

\subsubsection{Gravitational wave chirps from primordial black holes binaries}

The merger rates expected for the two illustrative PBH models considered here are represented on Fig.~\ref{fig:rates}, for the two binary formation channels detailed in the previous section.  From these rates, one can identify three types of GW sources that are relevant for LISA.

\paragraph{Intermediate-mass binaries:}  The LISA frequency range is ideal to detect IMBBH mergers with mass above $10^3 M_\odot$.  PBHs can produce such merger events with rates of order $\sim 10^{-2} {\rm yr}^{-1} {\rm Gpc}^{-3}$ for PBHs in clusters, which could therefore be detected by LISA given the expected astrophysical reach for such events, typically above Gpc. 
The rate of equal-mass primordial binaries roughly goes like $\sim M^{-1}$ for primordial binaires and is thus relatively low, while it is constant for binaries in clusters. Merger rates are also much larger for Model 1 than for Model 2, due to a much larger PBH density fraction in this range.  It must be pointed out that this fraction is restricted by CMB limits.  These are not calculated for such wide-mass distributions of highly clustered PBHs, and so they should not be directly applied to the models.

\paragraph{Extreme mass ratios:}  Another distinctive feature of PBHs with wide mass distributions is to predict large merger rates of binaries with extreme mass ratios, especially those  involving a PBH from the QCD peak, merging with an intermediate mass PBH, as shown in Fig.~\ref{fig:rates}.  For  Model 1 and tidal capture in clusters, these rates can be larger than $10^3 {\rm yr}^{-1} {\rm Gpc}^{-3}$.  Therefore LISA could detect such extreme mass ratio coalescences, because of the ideal sub-Hertz GW frequency of such mergers and the large merger rates that allow for the compensation of the reduced strain sensitivity of LISA compared to the one of ground-based detectors at higher frequencies. 

\paragraph{Continuous waves from galactic binaries:}  Models with an important PBH fraction at planetary-masses and below typically lead to very high merger rates at low mass, which could allow for the detection of such binaries within our galaxy.  This is especially motivated for the Model 2 and for primordial binaries.  But in the LISA range, such binaries are still far from the time of merger, when the strain evolution is almost constant over the duration of an observing run.  For these kind of signals the continuous wave searches (see e.g.~\cite{Miller:2020kmv}) look promising.
The sensitivity of LISA to planetary-mass PBHs using such search techniques is still to be determined.

\subsubsection{Stochastic gravitational wave background}

The overlap of GWs from PBH binaries that are close to merger form together a SGWB that could be detected by LISA.  The possible spectral shape of this SGWB for the two possible binary formation channels, the constraints coming from observations and their implications for PBH models, have been discussed in the recent literature \cite{Mandic:2016lcn,Clesse:2016ajp,Wang:2016ana,Raidal:2017mfl,Chen:2018rzo,Wang:2019kaf}.   Hereafter we review the basic principle and formula to compute this SGWB and present some predictions for particular PBH models.  

The amplitude of the SGWB is given by summing up the energy spectrum of each binary system by taking into account the merger rate distribution as well as its possible evolution with redshift.  For binary systems with a circular orbit and for a generic PBH mass function leading to merger rates $R(m_1,m_2)$, e.g.~given by Eq.~(\ref{eq:Rprimbinaries}) for primordial binaries and by Eq.~(\ref{eq:ratescatpure2}) for capture in halos, one gets~\cite{Clesse:2016ajp}
\begin{eqnarray}
  \Omega_{\rm GW}(f)
  &=& \frac{8\pi^{5/3}G^{5/3}}{9H_0^2}f^{2/3}
  \int_0^{z_{\rm max}} {\rm d} z ~ 
  \frac{1}{H(z)(1+z)^{4/3}} \nonumber\\
  &\times&
  \int {\rm d} \ln m_1
  \int {\rm d} \ln m_2 ~
  R (m_1,\,m_2, z)
  {\cal M}_c^{5/3}(m_1,m_2)
\end{eqnarray}
where ${\cal M}_c$ is the chirp mass.  The maximum frequency in the observer's frame is determined by the innermost stable circular orbit which is given by $f_{\rm ISCO}\approx 4.4 {\rm kHz} ~ {M_\odot}/{((m_1+m_2) (1+z)}$. 
This can be translated to the maximum redshift as $  z_{\rm max}={f}/{f_{\rm ISCO}}-1$.

As an example, we used a peaked mass distribution around $2.5 M_\odot$ (motivated by the QCD peak) and computed the corresponding SGWB, for binaries formed in halos.
This SGWB is clearly detectable by LISA but probably below the sensitivity of LIGO/Virgo.   
By probing the amplitude and spectral index of this SGWB, LISA would be able to distinguish between different mass functions, different clustering properties, as well as the dominant binary formation channel. Indeed, the lower limit of the frequency range covered by PBH binaries depends on their relative velocity when the binary is formed, which itself depends on the typical cluster mass and radius.

\begin{figure}
    \centering
	\includegraphics[width = 0.8\textwidth]{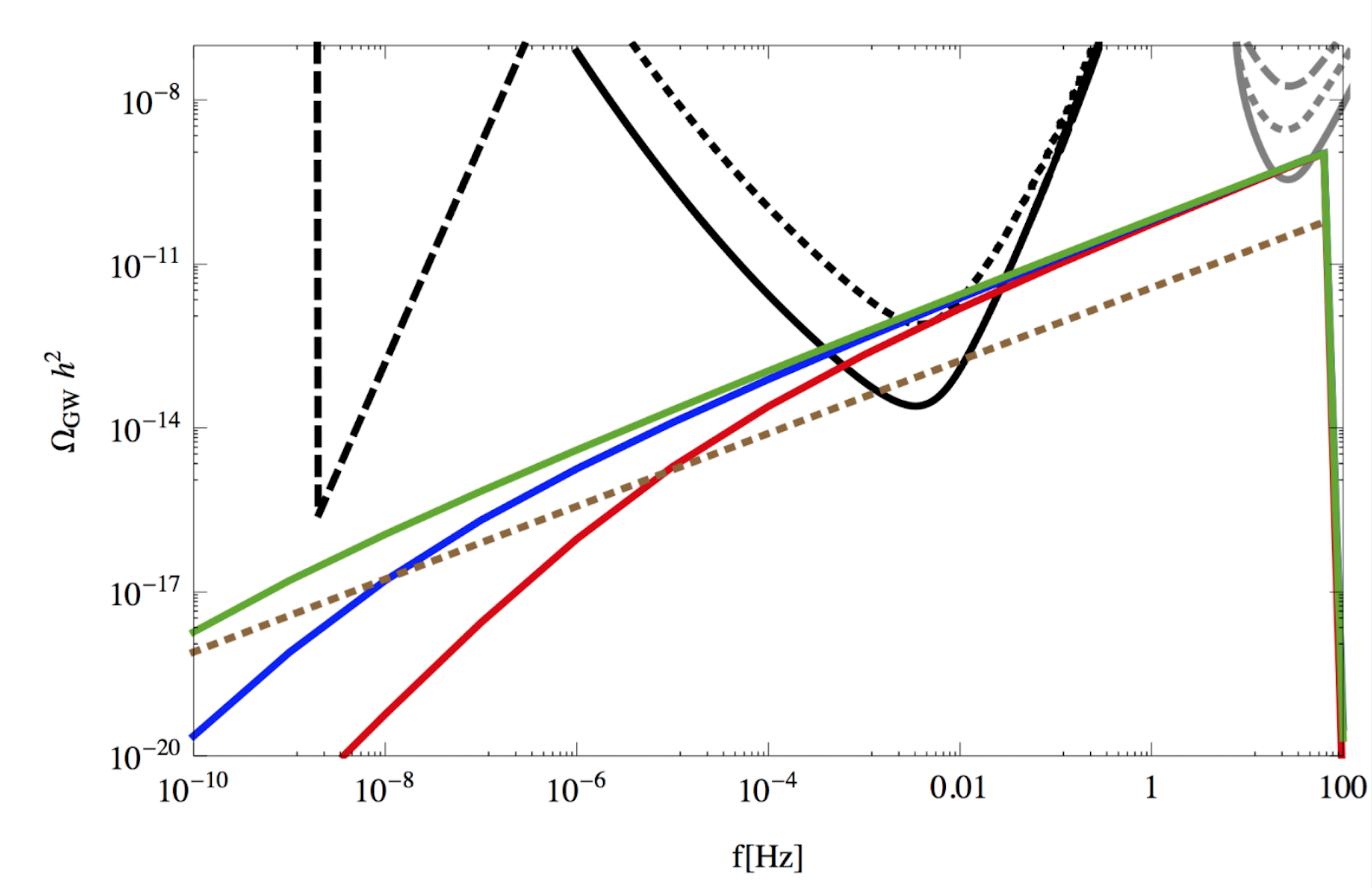}	
    \caption{\small Example of SGWB expected for PBH binaries formed by tidal capture in clusters, with rates given by Eq.~(\ref{eq:ratescatpure2}) and different virial velocities ($2$km/s in green, $20$km/s in blue and $200$km/s in red) and a peaked PBH mass distribution around $30 M_\odot$.  Figure taken from Ref.~\cite{Clesse:2016ajp}.  }
    \label{fig:SGWB_PBH_binaries}
\end{figure}

\subsubsection{Bursts from hyperbolic encounters}

If PBHs are at present grouped in clusters, a fraction of BH encounters will not
end up producing bound systems, but a single scattering event. This could happen e.g.~when the
relative velocity 
is too high for the capture to be possible. The GW signals from hyperbolic encounters can have frequencies, strain amplitudes and characteristic time durations that create GWs signals that can be detected by LISA. 
These events would have unique signatures~\cite{Garcia-Bellido:2017knh,Garcia-Bellido:2017qal} and would provide direct information about the orbital parameters and spatial distribution of these BHs, thereby providing complementary information to inspiral binaries, and strong evidence in favor of the scenario of clustered PBHs.

The waveforms of the GW emission in hyperbolic encounters are very different from
those of the inspiraling binaries, since the majority of the energy is released near the
point of closest approach, generating a burst of GWs with a characteristic "tear-drop" shape of the emission in the time-frequency
domain~\cite{Garcia-Bellido:2017knh,Garcia-Bellido:2017qal}.
The burst has a characteristic peak frequency 
\be
f_{\rm peak} = 0.32 \, {\rm mHz} \times \frac{\beta (e + 1)}{ (e -1) } \frac{{\rm AU}}{b } ,
\ee
which corresponds to the maximum GW amplitude and depends only on the impact parameter $b$, the total mass of the system $M$ and the eccentricity of the hyperbolic orbit $ e = \sqrt{1+b^2 v_0^4/G^2M^2}$), where $v_0$ is the asymptotic relative velocity of the encounter and $\beta \equiv v_0 / c$. 
The burst duration can be of the order of a few milliseconds to several hours (depending on the PBH masses and encounter parameters) and bright enough to be detected at distances up to several Gpc. The maximum strain amplitude of the GW burst is given by 
\bees
h^{\rm max}_{\rm c} &=& 3.24\times10^{-23}\frac{R_{\rm S} ({\rm km})}{d_{\rm L}({\rm Gpc})} \frac{q \beta^2 g_{\rm max}}{(1 + q)^2} \\
P_{max}(e) &=& 5.96\times 10^{26} L_\odot \frac{q^2 \beta^{10}}{(1 + q)^4} \frac{(e + 1)}{(e -1)^5},
\ees
where 
$L_\odot$ is the solar luminosity, 
$R_{\rm S}$ is the Schwarzschild radius, $q \equiv m_1/m_2$ is the black hole mass ratio, $m_1 = q m_2 \geq m_2$
and $g_{\rm max} = 2 \sqrt{18(e + 1) + 5e^2/(e -1)}$.

GWs detectable by LISA could be generated by close encounters of an intermediate-mass BH and a MBH as expected at the centers of galaxies, as well as from encounters of two MBHs that could occur during galactic collisions at low redshift.
In the first case, an intermediate-mass BH of mass $m_2 = 10^3 M_\odot$ and a MBH of mass $m_1 = 10^6 M_\odot$, with an impact parameter b = 1AU and velocity $v_0 = 0.05 \, c$, gives an eccentricity parameter $e = 1.031$  with  event duration of about 440\,s. 
The maximum
stress amplitude would be $h^{\rm max}_c=1.02\times10^{-19}$ at a distance $d_L = 1$ Gpc, with the peak at frequency $f_{peak} = 1.05$ mHz, well within the sensitivity band of LISA [17].
In a hyperbolic encounter between two MBHs of equal mass $m_1 = m_2 = 10^6 M_\odot$ with impact parameter $b = 10$ AU and relative velocity $v_0 = 0.015 c$,  the eccentricity is low $e = 1.01$ and the stress amplitude is huge $h^{\rm max}_c = 2.22\times10^{-17}$ at $f_{\rm peak} =1.51\times10^{-4} {\rm Hz}$, again right in the middle of the LISA observational band. Such an event would be clearly detectable.

\subsection{Massive black hole seeds from primordial black holes at high redshift} \label{sec:PBHhighz}

It is challenging to explain how MBHs can exist in only partially reionized environments at redshifts $z\gtrsim 7$~\cite{Banados:2017unc}.   The first populations of stars or the direct collapse of gas into BHs are two possible astrophysical mechanisms to generate the seeds of these MBHs.  But even if one invokes super-Eddington accretion, it is very challenging for these seeds to reach sufficiently large masses to explain observations.  PBHs are an alternative explanation to the existence of MBHs since they can provide seeds of intermediate-mass BHs at higher redshift than for the other astrophysical mechanisms~\cite{Duechting:2004dk,Kawasaki:2012kn,Clesse:2015wea,Bernal:2017nec}.  The easiest way to distinguish PBH seeds from other candidates is therefore to observe IMBBHs at  $z \gtrsim 20$, before star formation.  

\begin{figure}[t!]
\begin{centering}
\includegraphics[width = 1.
\textwidth]{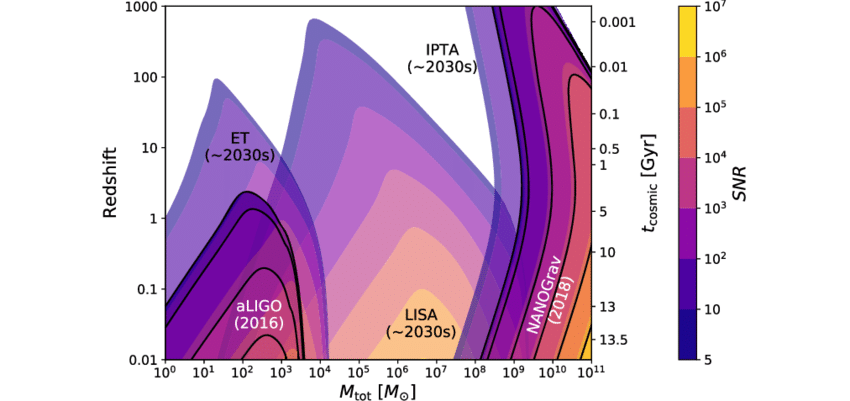}
\caption{\small Redshift range of LISA according to the analysis of \cite{Burke-Spolaor:2018bvk} for equal-mass BBH coalescences as a function of the total system mass, and comparison with the range of other detectors and pulsar timing arrays.  The color scale represents the expected SNR emerged from the study. Figure taken from Ref.~\cite{Burke-Spolaor:2018bvk}.}
 \label{fig:zrange}
 \end{centering}
\end{figure}

The astrophysical range of LISA will allow will allow for the observation of IMBBH mergers at redshifts $z>20$ with a SNR larger than five, for equal-mass mergers and progenitor masses between $10^3 M_\odot$ and $10^6 M_\odot$, as shown in Fig.~\ref{fig:zrange}.   The possible merger rates of PBHs for a broad mass function with the imprints of the thermal history, shown in Fig.~\ref{fig:rates},  can be larger than $\mathcal O(1) {\rm yr}^{-1}$ for primordial IMBBHs that would be formed in PBH clusters at high redshift.   The existence of these clusters is relevant since they would also form in the standard Press-Schechter theory.  LISA observations will be complementary to those of Earth-based GW detectors, like CE and ET, which will probe mergers with lower masses, and to future PTA limits from SKA, which will probe eventual mergers of MBHs at similar redshifts. 

\newpage
%%%%%%%%%%%%%%%%%%%%%%%%%%%%%%%%%
% Here Sec. 11 starts

\section{Tools/pipelines for the analysis of transient signal data in cosmology}
\label{sec:toolstrans}

\small \emph{Section coordinator: L.~Lombriser.
Contributors: T.~Baker, E.~Belgacem, G.~Calcagni, S.~Clesse, G.~Congedo,  J.M.~Ezquiaga, J.~Garcia-Bellido, D.~Laghi, L.~Lombriser, M.~Maggiore, A.~Raccanelli, M.~Sakellariadou, N.~Tamanini, M.~Zumalacarregui.
} \\

\noindent
This section presents a discussion of the tools available and in need of development for the analysis of the LISA transient signal data, including codes, pipelines, algorithms, and methodologies. Sec.~\ref{sec:tools_standard_sirens} focuses on standard sirens tests of the cosmic expansion (see Sec.~\ref{sec:standard_sirens}). Sec.~\ref{sec:toolsgwlensing} is devoted to the tools for GW lensing (Sec.~\ref{sec:gw_lensing}). Tools for tests of modified gravity (Sec.~\ref{sec:modified_gravity}) are presented in Sec.~\ref{sec:toolsmg}, and finally Sec.~\ref{sec:toolspbhs} discusses the tools for PBHs (see Sec.~\ref{sec:PBH}).

%%%
\subsection{Tools for testing the cosmic expansion with standard sirens} \label{sec:tools_standard_sirens}

As presented in Sec.~\ref{sec:standard_sirens}, GW observations are natural cosmological probes because their amplitude directly traces the luminosity distance as predicted by GR. 
When additional redshift information is obtained, GW events become standard sirens. Depending on whether the redshift information comes directly from an EM counterpart or from statistical inference, we will refer to them as bright and dark sirens respectively.
In both cases, modern GW data analysis techniques are based on Bayesian inference.
In the following we describe them independently.

\subsubsection{Bright sirens} \label{sec:tools_bright_sirens}

For bright sirens, the posterior for the cosmological parameters $\mathcal{H}=\{H_0,\Omega_m,\Omega_k,\Omega_\Lambda\}$ given GW data $\mathcal{D}_{\rm GW}$ and EM-wave data $\mathcal{D}_{\rm EM}$ can be written as~\cite{Chen:2020dyt}
\begin{equation}
 \label{eq:brightpos}
p(\mathcal{H}|\mathcal{D}_{\rm GW},\mathcal{D}_{\rm EM}) = p(\mathcal{H})
\frac{\displaystyle\int p(\mathcal{D}_{\rm GW}|\vec{\Theta})p(\mathcal{D}_{\rm EM}|\vec{\Theta})p_{\rm pop}(\vec{\Theta}|\mathcal{H})d\vec{\Theta}}{\displaystyle\int p_{\rm det}(\vec{\Theta})p_{\rm pop}(\vec{\Theta}|\mathcal{H})d\vec{\Theta}}\;,
\end{equation}
where $p(\mathcal{H})$ is the prior probability density on $\mathcal{H}$, and $\vec{\Theta}$ represents all the binary parameters. 
The population prior $p_{\rm pop}(\vec{\Theta}|\mathcal{H})$ is the probability density of binaries with parameters $\vec{\Theta}$ under our assumption of the rate evolution. The detection probability is given by
\begin{equation}
\label{eq:det}
p_{\rm det}(\vec{\Theta}) = \displaystyle \int\limits_{\substack{{\mathcal{D}_{\rm GW}>{\rm GW}_{\rm thr}} \\{\mathcal{D}_{\rm EM}>{\rm EM}_{\rm thr}}}} p(\mathcal{D}_{\rm GW}|\vec{\Theta})p(\mathcal{D}_{\rm EM}|\vec{\Theta})d\mathcal{D}_{\rm GW}d\mathcal{D}_{\rm EM}\;, 
\end{equation}  
in which the integration is only carried out over the data above the GW and EM detection thresholds, ${\rm GW}_{\rm thr}$ and ${\rm EM}_{\rm thr}$, respectively. 

GW and EM-wave observations can both introduce systematic uncertainties to the standard siren measurements. 
The measurements of the host redshifts can suffer from peculiar motions of the hosts~\cite{Howlett:2019mdh,Mukherjee:2019qmm,Nicolaou:2019cip}. The measurements of the binary luminosity distances 
are affected by the detector calibration uncertainty~\cite{2016RScI...87k4503K,Sun:2020wke}, lensing~\cite{Holz:2004xx,Hirata:2010ba}, and the accuracy of the waveforms~\cite{Abbott:2018wiz}. In addition, as shown in Eq.~\eqref{eq:brightpos}, our understanding of the possible observational selection effects~\cite{Chen:2020dyt} as well as the astrophysical rate evolution (see e.g.~Fig.~12 of Ref.~\cite{Finke:2021aom} for an application to LIGO-Virgo data)
are critical to the accuracy of the standard siren as well.

\subsubsection{Dark sirens} \label{sec:tools_dark_sirens}

In the absence of an EM counterpart, one may extract redshift information by putting a prior on potential hosts from a galaxy catalogue.\footnote{This method assumes that the true source is not of primordial origin.}
In this case, since the galaxy catalogue is compiled in redshift and does not carry any dependence on the parameters $\mathcal{H}$, the natural choice is to make a change of variables  $d_L \mapsto z(d_L, \mathcal{H})$ in Eq.~\eqref{eq:brightpos}, and use the redshift $z$ instead of the luminosity distance $d_L \in \vec{\Theta}.$\footnote{Note that the GW likelihood $p(\mathcal{D}_{\rm GW}|\vec{\Theta})$ is unaffected by this transformation since it is a PDF with respect to the data $\mathcal{D}_{\rm GW}$.} Denoting $\vec{\Theta} = \{ d_L, \hat{\Omega}, \vec{\Theta}' \}$, Eq.~\eqref{eq:brightpos} reads
\begin{equation}
    p(\mathcal{H}|\mathcal{D}_{\rm GW}) = 
\frac{p(\mathcal{H})}{\alpha(\mathcal{H})}  \displaystyle\int dz d\hat{\Omega} d\vec{\Theta}' \, p(\mathcal{D}_{\rm GW}|d_L(z, \mathcal{H}), \hat{\Omega}, \vec{\Theta}')p_{\rm pop}(z, \hat{\Omega}, \vec{\Theta}'|\mathcal{H})\, ,
\end{equation}
where $\alpha(\mathcal{H})\equiv\int p_{\rm det}(\vec{\Theta})p_{\rm pop}(\vec{\Theta}|\mathcal{H})d\vec{\Theta}$ and we set $p(\mathcal{D}_{\rm EM}|\vec{\Theta})=1$, being in the absence of a counterpart. Note that the GW likelihood now carries an explicit dependence on the parameters $\mathcal{H}$. 
The population prior on redshift and position is now naturally expressed as
\begin{equation}\label{eq:galCatPrior}
p_{\rm pop}(z, \hat{\Omega} | \mathcal{H} ) \propto \sum_i w_i   \,  p( z_{i}, \hat{\Omega}_{i} | z, \hat{\Omega}  )  \,,
\end{equation}
where $p(z_i, \hat{\Omega}_i | z, \hat{\Omega})$ is the likelihood of the redshift and position of the $i$-th galaxy in the catalogue, with observed values $z_{i}, \hat{\Omega}_{i}$.
The angular position can be assumed to be known with very good precision -- since the GW likelihood does not vary significantly over the scale of the galaxy location -- and the corresponding distribution is just a delta function about the observed values. As for the redshift, one can further put a prior that is usually taken as uniform in comoving volume~\cite{Chen:2017rfc,Soares-Santos:2019irc,Palmese:2020aof}, while the likelihood can be approximated by a Gaussian centered around the observed redshift $z_{i}$ and with standard deviation given by the redshift uncertainty.\footnote{However, the true PDF should be used when available, and this choice can have a non-negligible impact on the result~\cite{Palmese:2020aof}.}
Finally, the factors $w_i$ in Eq.~\eqref{eq:galCatPrior} represent the possibility of assigning different weights to the galaxies. Common choices are the B-band luminosity, which traces the star formation rate, or K-band luminosity that traces the total stellar mass.

One has then to deal with the fact that galaxy catalogues are in general not complete.
This requires one to include information on the missing galaxies, and the prior in Eq.~\eqref{eq:galCatPrior} must be supplemented by a suitable ``completion'' term~\cite{Chen:2017rfc,Fishbach:2018gjp,Gray:2019ksv,Finke:2021aom}, which requires two pieces of information. The first is the knowledge on the ``completeness'' of the catalogue, that is usually computed by comparing the luminosity distribution in the catalogue to the one given by a Schechter function~\cite{Chen:2017rfc,Fishbach:2018gjp,Finke:2021aom}.\footnote{A slightly different, though comparable, approach is to include the possibility that the true host is not in the catalogue in the formalism~\cite{Gray:2019ksv}.} The second is to specify a ``completion'' procedure, i.e.~to determine how the missing galaxies are spatially distributed. The simplest options can be to distribute them uniformly~\cite{Chen:2017rfc,Fishbach:2018gjp}, or to assume that the catalogue traces well the actual structures, hence assigning higher weight to the galaxies that are in low completeness regions~\cite{Finke:2021aom}, or even assuming that completion is not needed~\cite{Soares-Santos:2019irc,Palmese:2020aof, Laghi:2021pqk}. Eventually, the best choice depends on the galaxy survey, its sky coverage and on the completeness of the catalogue. A more refined procedure is to use the uniform, uninformative distribution in regions of low completeness, while switching to the second option where the galaxy survey traces the actual structures fairly well~\cite{Finke:2021aom}. 
This aspect will become more important as the precision on the localisation increases but will remain a limiting factor until catalogues with very large completeness are available. 
Since the completeness drops drastically with increasing redshift, this can be limiting for LISA sources at high redshift when using catalogues that have a large sky coverage but limited redshift range. Another option is to use deeper catalogues with limited sky coverage, but this requires to take into account the variation of the completeness with the angular position in the sky~\cite{Finke:2021aom}.
In any case, the availability of complete galaxy catalogues, accurate determination of the redshifts, and small GW localisation regions will be crucial in order for the statistical method to give competitive constraints on cosmological parameters.

Finally, a good understanding and an accurate computation of the selection bias is required~\cite{Chen:2020dyt, Finke:2021aom}. This is related to other relevant systematic uncertainties that deserve further investigation, related to our limited knowledge of the population properties of BBHs. In particular, the population prior should encode information on the mass distribution of the system's component and on the rate of BBH formation as a function of redshift. At the moment, very mild constraints exist on both in the case of SOBBH~\cite{Fishbach:2018edt, Abbott:2020gyp}, but their impact -- especially of the evolution of the BBH merger rate -- can be relevant~\cite{Finke:2021aom}.

A public \texttt{Python} code implementing the completion procedures described above, as well as calculation of selection effects, is available at 
\url{https://github.com/CosmoStatGW/DarkSirensStat}.
The code has been applied to the latest data release from the LIGO-Virgo collaboration for constraining $H_0$ and modified GW propagation and supports the combination with the GLADE, DES and GWENS galaxy catalogues, but it could easily be be adapted for LISA and other galaxy catalogues.

%%%
\subsection{Tools for gravitational wave lensing} \label{sec:toolsgwlensing}

GW lensing has attracted considerable attention recently, leading to different groups developing tools to compute amplification factors and lensed waveforms. 
In order to maximise the science yield of gravitational lensing with LISA, it is important to develop codes to compute amplification factors for general lenses and including wave effects.

\texttt{LensingGW} is a publicly available tool for GW lensing in geometric optics. It is a \texttt{Python} package designed to compute amplification factors for general lenses, adapted to the needs of GW observations~\cite{Pagano:2020rwj}. Currently, \texttt{LensingGW} operates in the geometric optics regime, using a selective sub-tiling to find microimages, their associated magnifications and time delays, cf.~Eq.~\eqref{eq:lensing_geometric_optics}. The code is based on the \texttt{lenstronomy} package~\cite{Birrer:2018xgm}, allowing it to draw from its library of algorithms and lens models. 

It is necessary to develop wave optics codes able to solve Eq.~\eqref{eq:lensing_wave optics}. The diffraction integral is highly oscillatory, which makes numerical solutions challenging. Proposed methods to address this problem use the time delay function~\cite{Ulmer:1994ij} and Fourier transforms to the frequency domain. Variants of these ideas have been implemented in Refs.~\cite{Diego:2019lcd,Cheung:2020okf} and others.
These codes are not publicly available.
An alternative approach employs Picard-Lefschetz theory~\cite{Feldbrugge:2019fjs}. This is valid only for lensing potentials that are meromorphic functions, which may be used to approximate non-analytic lensing potentials. 
Finally, corrections to geometric optics rely on higher derivatives of the Fermat potential around images~\cite{Takahashi:2004mc} and could be easily implemented in a geometric optics formalism.

Predictions on GW propagation beyond GR are notoriously complicated. Ref.~\cite{Ezquiaga:2020dao} develops a \texttt{Python} package to study GW birefringence. The code 
is able to (1) compute static spherical backgrounds in Horndeski theory; (2) compute the local propagation eigenstates and their speed as a function of position and direction; (3) solve the geodesics in the Born approximation to compute time delays and deflection angles for each eigenstate; and (4) obtain the waveform, including birefringence.
Future tools to study GW lensing should improve by (a) including higher-order corrections in the Wentzel-Kramers-Brillouin (WKB) expansion to compute the amplitude and diffraction effects (or optimally, arrive at a full wave optics expression); (b) include more general theories of gravity; and (c) work with more general backgrounds (for instance, performing GW ray tracing on modified gravity simulations). These codes could be interfaced with a statistical sampler to analyse real or mock data.
%%%

\subsection{Tools for testing modified gravity} \label{sec:toolsmg}

The sensitivity of LISA to modified GW propagation introduced by a frequency independent modification of the damping term was investigated in Ref.~\cite{Belgacem:2019pkk}. A MCMC analysis was conducted on LISA mock catalogues of MBBHs with EM counterparts, where luminosity distances were evaluated according to a fiducial $\Lambda$CDM cosmology. As discussed in  Refs.~\cite{Belgacem:2018lbp,Belgacem:2019pkk}, in the context of late-time modifications of GR invoked for DE studies, the most relevant parameters affecting GW propagation are the high-redshift ratio of GW to EM luminosity distances $\Xi_0$, defined in Eq.~\eqref{eq:fit}, the DE EoS $w_0$, the Hubble parameter $H_0$, and the current matter energy density fraction $\Omega_m$. The exponent $n$ of the parametrisation in Eq.~\eqref{eq:fit} plays a less important role and was fixed to a reference value.
In order to reduce the degeneracies among those parameters and assess the potential of LISA for cosmological parameter inference, in addition to LISA mock catalogues, the analysis in Ref.~\cite{Belgacem:2019pkk} also included further currently available datasets such as CMB measurements from Planck 2015, Type~Ia Supernovae (JLA catalog), and a collection of BAO measurements. The likelihood of the LISA mock data given the parameters \{$H_0,\Omega_m,w_0,\Xi_0$\} was assumed to follow
\begin{equation}\label{eq:GWlikelihood}
\ln(L(H_0,\Omega_m,w_0,\Xi_0))=-\frac{1}{2}\sum_{i=1}^{N_s}\frac{\left[d_L^{\,\rm GW}(z_i;H_0,\Omega_m,w_0,\Xi_0)-d_i\right]^2}{\sigma_i^2}\, ,
\end{equation}
where an additive constant in the logarithm of the likelihood from normalization is omitted.
Here $N_s$ is the number of mock sources in the catalogue, $d_L^{\,\rm GW}(z_i;H_0,\Omega_m,w_0,\Xi_0)$ is the theoretical value of the GW luminosity distance for the $i$-th source (at redshift $z_i$), $\sigma_i$ is the error on luminosity distance taking also into account the error on redshift determination, and $d_i$ is the ``measured" value of the luminosity distance of each event contained in the catalogue (obtained by scattering the fiducial $\Lambda$CDM prediction with a Gaussian distribution with variance $\sigma_i^2$). The MCMC code of Ref.~\cite{Belgacem:2019pkk} explores the cosmological parameter space, accepting or rejecting sampled points following a Metropolis-Hastings algorithm. The total likelihood used in the algorithm is obtained by multiplying the independent contributions from each of the four datasets employed, namely from CMB, Type~Ia supernovae, BAO, and LISA. The cosmological evolution, both at the level of the background and perturbations, is computed using the \texttt{CLASS} Boltzmann code. The modified version of \texttt{CLASS}, implementing the GW luminosity distance parametrisation in Eq.~\eqref{eq:fit}, is available at \url{https://github.com/enisbelgacem/class_public}. The MCMC code is built on \texttt{MontePython} with the inclusion of the LISA mock source catalogues and the likelihood from Eq.~\eqref{eq:GWlikelihood}, which uses the modified \texttt{CLASS} version. The code is available at \url{https://github.com/enisbelgacem/montepython_public}.

In addition to the test of the model-independent parameters $\{\Xi_0, n\}$, similar analyses have been carried out using the modified luminosity distances of specific gravity models, e.g.~Horndeski gravity~\cite{Lombriser:2015sxa,Baker:2020apq} or $f({\cal Q})$ gravity~\cite{Frusciante:2021sio}. These make use of the modified Boltzmann codes \texttt{hi$\_$CLASS}~\cite{Zumalacarregui:2016pph, Bellini:2019syt} and \texttt{EFTCAMB}~\cite{Hu:2013twa, Raveri:2014cka}, respectively. A wrapper to interface \texttt{hi$\_$CLASS} with the MCMC code \texttt{Cosmosis}~\cite{Zuntz:2014csq} is available at \url{https://github.com/itrharrison/hi_class}.

Current tools are predominantly specialised to probing the effects of frequency-independent modifications of the GW propagation, although frequency-dependent effects have been studied in the context of GW oscillations in bigravity~\cite{Belgacem:2019pkk}, where the associated numerical tools have however not been made publicly available yet.
The desire for UV-completion of modified gravity theories, whilst maintaining IR phenomenology, also motivates frequency-dependent modifications of the GW propagation equation~\cite{deRham:2018red}. Such modifications could be detectable in the LISA band whilst being suppressed to irrelevant levels in the band of ground-based detectors ($\sim$10 -- 1000~Hz). Their basic phenomenology corresponds to a frequency-dependent GW luminosity distance or a frequency-dependent GW propagation speed.

To study the detectability of these effects, one must first construct a motivated and manageable parameterisation for the frequency and redshift dependence of $c_T(f,z)$ and $\delta(f,z)$ that appear in Eq.~\eqref{prophmodgrav}. One then needs to solve for the amplitude and phase evolution of the waveform under this parameterisation, up to some appropriate post-Newtonian order. A Fisher forecast code, ideally calibrated against a smaller, more rigorous (but computationally expensive) MCMC forecast, can analyse the ability of LISA to constrain the frequency-dependent modifications to $c_T$ and $d_{\rm GW}$. The parameterisation must be carefully chosen to minimise degeneracies between the MG and standard source parameters. The creation of these tools is in progress~\cite{Baker:2022rhh}.

Finally, we note that sources of anisotropic stress in the universe act as source terms on the right-hand side of Eq.~\eqref{prophmodgrav}. Neutrinos are one such example, although their effects on GW propagation in the late-time universe are expected to be extremely small. However, their effects on GWs travelling through the early universe may be more significant. Analytical tools for solving GW evolution in the presence of neutrino anisotropic stress are presented in Ref.~\cite{Wren:2017bqz} and could likely be adapted for a generic imperfect fluid with anisotropic stress.

%%%
\subsection{Tools for PBHs} \label{sec:toolspbhs}

The recent developments in the field of PBHs have revealed a rather rich phenomenology, e.g.~related to PBH formation with a broad variety of primordial power spectra and new classes of models relying on non-trivial modifications of the statistical distribution of curvature fluctuations; related to the history of PBH clustering, accretion and mergings, etc.  The computations of GW observables therefore need to become more accurate and have to take into account model dependencies and various astrophysical uncertainties.  It becomes increasingly difficult to integrate all the recent development in new analyses.  Therefore, there is need in the community for an advanced numerical tool that progressively integrates all these recent developments at various levels (theoretical models, PBH formation, clustering, accretion, merging rates, SGWBs) in a unified and modular manner.   This is the main objective of an ongoing project of the LISA Cosmology Working Group that is developing the \texttt{PrimBholes} toolbox for the computation of model-dependent gravitational observables that are relevant for LISA and other GW experiments.  The \texttt{PrimBholes} toolbox is currently under development and will be made publicly available soon along with both a code companion paper and a review paper on PBHs.

The different computations performed with \texttt{PrimBholes} and the possible options and effects that will be implemented for selection are the following:

\begin{enumerate}

\item For the primordial curvature fluctuations this will include (a) several phenomenological models of power spectra (power-law, log-normal, Gaussian, broken power-law, etc.); (b) Gaussian and non-Gaussian perturbations with a generic distribution function and some specific realisations; and (c) generic power spectra and/or distribution functions imported from a file provided by the user.

\item The SGWB from second-order perturbations will be calculated for general as well a number of specific shapes of the primordial power spectrum.

\item The computation of the PBH formation and density distribution $\beta(m_{\rm PBH})$ at formation will be performed following (a) the standard formalism of Sec.~\ref{sec:PBHformstd} and (b) an advanced method and algorithm presented in Ref.~\cite{Musco:2020jjb} accounting for non-linear effects and the shape of the primordial power spectrum. This will include (c) effects of the evolution of the EoS through the thermal history of the universe and (d) a \textit{reversed} method to rescale the amplitude of primordial curvature fluctuations in order to get a fixed value of $f_{\rm PBH}$.

\item To obtain the late-time PBH mass function $f(m_{\rm PBH})$, \texttt{PrimBholes} will rely on the standard formalism and include additional effects from accretion and hierarchical mergers.

\item The distribution of merging rates $R(m_1,m_2)$ will be computed for (a) primordial binaries, accounting for the effects from the formation of early clusters due to the Poisson noise, from matter inhomogeneities and from nearby BHs, and (b) for tidal capture in clusters, accounting for different clustering histories and halo mass functions. This will include (c) the dependence on the redshift of the merging rates.

\item Finally, \texttt{PrimBholes} will include the spin distribution, based on Ref.~\cite{DeLuca:2019buf}; (7) the rate of hyperbolic encounters, based on Ref.~\cite{Garcia-Bellido:2017knh}; and (8) the SGWB from primordial binaries, for the two aforementioned formation channels.  

\end{enumerate}

\texttt{PrimBholes} will also provide plotting modules that will facilitate the production and exportation of key figures. Examples are given in Figs.~\ref{fig:fPBH} and~\ref{fig:rates}.

In a first step, the \texttt{PrimBholes} code will focus on the computation of GW observables without including a likelihood module for LISA. However, it will be made modular so that it can be used in combination with other codes such as the LISA analysis code for the SGWBs or the Botzmann code \texttt{CLASS}~\cite{Blas:2011rf}.  Later, likelihood modules for the \texttt{MontePython} code~\cite{Brinckmann:2018cvx} could be included for the  computation of the various astrophysical and cosmological limits on the PBH abundance.

\newpage
%%%%%%%%%%%%%%%%%%%%%%%%%%%%%%%%%
% Here Sec. 12 starts

\section{Tools/pipelines for the analysis of stochastic gravitational wave background data}
\label{sec:pipesgwb}

\small \emph{Section coordinators: N.~Karnesis, M.~Peloso, M.~Pieroni. Contributors: N.~Bartolo, G.~Boileau, N.~Christensen, C.~Contaldi, V.~Desjacques, R.~Flauger, N.~Karnesis, V.~Mandic, S.~Mataresse, M.~Peloso, M.~Pieroni, A.~Renzini, A.~Ricciardone, J.~Romano,  M.~Sakellariadou, L.~Sorbo, J.~Torrado.   
}\\ \normalsize

\noindent The SGWB to be measured by LISA consists mainly of the confusion noise created by the overlapping signals from unresolved astrophysical events, and possibly of a cosmological component due to one or more physical mechanisms taking place at different stages of the late or early universe (e.g.~the mechanisms discussed in Secs.\ref{sec:Inflation}, \ref{sec:CosmicStrings} and \ref{sec:PTs}). LISA will open a completely new window into the parameter space of each of these phenomena and has the potential to produce new discoveries about the physics of the early universe through the characterisation of this primordial SGWB component. A careful characterisation of the total background is also necessary to aid the detection and identification of individual events. In this section, we discuss the features of each contribution to the background signal, the instrumental noise, and the different tools we need to develop to separate and characterise all of of them.
	
	We start with Sec.~\ref{sec:Noise} which summarises the main characteristics of the noise in LISA. We comment on existing noise models, the assumptions on which they are built, and on possible improvements. In Sec.~\ref{sec:sgwb:foregrounds} we discuss foregrounds (i.e.~astrophysical sources of a SGWB) by describing the possible sources and the characteristics and main properties of the different signals. In Sec.~\ref{sec:sgwb:reconstruction} we then focus on frequency shape reconstruction methods i.e.~methods which aim at recovering the spectral profile of an unknown SGWB. We proceed by discussing anisotropy reconstruction i.e.~recovering the angular structure hidden in the signal. For this topic both theoretical and data oriented approaches are covered in Sec.~\ref{sec:anisotropy_rec} and in Sec.~\ref{sec:map_making} respectively. Finally, in Sec.~\ref{sec:global_fit} we frame the different topics touched in this section in the big picture of the so called \emph{global fit} problem. For all of these points the discussion covers both existing techniques and possible further developments.

	\subsection{Noise modelling}\label{sec:Noise}
\begin{figure}
		\centering
		\includegraphics[width = 0.48 \textwidth]{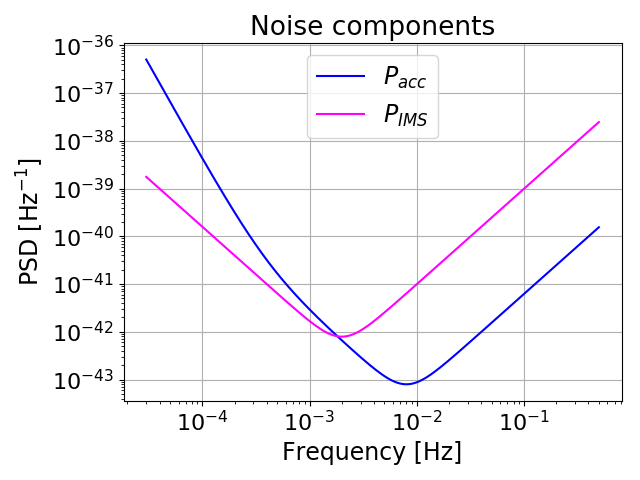}
		\includegraphics[width = 0.48 \textwidth]{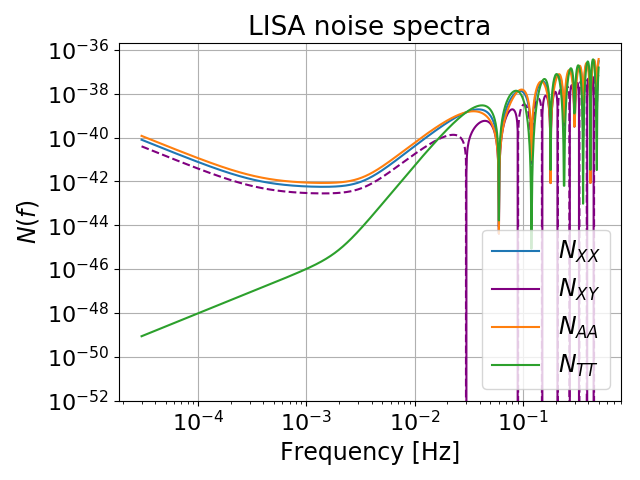}
		\caption{\small Left panel: IMS and acceleration noise power spectra expressed in the simplified scenario where all test-mass and laser noises are equal for all space-crafts. Right panel: LISA noise spectra in the $XYZ$ and $AET$ basis as given in e.g.~in Ref.~\cite{Flauger:2020qyi}. Plots taken from Ref.~\cite{Flauger:2020qyi}. \label{fig:PSD_strain} }
	\end{figure}

	The instrumental noise in LISA is expected to be non-stationary, and at the same time we expect noise transients that cause the statistical properties of the noise to depart from Gaussianity. At a first approximation we can assume that the short noise transients (i.e.~glitches) will be modelled, identified, and removed from the data streams~\cite{Robson:2018jly}. However, slow variations of the noise, like for example the slow decrease of the acceleration noise of the test-masses due to outgassing~\cite{Armano:2018kix}, are due to known effects that can be properly modelled and considered in the analysis. We can tackle those problems by adopting a piecewise analysis, where we can consider that each data segment can be assumed as stationary and Gaussian. In order to eliminate the dominant noise sources (i.e.~fluctuations of the laser central frequency and displacements of the of the optical benches), LISA will employ time domain interferometry (TDI)  techniques~\cite{Tinto:1999yr, Estabrook:2000ef, Tinto:2002de, Armstrong:2003ut, Tinto:2003vj, Tinto:2020fcc, Vallisneri:2020otf}. In the simplified scenario considered in most of the literature~\cite{Bartolo:2019oiq, Caprini:2019pxz, Contaldi:2020rht, Smith:2019wny, Pieroni:2020rob, Flauger:2020qyi, Orlando:2020oko} the residual noise for each arm link has two main components: the ``acceleration'' and the ``interferometric'' noise components. The acceleration component is associated with the random force noise acting on the test masses inside each of the three satellites, due for example to local environmental disturbances, and it dominates the low frequency part of the LISA band. The interferometric noise is directly connected to the interferometry metrology system (IMS). It describes the random readout noise of the optical system (mostly due to shot noise), and dominates the spectrum at high frequencies. For the 1.5 TDI variables~\cite{Bayle:2018hnm,Babak:2021mhe}, one can follow the recipe of Ref.~\cite{Flauger:2020qyi} and construct a likelihood based on the $\{ $X, Y, Z$\}$ TDI channels. In principle, analogous procedures can be defined for any kind of TDI variable combination.
	In order to remove noise correlations between different channels, it is customary to introduce an alternative TDI basis, typically dubbed the $AET$ basis, which diagonalises the noise matrix. The total noise spectra for both $XYZ$ and $AET$ basis are shown in the right panel of Fig.~\ref{fig:PSD_strain}. On the other hand the left panel of Fig.~\ref{fig:PSD_strain} shows the acceleration and IMS noise power spectra.
	
	Naturally, searching for SGWB signals requires a sufficient knowledge of the noise spectral shape, and it is possible that a simplified model such as the two-parameter model described above might not suffice for the signal search. Any instrument calibration mismatch will impact the joint fit of the noise and signal parameters, which may jeopardise our abilities to measure the underlying SGWB spectrum. Therefore, a possible solution to this problem would be to incorporate the calibration of the instrument in the analysis, by allowing a flexible model of the noise spectral shape. The development of methods (and of corresponding tools) which do not explicitly depend on analytical parameterisation for the noise spectra, will be an interesting research line to be explored over the years before the launch of LISA. 
	
	Another important aspect for searching for SGWB signals is assessing the quality of the data before they reach the designated pipelines. The use of signal subtraction, if implemented for LISA data, could result in imperfect subtraction and hence residuals, which in turn would affect the statistical properties of the noise~\cite{Thrane:2013kb,Romano:2016dpx,Smith:2017vfk,Ginat:2019aed}. The impact of theses residuals on the search for SGWBs with LISA is yet to be investigated. A better framing of this problem in the broader picture of the global fit problem for LISA is presented in Sec.~\ref{sec:global_fit}.

	\subsection{Expected astrophysical foregrounds}
	\label{sec:sgwb:foregrounds}
	
\begin{figure}[htb]
		\centering
		\includegraphics[width=\textwidth]{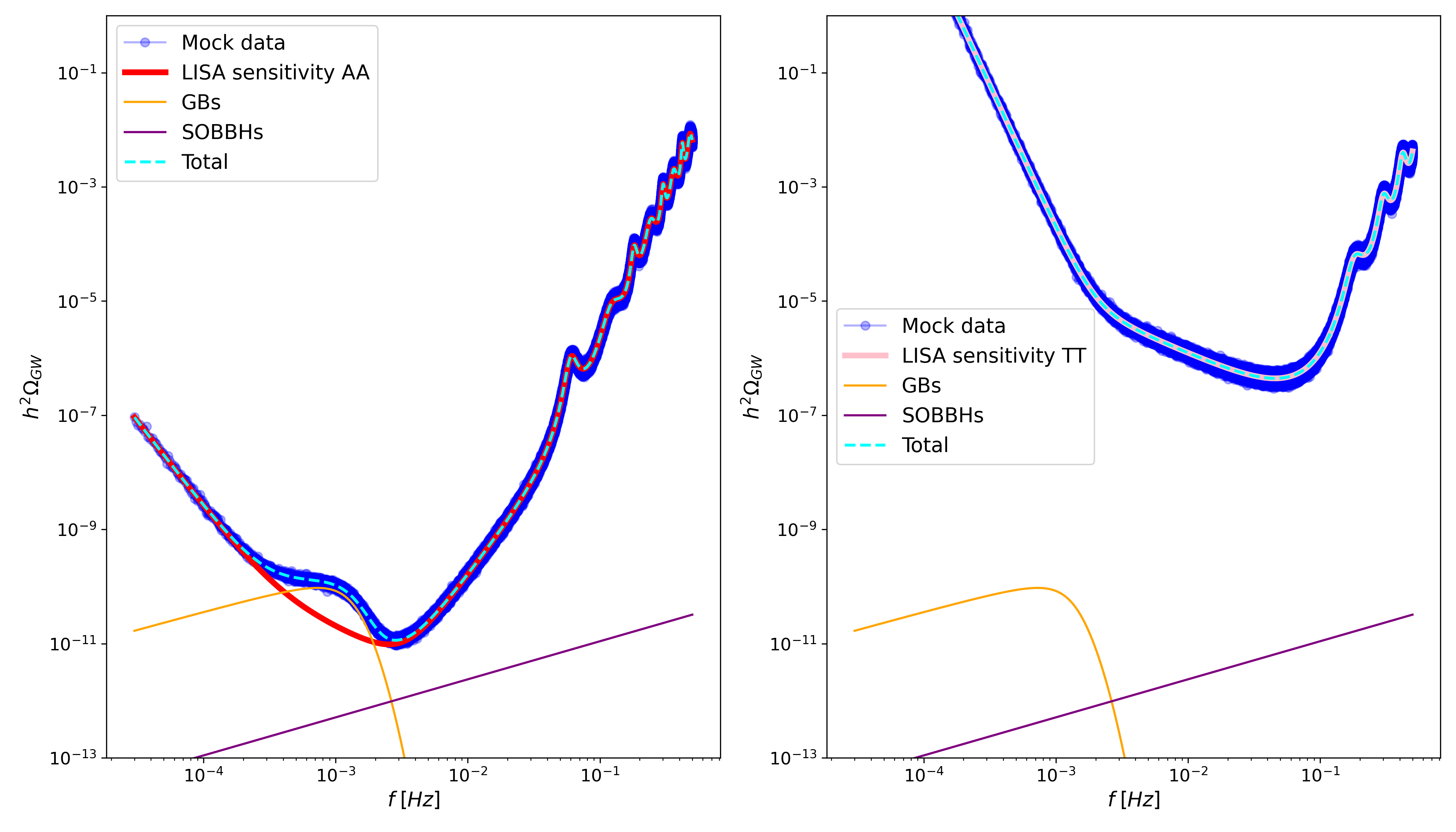}
		\caption{\small Theoretical models for the LISA sensitivity in the self-correlation of the $A$ and $T$ TDI data channels ($AA$ and $TT$, respectively) and for the SGWBs due to GBs and SOBBHs as specified in Ref.~\cite{Flauger:2020qyi}. Mock data for the sum of all these theoretical templates are shown for reference.}
		\label{fig:my_labelsloppy}
	\end{figure}

	Due to the richness of the sky in the milliHertz band LISA will be sensitive to a wide variety of sources such as  MBBHs with masses $ \sim 10^4-10^7 M_\odot$, SOBBHs,  EMRIs, and GBs. Beyond the resolvable sources, measurements by LISA will also be affected by a huge number of unresolvable events which will sum up incoherently, leading to the generation of a SGWB~\cite{Farmer:2003pa,Regimbau:2009rk,Regimbau:2011rp,Lamberts:2019nyk}. At least two SGWB components are guaranteed to be present in the LISA band: a contribution due to signals originating from compact GBs is expected to dominate the LISA band at low frequencies (up to $\sim 10^{-3} $~Hz), and a contribution from extragalactic BBH mergers is expected to be present at slightly larger frequencies ($\sim 10^{-3} - 10^{-2}$~Hz). A plot of the impact of these signals as described in Ref.~\cite{Flauger:2020qyi} on data measured in in the self correlations of the $A$ and $T$ channels (denoted by $AA$ and $TT$, respectively; we note that the self correltion $EE$ is identical to $AA$) is shown in Fig.~\ref{fig:my_labelsloppy}. While the latter is expected to be isotropic and stationary, the GB contribution is expected to be anisotropic (since the binaries  mostly lie on the galactic plane). Moreover, as discussed in Ref.~\cite{Adams:2013qma}, due to the yearly rotation of the satellite constellation, this signal is expected to present a yearly modulation. Both these features are thus expected to be, at least partially, present in the total SGWB measured by LISA.
	
	The characteristics of the GBs, i.e.~being almost monochromatic, and the majority of them being located in our galactic neighbourhood, will allow us to build reliable models for their residual foreground noise contribution~\cite{Crowder:2006eu,Sachdev:2020bkk}. The same applies to the case of the isotropic signal due to SOBBH events, where the spectral model of the residual confusion noise will be constructed based on the priors from ground observations~\cite{Abbott:2020niy} and the actual measurements with LISA. There is a possibility to measure a foreground signal component due to EMRIs~\cite{Barack:2005aj}, but a more detailed study on this type of source is needed in order to make robust predictions on their expected level. A generic method to make a first-level characterisation of the stochastic signals originating from compact binaries populations is studied in Ref.~\cite{Karnesis:2021tsh}.
	
	The classical central-limit theorem is violated when the single source variance is infinite, which is the case of e.g.~astrophysical backgrounds produced by compact mergers. As a consequence, the convergence to a Gaussian distribution can be much slower depending on the event rates (which determines the average number of sources overlapping in the detector frequency band(s)) and the properties of the source distribution~\cite{Ginat:2019aed}. If the sources are a spatial Poisson process, the resulting (non-Gaussian) distribution of the observed strain can be predicted (both in the time and frequency domains) using the techniques outlined in Ref.~\cite{Ginat:2019aed}. This approach can be extended to include the subtraction of bright sources, which (partly) gaussianises the signal~\cite{Timpano:2005gm}, and to furnish predictions for the distribution of the background of unresolved binary mergers. Quantifying deviations from Gaussianity is necessary to optimise searches for this confusion background~\cite{Thrane:2013kb,Smith:2017vfk} and backgrounds of cosmological origin~\cite{Bartolo:2019oiq}, and for their shape reconstruction. 
	
	Models for the residual foreground contributions can be provided when attempting a frequency shape reconstruction (see Sec.~\ref{sec:sgwb:reconstruction}), so that the reconstructed spectrum can be expected to consist of contributions from unknown, possibly cosmological sources. The energy spectrum of the foreground due to unresolved GBs can be modelled by a broken power law. Indeed, at sufficiently large frequencies the number of sources decreases, which produces a break in the $\alpha = 2/3$ power law behaviour. On the other hand, the foreground due to SOBBHs and BNSs is expected to be a power law with slope $ \alpha= 2/3$. Studies of SOBBH and BNS populations, such as Ref.~\cite{Perigois:2020ymr}, predict $\Omega_{\rm GW}(25~\text{Hz}) \simeq 4.97 \times 10^{-9} - 2.58 \times 10^{-8}$. Ground detectors~\cite{Abbott:2021xxi} set an upper limit for this foreground at $\Omega_{2/3}(f_{ref} = 25 ~ \text{ Hz}) \lesssim 5.8 \times 10^{-9} $. Clearly, these residual foreground models must account for the right amount of statistical uncertainty, or we risk overestimating the significance of a detection or of a spectral shape reconstruction. This includes, at descending levels in the Bayesian hierarchy, the possibility of different population models being used (mass or spin distribution, merger rate at different redshifts, etc), the uncertainty on the parameters of a given population when inferred from resolved events by LIGO/Virgo and LISA, and the possibility of residuals being left by the imperfect subtraction of these resolved events, either because of a sub-optimal SNR that leads to imperfect characterisation of the waveform to be subtracted, or to the very waveform not representing the actual events with perfect accuracy. Beyond these problems, as discussed in the previous paragraph, the statistical properties of the SGWBs of astrophysical origin (either before and after the removal of loud sources) may be dominated by strong non-Gaussianities which have to be appropriately modelled. A careful study is thus required to asses the impact of each of these sources of statistical uncertainty. A computational pipeline that, for each component of the residual confusion noise, can take a number of resolved events with their associated parameter uncertainties, and assuming a Bayesian-hierarchical population model (or family of them), can generate a probability distribution for each component of the foreground confusion noise needs to be developed. Preliminary work along this line has been developed for current- and next-generation ground-based detectors~\cite{Sachdev:2020bkk,Perigois:2020ymr,Abbott:2021xxi}, but a comprehensive study is needed in the framework of LISA.
	
	\subsection{Spectral shape reconstruction: existing tools and future developments}
	\label{sec:sgwb:reconstruction}
	
	The richness of sources in the LISA frequency band requires the development of techniques and tools  to disentangle the different contributions. These may include both a residual background of unresolved astrophysical sources (see Sec.~\ref{sec:sgwb:foregrounds}), a stochastic cosmological component sourced by topological defects,  phase transitions, or inflationary mechanisms, among others. Different cosmological sources are expected to produce SGWBs of characteristic frequency shapes~\cite{Bartolo:2016ami, Caprini:2015zlo, Caprini:2019egz, Auclair:2019wcv}. The reconstruction of the spectral shape of the SGWBs, regardless of its origin, is expected to play a special role in the separation of its different components. The correct estimation of the SGWB and its associated uncertainty is also essential for the characterisation and subtraction of resolved events, for which this background plays the role of an additional noise component, together with the instrumental noise.
	
\begin{figure}[htb!]
		\centering
		\includegraphics[width=\textwidth]{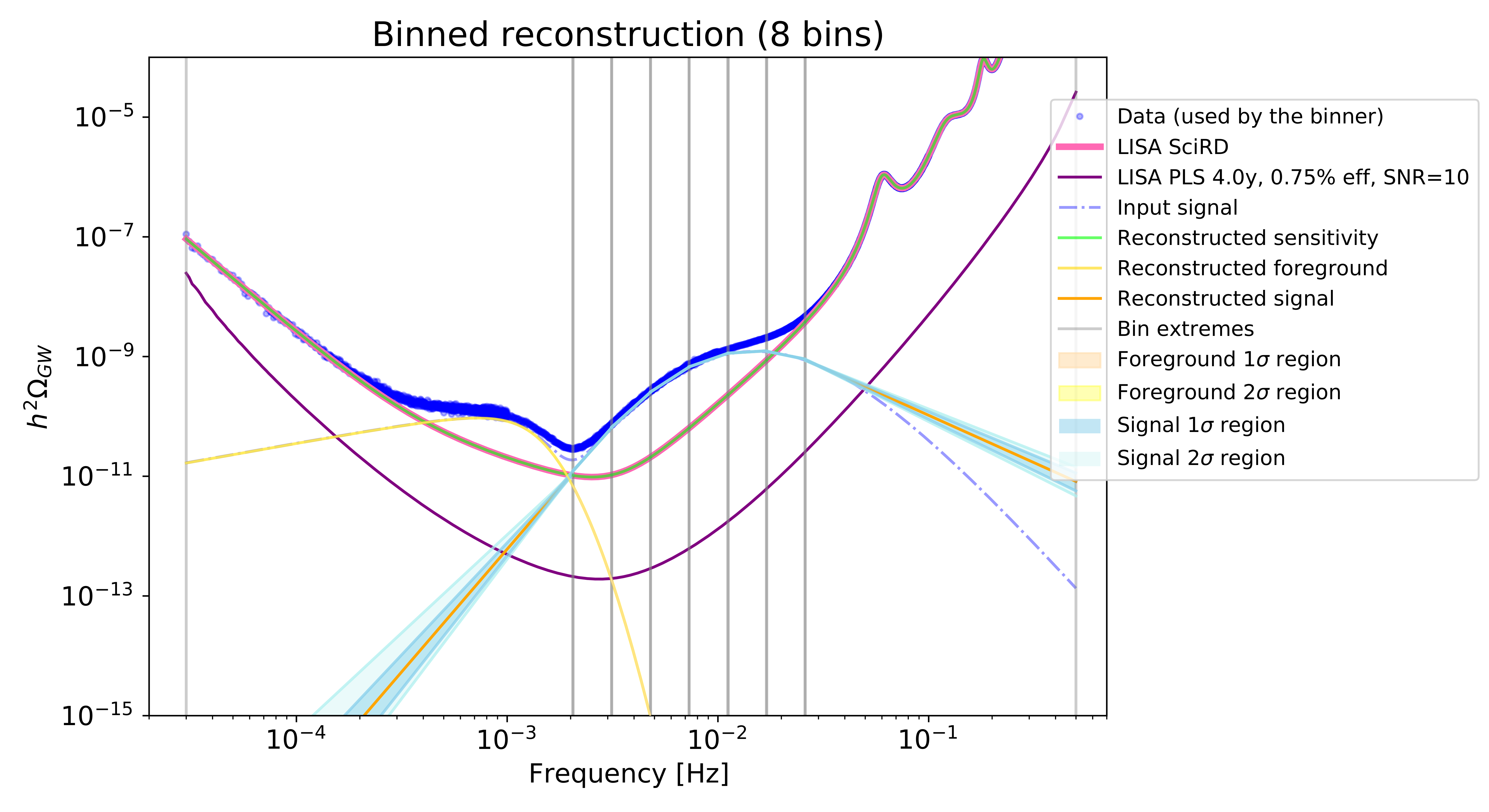}
		\caption{\small Reconstruction of a broken power law signal in presence of GB foreground. Figure taken from Ref.~\cite{Flauger:2020qyi}.}
		\label{fig:gal_MC_error_bars}
	\end{figure}

	Most of the SGWB detectability studies and ground-based searches have so far only focused on power-law templates~\cite{Thrane:2013oya, Adams:2013qma, Lentati:2015qwp, Arzoumanian:2018saf, LIGOScientific:2019vic} (and a few on more complicated, but fixed templates~\cite{Kuroyanagi:2018csn, Barish:2020vmy}). In the context of preparations for the LISA mission, some recent works~\cite{Karnesis:2019mph, Caprini:2019pxz, Pieroni:2020rob, Flauger:2020qyi} have attempted an agnostic template-free frequency shape reconstruction. One such reconstruction can take into account more complicated scenarios where the overall signal can be the superposition of several unknown signals, and can be extremely useful both for the disentanglement of the GW sources and for the constraining power that they have on astrophysical and cosmological parameters of the underlying theories (see previous sections). An example of reconstruction obtained with the method of Refs.~\cite{Caprini:2019pxz, Flauger:2020qyi} in the presence of the foreground due to GBs is shown in Fig.~\ref{fig:gal_MC_error_bars}.
	
	In Refs.~\cite{Karnesis:2019mph, Caprini:2019pxz, Flauger:2020qyi}, under the assumption of stationary, Gaussian, and isotropic signals, TDI data is processed into a set of frequency bins whose amplitudes are jointly estimated in the presence of noise. In the state-of-the art iteration~\cite{Flauger:2020qyi}, the number of bins is dynamically chosen using Bayesian criteria, and in each of the bins, together with the signal amplitude, a power-law tilt is also considered, leading to a more accurate reconstruction with fewer bins. On the other hand, the procedure employed in Ref.~\cite{Pieroni:2020rob} first expresses the signal as a linear combination of an arbitrarily dense basis of Gaussians and then selects the components associated with large eigenvalues of the Fisher information matrix. While technically different, this strategy follows the same philosophical approach set by Refs.~\cite{Karnesis:2019mph, Caprini:2019pxz, Flauger:2020qyi} and offers an interesting alternative to the methods developed therein. For example, in Ref.~\cite{Karnesis:2019mph}, after proper assumptions about the Gaussianity of the instrument noise, one can construct a posterior distribution for the level of the excess of spectral power, caused by an unmodelled SGWB signal at each frequency of the analysis. 
	
	All of the methods discussed in this section are focused on spectral shape reconstruction at the power spectrum level only. While in the most general set up this could be performed by only assuming the noise to be Gaussian, the techniques described here all assume (at the likelihood level) approximate Gaussianity of the SGWB as well. As long as the SNR per frequency mode is small, this is a good approximation but should be revisited for a bright signal. As discussed in Sec.~\ref{sec:sgwb:foregrounds} the assumption of Gaussianity can be violated for SGWBs of astrophysical origin. On the other hand, for SGWBs of cosmological origin any intrinsic non-Gaussianity in the metric perturbations 
	is expected to be washed out by propagation effects~\cite{Bartolo:2018evs, Bartolo:2018rku, Margalit:2020sxp}. Notice however that, as pointed out in Ref.~\cite{Bartolo:2019yeu}, non-Gaussianities in the energy density of the SGWB could be non-zero. For a better discussion of propagation effects see
	Sec.~\ref{Anisotropies and propagation effects}. If the backgrounds are expected to be significantly non-Gaussian, spectral shape reconstruction methods based on the observed strain distributions will have to be modified accordingly. 
	
	It is also worth mentioning that all the methods discussed in this section are applied to time-averaged data. While it could be difficult to induce any intrinsic time modulation in SGWBs of either cosmological or astrophysical origin, the motion of the detector with respect to the SGWB source frame is expected to naturally induce a yearly modulation in the amplitude of SGWB. While for isotropic SGWBs the effect is expected to be very small (of order of the velocity of the detector in units of the speed of light), for intrinsically anisotropic SGWBs (as for unresolved GBs) this effect could become sizeable~\cite{Adams:2013qma}. In order to model these effects it would be necessary to modify the existing pipelines to keep track (both at the generation and at the data analysis level) of the measurement time associated with each data segment. Such a modification in the existing tools would simultaneously improve their suitability  to be applied to real data and possibly their capability of disentangling the different components contributing to the SGWB. 
	
	The methods discussed in this section do not consider possible angular structure in the signal.	While techniques for reconstruction of anisotropies are more extensively discussed in the next section, it is worth mentioning that for most SGWB sources different angular structure is expected to be associated with different	spectra (namely, with a different frequency-dependence). As a consequence, by simultaneously tackling the frequency shape and the angular reconstruction it could be possible to break degeneracies between different components contributing to the observed signal. This could potentially lead to a significant improvement of the existing component separation techniques. For this reason embedding this possible method in the present pipelines may be an interesting possibility for future developments. 
		
	We conclude this section by commenting on the impact of foregrounds on the minimal SGWB intensity that would yield a detectable signal and allow us to attempt reconstruction.	
	While detection thresholds are clearly model dependent, i.e.~different amplitudes and SNRs are required in order to make quantitative statements for different SGWB shapes, some studies (see Sec.~\ref{sec:global_fit} for details) have already demonstrated the possibility of performing component separation with LISA. This implies that SGWB reconstruction can typically be performed even in the presence of foregrounds. However, it is worth mentioning that a crucial ingredient for component separation is the absence of degeneracies between the foregrounds and other SGWB components. Otherwise, unless this degeneracy is broken by the prior knowledge on the foreground parameters, the detection of the SGWB originating from cosmological sources could be negatively affected.
	
	\subsection{Anisotropy reconstruction}
	\label{sec:anisotropy_rec}
	
	The intensity of astrophysical and cosmological SGWBs are in general anisotropic. These anisotropies could allow one to differentiate between these two backgrounds. They can be generated both at the moment of the GW production and by the GW propagation in the perturbed universe.  In the case of the cosmological SGWB the anisotropies contain information about the primordial generation mechanism and moreover they can be a new probe of the primordial non-Gaussianity of the large-scale cosmological perturbations~\cite{Bartolo:2019oiq,Bartolo:2019yeu,Bartolo:2019zvb}. In the case of the astrophysical SGWB they contain information about the angular distribution of the sources and astrophysical properties, and they can be used as a tracer of large-scale structure; in both cases they also allow one to test the particle physics content of the universe~\cite{DallArmi:2020dar}. The LIGO/Virgo collaboration has already produced upper limits on the SGWB anisotropy in the 20-500 {\rm  Hz}  band~\cite{LIGOScientific:2019gaw}. The poor angular resolution of second-generation ground-based interferometers will probably not allow for the detection of the anisotropies of the SGWB, however future third-generation ground-based experiments like ET and CE will be sensitive to such a signal especially if it is characterised by a large monopole amplitude. For future space missions, like LISA, DECIGO, and BBO, prospects for reconstruction and measurement of the anisotropies in the SGWB, both astrophysical and cosmological, have been explored in the literature (see e.g.~Refs.~\cite{Giampieri:1997ie, Allen:1996gp, Cornish:2001hg, Seto:2004ji, Kudoh:2004he, Kudoh:2005as}). An updated  
	analysis for LISA, using the most up-to-date specifications, is ongoing~\cite{Bartolo:2022pez}, and in the next sections we report some main results. While in this section we discuss anisotropy reconstruction from a more theoretical point of view, in Sec.~\ref{sec:map_making} we present data analysis techniques to tackle this problem.

	\subsubsection{LISA response function}
	\label{sec:LISA_response}
	
	The reconstruction of an isotropic SGWB has been studied for instance in Ref.~\cite{Flauger:2020qyi}, combining measurements in the $A$,$E$,$T$ channels. This computation can be readily extended to an anisotropic SGWB. Details of this extension can be found in Ref.~\cite{Bartolo:2022pez}. Here we simply report the main result. 
	
	For a Gaussian SGWB, the object of our study is the two point correlation function of the GW signal. From the standard decomposition 
	\begin{equation} 
		h_{ab} \left( {\bf x} ,\, t \right) = \int_{-\infty}^{+\infty} df \int d \Omega_{\hat k} \, 
		{\rm e}^{2 \pi i f \left( t - {\hat k} \cdot {\bf x} \right)} \, \sum_\lambda {\tilde h}_\lambda \left( f ,\, {\hat k} \right) 
		e_{ab}^\lambda \left( {\hat k} \right) \;, 
	\end{equation} 
	where, for definiteness, $\lambda$ denotes the GW polarisation in the chiral basis. We are interested in the intensity of an unpolarised signal, for which 
	\begin{equation}
	    \label{eq:intensity_definition}
		\left\langle {\tilde h}_\lambda \left( f ,\, {\hat k} \right)  {\tilde h}_{\lambda'} \left( f' ,\,  {\hat k}' \right)  \right\rangle = \frac{1}{4 \pi} \, \delta \left( f + f' \right) \delta^{(2)} \left( {\hat k} - {\hat k}' \right) \, \delta_{\lambda,- \lambda'} \, I \left( f ,\, {\hat k} \right) \;. 
	\end{equation} 
	We further decompose 
	\begin{equation}
		I \left( f ,\, {\hat k} \right) \equiv   \sum_{\ell m} {\tilde I}_{\ell m} \left( f \right) \, {\tilde Y}_{\ell m} \left( {\hat k} \right) \;, 
		\label{anisotropic-I}
	\end{equation} 
	where ${\tilde Y}_{\ell m} \left( {\hat k} \right) \equiv \sqrt{4 \pi} \, Y_{\ell m} \left( {\hat k} \right)$, are rescaled spherical harmonics,  normalised so that ${\tilde Y}_{00} = 1 $. In this way the isotropic case corresponds to $I \left( f ,\, {\hat k} \right) = {\tilde I}_{00} \left( f \right)$. 
	
	The multipole coefficients ${\tilde I}_{\ell m}$ of the intensity  are related to those of the fractional density by~\cite{Bartolo:2022pez}
    \begin{equation}
    {\tilde I}_{\ell m} \left( f  \right)  = \frac{1}{\sqrt{4 \pi}} \, \frac{3 H_0^2}{4 \pi^2} \, \frac{\Omega_{\rm GW} \left( f \right)}{f^3} \delta_{\rm GW,\ell m} \;, 
    \label{It-delta}
    \end{equation} 
	where the latter are defined from
	\begin{equation}
	\frac{\omega_{\rm GW} \left( f ,\, {\hat k} \right) - \Omega_{\rm GW} \left( f \right)}{\Omega_{\rm GW} \left( f \right)} \equiv \sum_{\ell m} \delta_{\rm GW,\ell m} \left( f \right) \, Y_{\ell m} \left( {\hat k} \right) \;, 
    \end{equation}
    with $\omega_{\rm GW} \left( f ,\, {\hat k} \right)$ being the fractional energy density before angular integration, 
    \begin{equation}
    \Omega_{\rm GW} \left( f \right) = \int d^2 {\hat k} \,  \omega_{\rm GW} \left( f ,\, {\hat k} \right) / 4 \pi \;. 
    \end{equation}
	In Eq. (\ref{It-delta}), $H_0$ is the present value of the Hubble rate. 
	
\begin{figure}[htb]
		\centering
		\includegraphics[width=\textwidth]{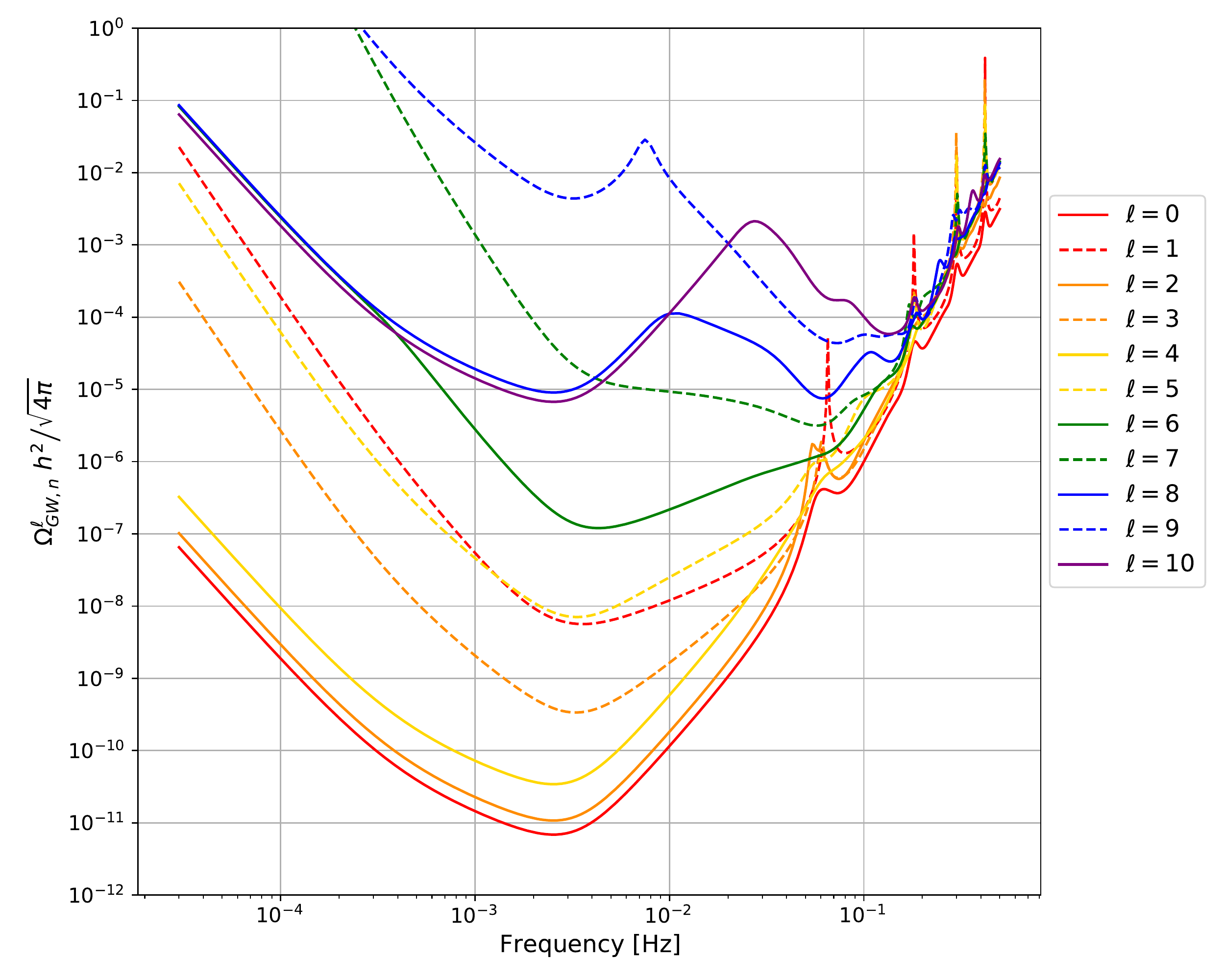}
		\caption{\small LISA Sensitivity (\ref{SNR-sensitvity}) to various multipoles of the SGWB. Figure taken from Ref.~\cite{Bartolo:2022pez}.}
		\label{fig:sensitivity-ell}
	\end{figure}

	For definiteness, we consider a statistically isotropic signal, so that the unbiased estimators ${\hat \delta}_{\rm GW,\ell m} $ for coefficients of the decomposition $\delta_{\rm GW,\ell m} $, have expectation value~\footnote{Notice that the brackets in Eq.~(\ref{eq:intensity_definition}) denote an ensemble average. Since we have a single realisation of the observable universe, an estimator $\hat{C}_\ell^{\rm GW}$ for $C_\ell^{\rm GW}$ is typically built by averaging the measured value of $\delta_{\rm GW,\ell m} \, \delta_{\rm GW,\ell' m'}^*$ over the different $m$ indices.}
	\begin{equation} 
	    \label{eq:pp_expectation}
		\left\langle \delta_{\rm GW,\ell m} \, \delta_{\rm GW,\ell' m'}^* \right\rangle = C_\ell^{\rm GW} \, \delta_{\ell \ell'} \, \delta_{m m'} \;. 
	\end{equation} 
	The expectation value of the SNR can be then expressed as the sum $\left\langle {\rm SNR} \right\rangle \equiv \sqrt{\sum_\ell \left\langle {\rm SNR}_\ell^2 \right\rangle } $ over multipoles. Each term is formally of the type 
	\begin{equation}
		\left\langle {\rm SNR}_\ell^2 \right\rangle = T_{\rm obs} \, 
		\int_0^\infty d f \, 
		\left[ \frac{\sqrt{C_\ell^{\rm GW}} \, \Omega_{\rm GW} \left( f \right) h^2 }{\Omega_{\rm GW,n}^\ell \left( f \right) \, h^2 } \right]^2 \,,
		\label{SNR-sensitvity}
	\end{equation}
where $T_{\rm obs}$ is the observation time, and 
 $\Omega_{\rm GW,n}^\ell \left( f \right) \, h^2$ 
	denotes the LISA sensitivity to the $\ell-$th multipole.
	The sensitivity to the first few multipoles are shown in Fig.~\ref{fig:sensitivity-ell}. The mathematical form of the sensitivity, and its derivation, are  given in Ref.~\cite{Bartolo:2022pez}. 
	
	\subsubsection{Cosmic dipole}
	
	One would  expect that LISA sensitivity degrades as the multipole order increases. While this is generally true, as can be also seen from Fig.~\ref{fig:sensitivity-ell}, due to the symmetries of the system, LISA turns out to be more sensitive to the quadrupole than to the dipole. Nevertheless, once the isotropic component of the SGWB is detected, a naturally expected value of the amplitude and the direction of the dipole will be determined by the velocity of the Earth with respect to the cosmic rest frame, providing a first target for the detection of SGWB anisotropies. 
	
	There are only two correlators sensitive to the dipole, namely the one between the $A$ and the $T$ channel, and the one between the $E$ and the $T$ channel. Remarkably, moreover, these $AT$ and $ET$ correlators vanish for the SGWB monopole. The SNR for the dipole induced by the Earth's motion with speed ${\beta}$ with respect to the cosmic rest frame is given, in the noise dominated regime, by the formula
	\begin{equation}
		{\rm SNR}^2_{A/ET}= 6 \,\beta^2\,\left(\frac{3\,H_0^2\,\Omega_{\rm GW}}{4\pi^2}\right)^2 \int_0^{T_{\rm obs}} dt \int\frac{df}{f^6}\, 
		\sum_{J=A,\,E}\frac{\left({\cal R}^{JT}_{\ell=1}\,\hat{n}(t)\cdot\hat{\beta}\right)^2}{N_J\,N_T}\,,
	\end{equation}
	where $N_J$ is the noise in mode $J$ (see the right panel of Fig.~\ref{fig:PSD_strain}) and ${\cal R}^{AT}_{\ell=1}={\cal R}^{ET}_{\ell=1}$ measures the amplitude of the $\ell=1$ mode of the angular response (see Ref.~\cite{Bartolo:2022pez}), evaluating to $\sim 10^{-2}$ for $f=0.1$~Hz and to $\sim 10^{-3}$ for $f=0.01$~Hz. Also, $\hat{\beta}$ is the direction of the dipole, whereas $\hat{n}(t)$ is the unit normal to the LISA configuration, whose direction changes over the course of the year.
	
	Assuming a flat energy spectrum in the LISA band, the SNR takes the approximate value 
	\begin{equation}\label{snr_dipole}
		{\rm SNR}\simeq 0.5\times\left(\frac{\beta}{10^{-3}}\right)\,\left(\frac{\Omega_{\rm GW}}{6\times 10^{-8}}\right)\,\sqrt{\frac{T_{\rm obs}}{1\ {\rm {year}}}}\,,
	\end{equation}
	where $\beta$ is normalised to its value with respect to the CMB rest frame, and $\Omega_{\rm GW}$ is normalised to its upper bound from LIGO/Virgo~\cite{LIGOScientific:2019vic}.  While Eq.~(\ref{snr_dipole}) suggests a small SNR already for a SGWB that saturates the current constraint, it is important to note that the bound~\cite{LIGOScientific:2019vic} holds at LIGO/Virgo frequencies, and there is no reason for the SGWB spectrum to be flat all the way to LISA frequencies.
	
	Remarkably, due to the higher sensitivity to the quadrupole, LISA will have comparable sensitivity to the dipole and to the quadruple induced by the motion of the Solar System with respect to the cosmic rest frame, despite the fact that the quadrupole has an amplitude that is a factor of $\beta\approx 10^{-3}$ smaller than the dipole.
	
	It is also worth pointing out that LISA will be much more sensitive to the dipole if the SGWB contains a chiral component. In this case, in fact, the AE correlator is not vanishing, leading to a SNR~\cite{Domcke:2019zls}
	\begin{equation}\label{snr_chiral_dipole}
		{\rm SNR} \simeq 10^3\,\delta\chi\left(\frac{\beta}{10^{-3}}\right)\,\left(\frac{\Omega_{\rm GW}}{6\times 10^{-8}}\right)\,\sqrt{\frac{T_{\rm obs}}{1\ {\rm {year}}}}\,,
	\end{equation}
	where $0\le \delta\chi\le 1$ measures the degree of chirality of the SGWB, see Sec.~\ref{inflation-chirality}. 

	\subsection{Gravitational wave map making with LISA}
	\label{sec:map_making}
	
	An available mapping algorithm tailored to the LISA detector is based on an optimal quadratic estimator~\cite{Contaldi:2020rht}. The mapper takes advantage of the time-dependent sky response of the LISA constellation to scan the sky over a long observation times. It reconstructs full-sky stochastic signals and their anisotropies. A prime example of anisotropic SGWB signal in the LISA band is the statistical signal from unresolved  white dwarf GBs. This signal is expected to trace out the distribution of white dwarfs in the Milky Way. Beyond this, the astrophysical SGWB from extragalactic stellar mass compact binaries is also expected to have some degree of angular anisotropy, as do many of the cosmological backgrounds reviewed in this paper. More details on anisotropic SGWB sources potentially detectable by LISA may be found in Ref.~\cite{Bartolo:2022pez}.
	
\begin{figure}[t]
		\centering
		\includegraphics[width = 0.31 \textwidth]{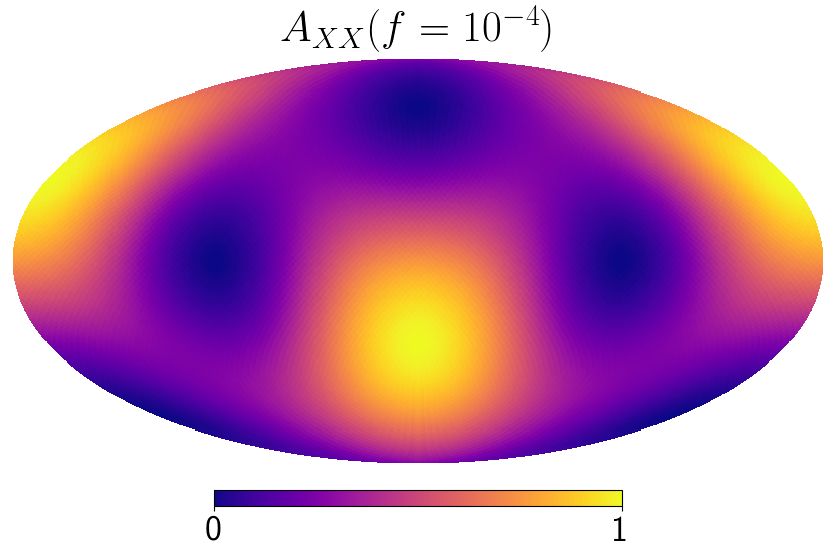}
		\hfill
		\includegraphics[width = 0.31 \textwidth]{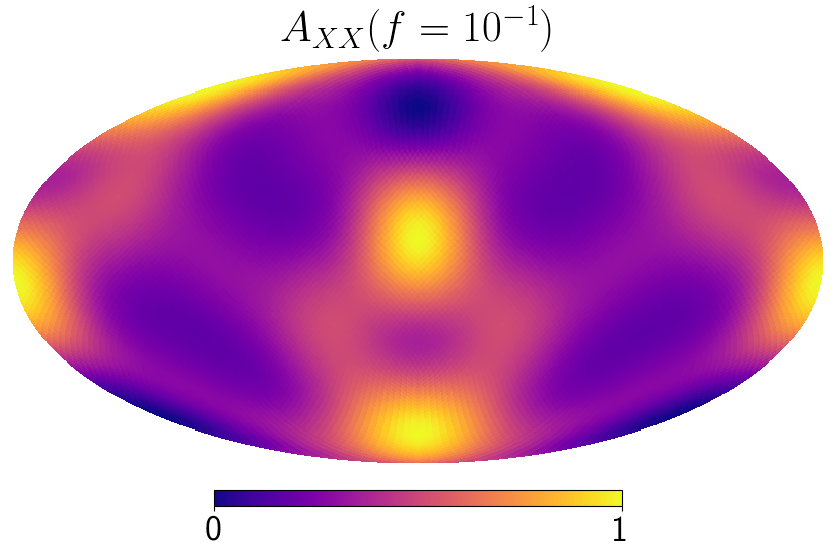}
		\hfill
		\includegraphics[width = 0.31 \textwidth]{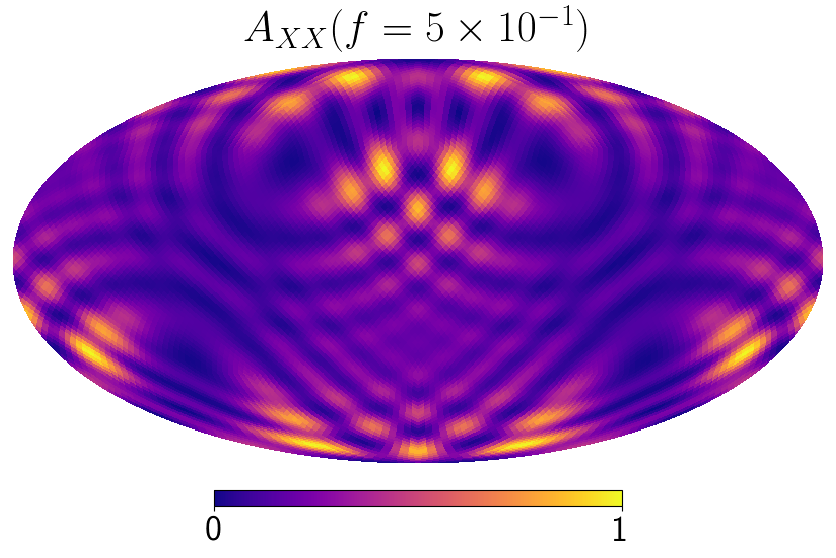}
		\caption{\small Normalised auto-correlated response of TDI channel X at time $t = 0$ in the Solar System barycentre reference frame, at frequencies $f = 10^{-4}$ Hz, $f = 10^{-1}$ Hz, $f = 5\times10^{-1}$ Hz from left to right respectively. Estimates based on Ref.~\cite{Contaldi:2020rht}.}
		\label{fig:freq_sky_response}
	\end{figure}

	The sky response used assumes equal arms and heliocentric circular orbits, which are considered to be good approximations until full, time-dependent flight solutions become available after launch. These are the same assumptions made throughout Sec.~\ref{sec:LISA_response}. The frequency transfer function included in the response induces a frequency dependence in the sky modulation, as may be seen in Fig.~\ref{fig:freq_sky_response}. This drives the effective angular resolution as a function of frequency sensitivity and determines the resolution of final maps. The map-making can be based on a broad-band integration of frequencies or the integration can be split into narrower frequency bands~\cite{Contaldi:2020rht}. 
	
	The estimator is based on the assumption that the signal and noise components in the timestream data are Gaussian and maximises the standard Gaussian likelihood. This approach is similar to optimal map-making steps in CMB analysis (see e.g.~Ref.~\cite{Bond:1998zw}). In this case, the likelihood for the data is parametrised by the sky-signal covariance as function of direction. The maximum-likelihood solution for the covariance (signal intensity) is obtained using a quadratic, iterative estimator which effectively inverts the time-integrated projection of the sky-signal onto the data. A noise model is added to the total covariance in the likelihood. The algorithm is tailored to and applied in the pixel domain but it can also be applied in the spherical harmonic domain. The two choices differ in the use of regularisation methods that need to be applied to the ill-conditioned problem.
	
	In practice, the data are segmented into short-duration frames, throughout which the sky response of the detector is considered to be constant. The time-frame length is a key element in the analysis: it sets the lower limit of frequency space probed, while also setting the maximum pixelisation scale to ensure a smooth transition on the sky, frame by frame. The covariance of the signal is initially estimated by subtracting the noise model from the data covariance, and is then noise-weighted over frequencies and projected onto the sky via the response operator to obtain a map. This operation is performed over all time-segments and subsequently averaged. The pixel-pixel Fisher matrix relative to the measurement is similarly calculated and inverted, using singular-value-decomposition techniques to regularise the inversion. The inverse Fisher matrix is then applied to the sky-and-noise weighted data to extract the optimal map solution. This is performed iteratively, until convergence requirements are satisfied.
	
	\begin{figure}
		\centering
		\includegraphics[width = 0.7\textwidth]{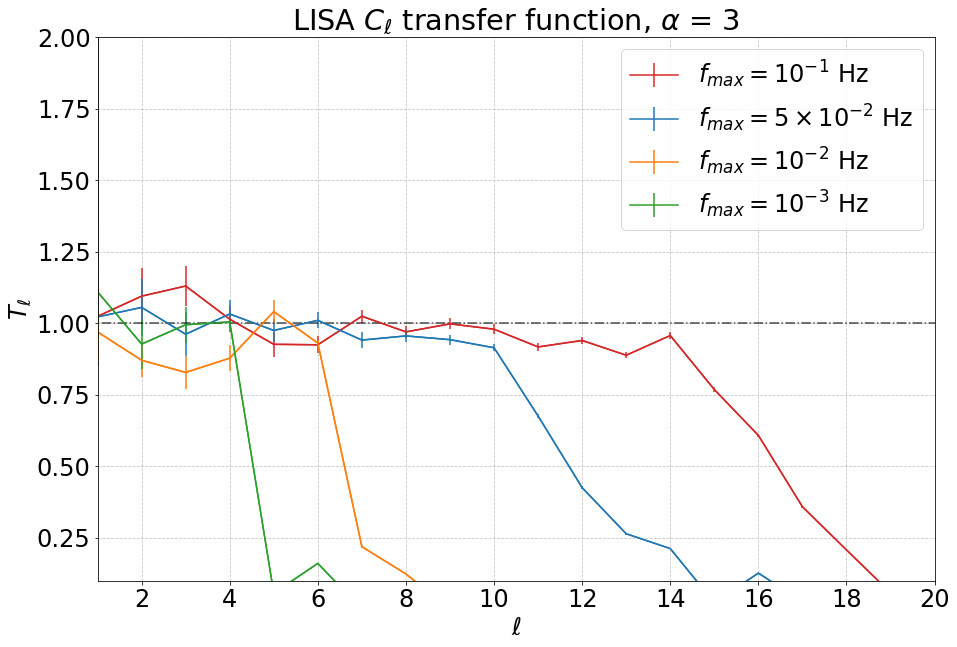}
		\caption{\small Transfer functions $T_{\ell}$ for the average reconstructed $C_{\ell}$s obtained with different frequency cutoffs $f_{max}$, drawn from Ref.~\cite{Contaldi:2020rht}. Each simulation set consists of 50 maps, each a different realisation of the same $C_{\ell}$ input  (dashed grey baseline). There appears to be a clear one-to-one relation between the resolution $\ell_{\rm max}$ of the instrument and the frequency cutoff.}
		\label{fig:mapping_transfer}
	\end{figure}

	Fig.~\ref{fig:mapping_transfer} shows the $\ell$-mode transfer function $T_\ell$ for sky-map reconstruction presented in Ref.~\cite{Contaldi:2020rht} in the case of very high SNR injections. This shows the effective angular resolution of LISA at different frequency pivot points. This is calculated by combining sets of 50 simulations for each frequency window examined. Each simulation consists of the injection and reconstruction of an $\ell^2C_\ell$-flat Gaussian realisation for the sky GW intensity. The procedure works under the assumption that the frequency and directional dependence can be separated:
	\begin{equation}
	\label{eq:spectral_angular_separation}
	\Omega_{\rm GW} (f, \textbf{n}) = \Omega (f) \mathcal{P}(\textbf{n}). 
    \end{equation}
	
	For simplicity, the spectral dependence of the signal is taken to be a simple power law, with spectral index $\alpha = 3$. The transfer function is defined as $T_\ell = C_\ell^{\rm out}/C_\ell^{\rm in}$. In this extremely high signal scenario, the solution converges after a single iteration. As shown in Fig.~\ref{fig:mapping_transfer}, the higher angular modes are well preserved when allowing the reconstructor to integrate up to higher frequencies, where there is finer structure in the response pattern, whereas they are aliased into lower modes when integrating over lower frequencies only. In the best case scenario of a strong signal at high frequencies, the map-maker can recover anisotropies up to scales of $\ell_{\rm max} \sim 15$. The reconstruction of a simplified model for the GB background is also tested; it is found that even though the signal has a relatively high SNR, it peaks above the noise around $10^{-3}$ Hz and hence may be recovered at best at $\ell_{\rm max}\sim 5$. 
	
	The estimator currently assumes a fixed instrumental noise model, however it is possible to include independent noise estimates for each time segment. A possible extension of this method would fit for both signal and noise components simultaneously. The current noise model employed in the mapping pipeline is the one presented in Sec.~\ref{sec:Noise}. It is also possible to extend the estimator to include different spectral shapes for the signal, beyond a simple power law, or alternatively solve for maps in narrow frequency bands to attempt a model-independent directional search. Furthermore, it is important to note that the Fisher matrix is highly singular and requires heavy conditioning before inversion; this is a delicate process which may bias the outcome of the mapping, and further investigation is required to assess its impact. The implications of the inversion for certain high and low SNR stationary SGWB signals are discussed in Ref.~\cite{Contaldi:2020rht}.
	
	 Going a step further, a recently developed Bayesian formalism~\cite{Banagiri:2021ovv} allows for simultaneous estimation of the frequency and directional content of a SGWB with LISA.\footnote{The code for this analysis pipeline can be found at \url{https://github.com/sharanbngr/blip}.} One again starts with the assumption in Eq.~\eqref{eq:spectral_angular_separation}, where a frequency dependence of choice could be inserted, for example a simple power law (as for example in~\cite{Adams:2010vc, Adams:2013qma, Caprini:2019pxz, Flauger:2020qyi}), while the angular dependence can be decomposed with respect to a basis on the sphere, such as spherical harmonics.   
	 
\begin{figure}
\begin{center}
\includegraphics[width=.4\textwidth]{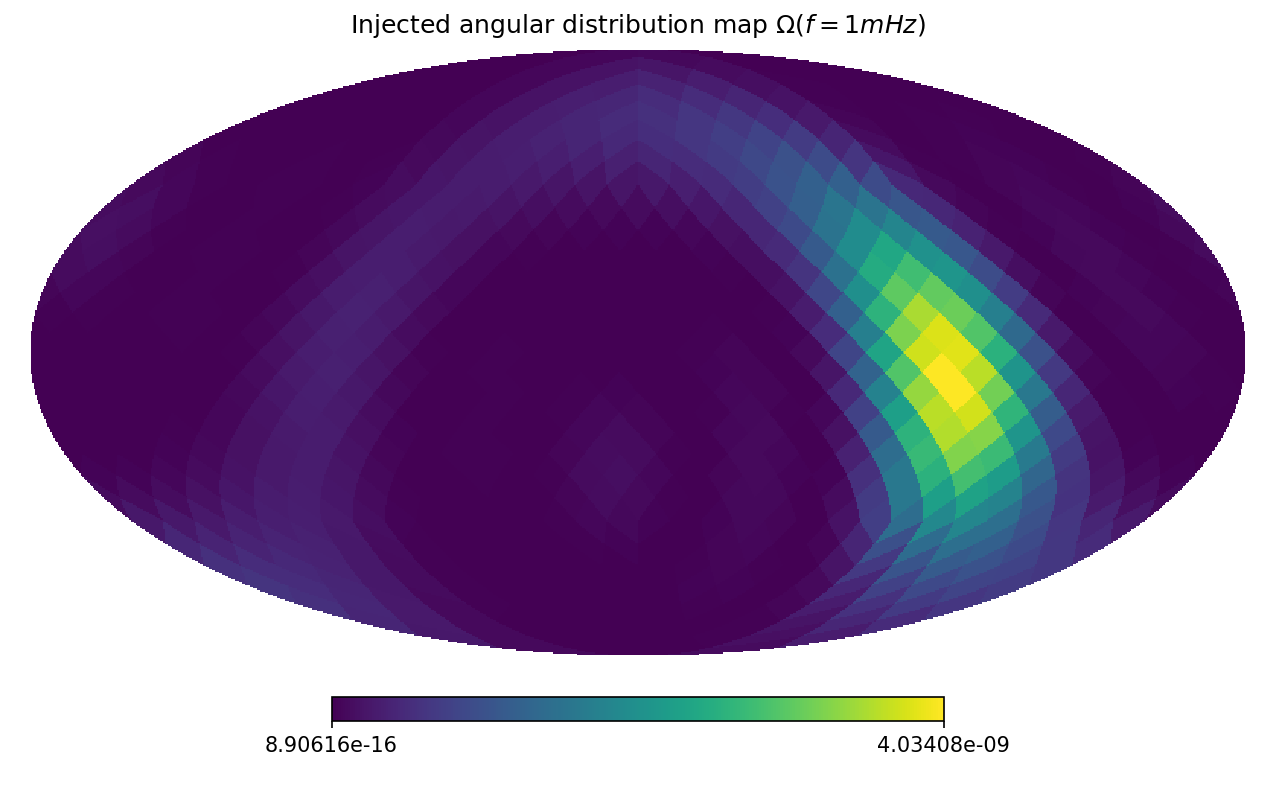}
\includegraphics[width=.4\textwidth]{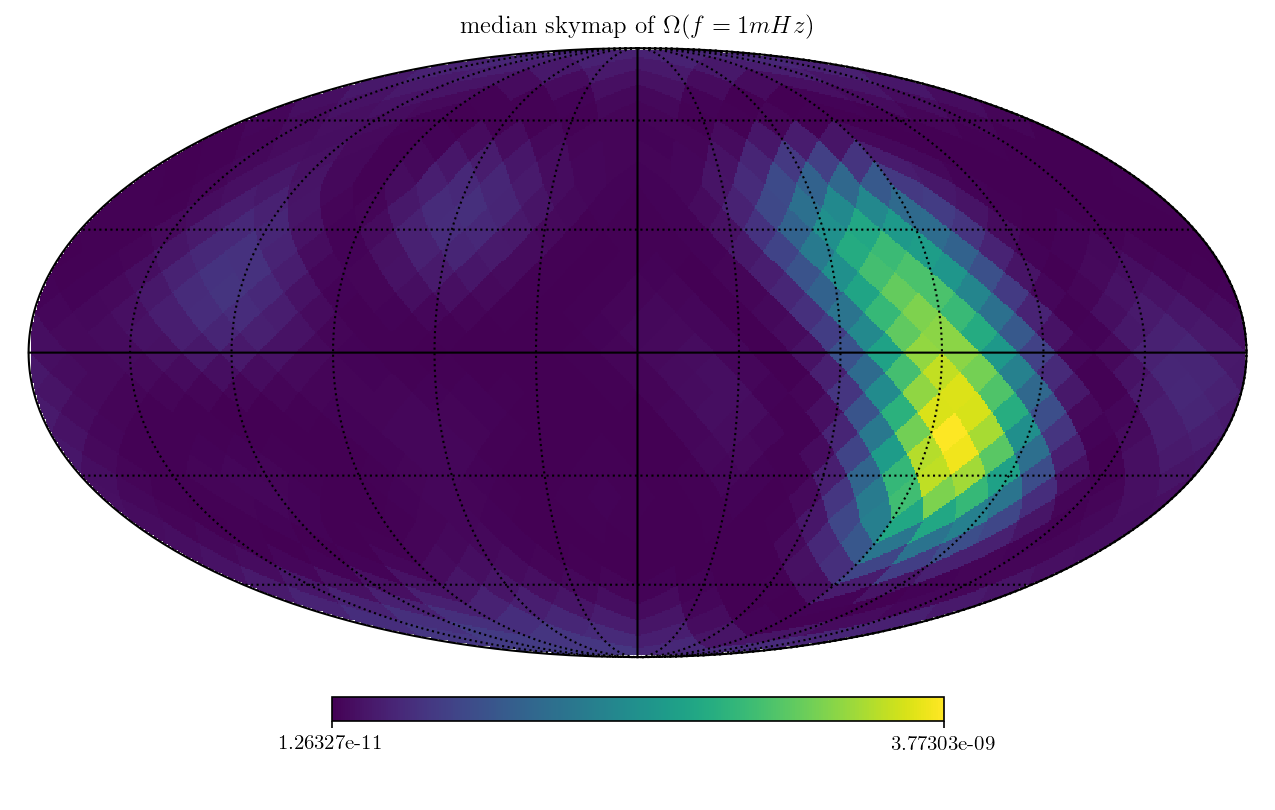}
\caption{\small Simulation (left) and recovery (right) of the  white dwarf GB foreground energy density in GWs at 1\,mHz, using one year of data in the Solar System barycentric frame. The bright spot in the map corresponds to the galactic central bulge. Figure taken from Ref.~\cite{Banagiri:2021ovv}.}
\label{Banagiri:sim2}
\end{center}
\end{figure}

One can then divide LISA data into segments of duration $T_{\rm seg}$ and compute the Fourier transforms of the LISA TDI channels in each segment (in either the X-Y-Z or A-E-T configuration). It is then straightforward to define the likelihood function
\begin{equation}
	\mathcal{L}(\tilde{d} | N_p, N_a, \alpha, \Omega_{\alpha}, \{b_{l,m}\}) =  \prod_{t, f} \frac{1}{2 \pi T_{\text{seg}} |C(t, f) |}  \times  \exp \left( - \frac{2 \, \tilde{d}^*_{t, f} \, C(t, f)^{-1} \, \tilde{d}_{t, f}}{T_{\text{seg}}} \right)\,,  
\end{equation} 
where $\tilde{d}_{t, f} = [ \tilde{d}_X (t, f), \tilde{d}_Y(t, f), \tilde{d}_Z (t, f)   ]$ is the array of data in the Fourier domain for the three LISA channels (in the time segment $t$ and at frequency $f$), and $C(t, f)$ is the $3 \times 3$ covariance matrix for the three channels. The covariance matrix can be modelled to include both instrumental noise (here described by position and acceleration noise parameters $N_p$ and $N_a$), as well as the astrophysical and cosmological contributions to the GWB. In this example, we use the amplitude $\Omega_\alpha$ and spectral index $\alpha$ to describe the frequency dependence of the astrophysical/cosmological contributions, and the $b_{l,m}$, coefficients of the decomposition onto spherical harmonics, to describe the directional dependence.
Using this likelihood in Bayesian parameter estimation, with suitable choices for prior distributions on the free parameters, allows for the recovery of the free parameters. In particular, the recovery of simulated noise and a cosmological background up to $l = 2$ is achievable~\cite{Banagiri:2021ovv}. The same technique can be applied to study the galactic foreground due to white dwarf binaries, as shown in Figure \ref{Banagiri:sim2}. 
Of course, additional model complexity could be added to this method: for example a more complex instrument noise model, or frequency dependence of $b_{l,m}$ coefficients, or polarisation dependence of the cosmological model.

\subsection{Global fit}
\label{sec:global_fit}
	One of the main data analysis challenges that LISA will face is to disentangle the various sources that overlap in time and in frequency. To tackle this problem, the LISA Consortium will employ a {\emph{global fit}} scheme~\cite{unpublishedGlobal, Littenberg:2020bxy} for detecting, separating and eventually identifying/classifying many overlapping signals of astrophysical and hopefully early-universe origin. Different strategies could be employed to tackle this compelling task. A fully consistent Bayesian framework, ideally the most accurate approach to this problem, would perform a simultaneous fit of all resolvable sources and SGWBs. From a computational point of view this would be extremely difficult due to very high dimensionality of the parameter space that would need to be explored.
	
	A possible approach (for a first step in this direction see Ref.~\cite{Karnesis:2021tsh}) would alternate between a step in which, given some current estimation of co-added background noise (instrumental noise, astrophysical confusion noise and a possible stochastic cosmological component), resolved events are detected, characterised and subtracted from the data streams, and a step in which the resulting ``residual'' data is used to update the knowledge on this background noise. In this step the residual is treated as a component which simultaneously models instrumental noise as well as the expected confusion noise inferred from the current catalogues of resolved sources (see Sec.~\ref{sec:sgwb:foregrounds}), and a free-form SGWB spectrum (see Sec.~\ref{sec:sgwb:reconstruction}). The advantage of this approach is that the dimensionality of the parameter inference problem at each of these steps is significantly lower (since technically we are performing conditional -- or Gibbs -- sampling at each step) and could be easily dealt with using standard techniques. Hereafter we discuss the state of the art on integrating SGWB detection, characterisation and component separation into a LISA \emph{global fit} pipeline, and the further developments which are crucial for LISA to match its scientific goals regarding this problem.
	
	The detection of an astrophysical SGWB 	with an accurate modelling, in the presence of the  instrumental noise has been established in e.g.~Refs.~\cite{Christensen:1996da, Adams:2010vc, Meyers:2020qrb}. In what concerns the simultaneous estimation of a cosmological component and one or more astrophysical confusion backgrounds, a number of approaches have been explored. We anticipate that their conclusions must be considered as preliminary since they all rely on some simplifying assumptions on the noise modelling, response functions and contamination from resolved-event  misreconstructions, among others. Properly addressing these aspects is one of the main priorities in the forthcoming years.      

Ref.~\cite{Flauger:2020qyi}  demonstrates the capability of LISA to separate between general astrophysical templates and a free-form cosmological component of sufficient intensity. In this work the three LISA TDI channels are fully exploited, using the $AET$ basis~\cite{Hogan:2001jn, Adams:2010vc}, in particular the ability of the closed-path $T$ channel to place strong, independent constraints on the instrumental noise, using the noise model of Sec.~\ref{sec:Noise}. The  proposed  pipeline  determines an optimal binning for the reconstruction of the cosmological signal using maximum likelihood estimation. This binning is then used in a fully Bayesian pipeline implemented in the Cosmological sampling framework Cobaya~\cite{Torrado:2020dgo}, to jointly estimate the parameters of the noise model, the amplitude of an astrophysical foreground template, and the free-form binned cosmological component. Using the Monte Carlo Nested Sampler PolyChord~\cite{Handley:2015vkr}, it turns out that a sufficiently high-SNR broken-power-law signal can be efficiently reconstructed in the presence of either a power-law model for extragalactic binaries consistent with LIGO/Virgo observations~\cite{LIGOScientific:2019vic}, or a foreground of GBs modelled as in Ref.~\cite{Robson:2018ifk}. On the contrary, SGWBs with SNR\,$\lesssim 10$ such as flat signal with $\Omega_{\rm GW}\lesssim 10^{-13}$ are likely to escape the LISA searches.

\begin{figure}
    \includegraphics[width=150mm]{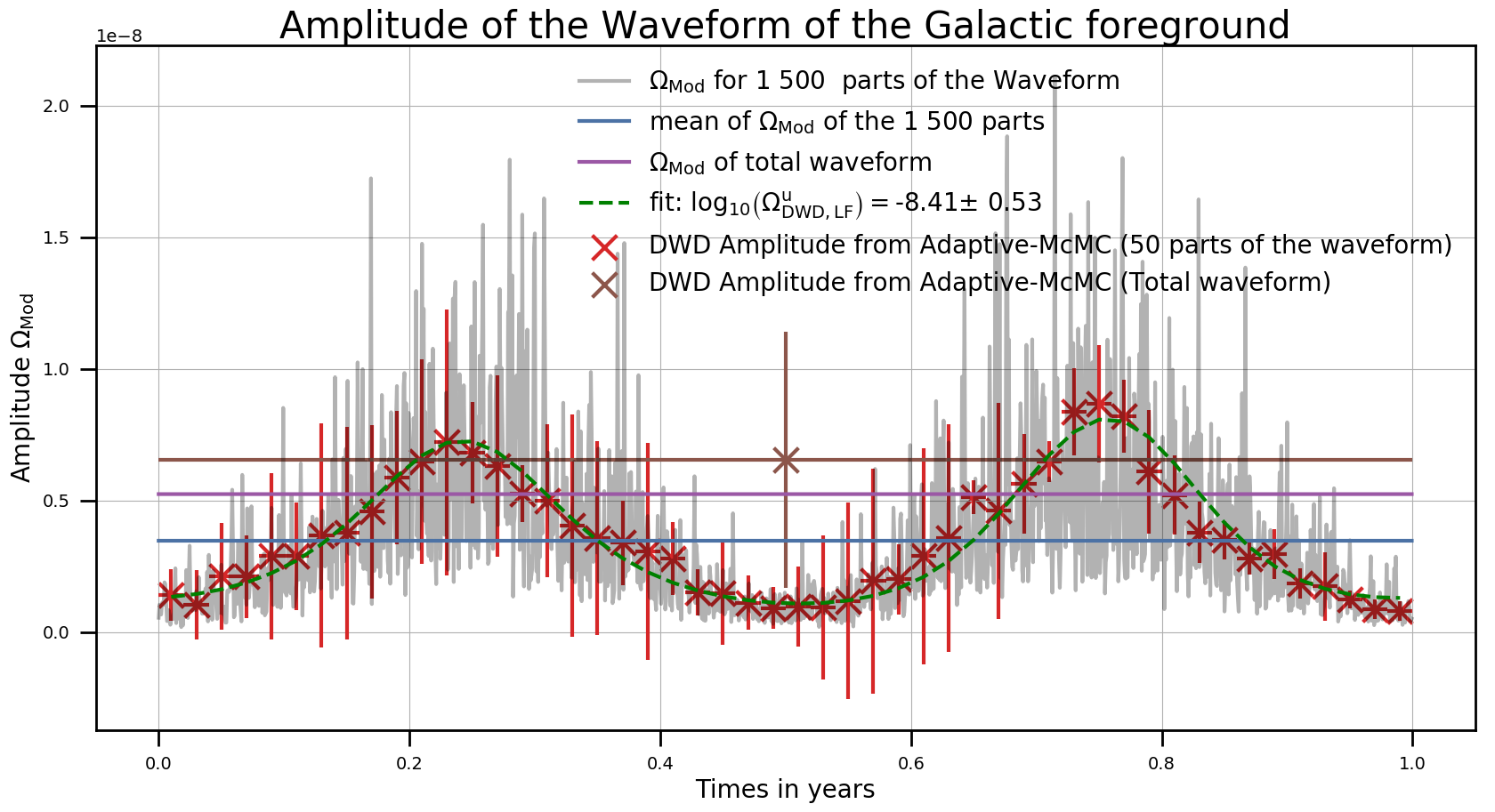}
    \caption{\small Measurement of the orbital modulation of the white dwarf binary foreground. In grey: 1500 estimates of $\Omega_{\textrm{Mod},i}=  \frac{4\pi^2}{3H_0} \left( \frac{c}{2\pi L}\right)^2 A_i^2$ ($A_i$ amplitude of the characteristic strain). In red: 50 MCMC results with 8 parameters (2 parameters for BBHs, 4 parameters for white dwarf binaries, 2 parameters for the LISA noise). In green, fit to the 50 MCMC run results to estimate the modulation from the LISA antenna pattern amplitude at 3 mHz. Modulation model: $\Omega_{\textrm{Mod},i} =\Omega_{\rm DWD,LF}^u\left(F^2_{+,i} + F^2_{\times,i} \right)$. Figure taken from Ref.~\cite{Boileau:2021sni}.}
    \label{fig:modulationpro}
    \end{figure}

\begin{figure}
    \includegraphics[width=150mm]{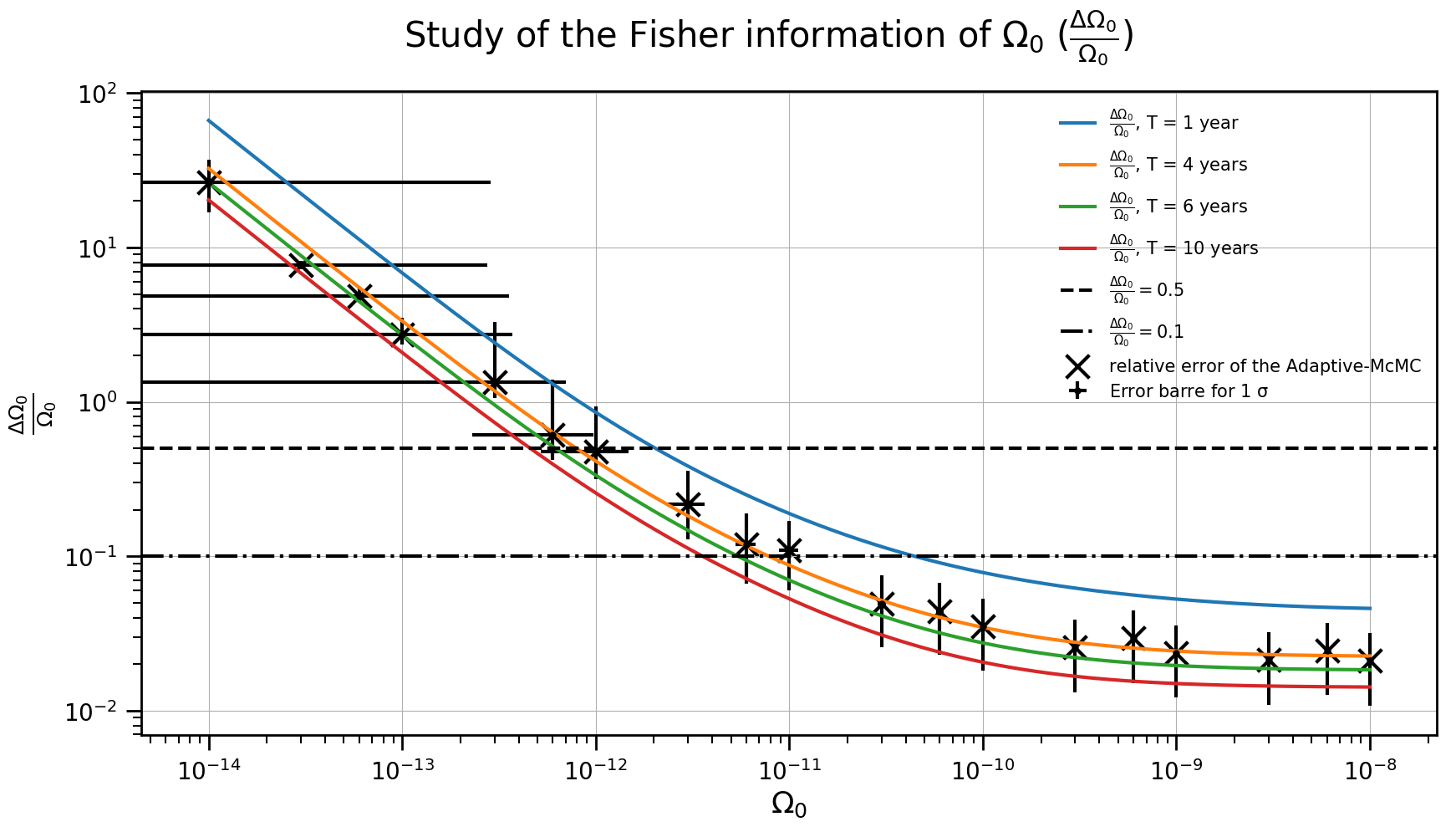}
    \caption{\small Cosmological amplitude uncertainty estimates from the Fisher Information study denoted by solid lines and from the MCMC by crosses. The upper horizontal dashed line represents the error level $50\%$. Figure taken from Ref.~\cite{Boileau:2021sni}.}
    \label{fig:Uncertaintypro}
    \end{figure}

Another step in this direction is taken in Ref.~\cite{Boileau:2020rpg} and Ref.~\cite{Boileau:2021sni} which progressively increase the complexity of the foreground. The analyses adopt a Bayesian strategy based on an Adaptive MCMC algorithm~\cite{Christensen:1998gf,Cornish:2007if} analysing the $A$, $E$ and $T$ channels given by the LISA model of Ref.~\cite{Smith:2019wny}. The Adaptive MCMC results are independently confirmed by an analysis from the Fisher information matrix. 
Also with these analyses, it turns out that the LISA $T$ channel helps to efficiently estimate the LISA noise parameters and to thus measure the SGWB in the $A$ and $E$ channels. Ultimately, it comes out that LISA can detect (power-law) cosmological backgrounds in the presence of astrophysical foregrounds. In particular,  given the expected LISA noise and the astrophysical foreground consistent with LIGO/Virgo observations, it is possible to observe a flat power-law SGWB with amplitude larger than $\Omega_{\rm GW, Cosmo} \approx (1 - 10) \times 10^{-13} $
after 4 years of LISA observations~\cite{Boileau:2020rpg}.
Ref.~\cite{Boileau:2021sni} makes this lower bound more robust by adding the foreground due to GBs simulated with the binary catalogues of Ref.~\cite{Lamberts:2019nyk}.
The study takes into account that the GB foreground measured by LISA has a yearly modulation (see  Fig.~\ref{fig:modulationpro}). It implements a Bayesian analysis for each week in the year, and the corresponding variation of the LISA pattern antenna.
Based on this analysis, the reconstruction of the amplitude of a flat SGWB has uncertainties below  50\%  when $ \Omega_{\rm GW,Cosmo} \gtrsim 8 \times 10^{-13}$ (see  Fig.~\ref{fig:Uncertaintypro}).

	Similar efforts is carried out in the context of ground-based experiments. Ref.~\cite{Martinovic:2020hru} explores a method for the simultaneous estimation of astrophysical and cosmological SGWBs in the frequency band of current and future  terrestrial interferometers. As an example, the detection of cosmological signals with GW fractional energy density $4.5\times 10^{-13}$ at 25 Hz, for cosmic strings, and $2.2\times 10^{-13}$, for a broken power-law model of an early universe phase transition, turns out to be possible in the presence of astrophysical confusion noise from compact binary coalescences, assuming a detector network containing CE and ET.
	In these approaches, reasonable levels of individual source subtraction are always assumed~\cite{Cutler:2005qq,Harms:2008xv,Sharma:2020btq,Sachdev:2020bkk}, given that future detector networks should be sensitive enough to accurately resolve a large fraction of the compact binary mergers in the visible universe.
	
	Ref.~\cite{Biscoveanu:2020gds} explores an alternative method to source subtraction. It proposes a Bayesian framework for simultaneously measuring individual compact binary mergers and a confusion foreground of these sources, together with a cosmological background. This method, in which the parameters of individual events are approximately marginalised-over, takes into account the non-Gaussian nature of the astrophysical signal (see Sec.~\ref{sec:sgwb:foregrounds}), increasing the sensitivity to the Gaussian cosmological background. The capability of the method is demonstrated with a combination of an astrophysical foreground of merging BBHs and a cosmological background with a power-law spectrum.
	
	The methods of both Refs.~\cite{Martinovic:2020hru,Biscoveanu:2020gds} could be adapted for the space-based detector LISA. In doing so, one should keep in mind that the  corresponding spectral separation study for LISA would be more involved due to the nature of the TDI and the necessity to simultaneously estimate the LISA noise. A successful contribution separation will give us valuable information on the one hand about merger rates and population models and the astrophysics of exotic objects, and on the other about phenomenology in early universe models. This is indeed a research line that should be incentivated  over the next years.

\newpage

\section*{Acknowledgement}
This work is partly supported by: 
A.G. Leventis Foundation;
Academy of Finland grants 328958 and 345070;
Alexander S. Onassis Foundation, Scholarship ID: FZO 059-1/2018-2019;
Amaldi Research Center funded by the MIUR program “Dipartimento di Eccellenza” (CUP: B81I18001170001); 
ASI Grants No. 2016-24-H.0 and No. 2016-24-H.1-2018;
Atracci\'{o}n de Talento grant 2019-T1/TIC-15784;
Atracción de Talento contract no. 2019-T1/TIC-13177 granted by the Comunidad de Madrid;
Ayuda `Beatriz Galindo Senior' by the Spanish `Ministerio de Universidades', grant BG20/00228;
Basque Government grant (IT-979-16);
Belgian Francqui Foundation;
Centre national d’Etudes spatiales;
Ben Gurion University Kreitman Fellowship, and the Israel Academy of Sciences and Humanities (IASH) \& Council for Higher Education (CHE) Excellence Fellowship Program for International Postdoctoral Researchers;
Centro de Excelencia Severo Ochoa Program SEV-2016-0597;
CERCA program of the Generalitat de Catalunya;
Cluster of Excellence “Precision Physics, Fundamental Interactions, and Structure of Matter” (PRISMA+ EXC 2118/1);
Comunidad de Madrid, Contrato de Atracción de Talento 2017-T1/TIC-5520; 
Czech Science Foundation GAČR, Grant No. 21-16583M;
Delta ITP consortium;
Department of Energy under Grant No. DE-SC0008541, DE-SC0009919 and DE-SC0019195;
Deutsche Forschungsgemeinschaft (DFG), Project ID 438947057;
Deutsche Forschungsgemeinschaft under Germany's Excellence Strategy - EXC 2121 Quantum Universe - 390833306;
European Structural and Investment Funds and the Czech Ministry of Education, Youth and Sports (Project CoGraDS - CZ.02.1.01/0.0/0.0/15 003/0000437);
European Union's H2020 ERC Consolidator Grant ``GRavity from Astrophysical to Microscopic Scales'' (Grant No.~GRAMS-815673);
European Union’s H2020 ERC, Starting Grant agreement no. DarkGRA–757480;
European Union's Horizon 2020 programme under the Marie Sklodowska-Curie grant agreement 860881 (ITN HIDDeN);
European Union's Horizon 2020 Research and Innovation Programme grant no.~796961, "AxiBAU" (K.S.);
European Union’s Horizon 2020 Research Council grant 724659 MassiveCosmo ERC-2016-COG;
FCT through national funds (PTDC/FIS-PAR/31938/2017) and through project ``BEYLA -- BEYond LAmbda" with ref. number PTDC/FIS-AST/0054/2021;
FEDER—Fundo Europeu de Desenvolvimento Regional through COMPETE2020 - Programa Operacional Competitividade e Internacionalização (POCI-01-0145-FEDER-031938) and research grants UIDB/04434/2020 and UIDP/04434/2020;
Fondation CFM pour la Recherche in France;
Foundation for Education and European Culture in Greece;
French ANR project MMUniverse (ANR-19-CE31-0020);
FRIA Grant No.1.E.070.19F of the Belgian Fund for Research, F.R.S.-FNRS
Funda{\c c}\~ao para a Ci\^encia e a Tecnologia (FCT) through contract no.~DL 57/2016/CP1364/CT0001;
Funda\c{c}\~{a}o para a  Ci\^{e}ncia e a Tecnologia (FCT) through grants UIDB/04434/2020, UIDP/04434/2020, PTDC/FIS-OUT/29048/2017, CERN/FIS-PAR/0037/2019  and ``CosmoTests -- Cosmological tests of gravity theories beyond General Relativity"  CEECIND/00017/2018;
Generalitat Valenciana grant PROMETEO/2021/083;
grant no.~758792, project GEODESI;
Government of Canada through the Department of Innovation, Science and Economic Development and Province of Ontario through the Ministry of Colleges and Universities;
Grants-in-Aid for JSPS Overseas Research Fellow (No. 201960698);
I+D grant PID2020-118159GB-C41 of the Spanish Ministry of Science and Innovation;
INFN \textit{iniziativa specifica} TEONGRAV;
Israel Science Foundation (grant no.~2562/20);
Japan Society for the Promotion of Science (JSPS) KAKENHI Grant no. 20H01899 and 20H05853;
IFT Centro de Excelencia Severo Ochoa Grant SEV-2;
Kavli Foundation and its founder Fred Kavli;
Minerva Foundation;
Ministerio de Ciencia e Innovacion grant PID2020-113644GB-I00;
NASA grant 80NSSC19K0318;
NASA Hubble Fellowship grants No.\ HST-HF2-51452.001-A awarded by the Space Telescope Science Institute with NASA contract NAS5-26555;
Netherlands Organisation for Science and Research (NWO) grant number 680-91-119;
new faculty seed start-up grant of the Indian Institute of Science, Bangalore, the Core Research Grant~CRG/2018/002200 of the Science and Engineering;
NSF grants PHY-1820675, PHY-2006645 and PHY-2011997;
Polish National Science Center grant 2018/31/D/ST2/02048;
Polish National Agency for Academic Exchange within the Polish Returns
Programme under agreement PPN/PPO/2020/1/00013/U/00001;
Pr\'o-Reitoria de Pesquisa of Universidade Federal de Minas Gerais (UFMG) under grant no.~28359;
Ram\'on y Cajal Fellowship contract RYC-2017-23493;
Research Project PGC2018-094773-B-C32 [MINECO-FEDER];
Research Project PGC2018-094773-B-C32 [MINECO-FEDER];
ROMFORSK grant Project.~No.~302640;
Royal Society grant URF/R1/180009 and ERC StG 949572: SHADE; 
Shota Rustaveli National Science Foundation (SRNSF) of Georgia (grant FR/18-1462);
Simons Foundation/SFARI 560536;
SNSF Ambizione grant;
SNSF professorship grant (No.~170547);
Spanish MINECO's ``Centro de Excelencia Severo Ochoa" Programme grants SEV-2016-0597 and PID2019-110058GB-C22;
Spanish Ministry MCIU/AEI/FEDER grant (PGC2018-094626-B-C21);
Spanish Ministry of Science and Innovation (PID2020-115845GB-I00/AEI/10.13039/501100011033);
Spanish Proyectos de I$+$D via grant PGC2018-096646-A-I00;
STFC Consolidated grant ST/T000732/1;
STFC Consolidated Grants ST/P000762/1 and ST/T000791/1;
STFC grant ST/S000550/1;
STFC grant ST/T000813/1;
STFC grants ST/P000762/1 and ST/T000791/1;
STFC under the research grant ST/P000258/1;
Swiss National Science Foundation (SNSF), project {\sl The Non-Gaussian Universe and Cosmological Symmetries}, project number: 200020-178787;
Swiss National Science Foundation Professorship grants No.~170547 and No.~191957;
SwissMap National Center for Competence in Research;
``The Dark Universe: A Synergic Multi-messenger Approach” number 2017X7X85K under the MIUR program PRIN 2017;
UK Space Agency;
UKSA Flagship Project, Euclid.

%%%%%%%%%%%%%%%%%%%%%%%%%%%%%%%
%\vfill
%\bibliographystyle{alpha}
%\bibliographystyle{h-elsevier}
%\bibliography{TOTAL_arxiv+byhand.bib}
%%%%%%%%%%%%%%%%%%%%%%%%%%%%%%%

\end{document}